\documentclass[twocolumn,american,aps, prb, floatfix, table, dvipsnames, superscriptaddress]{revtex4-2}
\usepackage[T1]{fontenc}
\usepackage[latin9]{inputenc}
\usepackage{color}
\usepackage{babel}
\usepackage{amsmath}
\usepackage{amssymb}
\usepackage{graphicx}
\usepackage{wasysym}
\usepackage{esint}
\usepackage[unicode=true,
 bookmarks=false,
 breaklinks=true,pdfborder={0 0 1},backref=false,colorlinks=true]
 {hyperref}
\hypersetup{
 pdfcreator={},pdfproducer={LaTeX with hyperref},linkcolor=blue,anchorcolor=blue,citecolor=blue,filecolor=red,menucolor=red,pagecolor=red,urlcolor=blue,pdfstartview=FitV,pdfhighlight=/I,pdfpagelayout=TwoColumns,hypertexnames=true}

\makeatletter
\usepackage{lmodern}
\usepackage{mathptmx, newtxtext, newtxmath, xspace}
\usepackage{bbm}

\usepackage[table, dvipsnames]{xcolor}
\definecolor{colorA}{cmyk}{0,0,0,0.05}
\definecolor{colorB}{cmyk}{0.14,0.04,0,0}
\definecolor{colorC}{cmyk}{0.02,0.0799,0,0}
\definecolor{colorD}{cmyk}{0.099,0.14,0,0}

\definecolor{TabGray}{cmyk}{0,0,0,0.15}
\definecolor{TabCyan}{cmyk}{0.09,0.02,0,0}
\definecolor{TabMagenta}{cmyk}{0.005,0.05,0,0}

\usepackage{natbib}
\setcitestyle{numbers}

\makeatother

\begin{document}
\title{Classification of electronic nematicity in three-dimensional crystals
and quasicrystals}
\author{Matthias Hecker }
\affiliation{School of Physics and Astronomy, University of Minnesota, Minneapolis
55455 MN, USA}
\author{Anant Rastogi }
\affiliation{School of Physics and Astronomy, University of Minnesota, Minneapolis
55455 MN, USA}
\author{Daniel F. Agterberg}
\affiliation{Department of Physics, University of Wisconsin-Milwaukee, Milwaukee
53201 WI, USA}
\author{Rafael M. Fernandes}
\affiliation{School of Physics and Astronomy, University of Minnesota, Minneapolis
55455 MN, USA}
\date{\today }
\begin{abstract}
Electronic nematic order has been reported in a rich landscape of
materials, encompassing not only a range of intertwined correlated
and topological phenomena, but also different underlying lattice symmetries.
Motivated by these findings, we investigate the behavior of electronic
nematicity as the spherical symmetry of three-dimensional (3D) space
is systematically lowered by the lattice environment. We consider
all $32$ crystallographic point groups as well as $4$ major classes
of quasicrystalline point groups, given the recent observations of
electronic phases of interest in quasicrystalline materials and artificial
twisted quasicrystals. Valuable insights are gained by establishing
a mapping between the five-component charge-quadrupolar nematic order
parameter of the electronic fluid and the 3D tensorial order parameter
of nematic liquid crystals. We find that a uniaxial nematic state
is only generically realized in polyhedral point groups (icosahedral and cubic), with the nematic director pointing along
different sets of rotational symmetry axes. Interestingly, icosahedral
point groups are the only ones in which the five nematic order parameter
components transform as the same irreducible representation, making
them the closest analog of 3D isotropic nematics. In axial point groups,
one of the nematic components is always condensed, whereas the other
four components decompose into an in-plane and an out-of-plane nematic
doublet, resulting in biaxial nematic ground states. Because these
two nematic doublets behave as $Z_{q}$-clock order parameters, this
allows us to identify the types of crystals and quasicrystals that
can host interesting electronic nematic phenomena enabled by the critical
properties of the $q\geq4$ clock model, such as emergent continuous
nematic fluctuations in 3D, critical phases with quasi-long-range
nematic order in 2D, and Ashkin-Teller nematicity in 2D.
\end{abstract}
\maketitle

\section{Introduction}

Fingerprints typical of electronic nematic behavior have been experimentally
seen in a wide range of settings, such as unconventional cuprates
\citep{Hinkov08,Davis10,Bozovic17,Matsuda17,Mukhopadhyay2019} and
iron-based superconductors \citep{Chuang2010,Chu2010,Yi2011,Chu12,Kuo16,Bohmer2022},
quantum Hall systems \citep{Eisenstein99,Feldman16}, correlated oxides
\citep{Mackenzie07}, doped topological insulators \citep{Tamegai2019,Cho2020},
twisted moiré systems \citep{Jiang2019,Cao_2021,Rubio2022,Zhang2022},
$f$-electron materials \citep{Ronning2017,Rosenberg2019,Seo2020,Massat2022},
kagome metals \citep{Drucker2024}, colossal magnetoresistance compounds
\citep{Beaudin2022}, triangular antiferromagnets \citep{Little2020,LiangWu2023,Sun2023,Hwangbo2023,Tan2023},
topological semimetals \citep{Siddiquee2022}, and optical lattices
\citep{cold_atoms_nem}. As its name suggests, this state of matter
is the quantum analog of classical nematic liquid crystals \citep{Chaikin2000,Kats1993},
as it causes the spontaneous breaking of a discrete rotational symmetry
of the system while preserving its properties under translations \citep{Kivelson1998}.
At first, the term electronic nematicity was coined to describe the
ordered state that emerges upon the partial melting of an underlying
charge or spin stripe state that restores the translational symmetry
of a correlated metal or insulator, in close analogy to the smectic-to-nematic
transition of liquid crystals \citep{Kivelson1998,Fradkin1999,Zaanen2004,Nie2014,Nie2017}.
More recently, this concept of a so-called vestigial nematic phase
was extended to the case of multi-component superconductors, which
were shown in certain circumstances to support a partially-melted
state that restores the U(1) gauge symmetry but keeps the rotational
symmetry of the pairing state broken \citep{Hecker2018,Hecker2023}.
Since the seminal work of Ref.~\citep{Kivelson1998}, the term electronic
nematicity has been broadly employed to refer to any spontaneous rotational-symmetry
breaking phase that is driven by electronic interactions (as opposed
to elastic interactions) and that does not break additional symmetries
\textendash{} for recent reviews, see \citep{Fradkin2010,Fernandes2014,Fradkin_intertwined,Fernandes2019}.
This includes the case of weakly-interacting Fermi liquids that undergo
a charge $l=2$ Pomeranchuk instability via a Stoner-like mechanism
\citep{Pomeranchuk1958,Oganesyan2001,Metzner2006,Zacharias2009,Maslov2010},
which is particularly favored when the electronic band structure is
close to a van Hove singularity and thus has an enhanced density of
states \citep{Valenzuela2008,Kiesel2013}. Another example of electronic
nematicity under this definition are insulating and metallic multi-orbital
systems that display ferroquadrupolar order \citep{Maharaj2017}.
Extensions of this concept to states that break rotational and time-reversal
or inversion symmetries have also been studied \citep{Wu2007,Zaanen2016,Xu2020,Mandal2023,Gali2024,Boettcher2018}.

\begin{figure*}[t]
\includegraphics[width=1\textwidth]{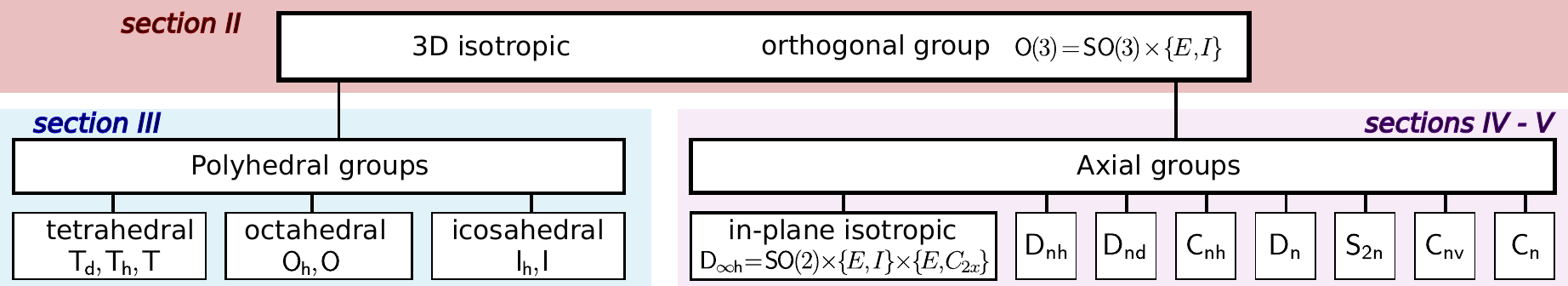}

\caption{Schematics of the systematic classification performed in this paper
of electronic nematicity in crystalline and quasicrystalline systems.
The analysis starts from the ideal isotropic three-dimensional system,
described by the orthogonal group $\mathsf{O(3)}$ (Sec.~\ref{sec:3D-Isotropic-nematicity}).
The spherical symmetry is systematically lowered, leading to two different
sets of groups: the seven polyhedral groups (Sec.~\ref{sec:Nematicity-in-Polyhedral})
and the seven classes of axial groups. The latter are subgroups of
$\mathsf{D_{\infty h}}$, which describes a two-dimensional isotropic
system. Axial groups include 27 crystallographic point groups (Sec.~\ref{sec:Nematicity-Axial})
as well as the non-crystallographic dodecagonal, decagonal, and octagonal
point groups that describe quasicrystals (Sec.~\ref{sec:Quasicrystal}).
\label{fig:descendant_tree}}
\end{figure*}

The main difference between electronic nematic materials and nematic
liquid crystals is that, in the former, rotational symmetry is already
explicitly broken by the underlying lattice, whereas the latter has
full rotational symmetry. As a result, the electronic nematic order
parameter always breaks a discrete symmetry that lowers the point
group of the crystal, and all nematic collective modes in the ordered
state are gapped. Historically, much of the theoretical investigations
on electronic nematicity have focused on strongly anisotropic layered
systems and ``planar'' nematic order parameters \citep{Fradkin1999,Oganesyan2001,HYKee2003,Metzner2006,Zacharias2009,Maslov2010,Maciejko2013,Fradkin_FQHE}.
Perhaps the most well recognized example is the electronically-driven
breaking of $C_{4z}$ symmetry (i.e. fourfold rotations about the
$z$-axis), in which case the electronic nematic order parameter is
Ising-like \citep{Carlson2006,Xu2008,Fang2008,Raghu2009,Fernandes2010,Metlitski2010,Fischer2011,Fernandes2012,Nie2014,Kontani2019,Wang2019}.
We emphasize that although such a nematic transition is, on symmetry
grounds, equivalent to a tetragonal-to-orthorhombic structural transition,
the microscopic mechanisms responsible for the spontaneously broken
symmetry are completely different \citep{Chu12}. The fact that electronic
nematic phenomena have been heavily studied in tetragonal lattices
is not surprising, since the initial experimental observations were
on tetragonal (or nearly tetragonal) compounds such as bilayer ruthenates
\citep{Mackenzie07}, cuprates \citep{Hinkov08}, and iron pnictides
\citep{Chuang2010,Chu2010}. More recently, experimental reports of
the spontaneous breaking of $C_{3z}$ symmetry (i.e. threefold rotations
about the $z$-axis) by electronic degrees of freedom in hexagonal
and trigonal systems such as doped $\mathrm{Bi_{2}Se_{3}}$ \citep{Tamegai2019,Cho2020},
twisted multi-layer graphene \citep{Jiang2019,Cao_2021,Rubio2022,Zhang2022},
and triangular antiferromagnets \citep{Little2020,LiangWu2023,Sun2023,Hwangbo2023,Tan2023}
have motivated a deeper theoretical investigation of 3-state Potts
electronic nematic order parameters \citep{Hecker2018,Fernandes2020,Xu2020,Buessen2021,Strockoz2022,Li2022,Kontani2022,Kimura2022,Nedic2023,Chakraborty2023}.

The diversity of systems in which electronic nematicity has been observed
indicates that a broader description and systematic classification
of this phenomenon is timely. Indeed, there is a rich landscape of
crystalline symmetries whose effects on electronic nematic order remain
to be explored in depth. This includes cubic systems such as $\mathrm{EuB_{6}}$
\citep{Beaudin2022} and $\mathrm{CaSn_{3}}$ \citep{Siddiquee2022},
which have been reported to display anisotropic properties consistent
with nematicity. Besides crystals, quasicrystals \citep{Shechtman1984,Levine1984}
have been recently found to display electronically ordered states
of interest such as superconductivity \citep{Kamiya2018,Uri2023},
magnetism \citep{Goldman1991}, and quantum criticality \citep{Deguchi2012},
raising interesting questions about how electronic nematicity would
be manifested in systems that lack periodicity but possess orientational
order. This includes not only quasicrystal materials with icosahedral,
dodecagonal, decagonal, and octagonal symmetries \citep{Socolar1989,Rabson1991,Luck1993,Lifshitz1996},
but also artificial quasicrystals that can be assembled in the laboratory
by twisting two periodic crystals by specific angles \citep{Ahn2018,Haenel2022,Yang2023_a,Yang2023_b,Kim2023,Uri2023}.
Moreover, besides the planar nematic order parameters studied in layered
systems, there are additional nematic channels available. Indeed,
in terms of the usual electronic operators $\hat{\psi}_{\boldsymbol{k}\sigma}$,
the nematic order parameters can be expressed as the expectation values
of the quadrupolar charge operator \citep{Oganesyan2001}, $d_{i}\equiv\sum\limits _{\boldsymbol{k},\sigma}\left\langle f_{i}\left(\hat{\boldsymbol{k}}\right)\hat{\psi}_{\boldsymbol{k}\sigma}^{\dagger}\hat{\psi}_{\boldsymbol{k}\sigma}\right\rangle $.
Since there are five $d$-wave form factors $f_{i}\left(\hat{\boldsymbol{k}}\right)$,
one must consider, in the most general case, a nematic order parameter
$\boldsymbol{d}$ with five components:

\begin{equation}
\boldsymbol{d}=\left(\begin{array}{c}
d_{\frac{1}{\sqrt{3}}(2z^{2}-x^{2}-y^{2})}\\
d_{x^{2}-y^{2}}\\
d_{2yz}\\
d_{2xz}\\
d_{2xy}
\end{array}\right),\label{eq:intro_d}
\end{equation}
where the subscript denotes the corresponding $d$-wave form factor
in Cartesian coordinates. This opens the possibility of realizing
``out-of-plane'' nematic order in materials that are not strongly
anisotropic, such as $d_{2xz}$ and $d_{2yz}$, beyond the widely
investigated ``in-plane'' nematic states corresponding to $d_{x^{2}-y^{2}}$
and $d_{2xy}$.

In this paper, we perform a systematic and thorough classification
of electronic nematic order in all crystalline and quasicrystalline
point groups. Specifically, we derive the Landau expansions and determine
the universal properties of the nematic transitions associated with
the different allowed types of in-plane and out-of-plane nematic order
parameters $d_{i}$ for a given point group. To gain further insight
into how the reduced rotational symmetry of the underlying crystalline
or quasicrystalline environment constrains nematicity, we establish
a direct relationship between the five-component charge-quadrupolar
nematic order parameter $\boldsymbol{d}$ in Eq.~(\ref{eq:intro_d}),
which describes the electronic fluid, and the full rank-2 traceless
symmetric tensor $Q_{\mu\mu^{\prime}}$ that describes nematic liquid
crystals in general \citep{Chaikin2000,Kats1993}. We find it illuminating
to write the latter in the eigenstate basis:

\begin{equation}
Q_{\mu\mu^{\prime}}=q_{1}\Big(n_{\mu}n_{\mu^{\prime}}-\frac{1}{3}\delta_{\mu\mu^{\prime}}\Big)-q_{2}\left(m_{\mu}m_{\mu^{\prime}}-l_{\mu}l_{\mu^{\prime}}\right),\label{eq:intro_Q}
\end{equation}
where $q_{1}$ and $q_{2}$ are scalars and $\boldsymbol{n}$, $\boldsymbol{m}$,
$\boldsymbol{l}$ form a complete set of orthonormal vectors in three-dimensional
space. When $q_{2}=0$ or $q_{2}=q_{1}$, $Q_{\mu\mu'}$ describes
a uniaxial nematic state with a uniquely defined nematic director
$\boldsymbol{n}$ or $\boldsymbol{m}$, respectively; in all other
cases, $Q_{\mu\mu'}$ describes a biaxial nematic state \citep{Fregoso2009}.
While in liquid crystals these axes determine the orientation of the
constituent molecules, in metallic systems they are manifested in
the Fermi surface of the nematic state. Thus, by establishing the
relationship between the five free parameters encompassed by $q_{1}$,
$q_{2}$, $\boldsymbol{n}$, $\boldsymbol{m}$, $\boldsymbol{l}$
in Eq.~(\ref{eq:intro_Q}) and the five charge-quadrupolar order
parameters encoded in $\boldsymbol{d}$ in Eq.~(\ref{eq:intro_d}),
we determine the characteristic Fermi surface distortion patterns
in the nematic states of every point group studied.

To perform the analysis in a transparent and insightful way, we start
from the isotropic three-dimensional system, described by the orthogonal
group $\mathsf{O(3)}=\mathsf{SO(3)}\times\left\{ E,I\right\} $, from
which all crystallographic and non-crystallographic point groups can
be obtained by systematically reducing the symmetry according to group
theory. Here, $\mathsf{SO(3)}$ contains arbitrary rotations about
any three-dimensional axis, $I$ denotes inversion, and $E$ denotes
the identity operation. Fig.~\ref{fig:descendant_tree} illustrates
the well-known path by which the $\mathsf{O(3)}$ symmetry is systematically
lowered to yield all point groups. First, one needs to distinguish
between two types of point groups: the polyhedral groups and the axial
groups. There are only $7$ polyhedral groups: $3$ tetrahedral and
$2$ octahedral groups, which form the cubic crystal system, and $2$
icosahedral groups, which describe a large number of quasicrystals. 

In contrast, there is an infinite number of axial groups, all of which
are subgroups of $\mathsf{D_{\infty h}=SO(2)}\times\{E,I\}\times\{E,C_{2x}\}$,
which describes a two-dimensional isotropic system and contains arbitrary
rotations about the $z$-axis, twofold rotations about any in-plane
axis, inversion, and reflection with respect to the plane. The axial
groups are further subdivided into the seven different groups outlined
in Fig.~\ref{fig:descendant_tree}: $\mathsf{D_{nh}}$, $\mathsf{D_{nd}}$,
$\mathsf{C_{nh}}$, $\mathsf{D_{n}}$, $\mathsf{S_{2n}}$, $\mathsf{C_{nv}}$,
and $\mathsf{C_{n}}$, which contain discrete $n$-fold rotations
around the axis perpendicular to the plane and, in some cases, twofold
rotations about an axis parallel to the plane. Enforcing the crystallographic
restriction theorem leads to $27$ crystallographic axial point groups,
which are then divided in six crystal systems: hexagonal ($7$ groups),
trigonal ($5$ groups), tetragonal ($7$ groups), orthorhombic ($3$
groups), monoclinic ($3$ groups), and triclinic ($2$ groups). Together
with the $5$ polyhedral cubic groups, they constitute the $32$ crystallographic
point groups and $7$ crystal systems. Importantly, the non-crystallographic
point groups can still describe materials that are non-periodic and
that can host non-trivial electronic states \citep{Socolar1989,Rabson1991}.
Thus, in this paper, we also focus on the non-crystallographic point
groups that describe quasicrystals, namely, the polyhedral icosahedral
groups as well as the axial dodecagonal, decagonal, and octagonal
point groups.

Of course, several of our results for the character of the nematic
transition in certain point groups recover well-established results
obtained in previous symmetry analyses of two-dimensional nematics
\citep{Fradkin2010,Fernandes2014,Fernandes2019}, structural and ferroelastic
transitions \citep{Cowley1976,Folk1976,Lookman2003}, multipolar order
\citep{Zaanen2016,Hayami2018}, and intertwined orders \citep{Li2021}.
We emphasize that our main goal here is to provide a complete, self-contained
classification of electronic nematicity in crystals \emph{and }quasicrystals
by systematically determining how the removal of symmetries impacts
the general structure of the tensorial nematic order parameter (\ref{eq:intro_Q})
via its relationship with the five-component charge-quadrupolar order
parameter (\ref{eq:intro_d}). Our results are summarized in Tables~\ref{tab:classification}
and \ref{tab:classification_quasicrystals} in Section~\ref{sec:All-crystals},
and their derivations are shown in Sections~\ref{sec:3D-Isotropic-nematicity}-\ref{sec:Quasicrystal}
and the Appendices~\ref{sec:cubic_appendix}-\ref{sec:tetragonal_Appendix}.
For the convenience of the reader not interested in the technical
details of our analysis, we outline here the main results.
\begin{itemize}
\item The icosahedral quasicrystalline point groups are the only ones for
which the five components of the nematic order parameter (\ref{eq:intro_d})
transform together as a single five-dimensional irreducible representation.
The nematic transition is first-order within mean-field due to a cubic
invariant in the Landau expansion. Depending on the signs of the Landau
coefficients, the nematic director that characterizes the uniaxial
nematic ground state aligns itself with either the $10$ axes of threefold
rotational symmetry or the $6$ axes of fivefold rotational symmetry
of the quasicrystal. In the former case, the residual point group
is crystalline (trigonal) whereas in the latter, it is quasicrystalline
(decagonal).
\item The components of the nematic order parameter $\boldsymbol{d}$ in
the cubic crystal groups split into separate triplet and doublet order
parameters. The triplet behaves as a $Z_{4}$-Potts order parameter,
giving rise to a uniaxial nematic state in which the director is parallel
to one of four axes of threefold rotational symmetry of the crystal
(i.e. the space diagonals). The doublet behaves as a $Z_{3}$-clock/Potts
order parameter, and the three nematic axes point along the coordinate
axes. In the cubic groups for which the coordinate axes are also axes
of fourfold rotational symmetry, the nematic ground state is uniaxial,
otherwise, it is biaxial. The nematic transitions are, again, first-order
within mean-field.
\item In any axial group, the nematic component $d_{\frac{1}{\sqrt{3}}(2z^{2}-x^{2}-y^{2})}$
is necessarily non-zero, which changes the shape of the Fermi surface
in the symmetry-unbroken phase from spherical to cylindrical. The
other four components are decomposed into at least two independent
nematic doublets, the in-plane $\boldsymbol{d}^{\mathrm{ip}}=\big(d_{x^{2}-y^{2}},\,d_{2xy}\big)^{T}$
and the out-of-plane $\boldsymbol{d}^{\mathrm{op}}=\left(d_{2yz},\,d_{2xz}\right)^{T}$.
The nematic ground state is always biaxial for the axial groups, unless
the Landau parameters are fine tuned. In most cases, at least one
of the nematic axes aligns with a high-symmetry in-plane direction. 
\item Each nematic doublet in the axial groups behaves as a $Z_{q}$-clock
order parameter. If the group has $\left(2n\right)$-fold rotational
symmetry, $q=n$ for $\boldsymbol{d}^{\mathrm{ip}}$ and $q=2n$ for
$\boldsymbol{d}^{\mathrm{op}}$. If the group has $\left(2n+1\right)$-fold
rotational symmetry, $q=2n+1$ for both $\boldsymbol{d}^{\mathrm{ip}}$
and $\boldsymbol{d}^{\mathrm{op}}$. In three dimensions, the $q$-state
clock model is known to undergo an XY transition for $q\geq4$, similarly
to the isotropic 2D nematic case. In two dimensions, the $Z_{q}$-clock
model displays an intermediate critical phase with quasi-long-range
order for $q\geq5$ and an Ashkin-Teller phase transition with non-universal
critical exponents for $q=4$. A first-order transition can only occur
for $q=3$ and above the upper critical dimension $d_{u}\apprge2$.
If the axial group lacks in-plane rotational symmetry axes, or if
$\boldsymbol{d}^{\mathrm{ip}}$ and $\boldsymbol{d}^{\mathrm{op}}$
have the same transformation properties, the clock term in the Landau
expansion acquires a non-universal offset, which we denote as the
$Z_{q}^{*}$-clock model.
\item The main difference between hexagonal and trigonal crystals is that,
in the latter, $\boldsymbol{d}^{\mathrm{ip}}$ and $\boldsymbol{d}^{\mathrm{op}}$
are not independent, which has a significant impact on the character
of the nematic transition. In tetragonal crystals, $\boldsymbol{d}^{\mathrm{ip}}$
is further decomposed into two one-component Ising-like order parameters.
Orthorhombic, monoclinic, and triclinic crystals are described by
Abelian point groups, which only admit one-dimensional irreducible
representations. As a result, any nematic transition must be Ising-like. 
\item In dodecagonal, decagonal, and octagonal quasicrystals, both nematic
doublets behave as $Z_{q}$-clock order parameters with $q\geq4$.
This makes axial quasicrystals interesting platforms to realize exotic
nematicity enabled by the critical properties of the $q\geq4$ clock
model, such as emergent XY nematic fluctuations and critical nematic
phases displaying quasi-long-range order. 
\end{itemize}
The organization of the paper, schematically shown in Fig.~\ref{fig:descendant_tree},
is as follows: Sec.~\ref{sec:3D-Isotropic-nematicity} introduces
the formalism and presents the results for nematicity in the three-dimensional
isotropic system. The case of polyhedral groups, i.e., cubic crystal
system and icosahedral quasicrystals, is discussed in Sec.~\ref{sec:Nematicity-in-Polyhedral}.
Sec.~\ref{sec:Nematicity-Axial} presents the properties of electronic
nematic order in the isotropic two-dimensional system and in the axial
crystallographic groups (hexagonal, trigonal, tetragonal, orthorhombic,
monoclinic, and triclinic crystal systems). The cases of dodecagonal,
decagonal, and octagonal quasicrystalline axial groups are presented
in Sec.~\ref{sec:Quasicrystal}. Sec.~\ref{sec:All-crystals} contains
our concluding remarks as well as a summary of our results for the
crystalline point groups in Table~\ref{tab:classification} and for
the quasicrystalline point groups in Table~\ref{tab:classification_quasicrystals}.
Appendix~\ref{sec:SO3_properties} contains additional details about
isotropic nematicity. Appendix~\ref{sec:Symmetrized-products} explains
the concept of symmetrized decomposition of products of irreducible
representations, whereas Appendix~\ref{sec:App_Ih} shows details
of the minimization of the nematic Landau expansion in icosahedral
systems. Appendix~\ref{sec:cubic_appendix} presents details of electronic
nematic order in cubic crystals without fourfold rotational symmetry
axes. Appendices~\ref{sec:hexagonal_Appendix}, \ref{sec:trigonal_appendix},
and \ref{sec:tetragonal_Appendix} derive the properties of electronic
nematicity, respectively, in hexagonal, trigonal, and tetragonal point
groups that lack in-plane rotational symmetry axes.

\section{Electronic nematicity in a 3D isotropic system\label{sec:3D-Isotropic-nematicity}}

\subsection{Representations of the nematic order parameter}

To set the stage for the remainder of the paper, we first review the
properties of electronic nematicity in 3D isotropic systems, i.e.
systems that are fully rotational invariant. In classical nematics,
the order parameter is a ``headless vector'' associated with the
orientation of the elongated molecules that form the liquid crystal.
It is conveniently expressed in terms of the symmetric traceless tensor
$Q_{\mu\mu^{\prime}}=\left\langle a_{\mu}a_{\mu^{\prime}}-\frac{1}{3}\delta_{\mu\mu^{\prime}}\boldsymbol{a}^{2}\right\rangle $,
with $\mu,\mu^{\prime}=1,2,3$ and the director $\boldsymbol{a}=(a_{1},a_{2},a_{3})$
\citep{Chaikin2000,Kats1993}. The generalization to the quantum (i.e.
electronic) case has been widely discussed in the literature \citep{Oganesyan2001,Fregoso2009}.
In terms of the electronic annihilation (creation) operators $\hat{\psi}_{\boldsymbol{k}}$
($\hat{\psi}_{\boldsymbol{k}}^{\dagger}$), the tensorial order parameter
is given by

\begin{align}
Q_{\mu\mu^{\prime}} & =\frac{1}{k_{F}^{2}}\sum_{\boldsymbol{k},\sigma}\left\langle \mathcal{F}_{\mu\mu^{\prime}}\left(\boldsymbol{k}\right)\hat{\psi}_{\boldsymbol{k}\sigma}^{\dagger}\hat{\psi}_{\boldsymbol{k}\sigma}\right\rangle ,\label{eq:Q_intro}
\end{align}
with momentum $\boldsymbol{k}$, Fermi momentum $k_{F}$, spin $\sigma=\uparrow,\downarrow$,
and the $d$-wave form factor $\mathcal{F}_{\mu\mu^{\prime}}\left(\boldsymbol{k}\right)=2\big(k_{\mu}k_{\mu^{\prime}}-\frac{1}{3}\delta_{\mu\mu^{\prime}}\boldsymbol{k}^{2}\big)$.
Generalizations to multi-orbital systems are straightforward, but
will not be covered here \citep{Fernandes2012,Hayami2018}. Thus,
the components of the electronic nematic order parameter correspond
to quadrupolar charge order which, in the theory of interacting Fermi
liquids, corresponds to an angular momentum $l=2$ Pomeranchuk instability
of the Fermi liquid in the singlet channel \citep{Pomeranchuk1958}.
The condensation of the tensor order parameter $Q$ leads to a distortion
of the otherwise spherically symmetric Fermi surface, thus breaking
the rotational invariance of the isotropic system:

\begin{align}
\epsilon_{\boldsymbol{k}} & =\frac{\boldsymbol{k}^{2}}{2m}-\mu_{0}+\frac{1}{2}\,\mathrm{tr}\left[Q\mathcal{F}\left(\boldsymbol{k}\right)\right].\label{eq:epsilon_k_nematic}
\end{align}
Here, $m$ is the effective electron mass and $\mu_{0}$, the chemical
potential. For concreteness, and to better visualize the fingerprints
of nematicity on the electronic degrees of freedom, our analysis will
focus on metals. Of course, the symmetry properties of the nematic
state would be the same in insulators.

Being a traceless symmetric tensor, $Q$ has five independent components.
Thus, it is also convenient to express nematic phenomena in terms
of a five-component ``vector'' $\boldsymbol{d}=\left(d_{1},d_{2},d_{3},d_{4},d_{5}\right)$.
To derive the symmetry properties of this vector, we note that a 3D
isotropic system is described by the continuous (orthogonal) group
$\mathrm{O(3)}=\mathrm{SO(3)}\times\left\{ E,I\right\} $, which combines
the full rotation group $\mathrm{SO(3)}$ with the inversion operation
$I$ (here, $E$ denotes the identity operation). Thus, in group-theory
notation, the five-component nematic vector $\boldsymbol{d}$ transforms
according to the five-dimensional irreducible representation (IR)
$\Gamma_{j=2}^{+}$ of the orthogonal group $\mathrm{O(3)}$. Note
that the superscript $+$ ($-$) indicates an inversion-even (inversion-odd)
IR.

Such a five-component order parameter can be obtained in a straightforward
way as a bilinear $\boldsymbol{M}$ constructed from the coordinate
vector $\boldsymbol{r}=(x,y,z)$, which in turn transforms according
to the three-dimensional vector IR $\Gamma_{j=1}^{-}$. That such
a bilinear $\boldsymbol{M}$ in the $\Gamma_{j=2}^{+}$-channel exists
follows from the product decomposition $\Gamma_{j=1}^{-}\otimes\Gamma_{j=1}^{-}=\Gamma_{j=0}^{+}\oplus\Gamma_{j=2}^{+}\oplus\Gamma_{j=1}^{+}$.
Writing the bilinear components as $M_{i}=r_{\mu}\lambda_{\mu\mu^{\prime}}^{j=2,i}r_{\mu^{\prime}}$
with $i\in\{1,\dots,5\}$ and $\mu\in\{1,2,3\}$, we obtain the five
matrices $\boldsymbol{\lambda}^{j=2}$ from the transformation condition
\begin{align}
\mathcal{R}_{-,j=1}^{T}(g)\lambda^{j=2,i}\mathcal{R}_{-,j=1}(g) & =\mathcal{R}_{+,j=2}(g)_{ii^{\prime}}\lambda^{j=2,i^{\prime}},\label{eq:SO3_trafo_condition}
\end{align}
where $\mathcal{R}_{\pm,j}(g)$ denotes the $5\times5$ transformation
matrix of a symmetry element $g$ associated with the IR $\Gamma_{j}^{\pm}$.
The symmetry elements $g=\big(\vartheta,\hat{\boldsymbol{\ell}},\mathcal{I}\big)$
are parametrized in terms of the rotation angle $\vartheta$ around
the unit rotation axis $\hat{\boldsymbol{\ell}}$ and the index $\mathcal{I}=\pm1$
for inversion being applied ($\mathcal{I}=-1$) or not ($\mathcal{I}=+1$).
Therefore, the transformation matrices are given by $\mathcal{R}_{+,j}(g)=\mathcal{R}_{j}(\vartheta,\hat{\boldsymbol{\ell}})$
and $\mathcal{R}_{-,j}(g)=\mathcal{I}\mathcal{R}_{j}(\vartheta,\hat{\boldsymbol{\ell}})$
where $\mathcal{R}_{j}(\vartheta,\hat{\boldsymbol{\ell}})=\exp\left(-\mathsf{i}\vartheta\,\boldsymbol{J}^{(j)}\cdot\hat{\boldsymbol{\ell}}\right)$
are the well-known $(2j+1)\times(2j+1)$-dimensional rotation matrices.
Using the $\boldsymbol{J}^{(j)}$ matrices outlined in Appendix~\ref{sec:SO3_properties},
one finds the five matrices $\boldsymbol{\lambda}^{j=2}$ in (\ref{eq:SO3_trafo_condition})
to be identical to the five symmetric Gell-Mann matrices, 
\begin{align}
\boldsymbol{\lambda}^{j=2} & =\left\{ \frac{1}{\sqrt{3}}\left(\begin{smallmatrix}\text{-}1 & 0 & 0\\
0 & \text{-}1 & 0\\
0 & 0 & 2
\end{smallmatrix}\right),\left(\begin{smallmatrix}1 & 0 & 0\\
0 & \text{-}1 & 0\\
0 & 0 & 0
\end{smallmatrix}\right),\left(\begin{smallmatrix}0 & 0 & 0\\
0 & 0 & 1\\
0 & 1 & 0
\end{smallmatrix}\right),\left(\begin{smallmatrix}0 & 0 & 1\\
0 & 0 & 0\\
1 & 0 & 0
\end{smallmatrix}\right),\left(\begin{smallmatrix}0 & 1 & 0\\
1 & 0 & 0\\
0 & 0 & 0
\end{smallmatrix}\right)\right\} .\label{eq:lambda_2}
\end{align}

We can now label the five nematic components of $\boldsymbol{d}$
as polynomials of the coordinate vector $\boldsymbol{r}=(x,y,z)$,
according to their symmetry properties encoded in the bilinear $\boldsymbol{M}$:
\begin{align}
\boldsymbol{d} & =\left(\begin{array}{c}
d_{1}\\
d_{2}\\
d_{3}\\
d_{4}\\
d_{5}
\end{array}\right)=\left(\begin{array}{c}
d_{\frac{1}{\sqrt{3}}(2z^{2}-x^{2}-y^{2})}\\
d_{x^{2}-y^{2}}\\
d_{2yz}\\
d_{2xz}\\
d_{2xy}
\end{array}\right).\label{eq:d_vector}
\end{align}
Unsurprisingly, the five components of $\boldsymbol{d}$ correspond
to the five $d$-wave form factors written as tesseral harmonics.
The symmetric Gell-Mann matrices (\ref{eq:lambda_2}) also conveniently
establish the relationship between the tensor notation $Q$ and the
vector notation $\boldsymbol{d}$: 
\begin{align}
Q & =\boldsymbol{d}\cdot\boldsymbol{\lambda}^{j=2}, & \boldsymbol{d} & =\mathrm{tr}\left[Q\boldsymbol{\lambda}^{j=2}\right]\,\big/2,\label{eq:Q_d_relation}
\end{align}
which, in explicit form, gives:

\begin{align}
Q & =\left(\begin{smallmatrix}d_{2}-\frac{1}{\sqrt{3}}d_{1} & d_{5} & d_{4}\\
d_{5} & -d_{2}-\frac{1}{\sqrt{3}}d_{1} & d_{3}\\
d_{4} & d_{3} & \frac{2}{\sqrt{3}}d_{1}
\end{smallmatrix}\right).\label{eq:Q_tensor}
\end{align}

In this work, we interchangeably use the tensor notation $Q$ (\ref{eq:Q_tensor})
and the five-component vector notation $\boldsymbol{d}$ (\ref{eq:d_vector}).
Moreover, throughout this work, we use the following transformation
relation for the nematic order parameter
\begin{align}
\boldsymbol{d} & \overset{g}{\longrightarrow}\mathcal{R}_{+,j=2}(g)\,\boldsymbol{d},\label{eq:d_5_trafo}
\end{align}
such that the transformation matrices are fixed by the set of given
symmetry elements $g$. This is particularly useful as it uniquely
defines the multi-component nematic channels within reduced symmetry
systems where $\mathcal{R}_{+,j=2}(g)$ is block-diagonal, such as
crystal lattices.

A convenient representation of the nematic tensor $Q$ (\ref{eq:Q_tensor})
is in terms of its eigenbasis. In this representation, which we denote
the $(\boldsymbol{n}\boldsymbol{m}\boldsymbol{l})$-representation,
the nematic tensor becomes 
\begin{align}
Q_{\mu\mu^{\prime}} & =\sqrt{3}|\boldsymbol{d}|\cos\alpha\Big(n_{\mu}n_{\mu^{\prime}}-\frac{1}{3}\delta_{\mu\mu^{\prime}}\Big)\nonumber \\
 & -|\boldsymbol{d}|\sin\alpha\left(m_{\mu}m_{\mu^{\prime}}-l_{\mu}l_{\mu^{\prime}}\right),\label{eq:Q_alt_parameterization}
\end{align}
where the angle $\alpha$ determines the eigenvalues while the unit
vectors $\boldsymbol{n}$, $\boldsymbol{m}$, $\boldsymbol{l}$ span
the eigenspace of the matrix (\ref{eq:Q_tensor}). As a set of orthonormal
eigenvectors, $\boldsymbol{n}$, $\boldsymbol{m}$, $\boldsymbol{l}$
obey the orthogonality relations $\boldsymbol{n}\cdot\boldsymbol{m}=\boldsymbol{n}\cdot\boldsymbol{l}=\boldsymbol{m}\cdot\boldsymbol{l}=0$,
as well as the completeness relation 
\begin{align}
n_{\mu}n_{\mu^{\prime}}+m_{\mu}m_{\mu^{\prime}}+l_{\mu}l_{\mu^{\prime}} & =\delta_{\mu\mu^{\prime}},\label{eq:nml_relation}
\end{align}
valid for any $\mu,\mu^{\prime}$. One reason why the $(\boldsymbol{n}\boldsymbol{m}\boldsymbol{l})$-representation
is particularly useful is because the tensor (\ref{eq:Q_alt_parameterization})
is readily diagonalized using the orthogonal matrix $U=\left(\boldsymbol{n},\boldsymbol{m},\boldsymbol{l}\right)$,
\begin{align}
Q^{d} & =U^{T}QU=\frac{2|\boldsymbol{d}|}{\sqrt{3}}\left(\begin{smallmatrix}\cos\left(\alpha\right) & 0 & 0\\
0 & \cos\left(\alpha+\frac{2\pi}{3}\right) & 0\\
0 & 0 & \cos\left(\alpha+\frac{4\pi}{3}\right)
\end{smallmatrix}\right).\label{eq:Q_d}
\end{align}
Moreover, in this representation, we can use Eq.~(\ref{eq:Q_d_relation})
to write the five-component vector $\boldsymbol{d}$ (\ref{eq:d_vector})
as
\begin{align}
\boldsymbol{d} & =\frac{|\boldsymbol{d}|}{\sqrt{3}}\Big\{\cos\left(\alpha\right)\left(n_{\mu}\boldsymbol{\lambda}_{\mu\mu^{\prime}}^{j=2}n_{\mu^{\prime}}\right)+\cos\left(\alpha+\frac{2\pi}{3}\right)\left(m_{\mu}\boldsymbol{\lambda}_{\mu\mu^{\prime}}^{j=2}m_{\mu^{\prime}}\right)\nonumber \\
 & \quad+\cos\left(\alpha+\frac{4\pi}{3}\right)\left(l_{\mu}\boldsymbol{\lambda}_{\mu\mu^{\prime}}^{j=2}l_{\mu^{\prime}}\right)\Big\},\label{eq:d_nml}
\end{align}
where summation over $\mu$,$\mu^{\prime}$ is implied. The parameterization
of the orthonormal eigenvectors $\boldsymbol{n}$,$\boldsymbol{m}$,$\boldsymbol{l}$
involves three angles and it is not unique, see Appendix~\ref{sec:SO3_properties}
for details. Importantly, to ensure a one-to-one mapping between Eq.~(\ref{eq:Q_tensor})
and Eq.~(\ref{eq:Q_alt_parameterization}), the eigenvalue-angle
$\alpha$ must be restricted to the range $[0,\pi/3]$. Plotting the
three eigenvalues as a function of $\alpha$ in Fig.~\ref{fig:cosines}(a)
makes it clear that the smallest eigenvalue is always related to $\boldsymbol{m}$
whereas the largest one is associated with $\boldsymbol{n}$, see
also Eq.~(\ref{eq:d_nml}). Moreover, the figure also reveals two
special points, $\alpha=0,\,\pi/3$, for which two eigenvalues are
degenerate \textendash{} either the smallest eigenvalue (for $\alpha=0$)
or the largest eigenvalue (for $\alpha=\pi/3$), corresponding to
an uniaxial nematic state. For later convenience, we note that in
the $(\boldsymbol{n}\boldsymbol{m}\boldsymbol{l})$-representation
of Eq.~(\ref{eq:d_nml}), a sign change in $\boldsymbol{d}$ corresponds
to changing $\alpha\rightarrow\frac{\pi}{3}-\alpha$ and swapping
$\boldsymbol{n}\leftrightarrow\boldsymbol{m}$:
\begin{align}
-\boldsymbol{d}\Big[\!\left|\boldsymbol{d}\right|,\alpha,\boldsymbol{n},\boldsymbol{m},\boldsymbol{l}\Big] & =\boldsymbol{d}\left[\left|\boldsymbol{d}\right|,\frac{\pi}{3}-\alpha,\boldsymbol{m},\boldsymbol{n},\boldsymbol{l}\right].\label{eq:negative_d_nml}
\end{align}

\subsection{Minimization of the nematic free energy}

The value of the angle $\alpha$ that determines the eigenvalues of
the nematic order parameter $Q$ or $\boldsymbol{d}$, see Eqs.~(\ref{eq:Q_d})
and (\ref{eq:d_nml}), can be obtained by minimizing the corresponding
Landau expansion. The symmetry-allowed terms in the Landau expansion
can be obtained by decomposing the products of the non-trivial IR
that defines the order parameter and then picking the terms in the
decomposition that transform trivially under the group operations.
One way to unambiguously determine these terms, which will be very
useful once we consider point groups, is through the so-called decomposition
of the symmetrized product, see Appendix~\ref{sec:Symmetrized-products}
for details. This special decomposition removes any redundancy related
to the anti-symmetric channels and avoids the double-counting that
one would encounter by considering the non-symmetrized product. With
the order parameter $\boldsymbol{d}$ transforming according to the
IR $\Gamma_{j=2}^{+}$, the decomposition of the symmetrized products
for each Landau expansion order becomes

\begin{align}
\big[\otimes_{l=1}^{2}\Gamma_{2}^{+}\big]_{s} & =\Gamma_{0}^{+}\oplus\Gamma_{2}^{+}\oplus\Gamma_{4}^{+},\label{eq:symm_decomp1}\\
\big[\otimes_{l=1}^{3}\Gamma_{2}^{+}\big]_{s} & =\Gamma_{0}^{+}\oplus\Gamma_{2}^{+}\oplus\Gamma_{3}^{+}\oplus\Gamma_{4}^{+}\oplus\Gamma_{6}^{+},\label{eq:symm_decomp2}\\
\big[\otimes_{l=1}^{4}\Gamma_{2}^{+}\big]_{s} & =\Gamma_{0}^{+}\oplus2\Gamma_{2}^{+}\oplus2\Gamma_{4}^{+}\oplus\Gamma_{5}^{+}\oplus\Gamma_{6}^{+}\oplus\Gamma_{8}^{+}.\label{eq:symm_decomp3}
\end{align}
Here, we use the abbreviated tensor product notation, e.g. $\otimes_{l=1}^{3}\Gamma_{2}^{+}=\Gamma_{2}^{+}\otimes\Gamma_{2}^{+}\otimes\Gamma_{2}^{+}$,
and the subscript $s$ to indicate that only the symmetrized product
is considered. The symmetrized decompositions (\ref{eq:symm_decomp1})-(\ref{eq:symm_decomp3})
imply the existence of one Landau invariant per expansion order, since
there is exactly one trivial channel ($\Gamma_{0}^{+}$) per expansion
order. To systematically identify these invariants, it is useful to
first determine the bilinears associated with each IR in Eq.~(\ref{eq:symm_decomp1}),
which we denote by $D^{j=0}=|\boldsymbol{d}|^{2}$ ($\Gamma_{0}^{+}$),
$\boldsymbol{D}^{j=2}$ ($\Gamma_{2}^{+}$) and $\boldsymbol{D}^{j=4}$
($\Gamma_{4}^{+}$). Solving the respective transformation conditions,
similar to Eq.~(\ref{eq:SO3_trafo_condition}), leads to the five-component
vector 
\begin{align}
\boldsymbol{D}^{j=2} & =\left(\begin{smallmatrix}d_{1}^{2}-d_{2}^{2}+\frac{1}{2}\left(d_{3}^{2}+d_{4}^{2}-2d_{5}^{2}\right)\\
-2d_{1}d_{2}-\frac{\sqrt{3}}{2}\left(d_{3}^{2}-d_{4}^{2}\right)\\
\left(d_{1}-\sqrt{3}d_{2}\right)d_{3}+\sqrt{3}d_{4}d_{5}\\
\left(d_{1}+\sqrt{3}d_{2}\right)d_{4}+\sqrt{3}d_{3}d_{5}\\
-2d_{1}d_{5}+\sqrt{3}d_{3}d_{4}
\end{smallmatrix}\right),\label{eq:bilinears_j02}
\end{align}
and the nine-component vector $\boldsymbol{D}^{j=4}$ given in Eq.~(\ref{eq:Dj4})
in Appendix~\ref{sec:SO3_properties}. Since $\boldsymbol{D}^{j=2}$
is quadratic in $d_{i}$, it is straightforward to construct the cubic
invariant that appears in the symmetrized decomposition in Eq.~(\ref{eq:symm_decomp2}).
Specifically, since $\boldsymbol{d}$ and $\boldsymbol{D}^{j=2}$
transform as the same IR $\Gamma_{j=2}^{+}$, their scalar product
must transform trivially and thus appear in the free-energy expansion:
\begin{align}
\boldsymbol{d}\cdot\boldsymbol{D}^{j=2} & =\frac{\sqrt{3}}{2}\mathrm{tr}\left[Q^{3}\right]=\left|\boldsymbol{d}\right|^{3}\cos\left(3\alpha\right).\label{eq:cubic_term_SO3}
\end{align}
In the last step, we employed the $(\boldsymbol{n}\boldsymbol{m}\boldsymbol{l})$-representation
(\ref{eq:Q_alt_parameterization})-(\ref{eq:d_nml}) to further simplify
the expression. As for the quartic decomposition in Eq.~(\ref{eq:symm_decomp3}),
since there is only one term in the symmetrized product that transforms
trivially, we can readily identify it as $D^{j=0}D^{j=0}=|\boldsymbol{d}|^{4}$.
Therefore, the resulting nematic Landau expansion is given by the
action 
\begin{align}
\mathcal{S}\left[\left|\boldsymbol{d}\right|,\alpha\right] & =\int_{\mathsf{x}}\left\{ r_{0}\left|\boldsymbol{d}\right|^{2}+g\left|\boldsymbol{d}\right|^{3}\cos\left(3\alpha\right)+u\left|\boldsymbol{d}\right|^{4}\right\} ,\label{eq:action_SO3-1}
\end{align}
where $\mathsf{x}=(\boldsymbol{r},\tau)$ comprises position and imaginary
time, $\int_{\mathsf{x}}\equiv\int d^{3}r\int_{0}^{1/T}d\tau$, $r_{0}$
is the control parameter that tunes the system across a nematic transition,
and $g$, $u$ are cubic and quartic Landau parameters, respectively.
We opted to represent the Landau expansion in terms of an action rather
than a free energy in order to explicitly account for the temporal
dependence of the nematic order parameter, which is necessary in the
case of a quantum phase transition. For a thermal transition, we can
write $r_{0}=a_{0}(T-T_{0})$ with $a_{0}>0$ and $T_{0}$ a reference
temperature, such that the free energy is given by $F=ST_{0}$. 

We note that the action (\ref{eq:action_SO3-1}) does not depend on
the orientation of the nematic axes $\boldsymbol{n}$, $\boldsymbol{m}$,
and $\boldsymbol{l}$, which is a manifestation of the full rotational
invariance of the 3D isotropic system. Although the expansion (\ref{eq:action_SO3-1})
resembles that of a 3-state Potts/clock model ($Z_{3}$ model), it
is important to emphasize that $\alpha\in[0,\frac{\pi}{3}]$ and that
$\boldsymbol{d}$ is a five-component vector. Minimization with respect
to $\alpha$ gives the mean-field ground-state angle:
\begin{align}
\alpha_{0} & =\frac{\pi}{3}\left(\frac{1+\mathrm{sign}\left(g\right)}{2}\right),\label{eq:SO3_alpha0}
\end{align}
i.e. the angle is either $\alpha_{0}=0$ (for $g<0$) or $\alpha_{0}=\pi/3$
(for $g>0$). In either case, the effective action in terms of $\left|\boldsymbol{d}\right|$
alone assumes the form:

\begin{equation}
\mathcal{S}_{\mathrm{eff}}\left[\left|\boldsymbol{d}\right|\right]=\int_{\mathsf{x}}\left\{ r_{0}\left|\boldsymbol{d}\right|^{2}-\left|g\right|\left|\boldsymbol{d}\right|^{3}+u\left|\boldsymbol{d}\right|^{4}\right\} .
\end{equation}
The existence of a negative cubic term implies that, within a mean-field
solution, the isotropic nematic transition is first order. Another
important property of the $\alpha_{0}=0,\,\pi/3$ solutions is that
they correspond to uniaxial nematic states, i.e., states that only
depend on one eigenvector and for which two eigenvalues are degenerate
\citep{Fregoso2009}. This can be seen directly from Eq.~(\ref{eq:Q_alt_parameterization}):
when $\alpha=0$, the second term vanishes and we obtain

\begin{equation}
Q_{\mu\mu^{\prime}}\left[\alpha=0\right]=\sqrt{3}\left|\boldsymbol{d}\right|\Big(n_{\mu}n_{\mu^{\prime}}-\frac{1}{3}\delta_{\mu\mu^{\prime}}\Big),
\end{equation}
which corresponds to a nematic director along $\boldsymbol{n}$. Correspondingly,
for $\alpha=\pi/3$, we have $\sqrt{3}\cos\alpha=\sin\alpha=\sqrt{3}/2$;
using the completeness relation (\ref{eq:nml_relation}) we find:

\begin{equation}
Q_{\mu\mu^{\prime}}\left[\alpha=\pi/3\right]=-\sqrt{3}\left|\boldsymbol{d}\right|\Big(m_{\mu}m_{\mu^{\prime}}-\frac{1}{3}\delta_{\mu\mu^{\prime}}\Big)
\end{equation}
corresponding to a nematic director along $\boldsymbol{m}$. Similarly,
inserting the two $\alpha_{0}$ values into the nematic order parameter
(\ref{eq:d_nml}) gives:
\begin{align}
\boldsymbol{d}\left[\alpha=0\right] & =\big(\sqrt{3}\big/2\big)\,|\boldsymbol{d}|\;n_{\mu}\boldsymbol{\lambda}_{\mu\mu^{\prime}}^{j=2}n_{\mu^{\prime}},\label{eq:Q_alt_parameterization-alpha_0}\\
\boldsymbol{d}\left[\alpha=\pi/3\right] & =-\big(\sqrt{3}\big/2\big)\,|\boldsymbol{d}|\;m_{\mu}\boldsymbol{\lambda}_{\mu\mu^{\prime}}^{j=2}m_{\mu^{\prime}},\label{eq:Q_alt_parameterization-alpha_pi3}
\end{align}
where we used, once again, the completeness relation (\ref{eq:nml_relation})
as well as the fact that $\mathrm{tr}\left(\lambda^{j=2,i}\right)=0$.
Note that (\ref{eq:Q_alt_parameterization-alpha_pi3}) is just the
negative of (\ref{eq:Q_alt_parameterization-alpha_0}), in accordance
with (\ref{eq:negative_d_nml}). Clearly, the continuous rotational
symmetry of the isotropic system is spontaneously broken in the nematic
ground state, as the nematic director $\boldsymbol{n}$ (or $\boldsymbol{m}$)
can point in any direction. 

In the case of a metallic system, these distortion patterns are manifested
in the Fermi surface of the nematic state, and the nematic instability
is nothing but an $l=2$ Pomeranchuk instability in the charge channel.
The corresponding electronic dispersion (\ref{eq:epsilon_k_nematic})
can be conveniently rewritten in the $(\boldsymbol{n}\boldsymbol{m}\boldsymbol{l})$-representation.
Using Eq.~(\ref{eq:Q_d}), we find 

\begin{equation}
\epsilon_{\boldsymbol{k}}=\frac{\boldsymbol{k}^{2}}{2m}-\mu_{0}+\frac{1}{2}\,\mathrm{tr}\left[Q_{d}U^{T}\mathcal{F}\left(\boldsymbol{k}\right)U\right].
\end{equation}
The transformed matrix $U^{T}\mathcal{F}\left(\boldsymbol{k}\right)U$
is given by:

\begin{align}
U^{T}\mathcal{F}\left(\boldsymbol{k}\right)U & =\left(\begin{smallmatrix}\boldsymbol{n}^{T}\\
\boldsymbol{m}^{T}\\
\boldsymbol{l}^{T}
\end{smallmatrix}\right)\mathcal{F}\left(\boldsymbol{k}\right)\left(\boldsymbol{n},\boldsymbol{m},\boldsymbol{l}\right)\nonumber \\
 & =\left(\begin{smallmatrix}\boldsymbol{n}^{T}\mathcal{F}\left(\boldsymbol{k}\right)\boldsymbol{n} & \boldsymbol{n}^{T}\mathcal{F}\left(\boldsymbol{k}\right)\boldsymbol{m} & \boldsymbol{n}^{T}\mathcal{F}\left(\boldsymbol{k}\right)\boldsymbol{l}\\
\boldsymbol{m}^{T}\mathcal{F}\left(\boldsymbol{k}\right)\boldsymbol{n} & \boldsymbol{m}^{T}\mathcal{F}\left(\boldsymbol{k}\right)\boldsymbol{m} & \boldsymbol{m}^{T}\mathcal{F}\left(\boldsymbol{k}\right)\boldsymbol{l}\\
\boldsymbol{l}^{T}\mathcal{F}\left(\boldsymbol{k}\right)\boldsymbol{n} & \boldsymbol{l}^{T}\mathcal{F}\left(\boldsymbol{k}\right)\boldsymbol{m} & \boldsymbol{l}^{T}\mathcal{F}\left(\boldsymbol{k}\right)\boldsymbol{l}
\end{smallmatrix}\right).
\end{align}
Using the fact that

\begin{equation}
\boldsymbol{v}^{T}\mathcal{F}\left(\boldsymbol{k}\right)\boldsymbol{w}=2\big(\boldsymbol{k}\cdot\boldsymbol{v}\big)\big(\boldsymbol{k}\cdot\boldsymbol{w}\big)-\frac{2}{3}\boldsymbol{k}^{2}\big(\boldsymbol{v}\cdot\boldsymbol{w}\big),
\end{equation}
and performing the matrix product, we find:

\begin{align}
\epsilon_{\boldsymbol{k}} & =\frac{\boldsymbol{k}^{2}}{2m}-\mu_{0}+\frac{2|\boldsymbol{d}|\boldsymbol{k}^{2}}{\sqrt{3}}\Big[\big(\hat{\boldsymbol{k}}\cdot\boldsymbol{n}\big)^{2}\cos\left(\alpha\right)\nonumber \\
 & +\big(\hat{\boldsymbol{k}}\cdot\boldsymbol{m}\big)^{2}\cos\Big(\alpha+\frac{2\pi}{3}\Big)+\big(\hat{\boldsymbol{k}}\cdot\boldsymbol{l}\big)^{2}\cos\Big(\alpha+\frac{4\pi}{3}\Big)\Big],\label{eq:epsilon_fin}
\end{align}
with $\hat{\boldsymbol{k}}=\boldsymbol{k}/\left|\boldsymbol{k}\right|$.
Here, we used $\sum_{\nu=0}^{2}\cos\left(\alpha+\nu\frac{2\pi}{3}\right)=0$.

\begin{figure}
\begin{centering}
\includegraphics[width=1\columnwidth]{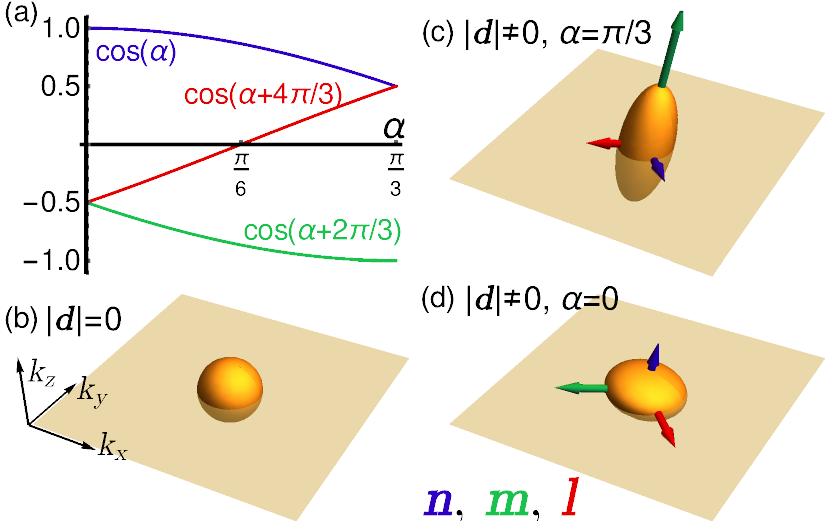}
\par\end{centering}
\caption{(a) Eigenvalues of the tensorial nematic order parameter $Q$ (\ref{eq:Q_d}),
in units of $2|\boldsymbol{d}|/\sqrt{3}$, as a function of the parameter
$\alpha$. The eigenvalues are colored according to their corresponding
eigenvector $\boldsymbol{n}$ (blue), $\boldsymbol{m}$ (green), and
$\boldsymbol{l}$ (red), see Eq.~(\ref{eq:d_nml}). Note that for
$\alpha=0,\,\pi/3$, two eigenvalues are degenerate and the corresponding
nematic state is uniaxial. (b)-(d) Fermi surfaces (\ref{eq:kF_khat})
of the 3D isotropic system in the absence (panel (b)) or presence
of nematic order in the ground state $\alpha=\pi/3$ (panel (c), corresponding
to a tensile nematic distortion) and $\alpha=0$ (panel (d), corresponding
to a compressive nematic distortion). The nematic axes $\boldsymbol{n}$,$\boldsymbol{m}$,$\boldsymbol{l}$
are also shown; for visual purposes, they are rescaled by the corresponding
value of the Fermi momentum along that direction, $2.25\,k_{F}\left(\left\{ \boldsymbol{n},\boldsymbol{m},\boldsymbol{l}\right\} \right)$.
\label{fig:cosines}}
\end{figure}

The Fermi wave-vector of the dispersion (\ref{eq:epsilon_fin}) can
be readily obtained as a function of $\hat{\boldsymbol{k}}$:
\begin{align}
k_{F}\left(\hat{\boldsymbol{k}}\right) & =k_{F,0}\,\Big\{1+\frac{4m|\boldsymbol{d}|}{\sqrt{3}}\Big[\big(\hat{\boldsymbol{k}}\cdot\boldsymbol{m}\big)^{2}\cos\Big(\alpha+\frac{2\pi}{3}\Big)\nonumber \\
 & +\big(\hat{\boldsymbol{k}}\cdot\boldsymbol{l}\big)^{2}\cos\Big(\alpha+\frac{4\pi}{3}\Big)+\big(\hat{\boldsymbol{k}}\cdot\boldsymbol{n}\big)^{2}\cos\left(\alpha\right)\Big]\Big\}^{-\frac{1}{2}},\label{eq:kF_khat}
\end{align}
where we defined the isotropic Fermi momentum $k_{F,0}=\sqrt{2m\mu_{0}}$.
Using the results of Fig.~\ref{fig:cosines}(a), we conclude that
the Fermi wave-vector is longest along the $\boldsymbol{m}$ axis,
$k_{F,\mathrm{max}}=k_{F}\big(\hat{\boldsymbol{k}}=\pm\boldsymbol{m}\big)$,
and shortest along the $\boldsymbol{n}$ directions, $k_{F,\mathrm{min}}=k_{F}\big(\hat{\boldsymbol{k}}=\pm\boldsymbol{n}\big)$.
Therefore, we identify $\boldsymbol{m}$ and $\boldsymbol{n}$ as
the long and short nematic axes of the distorted Fermi surface, respectively.
Moreover, when $\alpha=0$ ($g<0$), the Fermi momentum along the
$\boldsymbol{l}$ direction is as large as the Fermi momentum along
$\boldsymbol{m}$, resulting in a Fermi surface with the shape of
an oblate spheroid; we dub this a ``compressive'' nematic deformation,
and associate the nematic director to $\boldsymbol{n}$. In contrast,
when $\alpha=\pi/3$ ($g>0$), the Fermi momenta along $\boldsymbol{l}$
and along the short axes $\boldsymbol{n}$ are equivalent, resulting
in a Fermi surface with the shape of a prolate spheroid, which we
associate with a ``tensile'' nematic deformation. The nematic director
in this case is parallel to $\boldsymbol{m}$.

We plot the Fermi surface (\ref{eq:kF_khat}) associated with the
two mean-field nematic ground-states $\alpha=\pi/3,\,0$ in Figs.~\ref{fig:cosines}(c)-(d)
with nematic magnitude $|\boldsymbol{d}|=1/(3m)$, together with the
undistorted Fermi surface in Fig.~\ref{fig:cosines}(b). The same
nematic magnitude is employed in all figures in this work. Additionally,
we also plot the corresponding eigenvectors $\boldsymbol{n}$, $\boldsymbol{m}$,
$\boldsymbol{l}$ rescaled by $2.25\,k_{F}(\boldsymbol{n})$, $2.25\,k_{F}(\boldsymbol{m})$,
$2.25\,k_{F}(\boldsymbol{l})$, respectively, to better visualize
the long and short axes in each case. Regardless of the value of $g$,
the resulting Fermi surface is always a uniaxial ellipsoid, which
has the shape of either an oblate spheroid ($\alpha=0$ , $g<0$,
resulting in the compressive distortion of panel (d)) or a prolate
spheroid ($\alpha=\pi/3$ , $g>0$, resulting in the tensile distortion
of panel (c)). Importantly, the uniaxial vector $\boldsymbol{n}$
or $\boldsymbol{m}$ can point in any direction, reflecting the spontaneous
breaking of the continuous rotational symmetry below the nematic transition.

\section{Electronic nematicity in polyhedral point groups\label{sec:Nematicity-in-Polyhedral}}

Once full rotational symmetry is explicitly broken, the system is
described in terms of point groups, which, in contrast to the continuous
orthogonal group $\mathsf{O(3)}$, are finite groups. When investigating
electronically ordered states, it is customary to focus on the $32$
crystallographic point groups, which have at most twofold, threefold,
fourfold, or sixfold rotation symmetry. However, quasicrystals \citep{Shechtman1984,Levine1984}
have been recently shown to realize various electronically-driven
phenomena observed in periodic crystals \citep{Luck1993,Lifshitz1996},
such as superconductivity \citep{Kamiya2018,Uri2023}, magnetism \citep{Goldman1991},
and quantum criticality \citep{Deguchi2012}. While nematicity has
not yet been observed in quasicrystalline environments, it is interesting
to analyze this possibility not only to establish valuable predictions
for future experiments, but also to gain deep insights about the structure
of electronic nematicity as the symmetries of the non-isotropic system
are systematically reduced. Therefore, in this paper, we will not
restrict our analysis to crystallographic point groups only, but will
also consider electronic nematicity in point groups that describe
icosahedral, dodecagonal, decagonal, and octagonal quasicrystals,
as well as twisted quasicrystals \citep{Ahn2018,Haenel2022,Yang2023_a,Yang2023_b,Kim2023,Uri2023}.

There are two different classes of point groups, namely, polyhedral
groups (which do not have axial symmetry) and axial groups (which,
as the name implies, have cylindrical symmetry). As we show below,
in what concerns nematicity, the polyhedral groups are special, since
none of the five elements of the vector $\boldsymbol{d}$ transform
as a trivial IR. In fact, they always transform as multi-dimensional
IRs, which can be either 2-dimensional, 3-dimensional, or 5-dimensional.
This makes a description of electronic nematicity in the $(\boldsymbol{n}\boldsymbol{m}\boldsymbol{l})$-representation
particularly insightful and convenient. In contrast, in the axial
groups, at least one of the elements of $\boldsymbol{d}$ transforms
trivially, and the other components transform either as 2-dimensional
IRs or as 1-dimensional IRs. As we will see, this does not preclude
a description in terms of the $(\boldsymbol{n}\boldsymbol{m}\boldsymbol{l})$-representation.

The seven existing polyhedral groups, which are the focus of this
section, are depicted in Fig.~\ref{fig:descendant_tree} and include
2 octahedral, 3 tetrahedral, and 2 icosahedral point groups. The octahedral
and tetrahedral groups define the cubic crystal system, whereas the
icosahedral groups describe a large class of quasicrystals \citep{Rabson1991}.

The two icosahedral groups $\mathsf{I}$ and $\mathsf{I_{h}}=\mathsf{I}\times\left\{ E,I\right\} $
have a total of $60$ and $120$ symmetry elements. Those are constructed
from twofold ($15$ axes), threefold ($10$ axes), and fivefold ($6$
axes) rotations. For later convenience, we explicitly list the sets
of $6$ fivefold and the $10$ threefold rotation axes:
\begin{align}
\mathcal{V}_{5}^{\mathrm{ico}} & =\Bigg\{\left(\begin{smallmatrix}\pm\alpha_{+}^{(5)}\\
\alpha_{-}^{(5)}\\
0
\end{smallmatrix}\right),\left(\begin{smallmatrix}0\\
\pm\alpha_{+}^{(5)}\\
\alpha_{-}^{(5)}
\end{smallmatrix}\right),\left(\begin{smallmatrix}\alpha_{-}^{(5)}\\
0\\
\pm\alpha_{+}^{(5)}
\end{smallmatrix}\right)\Bigg\},\label{eq:n_C5s}\\
\mathcal{V}_{3}^{\mathrm{ico}} & =\Bigg\{\left(\begin{smallmatrix}\pm\alpha_{-}^{(3)}\\
\alpha_{+}^{(3)}\\
0
\end{smallmatrix}\right),\left(\begin{smallmatrix}0\\
\pm\alpha_{-}^{(3)}\\
\alpha_{+}^{(3)}
\end{smallmatrix}\right),\left(\begin{smallmatrix}\alpha_{+}^{(3)}\\
0\\
\pm\alpha_{-}^{(3)}
\end{smallmatrix}\right),\mathcal{V}_{111}\Bigg\}.\label{eq:n_C3s}
\end{align}
Here, we defined $\alpha_{\pm}^{(3)}=\frac{1}{\sqrt{6}}\sqrt{3\pm\sqrt{5}}$,
$\alpha_{\pm}^{(5)}=\frac{1}{\sqrt{10}}\sqrt{5\pm\sqrt{5}}$, as well
as the set
\begin{align}
\mathcal{V}^{111} & =\frac{1}{\sqrt{3}}\left\{ \left(\begin{smallmatrix}1\\
1\\
1
\end{smallmatrix}\right),\left(\begin{smallmatrix}\text{-}1\\
\text{-}1\\
1
\end{smallmatrix}\right),\left(\begin{smallmatrix}1\\
\text{-}1\\
\text{-}1
\end{smallmatrix}\right),\left(\begin{smallmatrix}\text{-}1\\
1\\
\text{-}1
\end{smallmatrix}\right)\right\} ,\label{eq:V111}
\end{align}
containing the four corners of a tetrahedron. The five octahedral
and tetrahedral point groups, which form the cubic crystal system,
have much fewer elements, and can be conveniently expressed as 

\begin{align}
\mathsf{T} & =\{E,C_{3\alpha}^{\pm1}\}\times\{E,C_{2z}\}\times\{E,C_{2x}\},\label{eq:group_T}\\
\mathsf{T_{h}} & =\mathsf{T}\times\{E,I\},\label{eq:ggrgoup_Thh}\\
\mathsf{T_{d}} & =\mathsf{T}\times\{E,IC_{4z}\},\label{eq:group_Td}\\
\mathsf{O} & =\mathsf{T}\times\{E,C_{4z}\},\label{eq:group_O}\\
\mathsf{O_{h}} & =\mathsf{T}\times\{E,C_{4z}\}\times\{E,I\}.\label{eq:group_Oh}
\end{align}
Within the cubic crystal system (\ref{eq:group_T})-(\ref{eq:group_O}),
$\mathsf{O_{h}}$ is the supergroup, since all the other ones are
subgroups to $\mathsf{O_{h}}$. The characteristic elements of the
$\mathsf{T}$-group comprise $4$ threefold rotation axes $\hat{\boldsymbol{\ell}}\in\mathcal{V}^{111}$,
see Eq.~(\ref{eq:V111}), as well as $3$ twofold rotation axes $\hat{\boldsymbol{\ell}}\in\mathcal{V}^{100}$,
where 
\begin{align}
\mathcal{V}^{100} & =\left\{ \hat{\boldsymbol{e}}_{x},\hat{\boldsymbol{e}}_{y},\hat{\boldsymbol{e}}_{z}\right\} .\label{eq:V100}
\end{align}
For instance, the element $C_{3\alpha}^{\pm1}$ denotes a rotation
by an angle $\vartheta=\pm\frac{2\pi}{3}$ about the axis $\hat{\boldsymbol{\ell}}=(1,1,1)/\sqrt{3}$.
Similarly, $C_{2z}$ corresponds to a $\vartheta=\frac{2\pi}{2}$
rotation about $\hat{\boldsymbol{\ell}}=\hat{\boldsymbol{e}}_{z}$.

\subsection{Icosahedral quasicrystals\label{sec:Nematicity-Icosahedral}}

Within the polyhedral point groups, the icosahedral class contains
the largest number of symmetry elements. In fact, the large number
of symmetry elements forces the five-component order parameter $\boldsymbol{d}$
to still transform as a five-dimensional IR. In other words, in an
icosahedral environment the nematic order parameter is not symmetry-decomposed
but remains the five-dimensional expression (\ref{eq:d_vector}),
i.e. 
\begin{align}
\boldsymbol{d} & \quad\overset{\mathrm{icosahedral}}{\longrightarrow}\quad\boldsymbol{d}.\label{eq:d_Ih}
\end{align}
This property is unique to icosahedral quasicrystals as no other crystallographic
or non-crystallographic point group allows for a five-component IR.
For concreteness, in this section we focus on the larger group $\mathsf{I_{h}}$,
but our results also apply to the group $\mathrm{I}$. Within $\mathsf{I_{h}}$,
the order parameter transforms according to the five-dimensional IR
$H_{g}$. While both the isotropic $\mathsf{O(3)}$ and the icosahedral
$\mathsf{I_{h}}$ environments host five-component nematic order parameters,
the preferred directions of the nematic director in the latter should
be severely restricted due to the finite number of symmetry axes.
Therefore, because the system lacks a continuous rotational symmetry,
the nematic ground state of an icosahedral quasicrystal should be
qualitatively different from the isotropic case. Mathematically, this
difference should be reflected in the corresponding nematic Landau
expansion, which we now derive.

\begin{figure}
\begin{centering}
\includegraphics[width=1\columnwidth]{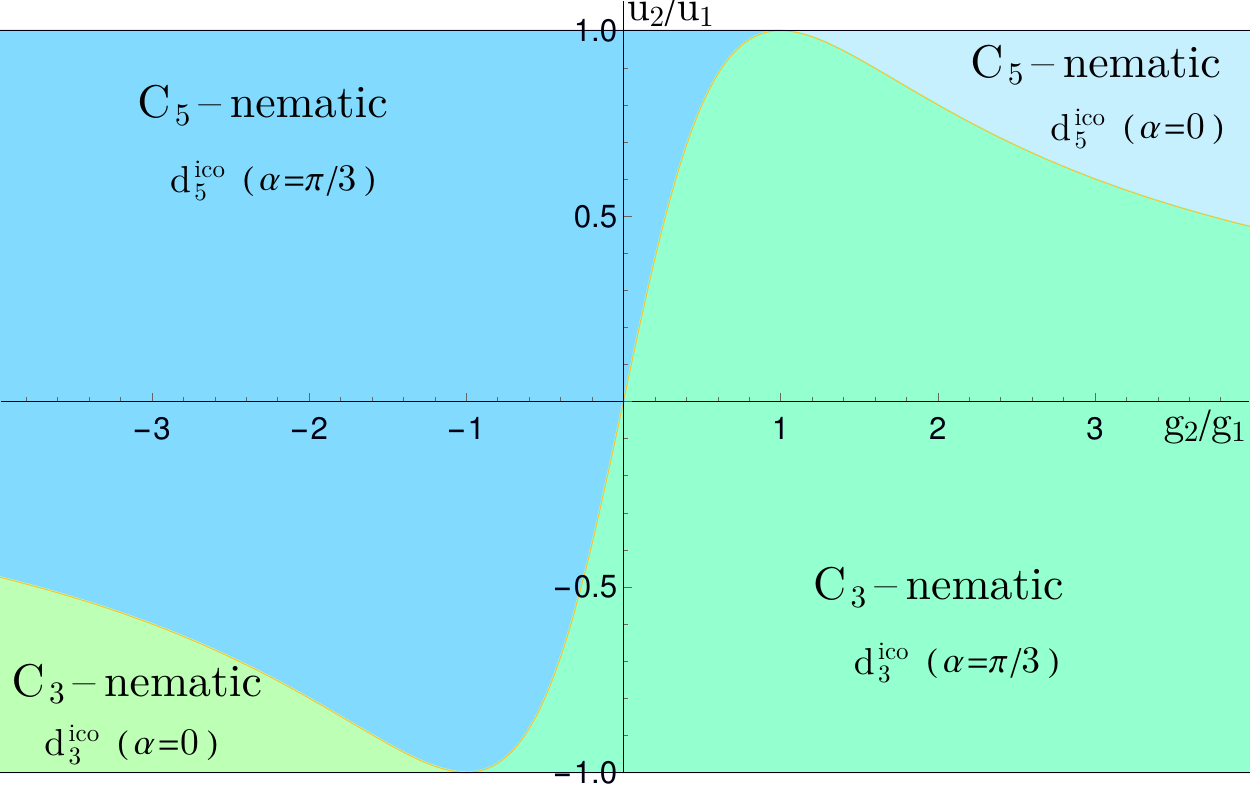}
\par\end{centering}
\caption{Nematic mean-field phase diagram of an icosahedral quasicrystal obtained
by a mean-field minimization of the action (\ref{eq:S_Hg}). The parameters
$u_{1},\,u_{2}$ and $g_{1},\:g_{2}$ are the quartic and cubic Landau
coefficients, respectively. Each state is characterized by the angle
$\alpha$ denoting either a tensile nematic state ($\alpha=\pi/3)$
or a compressive nematic state ($\alpha=0$), see Eq.~(\ref{eq:d_nml}).
In this figure, we set $g_{1}>0$. The case $g_{1}<0$ is recovered
by switching $(\alpha=0)\leftrightarrow(\alpha=\pi/3)$. \label{fig:phase_diag}}
\end{figure}

Here, and in all subsequent sections of this paper, we follow the
same strategy to derive the Landau expansion. First, we compute the
decomposition, for each expansion order, of the symmetrized products
of the IR according to which the nematic order parameter transforms
(see Appendix~\ref{sec:Symmetrized-products}). We find:
\begin{align}
\big[\!\otimes_{j=1}^{2}H_{g}\big]_{s} & =A_{g}\oplus G_{g}\oplus2H_{g},\label{eq:Hg2_decomp}\\
\big[\!\otimes_{j=1}^{3}H_{g}\big]_{s} & =2A_{g}\oplus T_{1g}\oplus T_{2g}\oplus3G_{g}\oplus3H_{g},\label{eq:Hg3_decomp}\\
\big[\!\otimes_{j=1}^{4}H_{g}\big]_{s} & =2A_{g}\!\oplus\!2T_{1g}\!\oplus\!2T_{2g}\!\oplus4G_{g}\!\oplus8H_{g}.\label{eq:Hg4_decomp}
\end{align}
Therefore, the Landau expansion up to fourth order contains a total
of $5$ invariants (i.e. $5$ terms that transform as the trivial
IR $A_{g}$), in contrast to the $3$ invariants of the isotropic
case, see Eqs.~(\ref{eq:symm_decomp1})-(\ref{eq:symm_decomp3}).
To construct these five invariants it is convenient to determine the
bilinear combinations associated with Eq.~(\ref{eq:Hg2_decomp}),
which we denote by $D^{A_{g}}=\left|\boldsymbol{d}\right|^{2}$, $\boldsymbol{D}^{G_{g}}$,
$\boldsymbol{D}^{H_{g},1}$ and $\boldsymbol{D}^{H_{g},2}$. Note
that the two $H_{g}$-bilinears are degenerate. As explained in Appendix~\ref{sec:App_Ih},
we choose a representation where $\boldsymbol{D}^{H_{g},1}=\boldsymbol{D}^{j=2}$
, as defined in Eq.~(\ref{eq:bilinears_j02}), and
\begin{align}
\boldsymbol{D}^{H_{g},2}= & \mathcal{R}_{5}\left(-\phi_{0}\right)\left(\begin{smallmatrix}\tilde{d}_{1}^{2}-\tilde{d}_{2}^{2}+\frac{1}{2}\left(\tilde{d}_{3}^{2}+\tilde{d}_{4}^{2}-2\tilde{d}_{5}^{2}\right)\\
-2\tilde{d}_{1}\tilde{d}_{2}-\frac{\sqrt{3}}{2}\left(\tilde{d}_{3}^{2}-\tilde{d}_{4}^{2}\right)\\
\left(\tilde{d}_{1}-\sqrt{3}\tilde{d}_{2}\right)\tilde{d}_{3}+\sqrt{3}\tilde{d}_{4}\tilde{d}_{5}\\
\left(\tilde{d}_{1}+\sqrt{3}\tilde{d}_{2}\right)\tilde{d}_{4}+\sqrt{3}\tilde{d}_{3}\tilde{d}_{5}\\
-2\tilde{d}_{1}\tilde{d}_{5}+\sqrt{3}\tilde{d}_{3}\tilde{d}_{4}
\end{smallmatrix}\right).\label{eq:D_Hg2_ico}
\end{align}
Here, we introduced $\tilde{\boldsymbol{d}}=\mathcal{R}_{5}(\phi_{0})\boldsymbol{d}$
with angle $\phi_{0}=\arccos\left(-1/4\right)$ and the particular
rotation matrix 
\begin{align}
\mathcal{R}_{5}(\phi) & =\left(\begin{smallmatrix}\cos\phi & \sin\phi & 0 & 0 & 0\\
-\sin\phi & \cos\phi & 0 & 0 & 0\\
0 & 0 & 1 & 0 & 0\\
0 & 0 & 0 & 1 & 0\\
0 & 0 & 0 & 0 & 1
\end{smallmatrix}\right).\label{eq:R5-1}
\end{align}
Correspondingly, the two $H_{g}$-bilinears are related through the
equation $\boldsymbol{D}^{H_{g},2}=\mathcal{R}_{5}(-\phi_{0})\boldsymbol{D}^{H_{g},1}\big|_{\boldsymbol{d}\rightarrow\tilde{\boldsymbol{d}}}$,
and thus have the same magnitude $\left|\boldsymbol{D}^{H_{g},1}\right|=\left|\boldsymbol{D}^{H_{g},2}\right|=\left|\boldsymbol{d}\right|^{2}$.
For our purposes, we do not need $\boldsymbol{D}^{G_{g}}$, although
it can be derived in a straightforward way. With the aid of the two
$H_{g}$-bilinears, the five $A_{g}$ invariants in Eqs.~(\ref{eq:Hg2_decomp})-(\ref{eq:Hg4_decomp})
can be constructed as $|\boldsymbol{d}|^{2}$, $|\boldsymbol{d}|^{4}$,
$\boldsymbol{d}\cdot\boldsymbol{D}^{H_{g},1}$, $\boldsymbol{d}\cdot\boldsymbol{D}^{H_{g},2}$,
and $\boldsymbol{D}^{H_{g},1}\cdot\boldsymbol{D}^{H_{g},2}$. The
resulting nematic Landau action becomes
\begin{align}
\mathcal{S} & =\int_{\mathsf{x}}\Big\{ r_{0}\left|\boldsymbol{d}\right|^{2}+\left|\boldsymbol{d}\right|^{3}\Big(g_{1}\hat{\boldsymbol{d}}\cdot\hat{\boldsymbol{D}}^{H_{g},1}+g_{2}\hat{\boldsymbol{d}}\cdot\hat{\boldsymbol{D}}^{H_{g},2}\Big)\nonumber \\
 & \qquad\qquad\qquad+\left|\boldsymbol{d}\right|^{4}\Big(u_{1}+u_{2}\,\hat{\boldsymbol{D}}^{H_{g},1}\cdot\hat{\boldsymbol{D}}^{H_{g},2}\Big)\Big\},\label{eq:S_Hg}
\end{align}
with the unit vectors $\hat{\boldsymbol{d}}=\boldsymbol{d}/|\boldsymbol{d}|$
and $\hat{\boldsymbol{D}}^{H_{g},1/2}=\boldsymbol{D}^{H_{g},1/2}/|\boldsymbol{d}|^{2}$,
the cubic coefficients $g_{1,2}$, and the quartic coefficients $u_{1,2}$,
which are restricted to $0\leq|u_{2}|<u_{1}$ in order for the action
to be bounded. Since the action (\ref{eq:S_Hg}) has not been explored
in the literature, we first establish the corresponding mean-field
ground-state phase diagram. The details of the derivation are given
in Appendix~\ref{sec:App_Ih}; here, we focus on the numerically-obtained
phase diagram shown in Fig.~\ref{fig:phase_diag} in the $\left(g_{2}/g_{1},\,u_{2}/u_{1}\right)$
parameter-space. Clearly, the Landau expansion (\ref{eq:S_Hg}) has
two distinct ground-state phases, which we dub $C_{3}$-nematic and
$C_{5}$-nematic, based on the residual rotational symmetry of each
state.

\begin{figure}
\begin{centering}
\includegraphics[width=1\columnwidth]{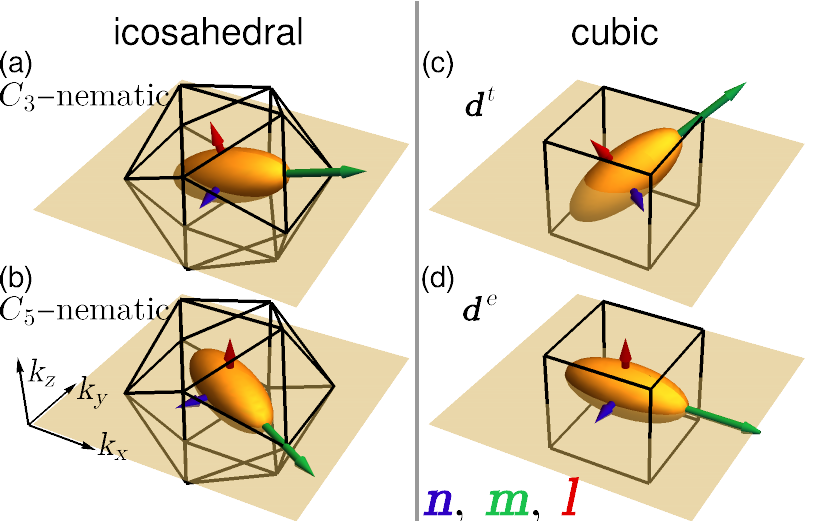}
\par\end{centering}
\caption{Distorted Fermi surface in the nematic phase, as given by Eq.~(\ref{eq:kF_khat}),
plotted together with the nematic axes $\boldsymbol{n}$,$\boldsymbol{m}$,$\boldsymbol{l}$.
(a) Icosahedral $C_{3}$-nematic state (\ref{eq:d_ico3_1}); (b) icosahedral
$C_{5}$-nematic state (\ref{eq:d_ico5_1}); (c) cubic triplet ($T_{2g}$)
state (\ref{eq:d_cubic_t_1}); (d) cubic doublet ($E_{g}$) state
(\ref{eq:d_cubic_e_1}). In all cases, the $\alpha=\pi/3$ (tensile)
state is plotted, and the nematic director is parallel to $\boldsymbol{m}$.
The unit vectors $\left\{ \boldsymbol{n},\boldsymbol{m},\boldsymbol{l}\right\} $
are rescaled by $2.2\,k_{F}\left(\left\{ \boldsymbol{n},\boldsymbol{m},\boldsymbol{l}\right\} \right)$,
respectively. \label{fig:FS_ico_cubic}}
\end{figure}

We first discuss the $C_{3}$-nematic phase, by focusing on the parameter
regime $\mathrm{sign}\,g_{1}=\mathrm{sign}\,g_{2}$ and $u_{2}<0$
(bottom right quadrant in Fig.~\ref{fig:phase_diag}). In this region,
the condition $\hat{\boldsymbol{D}}^{H_{g},1}=\hat{\boldsymbol{D}}^{H_{g},2}=-\mathrm{sign}\,g_{1}\hat{\boldsymbol{d}}$
minimizes not only the two cubic terms of Eq.~(\ref{eq:S_Hg}), but
also the anisotropic quartic term, i.e. all three direction-dependent
terms are simultaneously minimized. This condition is satisfied for
$10$ directions of the nematic order parameter $\hat{\boldsymbol{d}}$,
which are associated with the $10$ threefold symmetry axes $\mathcal{V}_{3}^{\mathrm{ico}}$
in Eq.~(\ref{eq:n_C3s}). Indeed, in the $(\boldsymbol{n}\boldsymbol{m}\boldsymbol{l})$-representation
(\ref{eq:d_nml}), the $C_{3}$-nematic ground state is given by 
\begin{align}
g_{1} & >0: & \boldsymbol{d}_{3,i}^{\mathrm{ico}} & =\boldsymbol{d}\left[\left|\boldsymbol{d}\right|,\alpha=\pi/3,\boldsymbol{n}_{i},\boldsymbol{m}_{i}=\mathcal{V}_{3,i}^{\mathrm{ico}},\boldsymbol{l}_{i}\right],\label{eq:d_ico3_1}\\
g_{1} & <0: & \boldsymbol{d}_{3,i}^{\mathrm{ico}} & =\boldsymbol{d}\left[\left|\boldsymbol{d}\right|,\alpha=0,\boldsymbol{n}_{i}=\mathcal{V}_{3,i}^{\mathrm{ico}},\boldsymbol{m}_{i},\boldsymbol{l}_{i}\right],\label{eq:d_ico3_2}
\end{align}
with $i=1,\ldots,10$. Since $\alpha=\pi/3$ or $\alpha=0$, both
ground states are uniaxial, see Eqs.~(\ref{eq:Q_alt_parameterization-alpha_0})-(\ref{eq:Q_alt_parameterization-alpha_pi3}),
with the state described by Eq.~(\ref{eq:d_ico3_1}) being a tensile
nematic state ($\alpha=\pi/3$) and the state described by Eq.~(\ref{eq:d_ico3_2})
being a compressive nematic state ($\alpha=0$). In Fig.~\ref{fig:FS_ico_cubic}(a),
we plot the Fermi surface distortion associated with the tensile $C_{3}$-nematic
ground state (\ref{eq:d_ico3_1}) using Eq.~(\ref{eq:epsilon_fin}).
Note that while strictly speaking the crystal momentum is not a good
quantum number for a quasicrystal, plotting Eq.~(\ref{eq:kF_khat})
is useful to visualize, even if perturbatively only, the effect of
the broken rotational symmetry on the otherwise spherically symmetric
electronic charge distribution. To visualize the allowed directions
of the nematic director $\boldsymbol{m}$, we draw in the same figure
an icosahedron concentric to the Fermi surface, which consists of
$20$ identical equilateral triangles. Clearly, the long axis $\boldsymbol{m}$
points towards the center of one these equilateral triangles. Since
there are $20$ such triangles, and because $\pm\boldsymbol{m}$ are
identical nematic states, there are indeed $10$ degenerate states.
Because $\boldsymbol{n}_{i}$ or $\boldsymbol{m}_{i}$ can only point
along 10 symmetry-related directions, the broken symmetry is discrete.
Note that the resulting point group in this case is $\mathsf{D_{3d}}$.
While it belongs to the crystallographic trigonal system, this does
not imply that the quasicrystal will become a crystal rather than
an incommensurate crystal. Nevertheless, it is interesting that a
nematic transition changes the point group from non-crystallographic
to crystallographic.

To discuss the $C_{5}$-nematic phase, we consider the parameter range
$\mathrm{sign}\,g_{1}=-\mathrm{sign}\,g_{2}$ and $u_{2}>0$, corresponding
to the top left quadrant in the phase diagram of Fig.~\ref{fig:phase_diag}.
In this case, the simultaneous minimization of the three anisotropic
terms of the action (\ref{eq:S_Hg}) is achieved by the condition
$\hat{\boldsymbol{D}}^{H_{g},1}=-\hat{\boldsymbol{D}}^{H_{g},2}=-\mathrm{sign}\,g_{1}\hat{\boldsymbol{d}}$.
This condition, in turn, is satisfied for six directions of $\hat{\boldsymbol{d}}$
that are associated with the $6$ fivefold symmetry axes $\mathcal{V}_{5}^{\mathrm{ico}}$
in Eq.~(\ref{eq:n_C5s}). In analogy to the $C_{3}$-nematic phase,
the $C_{5}$-nematic ground state is uniaxial and parametrized as:
\begin{align}
g_{1} & >0: & \boldsymbol{d}_{5,i}^{\mathrm{ico}} & =\boldsymbol{d}\left[\left|\boldsymbol{d}\right|,\alpha=\pi/3,\boldsymbol{n}_{i},\boldsymbol{m}_{i}=\mathcal{V}_{5,i}^{\mathrm{ico}},\boldsymbol{l}_{i}\right],\label{eq:d_ico5_1}\\
g_{1} & <0: & \boldsymbol{d}_{5,i}^{\mathrm{ico}} & =\boldsymbol{d}\left[\left|\boldsymbol{d}\right|,\alpha=0,\boldsymbol{n}_{i}=\mathcal{V}_{5,i}^{\mathrm{ico}},\boldsymbol{m}_{i},\boldsymbol{l}_{i}\right],\label{eq:d_ico5_2}
\end{align}
where $i=1,\ldots,6$. We visualize the tensile Fermi surface distortion
associated with the tensile state (\ref{eq:d_ico5_1}) in Fig.~\ref{fig:FS_ico_cubic}(b).
In the $C_{5}$-nematic phase, the nematic director $\boldsymbol{m}$
aligns with the corners of the icosahedron. Since there are $12$
such corners, we find indeed $6$ degenerate $C_{5}$-nematic states.
The residual symmetries inside the nematic phase result in the point
group $\mathsf{D_{5d}}$, which is also a non-crystallographic point
group.

The nematic phase diagram in Fig.~\ref{fig:phase_diag} shows that
the $C_{3}$-nematic state (\ref{eq:d_ico3_1}) and the $C_{5}$-nematic
state (\ref{eq:d_ico5_1}) actually occupy the entire phase space,
including regions where not all three anisotropic terms in (\ref{eq:S_Hg})
can be satisfied simultaneously. As shown in the top-right and bottom-left
corner, a tensile ground state can become a compressive one upon traversing
the phase diagram.

\subsection{Cubic crystals\label{sec:Nematicity-Cubic}}

The cubic crystal system consists of two octahedral point groups ($\mathsf{O_{h}}$,
$\mathsf{O}$) and three tetrahedral point groups ($\mathsf{T_{h}}$,
$\mathsf{T_{d}}$, $\mathsf{T}$). In all five cases, the reduced
symmetry with respect to the icosahedral point group leads to a symmetry
decomposition of the five-component nematic order parameter $\boldsymbol{d}$
(\ref{eq:d_vector}) according to: 
\begin{align}
\boldsymbol{d} & \quad\overset{\mathrm{cubic}}{\longrightarrow}\quad\left(\boldsymbol{d}^{e},\,\boldsymbol{d}^{t}\right)^{T},\label{eq:d_Oh}
\end{align}
where the doublet $\boldsymbol{d}^{e}$ and the triplet $\boldsymbol{d}^{t}$
nematic vectors are given by:
\begin{align}
\boldsymbol{d}^{e} & =\left(\begin{array}{c}
d_{1}\\
d_{2}
\end{array}\right)=\left(\begin{array}{c}
d_{\frac{1}{\sqrt{3}}(2z^{2}-x^{2}-y^{2})}\\
d_{x^{2}-y^{2}}
\end{array}\right)=\left|\boldsymbol{d}^{e}\right|\left(\begin{array}{c}
\cos\gamma_{e}\\
\sin\gamma_{e}
\end{array}\right),\label{eq:de_cubic}\\
\boldsymbol{d}^{t} & =\left(d_{3},d_{4},d_{5}\right)^{T}=\left(d_{2yz},d_{2xz},d_{2xy}\right)^{T}.\label{eq:dt_cubic}
\end{align}
In terms of the IRs of the five groups ($\mathsf{O_{h}}$, $\mathsf{O}$,
$\mathsf{T_{h}}$, $\mathsf{T_{d}}$, $\mathsf{T}$), $\boldsymbol{d}^{t}$
always transforms as a three-dimensional IR ($T_{2g}$, $T_{2}$,
$T_{g}$, $T_{2}$, and $T$, respectively) whereas $\boldsymbol{d}^{e}$
transforms as a two-dimensional IR ($E_{g}$, $E$, $E_{g}\oplus\bar{E}_{g}$,
$E$, and $E\oplus\bar{E}$, respectively). Our goal is to derive
the Landau expansions for both types of nematics and then determine
the mean-field ground states in the $(\boldsymbol{n}\boldsymbol{m}\boldsymbol{l})$-representation
(\ref{eq:d_nml}). For concreteness, we focus on the most symmetric
cubic point group, $\mathsf{O_{h}}$, and comment on the extension
to the other four cubic point groups in the end of this section.

We begin with the nematic triplet order parameter $\boldsymbol{d}^{t}$
(\ref{eq:dt_cubic}), which transforms as the $T_{2g}$ IR of $\mathsf{O_{h}}$.
The decomposition of the symmetrized product for each expansion order
gives 

\begin{align}
\big[\otimes_{j=1}^{2}T_{2g}\big]_{s} & =A_{1g}\oplus E_{g}\oplus T_{2g},\label{eq:symm_decomp_T2gA}\\
\big[\otimes_{j=1}^{3}T_{2g}\big]_{s} & =A_{1g}\oplus T_{1g}\oplus2T_{2g},\label{eq:symm_decomp_T2gB}\\
\big[\otimes_{j=1}^{4}T_{2g}\big]_{s} & =2A_{1g}\oplus2E_{g}\oplus T_{1g}\oplus2T_{2g},\label{eq:symm_decomp_T2gC}
\end{align}
which implies the existence of $4$ invariants in the Landau expansion
up to fourth order. To systematically construct them, it is convenient
to exploit the bilinears associated with the quadratic decomposition,
Eq.~(\ref{eq:symm_decomp_T2gA}). We find that $D^{A_{1g}}=|\boldsymbol{d}^{t}|^{2}$
and
\begin{align}
\boldsymbol{D}^{E_{g}} & =\left(\begin{smallmatrix}\frac{1}{\sqrt{3}}\left(2d_{5}^{2}-d_{3}^{2}-d_{4}^{2}\right)\\
\left(d_{3}^{2}-d_{4}^{2}\right)
\end{smallmatrix}\right), & \boldsymbol{D}^{T_{2g}} & =\left(\begin{smallmatrix}2d_{4}d_{5}\\
2d_{3}d_{5}\\
2d_{3}d_{4}
\end{smallmatrix}\right).\label{eq:bilinears_j02-1-1}
\end{align}
In terms of these bilinears, the four invariants become $\left|\boldsymbol{d}^{t}\right|^{2}$
to quadratic order, $\boldsymbol{d}^{t}\cdot\boldsymbol{D}^{T_{2g}}$
to cubic order, and $\left|\boldsymbol{d}^{t}\right|^{4}$ and $\boldsymbol{D}^{T_{2g}}\cdot\boldsymbol{D}^{T_{2g}}$
to quartic order. Note that the quartic term could equally be written
in terms of $\boldsymbol{D}^{E_{g}}\cdot\boldsymbol{D}^{E_{g}}$ by
using the Fierz identity $\boldsymbol{D}^{E_{g}}\cdot\boldsymbol{D}^{E_{g}}=\frac{4}{3}\left|\boldsymbol{d}^{t}\right|^{4}-\boldsymbol{D}^{T_{2g}}\cdot\boldsymbol{D}^{T_{2g}}$.
Therefore, the resulting Landau expansion is
\begin{align}
\mathcal{S} & =\int_{\mathsf{x}}\Big\{ r_{0}\left|\boldsymbol{d}^{t}\right|^{2}+g\,d_{3}d_{4}d_{5}+u_{1}\left|\boldsymbol{d}^{t}\right|^{4}\nonumber \\
 & +u_{2}\left(d_{3}^{2}d_{4}^{2}+d_{3}^{2}d_{5}^{2}+d_{4}^{2}d_{5}^{2}\right)\Big\},\label{eq:S_T2g}
\end{align}
with the cubic and quartic Landau coefficients $g$, $u_{1}$, $u_{2}$.
The Landau expansion (\ref{eq:S_T2g}) is the the same as that of
the 4-state Potts model ($Z_{4}$-Potts) \citep{Potts_Wallace}. It
is well established that the the upper critical dimension of this
model is below $3$, $d_{u}<3$ \citep{Potts_Wu}. Consequently, a
mean-field solution of Eq.~(\ref{eq:S_T2g}) is appropriate, resulting
in a first-order transition and in a fourfold degenerate ground-state
manifold given by $\boldsymbol{d}_{0,i}^{t}=\mathrm{sign}\left(-g\right)\left|\boldsymbol{d}^{t}\right|\mathcal{V}_{i}^{Z_{4}}$
with $i=1,\ldots,4$ and:
\begin{align}
\mathcal{V}^{Z_{4}} & =\frac{1}{\sqrt{3}}\left\{ \left(\begin{smallmatrix}1\\
1\\
1
\end{smallmatrix}\right),\left(\begin{smallmatrix}\text{-}1\\
\text{-}1\\
1
\end{smallmatrix}\right),\left(\begin{smallmatrix}1\\
\text{-}1\\
\text{-}1
\end{smallmatrix}\right),\left(\begin{smallmatrix}\text{-}1\\
1\\
\text{-}1
\end{smallmatrix}\right)\right\} .\label{eq:V_Z4}
\end{align}
This fourfold degeneracy is associated with the $4$ threefold symmetry
axes $\mathcal{V}^{111}$ defined in Eq.~(\ref{eq:V111}). Indeed,
in the $(\boldsymbol{n}\boldsymbol{m}\boldsymbol{l})$-representation
of Eq.~(\ref{eq:d_nml}), these four states are given by 
\begin{align}
g & >0: & \boldsymbol{d}_{0,i}^{t} & =\boldsymbol{d}\left[\left|\boldsymbol{d}^{t}\right|,\alpha=\pi/3,\boldsymbol{n}_{i},\boldsymbol{m}_{i}=\mathcal{V}_{i}^{111},\boldsymbol{l}_{i}\right],\label{eq:d_cubic_t_1}\\
g & <0: & \boldsymbol{d}_{0,i}^{t} & =\boldsymbol{d}\left[\left|\boldsymbol{d}^{t}\right|,\alpha=0,\boldsymbol{n}_{i}=\mathcal{V}_{i}^{111},\boldsymbol{m}_{i},\boldsymbol{l}_{i}\right].\label{eq:d_cubic_t_2}
\end{align}
Since $\alpha=\pi/3$ or $\alpha=0$, the nematic states (\ref{eq:d_cubic_t_1})-(\ref{eq:d_cubic_t_2})
are uniaxial {[}recall Eqs.~(\ref{eq:Q_alt_parameterization-alpha_0})-(\ref{eq:Q_alt_parameterization-alpha_pi3}){]},
and the nematic director is determined by either $\boldsymbol{m}$
or $\boldsymbol{n}$, respectively. In Fig.~\ref{fig:FS_ico_cubic}(c),
we show the distorted Fermi surface associated with the tensile (i.e.
$\alpha=\pi/3$) nematic state, Eq.~(\ref{eq:d_cubic_t_1}). Note
that the nematic director $\boldsymbol{m}$ points towards the corners
of the cube that is concentric to the Fermi surface. Since there are
$8$ such corners, the number of degenerate states is $4$, since
$\pm\boldsymbol{m}$ correspond to the same state.

Next, we consider the nematic doublet order parameter $\boldsymbol{d}^{e}$
(\ref{eq:de_cubic}), which transforms as the $E_{g}$ IR of $\mathsf{O_{h}}$.
Following the same procedure as above, we first perform the decomposition
of the symmetrized product for each Landau-expansion order:
\begin{align}
\big[\!\otimes_{j=1}^{2}E_{g}\big]_{s} & =A_{1g}\oplus E_{g},\label{eq:symm_decomp_Eg2-1}\\
\big[\!\otimes_{j=1}^{3}E_{g}\big]_{s} & =A_{1g}\oplus A_{2g}\oplus E_{g},\label{eq:symm_decomp_Eg3-1}\\
\big[\!\otimes_{j=1}^{4}E_{g}\big]_{s} & =A_{1g}\!\oplus\!2E_{g},\label{eq:symm_decomp_Eg4-1}
\end{align}
which reveals the existence of one $A_{1g}$ invariant per expansion
order. The bilinear combinations in Eq.~(\ref{eq:symm_decomp_Eg2-1})
can be readily obtained:
\begin{align}
D^{A_{1g}} & =\left|\boldsymbol{d}^{e}\right|^{2}, & \boldsymbol{D}^{E_{g}} & =\left|\boldsymbol{d}^{e}\right|^{2}\left(\begin{array}{c}
\cos2\gamma_{e}\\
-\sin2\gamma_{e}
\end{array}\right).\label{eq:bilinears_j02-1}
\end{align}
Using these bilinears, we can express the three invariants as $\left|\boldsymbol{d}^{e}\right|^{2}$,
$\left|\boldsymbol{d}^{e}\right|^{4}$ and $\boldsymbol{d}^{e}\cdot\boldsymbol{D}^{E_{g}}$,
such that the nematic Landau expansion becomes:
\begin{align}
\mathcal{S} & =\int_{\mathsf{x}}\Big\{ r_{0}\left|\boldsymbol{d}^{e}\right|^{2}+g\left|\boldsymbol{d}^{e}\right|^{3}\,\cos\left(3\gamma_{e}\right)+u\left|\boldsymbol{d}^{e}\right|^{4}\Big\}.\label{eq:S_Eg}
\end{align}
The nematic action (\ref{eq:S_Eg}) has the same Landau expansion
as the $3$-state Potts model ($Z_{3}$-Potts) \citep{Potts_Wu}.
We note that the $3$-state Potts model is in the same universality
class as the $3$-state clock model ($Z_{3}$-clock), so we will use
these terms interchangeably in the remainder of the paper. We will
discuss further the differences between $Z_{q}$-Potts and $Z_{q}$-clock
models in Sec.~\ref{sec:Nematicity-Axial}. The key point is that
the upper critical dimension of the $Z_{3}$-Potts/clock model is
$d_{u}<3$. Therefore, as in the case of the triplet nematic, a mean-field
solution of Eq.~(\ref{eq:S_Eg}) is appropriate. We find a threefold
degenerate ground-state manifold given by $\boldsymbol{d}_{0,i}^{e}=\mathrm{sign}\left(-g\right)\left|\boldsymbol{d}^{e}\right|\mathcal{V}_{i}^{Z_{3}}$
with $i=1,2,3$ and
\begin{align}
\mathcal{V}^{Z_{3}} & =\left\{ \left(\begin{smallmatrix}1\\
0
\end{smallmatrix}\right),\frac{1}{2}\left(\begin{smallmatrix}-1\\
\sqrt{3}
\end{smallmatrix}\right),\frac{1}{2}\left(\begin{smallmatrix}-1\\
-\sqrt{3}
\end{smallmatrix}\right)\right\} .\label{eq:V_Z3}
\end{align}
This threefold degeneracy is related to the $3$ fourfold rotation
axes $\mathcal{V}^{100}$ in Eq.~(\ref{eq:V100}), which are parallel
to the three coordinate axes. This can be directly seen upon writing
the nematic ground state in the $(\boldsymbol{n}\boldsymbol{m}\boldsymbol{l})$-representation
(\ref{eq:d_nml}): 
\begin{align}
g & >0: & \boldsymbol{d}_{0,i}^{e} & =\boldsymbol{d}\left[\left|\boldsymbol{d}^{e}\right|,\alpha=\pi/3,\boldsymbol{n}_{i},\boldsymbol{m}_{i}=\mathcal{V}_{i}^{100},\boldsymbol{l}_{i}\right],\label{eq:d_cubic_e_1}\\
g & <0: & \boldsymbol{d}_{0,i}^{e} & =\boldsymbol{d}\left[\left|\boldsymbol{d}^{e}\right|,\alpha=0,\boldsymbol{n}_{i}=\mathcal{V}_{i}^{100},\boldsymbol{m}_{i},\boldsymbol{l}_{i}\right].\label{eq:d_cubic_e_2}
\end{align}
As in the case of the $T_{2g}$ (triplet) nematic, the $E_{g}$ (doublet)
nematic state is uniaxial with the nematic director $\boldsymbol{m}$
or $\boldsymbol{n}$ pointing along one of the three coordinate axes.
This is illustrated in Fig.~\ref{fig:FS_ico_cubic}(d), where we
plot the Fermi-surface distortion associated with the tensile (i.e.
$\alpha=\pi/3$) state given by Eq.~(\ref{eq:d_cubic_e_1}). 

While the analysis performed above focused on the $\mathsf{O_{h}}$
group, the same general results hold for the doublet and triplet nematics
of the other four cubic point groups, as shown in Table~\ref{tab:classification}.
It is important to note, however, that the case of the nematic doublet
order parameter $\boldsymbol{d}^{e}$ in the tetrahedral point groups
$\mathsf{T_{h}}$ and $\mathsf{T}$ needs to be treated slightly differently,
because $\boldsymbol{d}^{e}$ transforms according to a \emph{complex}
IR, which allows for additional invariants in the Landau expansion,
which in turn make the nematic state biaxial rather than uniaxial.
A detailed derivation of this case is presented in Appendix~\ref{sec:cubic_appendix}. 

\section{Electronic nematicity in crystallographic axial point groups\label{sec:Nematicity-Axial}}

In contrast to the family of polyhedral point groups, which encompasses
the seven point groups discussed in Sec.~\ref{sec:Nematicity-in-Polyhedral},
there is an infinite number of axial groups. Their defining property,
as the name indicates, is their underlying cylindrical symmetry, which
implies that the system is invariant under some $n$-fold proper or
improper rotation with respect to the $z$-axis. This symmetry can
be accompanied by additional symmetries related to inversion or twofold
rotations with respect to in-plane axes, resulting in a total of seven
series of axial groups, as shown in Fig.~\ref{fig:descendant_tree}.
They can be constructed from two axial group series corresponding
to the cyclic group $\mathsf{C_{n}}$ and the group $\mathsf{S_{2n}}$:
\begin{align}
\mathsf{C_{n}} & =\left\{ E,C_{nz},C_{nz}^{2},\dots,C_{nz}^{n-1}\right\} , & n & \in[2,3,\dots,\infty],\label{eq:group_Cn}\\
\mathsf{S_{2n}} & =\left\{ E,S_{2nz},S_{2nz}^{2},\dots,S_{2nz}^{2n-1}\right\} , & n & \in[2,3,\dots,\infty].\label{eq:group_Sn}
\end{align}
Here, $E$ is the identity operation, $C_{nz}$ corresponds to an
$n$-fold rotation with respect to the $z$ axis, and the improper
rotation (or rotoinversion) operation $S_{nz}$ is defined as a proper
rotation $C_{nz}$ followed by a reflection with respect to the horizontal
mirror, $S_{nz}=IC_{2z}C_{nz}$. Note that we can also extend these
definitions to include the groups $\mathsf{C_{1}}=\left\{ E\right\} $
and $\mathsf{S_{2}}=\mathsf{C_{i}}=\left\{ E,I\right\} $. The remaining
five infinite series of axial groups are obtained from $\mathsf{C_{n}}$
or $\mathsf{S_{2n}}$ in the following way: 
\begin{align}
\mathsf{D_{n}} & =\mathsf{C_{n}}\times\left\{ E,C_{2x}\right\} ,\label{eq:group_Dn}\\
\mathsf{C_{nv}} & =\mathsf{C_{n}}\times\left\{ E,IC_{2x}\right\} ,\label{eq:group_Cnv}\\
\mathsf{C_{nh}} & =\mathsf{C_{n}}\times\left\{ E,IC_{2z}\right\} ,\label{eq:group_Cnh}\\
\mathsf{D_{nd}} & =\mathsf{S_{2n}}\times\left\{ E,C_{2x}\right\} ,\label{eq:group_Dnd}\\
\mathsf{D_{nh}} & =\mathsf{C_{n}}\times\left\{ E,IC_{2z}\right\} \times\left\{ E,C_{2x}\right\} ,\label{eq:group_Dnh}
\end{align}
where $I$ denotes inversion and $C_{2x}$, twofold rotation with
respect to the in-plane $x$-axis. Note that the compositions $IC_{2z}$
and $IC_{2x}$ correspond to reflections with respect to the horizontal
and vertical mirrors (denoted $\sigma_{h}$ and $\sigma_{v}$), whereas
$IC_{2y}C_{4z}=S_{4z}C_{2x}$ corresponds to a reflection with respect
to the diagonal mirror (denoted $\sigma_{d}$).

Upon imposing the crystallographic restriction theorem, one finds
$27$ crystallographic axial point groups corresponding to the following
six crystal systems: hexagonal, trigonal, tetragonal, orthorhombic,
monoclinic, and triclinic. Together with the $5$ crystallographic
polyhedral point groups discussed in Sec.~\ref{sec:Nematicity-in-Polyhedral}
that form the cubic crystal system, one finds $32$ crystallographic
point groups. While this section focuses on the $27$ crystallographic
axial groups, we will discuss non-crystallographic axial point groups
that are relevant to quasicrystalline materials \citep{Rokhsar1987,Socolar1989,Goldman1991,Rabson1991,Lifshitz1996}
(octagonal, decagonal, and dodecagonal) and to twisted quasicrystals
\citep{Haenel2022,Yang2023_a,Yang2023_b,Liu2024nematic} in Sec.~\ref{sec:Quasicrystal}. 

\subsection{2D isotropic systems\label{subsec:Dihedral-group_Dih}}

To set the stage for the analysis of electronic nematicity in axial
groups, it is instructive to analyze the case in which the system
has full in-plane rotational symmetry. To describe such a system,
one possibility would be to consider the continuous $\mathsf{SO(2)}$
group, which describes 2D rotations and is equivalent to $\mathsf{C_{\infty}}$
defined in Eq.~(\ref{eq:group_Cn}). However, for our purposes, it
is more convenient to consider the continuous dihedral group $\mathsf{D_{\infty h}}$,
since all other axial groups are subgroups of $\mathsf{D_{\infty h}}$.
Formally, the elements of this group include continuous proper and
improper 2D rotations as well as a horizontal mirror, i.e. $\mathsf{D_{\infty h}=SO(2)}\times\{E,I\}\times\{E,C_{2x}\}$.
Hereafter, we will simply refer to such a system as a 2D isotropic
system.

In the 2D isotropic case, the five-dimensional nematic order parameter
$\boldsymbol{d}$ (\ref{eq:d_vector}) decomposes into three separate
channels,
\begin{align}
\boldsymbol{d} & \quad\overset{\mathrm{in\text{-}plane}\,\,\mathrm{isotropic}}{\longrightarrow}\quad\Big(\,\underline{d_{1}},\,\boldsymbol{d}^{\mathrm{ip}},\,\boldsymbol{d}^{\mathrm{op}}\,\Big)^{T}.\label{eq:d_Dih}
\end{align}
The first component $d_{1}=d_{\frac{1}{\sqrt{3}}(2z^{2}-x^{2}-y^{2})}$
transforms as the trivial IR $A_{1g}$ of $\mathsf{D_{\infty h}}$,
which we indicate by the underline in the expression above. The other
components $\boldsymbol{d}^{\mathrm{ip}}$ and $\boldsymbol{d}^{\mathrm{op}}$
are two-component ``vectors'' (i.e. doublets) that transform as
different two-dimensional IRs of $\mathsf{D_{\infty h}}$, $E_{2g}$
and $E_{1g}$, respectively. They are given by:

\begin{align}
\boldsymbol{d}^{\mathrm{ip}} & =\left(\begin{array}{c}
d_{2}\\
d_{5}
\end{array}\right)=\left(\begin{array}{c}
d_{x^{2}-y^{2}}\\
d_{2xy}
\end{array}\right)=\left|\boldsymbol{d}^{\mathrm{ip}}\right|\left(\begin{array}{c}
\cos\gamma_{\mathrm{ip}}\\
\sin\gamma_{\mathrm{ip}}
\end{array}\right),\label{eq:d_ip_Dih}\\
\boldsymbol{d}^{\mathrm{op}} & =\left(\begin{array}{c}
d_{3}\\
d_{4}
\end{array}\right)=\left(\begin{array}{c}
d_{2yz}\\
d_{2xz}
\end{array}\right)=\left|\boldsymbol{d}^{\mathrm{op}}\right|\left(\begin{array}{c}
\cos\gamma_{\mathrm{op}}\\
\sin\gamma_{\mathrm{op}}
\end{array}\right),\label{eq:d_op_Dih}
\end{align}
where we introduced a polar parameterization in terms of the angles
$\gamma_{\mathrm{ip}}$ and $\gamma_{\mathrm{op}}$. The superscripts
$\mathrm{ip}$ and $\mathrm{op}$ are used to indicate that the condensation
of the order parameter promotes an in-plane or out-of-plane distortion
of the Fermi surface, respectively. Importantly, as explained in Sec.~\ref{sec:3D-Isotropic-nematicity},
we are using the transformation rule in Eq.~(\ref{eq:d_5_trafo})
to define the nematic order parameter for all point groups, which
results in the definitions (\ref{eq:d_ip_Dih})-(\ref{eq:d_op_Dih}).
In the literature, one often uses a point-group-specific basis to
define the nematic order parameter, resulting in definitions such
as $\tilde{\boldsymbol{d}}^{\mathrm{ip}}=\big(d_{x^{2}-y^{2}},\,-d_{2xy}\big)^{T}$
and $\tilde{\boldsymbol{d}}^{\mathrm{op}}=\left(d_{2yz},\,-d_{2xz}\right)^{T}$
\citep{Hecker2022}. Further implications arising from different definitions
are discussed in Sec.~\ref{sec:All-crystals}. Of course, the final
results do not depend on the basis used to express the nematic order
parameter.

We emphasize that, because the nematic component $d_{1}$ transforms
trivially in $\mathsf{D_{\infty h}}$, it will do so for all axial
groups. Therefore, $d_{\frac{1}{\sqrt{3}}(2z^{2}-x^{2}-y^{2})}$ should
not be interpreted as a nematic order parameter, since it is generically
non-zero for any temperature or tuning parameter. Physically, from
Eq.~(\ref{eq:kF_khat}), $d_{1}\neq0$ corresponds to a distortion
of the isotropic Fermi surface along the $k_{z}$-axis, lowering its
symmetry from spherical to cylindrical \textendash{} as expected for
an axial group. As we will argue below, while $d_{1}$ does not impact
the critical properties of the nematic instability, it needs to be
included if one chooses to express the nematic order parameter $\boldsymbol{d}$
in the $(\boldsymbol{n}\boldsymbol{m}\boldsymbol{l})$-representation
of Eq.~(\ref{eq:d_nml}).

We now proceed to derive the Landau expansions for $\boldsymbol{d}^{\mathrm{ip}}$
and $\boldsymbol{d}^{\mathrm{op}}$. Since they transform according
to the IRs $E_{2g}$ and $E_{1g}$ of the infinite dihedral group
$\mathsf{D_{\infty h}}$, respectively, the number of invariants for
each expansion order can be read off from the symmetrized-product
decomposition:
\begin{align}
\big[\otimes_{j=1}^{2N}E_{2g}\big]_{s} & =A_{1g}\oplus_{n=1}^{N}E_{(4n)g},\label{eq:Dih_decomp1}\\
\big[\otimes_{j=1}^{2N-1}E_{2g}\big]_{s} & =\oplus_{n=1}^{N}E_{(4n-2)g},\label{eq:Dih_decomp2}\\
\big[\otimes_{j=1}^{2N}E_{1g}\big]_{s} & =A_{1g}\oplus_{n=1}^{N}E_{(2n)g},\label{eq:Dih_decomp3}\\
\big[\otimes_{j=1}^{2N-1}E_{1g}\big]_{s} & =\oplus_{n=1}^{N}E_{(2n-1)g},\label{eq:Dih_decomp4}
\end{align}
with $N=1,2,3$ and where we used a tensor summation notation, e.g.
$\oplus_{n=1}^{2}E_{(4n)g}=E_{4g}\oplus E_{8g}$. The decompositions
(\ref{eq:Dih_decomp1})-(\ref{eq:Dih_decomp4}) are carried out up
to sixth order in order to emphasize that no $A_{1g}$ invariants
occur other than $\left|\boldsymbol{d}^{\mathrm{ip}}\right|^{2N}$
and $\left|\boldsymbol{d}^{\mathrm{op}}\right|^{2N}$, i.e. the invariants
only occur at even order. Consequently, the Landau expansions for
the in-plane and out-of-plane nematic order parameters become

\begin{align}
\mathcal{S}_{\mathrm{ip}} & =\int_{\mathsf{x}}\Big\{ r_{0}|\boldsymbol{d}^{\mathrm{ip}}|^{2}+u|\boldsymbol{d}^{\mathrm{ip}}|^{4}\Big\}, & \mathcal{S}_{\mathrm{op}} & =\int_{\mathsf{x}}\Big\{ r_{0}|\boldsymbol{d}^{\mathrm{op}}|^{2}+u|\boldsymbol{d}^{\mathrm{op}}|^{4}\Big\}.\label{eq:S_Dih_E2g}
\end{align}
Therefore, since $\boldsymbol{d}^{\mathrm{ip}}$ and $\boldsymbol{d}^{\mathrm{op}}$
are two-component order parameters, their actions have the same Landau
expansion as the well-known XY model. Their condensation leads to
a continuous symmetry breaking, since the nematic angles $\gamma_{\mathrm{ip}},\gamma_{\mathrm{op}}\in[0,2\pi]$
are not constrained by symmetry and can point in any direction. 

It is illustrative to rewrite the fully-isotropic nematic order parameters
(\ref{eq:d_ip_Dih})-(\ref{eq:d_op_Dih}) in the $(\boldsymbol{n}\boldsymbol{m}\boldsymbol{l})$-representation
of Eq.~(\ref{eq:d_nml}). For the description of the in-plane nematic
order parameter $\boldsymbol{d}^{\mathrm{ip}}$, it is convenient
to introduce the three orthonormal vectors in cylindrical coordinates

\begin{align}
\boldsymbol{e}_{\parallel}^{A} & =\left(\begin{smallmatrix}\cos\left(\gamma_{\mathrm{ip}}/2\right)\\
\sin\left(\gamma_{\mathrm{ip}}/2\right)\\
0
\end{smallmatrix}\right), & \boldsymbol{e}_{\parallel}^{B} & =\left(\begin{smallmatrix}-\sin\left(\gamma_{\mathrm{ip}}/2\right)\\
\cos\left(\gamma_{\mathrm{ip}}/2\right)\\
0
\end{smallmatrix}\right), & \boldsymbol{e}_{z} & =\left(\begin{smallmatrix}0\\
0\\
1
\end{smallmatrix}\right).\label{eq:Dih_ip_vectors}
\end{align}
The magnitude of the nematic order parameter is given by $\left|\boldsymbol{d}\right|=\sqrt{\left|\boldsymbol{d}^{\mathrm{ip}}\right|^{2}+\left(d_{1}\right)^{2}}$,
where $d_{1}$, as explained above, is always non-zero, and can thus
be considered as an intrinsic parameter characterizing the 2D isotropic
system even in the absence of symmetry-breaking nematic order. Depending
on the value of the ratio $-1<d_{1}/\left|\boldsymbol{d}\right|<1$,
we find three different regimes of parameters that characterize $\boldsymbol{d}$
in the $(\boldsymbol{n}\boldsymbol{m}\boldsymbol{l})$-representation: 

\begin{align}
\frac{\left|d_{1}\right|}{\left|\boldsymbol{d}\right|} & \!\le\!\frac{1}{2}\,: & \!\!\alpha & =\frac{\pi}{6}+\arcsin\left(\frac{d_{1}}{\left|\boldsymbol{d}\right|}\right), & \!\boldsymbol{n} & =\boldsymbol{e}_{\parallel}^{A},\;\boldsymbol{m}=\boldsymbol{e}_{\parallel}^{B},\;\boldsymbol{l}=\boldsymbol{e}_{z},\nonumber \\
\frac{d_{1}}{\left|\boldsymbol{d}\right|} & \!<\!\frac{\text{-}1}{2}\!: & \!\!\alpha & =\arcsin\left(\frac{\left|d_{1}\right|}{\left|\boldsymbol{d}\right|}\right)-\frac{\pi}{6}, & \!\boldsymbol{n} & =\boldsymbol{e}_{\parallel}^{A},\;\boldsymbol{m}=\boldsymbol{e}_{z},\;\boldsymbol{l}=\boldsymbol{e}_{\parallel}^{B},\nonumber \\
\frac{d_{1}}{\left|\boldsymbol{d}\right|} & \!>\!\frac{1}{2}\,: & \!\!\alpha & =\frac{\pi}{2}-\arcsin\left(\frac{\left|d_{1}\right|}{\left|\boldsymbol{d}\right|}\right), & \!\boldsymbol{n} & =\boldsymbol{e}_{z},\;\boldsymbol{m}=\boldsymbol{e}_{\parallel}^{B},\;\boldsymbol{l}=\boldsymbol{e}_{\parallel}^{A}.\label{eq:D_ih_ip_gs_nml}
\end{align}
As for the out-of-plane nematic order parameter, we have $\left|\boldsymbol{d}\right|=\sqrt{\left|\boldsymbol{d}^{\mathrm{op}}\right|^{2}+\left(d_{1}\right)^{2}}$
and: 
\begin{align}
\alpha & =\frac{\pi}{6}-\arcsin\left[d_{1}/\left(2\left|\boldsymbol{d}\right|\right)\right],\nonumber \\
\boldsymbol{n} & =\left(\begin{smallmatrix}\sin\gamma_{\mathrm{op}}\cos\eta_{\mathrm{op}}\\
\cos\gamma_{\mathrm{op}}\cos\eta_{\mathrm{op}}\\
\sin\eta_{\mathrm{op}}
\end{smallmatrix}\right),\;\;\boldsymbol{m}=\left(\begin{smallmatrix}\sin\gamma_{\mathrm{op}}\sin\eta_{\mathrm{op}}\\
\cos\gamma_{\mathrm{op}}\sin\eta_{\mathrm{op}}\\
-\cos\eta_{\mathrm{op}}
\end{smallmatrix}\right),\;\;\boldsymbol{l}=\left(\begin{smallmatrix}\cos\gamma_{\mathrm{op}}\\
-\sin\gamma_{\mathrm{op}}\\
0
\end{smallmatrix}\right),\label{eq:D_ih_op_gs_nml}
\end{align}
where we defined the polar angle:
\begin{align}
\eta_{\mathrm{op}} & =-\frac{1}{2}\mathrm{sign}\left(d_{1}\right)\arcsin\Bigg(\left|\boldsymbol{d}^{\mathrm{op}}\right|\Big/\sqrt{\left|\boldsymbol{d}^{\mathrm{op}}\right|^{2}+\frac{3}{4}\left(d_{1}\right)^{2}}\;\Bigg)\nonumber \\
 & +\frac{\pi}{2}\frac{1+\mathrm{sign}\left(d_{1}\right)}{2}.\label{eq:eta_op}
\end{align}
Both the in-plane (\ref{eq:D_ih_ip_gs_nml}) and the out-of-plane
(\ref{eq:D_ih_op_gs_nml}) nematic states are generically biaxial,
since $\alpha\neq0,\,\pi/3$. Note that, if the contribution from
the trivial nematic component is negligible, $\left|d_{1}\right|\ll\left|\boldsymbol{d}^{\mathrm{ip}}\right|,\left|\boldsymbol{d}^{\mathrm{op}}\right|$,
both states are ``maximally'' biaxial with $\alpha=\pi/6$ {[}cf.
Fig.~\ref{fig:cosines}(a){]}. In this regard, a non-zero $d_{1}$
brings the states (\ref{eq:D_ih_ip_gs_nml}) and (\ref{eq:D_ih_op_gs_nml})
closer to the uniaxial regime. Indeed, in the limit $\left|d_{1}\right|\gg\left|\boldsymbol{d}^{\mathrm{ip}}\right|,\left|\boldsymbol{d}^{\mathrm{op}}\right|$,
$\alpha\rightarrow0,\,\pi/3$ in both cases, with the nematic director
$\boldsymbol{n}$ or $\boldsymbol{m}$ pointing along the $k_{z}$-axis.
Interestingly, the nematic in-plane order parameter (\ref{eq:D_ih_ip_gs_nml})
also establishes a uniaxial state when $\left|d_{1}\right|=\left|\boldsymbol{d}^{\mathrm{ip}}\right|/\sqrt{3}$,
corresponding to $\left|d_{1}\right|\big/\left|\boldsymbol{d}\right|=1/2$.
In this fine-tuned parameter regime, there is a swap between the nematic
axes that point along the $k_{z}$-axis, since for $\left|d_{1}\right|>\left|\boldsymbol{d}^{\mathrm{ip}}\right|/\sqrt{3}$
either the long or the short nematic axis ($\boldsymbol{m}$ or $\boldsymbol{n}$)
points out-of-plane, while for $\left|d_{1}\right|<\left|\boldsymbol{d}^{\mathrm{ip}}\right|/\sqrt{3}$
the long and short nematic axes both lie within the plane. Since $\boldsymbol{d}^{\mathrm{ip}}$
and $\boldsymbol{d}^{\mathrm{op}}$ also emerge in the description
of hexagonal and tetragonal lattices, we will defer plotting the Fermi
surface distortion patterns triggered by the condensation of these
nematic order parameters to the subsequent subsections.

An alternative representation of the 2D isotropic in-plane nematic
order parameter $\boldsymbol{d}^{\mathrm{ip}}$ has also been widely
employed in the literature \citep{Fradkin1999,Oganesyan2001,HYKee2003,Metzner2006,Zacharias2009,Maslov2010,Maciejko2013,Fradkin_FQHE},
building on the analogy with the classical \emph{two-dimensional }tensorial
nematic order parameter $Q_{\mu\mu^{\prime}}=\left\langle a_{\mu}a_{\mu^{\prime}}-\frac{1}{2}\delta_{\mu\mu^{\prime}}\boldsymbol{a}^{2}\right\rangle $,
with director $\boldsymbol{a}=\left(a_{x},\,a_{y}\right)$. In terms
of the quadrupolar order parameters, the electronic nematic order
parameter is given by:

\begin{equation}
Q=\left(\begin{array}{cc}
d_{x^{2}-y^{2}} & d_{2xy}\\
d_{2xy} & -d_{x^{2}-y^{2}}
\end{array}\right)=\boldsymbol{d}^{\mathrm{ip}}\cdot\boldsymbol{\tau}^{\mathrm{ip}},\label{eq:aux_Q_2D}
\end{equation}
where $\boldsymbol{\tau}^{\mathrm{ip}}=(\tau^{3},\tau^{1})$ with
Pauli matrices $\tau^{i}$. This form establishes a straightforward
connection between the two-component ``vector'' $\boldsymbol{d}^{\mathrm{ip}}=\left(d_{2},\,d_{5}\right)^{T}=\big(d_{x^{2}-y^{2}},\,d_{2xy}\big)^{T}$
and $Q$ similar to the 3D case in Eq.~(\ref{eq:Q_d_relation}),
with the symmetric Pauli matrices replacing the symmetric Gell-Mann
matrices. To preserve rotational invariance, the Landau expansion
can only depend on traces of powers of $Q$:

\begin{equation}
\mathcal{S}_{\mathrm{ip}}=\int_{\mathsf{x}}\sum_{n}\frac{\kappa_{n}}{2}\,\mathrm{tr}\left(Q^{n}\right),
\end{equation}
where $\kappa_{n}$ are Landau coefficients. Using the facts that
$\left(\boldsymbol{d}^{\mathrm{ip}}\cdot\boldsymbol{\tau}^{\mathrm{ip}}\right)^{2n}=\left|\boldsymbol{d}^{\mathrm{ip}}\right|^{2n}\tau^{0}$,
and thus $\mathrm{tr}\left[(\boldsymbol{d}^{\mathrm{ip}}\cdot\boldsymbol{\tau}^{\mathrm{ip}})^{2n+1}\right]=\left|\boldsymbol{d}^{\mathrm{ip}}\right|^{2n}\mathrm{tr}\left[\boldsymbol{d}^{\mathrm{ip}}\cdot\boldsymbol{\tau}^{\mathrm{ip}}\right]=0$,
we find:

\begin{align}
\mathcal{S}_{\mathrm{ip}} & =\int_{\mathsf{x}}\sum_{n}\kappa_{2n}\,|\boldsymbol{d}^{\mathrm{ip}}|^{2n}=\int_{\mathsf{x}}\Big\{ r_{0}|\boldsymbol{d}^{\mathrm{ip}}|^{2}+u|\boldsymbol{d}^{\mathrm{ip}}|^{4}+\mathcal{O}\left(|\boldsymbol{d}^{\mathrm{ip}}|^{6}\right)\Big\},
\end{align}
which is identical to Eq.~(\ref{eq:S_Dih_E2g}). When comparing to
the Landau expansion of the 3D isotropic case, Eq.~(\ref{eq:action_SO3-1}),
the main difference is the absence of the cubic term. Ultimately,
this can be traced back to the fundamental differences in the $\mathrm{SU(2)}$
and $\mathrm{SU(3)}$ Lie algebras, reflected in the fact that different
Pauli matrices necessarily anti-commute whereas different Gell-Mann
matrices do not necessarily anti-commute.

As discussed above in the context of Eq.~(\ref{eq:S_Dih_E2g}), electronic
nematicity in the 2D isotropic system (either in-plane or out-of-plane)
belongs to the XY universality class, implying the existence of a
Goldstone mode in the nematically ordered state. As discussed in Refs.~\citep{Oganesyan2001,Watanabe2014},
the coupling between this Goldstone mode and low-energy electronic
degrees of freedom can promote interesting phenomena, such as non-Fermi-liquid
behavior. As we will see below, the main effect of the explicit breaking
of the full rotational symmetry in a lattice described by an axial
point group is the emergence of an anisotropic term in the Landau
expansion:

\begin{equation}
\mathcal{S}_{\mathrm{axial}}=\int_{\mathsf{x}}\Big\{ r_{0}|\boldsymbol{d}^{\mathrm{\alpha}}|^{2}+u|\boldsymbol{d}^{\mathrm{\alpha}}|^{4}+h_{q}|\boldsymbol{d}^{\mathrm{\alpha}}|^{q}\cos\left(q\gamma_{\alpha}\right)\Big\},\label{eq:S_axial_clock}
\end{equation}
where $q$ is a positive integer that is the same within the same
crystal system (hexagonal, trigonal, tetragonal, orthorhombic, monoclinic,
and triclinic), $\alpha$ can refer to $\mathrm{ip}$ or $\mathrm{op}$,
and $h_{q}$ is a Landau coefficient. For the (crystallographic and
non-crystallographic) axial groups that we investigated here, we found
the following general result: if the group has $(2n)$-fold symmetry,
then $q=n$ for $\alpha=\mathrm{ip}$ and $q=2n$ for $\alpha=\mathrm{op}$.
On the other hand, if the axial group has $(2n+1)$-fold symmetry,
$q=2n+1$ for both $\alpha=\mathrm{ip}$ and $\alpha=\mathrm{op}$. 

We note that Eq.~(\ref{eq:S_axial_clock}) corresponds to the Landau
expansion of the $q$-state ``soft'' clock model ($Z_{q}$-clock),
whose classical critical properties are well understood \citep{Clock_Jose}.
In three dimensions (and non-zero temperatures), which is the case
of most relevance here, the character of the transition is (of course,
$q=1$ implies explicitly broken symmetry): 3D-Ising for $q=2$, first-order
for $q=3$, and 3D-XY for $q\geq4$. The latter result is a consequence
of the fact that $h_{q}$ is a dangerously irrelevant perturbation
(in the renormalization-group sense) for $q\geq4$, i.e. it is irrelevant
at the nematic critical point but relevant inside the nematically
ordered state \citep{Clock_Oshikawa,Clock_Carmona,Clock_Sandvik}.
Thus, even though the transition is XY-like, the ordered state does
not display a Goldstone mode, but a gapped pseudo-Goldstone mode. 

It is important to point out that crystals described by an axial point
group can be very anisotropic and display behavior intermediate between
2D and 3D. Moreover, as we will discuss in Sec.~\ref{subsec:twisted-quasicrystals},
twisted quasicrystals are 2D systems. In this regard, it is interesting
to note that the $Z_{q}$-clock model has rather unique properties
in 2D \citep{Clock_Jose}. While the $q=2$ model undergoes a standard
2D-Ising transition, the $q=3$ model undergoes a second-order transition
despite the existence of a cubic term in the Landau expansion. More
surprisingly, for $q\geq5$, the 2D $Z_{q}$-clock model undergoes
two Berezinskii-Kosterlitz-Thouless (BKT) transitions: a higher-temperature
BKT transition towards a critical phase with quasi-long-range order,
analogous to that displayed by the XY model at non-zero temperatures,
and a lower-temperature BKT transition towards a state with long-range
order where the discrete clock symmetry is broken. Interestingly,
for $q=4$, these two BKT transitions merge into a single line. As
a result, the critical behavior of the 2D $Z_{4}$-clock model, which
maps onto the 2D Ashkin-Teller model, is only weakly-universal \citep{Suzuki1974},
since while the anomalous critical exponent is fixed to $\eta=1/4$,
the other exponents depend on the value of $h_{4}$ \citep{Jin2013}.
The crossover from 2D to 3D $Z_{q}$-clock behavior has been recently
discussed numerically in Ref.~\citep{Clock_Sandvik} and also indirectly
in the context of quantum critical points in 2D $Z_{q}$-clock systems
\citep{Podolsky2016,Xu2020,Arnold2022,Mandal2023,Gali2024}. Finally,
we emphasize that the critical properties of the $Z_{q}$-clock model
are generically different from those of the $Z_{q}$-Potts model,
except for the special case $q=3$, where they share the same universality
class \citep{Potts_Wu}. For this reason, as explained in the previous
section, we use the terms $Z_{3}$-clock and $Z_{3}$-Potts interchangeably. 

Finally, we will show in the Appendices~\ref{sec:hexagonal_Appendix}-\ref{sec:tetragonal_Appendix}
that, in the cases where the nematic order parameter transforms as
a complex IR, which reflects the lack of twofold in-plane rotation
axes in the crystal, the clock term in the action (\ref{eq:S_axial_clock})
acquires an offset angle, which can be traced back to the existence
of another anisotropic term in the Landau expansion, $\tilde{h}_{q}|\boldsymbol{d}^{\mathrm{\alpha}}|^{q}\sin\left(q\gamma_{\alpha}\right)$,
(see also Ref.~\citep{Li2021}):

\begin{equation}
\mathcal{S}_{\mathrm{axial}}=\int_{\mathsf{x}}\Big\{ r_{0}|\boldsymbol{d}^{\mathrm{\alpha}}|^{2}+u|\boldsymbol{d}^{\mathrm{\alpha}}|^{4}+h_{q}|\boldsymbol{d}^{\mathrm{\alpha}}|^{q}\cos\left(q\gamma_{\alpha}-\delta_{0}\right)\Big\}.\label{eq:S_axial_clock_mod}
\end{equation}
To distinguish it from Eq.~(\ref{eq:S_axial_clock_mod}), we denote
the Landau expansion above a $Z_{q}^{*}$-clock model. An offset angle
is also found when the two nematic doublets are ``degenerate,''
i.e., transform as the same IR, as is the case in trigonal point groups.

\subsection{Hexagonal crystals\label{subsec:Nematicity_hexagonal}}

The hexagonal crystal system consists of seven crystallographic axial
point groups: $\mathsf{D_{6h}}$, $\mathsf{D_{6}}$, $\mathsf{D_{3h}}$,
$\mathsf{C_{6v}}$, $\mathsf{C_{6h}}$, $\mathsf{C_{3h}}$, and $\mathsf{C_{6}}$.
In all cases, the nematic order parameter $\boldsymbol{d}$ in Eq.~(\ref{eq:d_vector})
decomposes into the three channels in the same way as in the 2D isotropic
case, 
\begin{align}
\boldsymbol{d} & \quad\overset{\mathrm{hexagonal}}{\longrightarrow}\quad\Big(\,\underline{d_{1}},\,\boldsymbol{d}^{\mathrm{ip}},\,\boldsymbol{d}^{\mathrm{op}}\,\Big)^{T}.\label{eq:d_hexagonal}
\end{align}
We recall that an underline means that the corresponding nematic component
transforms as the trivial IR of the point group. The most notable
difference with respect to the isotropic case is, of course, the limited
number of $C_{nz}$ or $IC_{nz}$ symmetry elements, which reduces
the continuous XY-degeneracy of the nematic angles $\gamma_{\mathrm{ip}}$
and $\gamma_{\mathrm{op}}$ down to either threefold or sixfold degeneracies,
as we demonstrate below. Within the hexagonal crystal system, one
needs to distinguish between the sets of point groups \{$\mathsf{D_{6h}}$,
$\mathsf{D_{6}}$, $\mathsf{D_{3h}}$, $\mathsf{C_{6v}}$\} and \{$\mathsf{C_{6h}}$,
$\mathsf{C_{3h}}$, $\mathsf{C_{6}}$\}, since only the former possess
the characteristic $6$ in-plane symmetry axes 
\begin{align}
\mathcal{V}_{1}^{\mathrm{hex}} & =\left\{ \left(\begin{smallmatrix}1\\
0\\
0
\end{smallmatrix}\right),\left(\begin{smallmatrix}\text{-}\frac{1}{2}\\
\frac{\sqrt{3}}{2}\\
0
\end{smallmatrix}\right),\left(\begin{smallmatrix}\text{-}\frac{1}{2}\\
\text{-}\frac{\sqrt{3}}{2}\\
0
\end{smallmatrix}\right)\right\} , & \mathcal{V}_{2}^{\mathrm{hex}} & =\left\{ \left(\begin{smallmatrix}0\\
1\\
0
\end{smallmatrix}\right),\left(\begin{smallmatrix}\text{-}\frac{\sqrt{3}}{2}\\
\text{-}\frac{1}{2}\\
0
\end{smallmatrix}\right),\left(\begin{smallmatrix}\frac{\sqrt{3}}{2}\\
\text{-}\frac{1}{2}\\
0
\end{smallmatrix}\right)\right\} ,\label{eq:V_hex}
\end{align}
see Eqs.~(\ref{eq:group_Cn}),(\ref{eq:group_Cnh}). Mathematically,
the absence of in-plane symmetry axes is manifested in the nematic
doublets $\boldsymbol{d}^{\mathrm{ip}}$ and $\boldsymbol{d}^{\mathrm{op}}$
transforming according to complex IRs, which causes additional Landau
invariants to emerge when compared to the generic $Z_{q}$-clock Landau
expansion of Eq.~(\ref{eq:S_axial_clock}). As we show in detail
in Appendix~\ref{sec:hexagonal_Appendix}, this additional Landau
invariant for the groups \{$\mathsf{C_{6h}}$, $\mathsf{C_{3h}}$,
$\mathsf{C_{6}}$\} can be recast as an offset angle in the clock
term, see Eq.~(\ref{eq:S_axial_clock_mod}). In this section, we
focus instead on the first set of groups, \{$\mathsf{D_{6h}}$, $\mathsf{D_{6}}$,
$\mathsf{D_{3h}}$, $\mathsf{C_{6v}}$\}, for which the nematic doublets
$\boldsymbol{d}^{\mathrm{ip}}$ and $\boldsymbol{d}^{\mathrm{op}}$
transform as real two-dimensional IRs. For concreteness, we show the
derivation for the point group $\mathsf{D_{6h}}$, where $\boldsymbol{d}^{\mathrm{ip}}$
and $\boldsymbol{d}^{\mathrm{op}}$ transform according to the IRs
$E_{2g}$ and $E_{1g}$, respectively. The results equally apply to
the other groups of the set, \{$\mathsf{D_{6}}$, $\mathsf{D_{3h}}$,
$\mathsf{C_{6v}}$\}.

Consider first the in-plane nematic order parameter $\boldsymbol{d}^{\mathrm{ip}}=\left(d_{2},\,d_{5}\right)^{T}=\big(d_{x^{2}-y^{2}},\,d_{2xy}\big)^{T}$.
The symmetrized product decomposition for each expansion order is
given by 
\begin{align}
\big[\otimes_{j=1}^{2}E_{2g}\big]_{s} & =A_{1g}\oplus E_{2g},\label{eq:E2g_bil_hex}\\
\big[\otimes_{j=1}^{3}E_{2g}\big]_{s} & =A_{1g}\oplus A_{2g}\oplus E_{2g},\label{eq:hex_E2g_decomp3}\\
\big[\otimes_{j=1}^{4}E_{2g}\big]_{s} & =A_{1g}\!\oplus2E_{2g}.\label{eq:hex_E2g_decomp4}
\end{align}
To construct the three $A_{1g}$ Landau invariants, we derive the
bilinear combinations associated with Eq.~(\ref{eq:E2g_bil_hex}):
\begin{align}
D_{\mathrm{ip}}^{A_{1g}} & =\left|\boldsymbol{d}^{\mathrm{ip}}\right|^{2}, & \boldsymbol{D}_{\mathrm{ip}}^{E_{2g}} & =\left|\boldsymbol{d}^{\mathrm{ip}}\right|^{2}\left(\cos\left(2\gamma_{\mathrm{ip}}\right),\text{-}\sin\left(2\gamma_{\mathrm{ip}}\right)\right)^{T}.\label{eq:hex_E2g_bilinears}
\end{align}
Then, the three invariants become $\left|\boldsymbol{d}^{\mathrm{ip}}\right|^{2}$,
$\left|\boldsymbol{d}^{\mathrm{ip}}\right|^{4}$, and $\boldsymbol{d}^{\mathrm{ip}}\cdot\boldsymbol{D}_{\mathrm{ip}}^{E_{2g}}$,
since both $\boldsymbol{d}^{\mathrm{ip}}$ and $\boldsymbol{D}_{\mathrm{ip}}^{E_{2g}}$
transform as $E_{2g}$. The corresponding Landau expansion, 
\begin{align}
\mathcal{S}_{\mathrm{ip}} & =\int_{\mathsf{x}}\Big\{ r_{0}\left|\boldsymbol{d}^{\mathrm{ip}}\right|^{2}+g_{\mathrm{ip}}\left|\boldsymbol{d}^{\mathrm{ip}}\right|^{3}\cos\left(3\gamma_{\mathrm{ip}}\right)+u\left|\boldsymbol{d}^{\mathrm{ip}}\right|^{4}\Big\},\label{eq:S_hex_ip}
\end{align}
has the same form as that of the $Z_{3}$-clock model, Eq.~(\ref{eq:S_axial_clock}),
as derived elsewhere \citep{Hecker2018,Fernandes2020,Xu2020,Kimura2022}.
Correspondingly, the threefold-degenerate mean-field ground state
is given by 
\begin{align}
\gamma_{\mathrm{ip}}^{0} & =\frac{\pi}{3}\left(\frac{1+\mathrm{sign}\,g_{\mathrm{ip}}}{2}\right)+\frac{2\pi}{3}n, & n & =\left\{ 0,1,2\right\} .\label{eq:angles_ip_Z3}
\end{align}

\begin{figure}
\begin{centering}
\includegraphics[width=1\columnwidth]{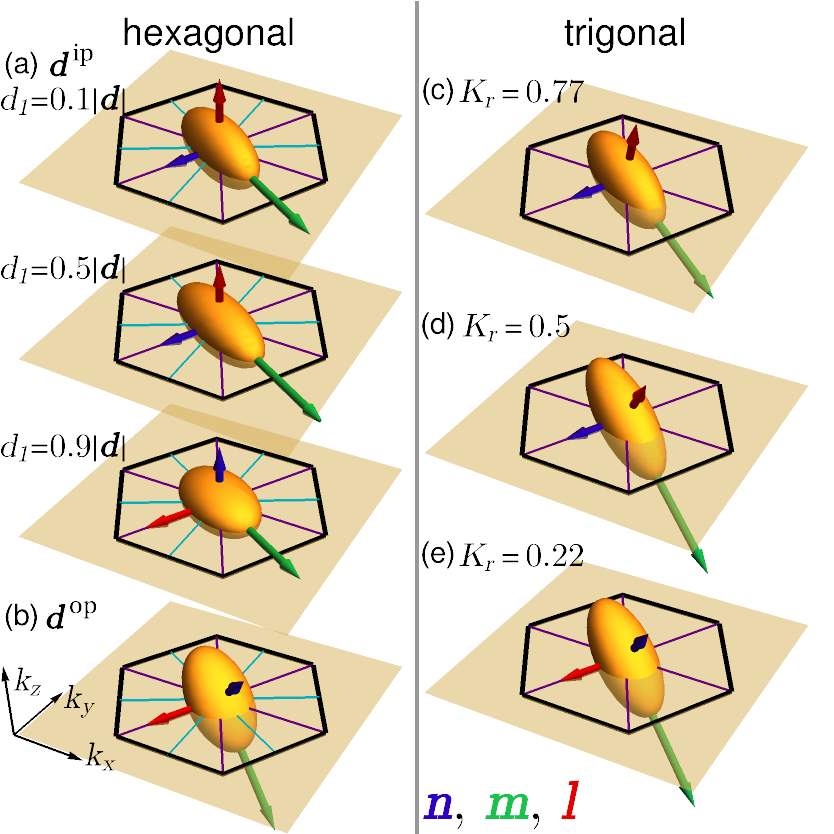}
\par\end{centering}
\caption{Distorted Fermi surface in the nematic phase, as given by Eq.~(\ref{eq:kF_khat}),
plotted together with the nematic axes $\boldsymbol{n}$,$\boldsymbol{m}$,$\boldsymbol{l}$
for the cases of: (a) hexagonal in-plane doublet state (\ref{eq:angles_ip_Z3})
for three values of $d_{1}/\left|\boldsymbol{d}\right|$, and (b)
out-of-plane doublet state (\ref{eq:angles_op_Z6}); (c)-(e) trigonal
nematic state (\ref{eq:sol_tri}) for three values of $K_{r}$ (\ref{eq:Kr_tri}).
The unit vectors $\left\{ \boldsymbol{n},\boldsymbol{m},\boldsymbol{l}\right\} $
are rescaled by $2.5\,k_{F}\left(\left\{ \boldsymbol{n},\boldsymbol{m},\boldsymbol{l}\right\} \right)$,
respectively. Unless specified otherwise, we chose $d_{1}/\left|\boldsymbol{d}\right|=0.1$.
\label{fig:FS_hex_trig}}
\end{figure}

In the $(\boldsymbol{n}\boldsymbol{m}\boldsymbol{l})$-representation
(\ref{eq:d_nml}), this ground state is parametrized according to
Eq.~(\ref{eq:D_ih_ip_gs_nml}); recall that $d_{1}$ is the symmetry-conforming
nematic component, which is non-zero at any temperature or tuning
parameter range. We plot the corresponding Fermi surface distortion
in Fig.~\ref{fig:FS_hex_trig}(a) by using the general expression
(\ref{eq:kF_khat}). In this figure, we chose $g_{\mathrm{ip}}<0$
and three values of the trivial component $d_{1}/\left|\boldsymbol{d}\right|=\left\{ 0.1,\,0.5,\,0.9\right\} $.
In the top panel where $\left|d_{1}\right|<\left|\boldsymbol{d}^{\mathrm{ip}}\right|/\sqrt{3}$,
the short and long nematic axes $\boldsymbol{n}$ and $\boldsymbol{m}$
align, respectively, with the in-plane symmetry axes $\mathcal{V}_{1}^{\mathrm{hex}}$
and $\mathcal{V}_{2}^{\mathrm{hex}}$ of Eq.~(\ref{eq:V_hex}), highlighted
in purple and light-blue in the figure. The three degenerate states
correspond to $\boldsymbol{m}$ being aligned with one of the three
light-blue axes, corresponding to $\mathcal{V}_{2}^{\mathrm{hex}}$
in Eq.~(\ref{eq:V_hex}). For the opposite sign $g_{\mathrm{ip}}>0$,
the long nematic axis $\boldsymbol{m}$ aligns with the purple axes,
which correspond to $\mathcal{V}_{1}^{\mathrm{hex}}$ in Eq.~(\ref{eq:V_hex}).
The biaxial nature of the nematic order parameter is clear from the
shape of the distorted Fermi surface. Upon increasing $\left|d_{1}\right|/\left|\boldsymbol{d}\right|$,
we demonstrate in the second panel of Fig.~\ref{fig:FS_hex_trig}(a)
how the nematic state passes through a fine-tuned uniaxial point at
$\left|d_{1}\right|=\left|\boldsymbol{d}\right|/2$, according to
Eq.~(\ref{eq:D_ih_ip_gs_nml}). In the parameter range $\left|d_{1}\right|>\left|\boldsymbol{d}\right|/2$
(third panel), two nematic axes are interchanged, and the nematic
state becomes gradually more in-plane isotropic. 

We now discuss the case of the out-of-plane nematic doublet $\boldsymbol{d}^{\mathrm{op}}=\left(d_{3},\,d_{4}\right)^{T}=\left(d_{2yz},\,d_{2xz}\right)^{T}$,
which transforms as the $E_{1g}$ IR of $\mathsf{D_{6h}}$. The decomposition
of the symmetrized product gives:
\begin{align}
\big[\otimes_{j=1}^{2}E_{1g}\big]_{s} & =A_{1g}\oplus E_{2g},\label{eq:hex_decomp_op_2}\\
\big[\otimes_{j=1}^{3}E_{1g}\big]_{s} & =B_{1g}\oplus B_{2g}\oplus E_{1g},\label{eq:hex_decomp_op_3}\\
\big[\otimes_{j=1}^{4}E_{1g}\big]_{s} & =A_{1g}\!\oplus2E_{2g},\label{eq:hex_decomp_op_4}\\
\big[\otimes_{j=1}^{5}E_{1g}\big]_{s} & =B_{1g}\!\oplus B_{2g}\oplus2E_{1g},\label{eq:hex_decomp_op_5}\\
\big[\otimes_{j=1}^{6}E_{1g}\big]_{s} & =2A_{1g}\!\oplus A_{2g}\oplus2E_{2g}.\label{eq:hex_decomp_op_6}
\end{align}
Here, we extended the expansion to sixth-order to include the leading
anisotropic invariant, i.e. the leading-order invariant that depends
on the angle $\gamma_{\mathrm{op}}$. Using the bilinears associated
with Eq.~(\ref{eq:hex_decomp_op_2}),
\begin{align}
D_{\mathrm{op}}^{A_{1g}} & =\left|\boldsymbol{d}^{\mathrm{op}}\right|^{2}, & \boldsymbol{D}_{\mathrm{op}}^{E_{2g}} & =\left|\boldsymbol{d}^{\mathrm{op}}\right|^{2}\left(\cos\left(2\gamma_{\mathrm{op}}\right),-\sin\left(2\gamma_{\mathrm{op}}\right)\right)^{T},\label{eq:hex_op_bilinears}
\end{align}
the four invariants can be constructed as $\left|\boldsymbol{d}^{\mathrm{op}}\right|^{2}$,
$\left|\boldsymbol{d}^{\mathrm{op}}\right|^{4}$, $\left|\boldsymbol{d}^{\mathrm{op}}\right|^{6}$,
$(\boldsymbol{d}^{\mathrm{op}}\cdot\boldsymbol{D}_{\mathrm{op}}^{E_{2g}})^{2}$,
where we used the fact that $\boldsymbol{d}^{\mathrm{op}}\cdot\boldsymbol{D}_{\mathrm{op}}^{E_{2g}}$
transforms as $B_{1g}$. The resulting nematic Landau expansion can
be written as
\begin{align}
\mathcal{S}_{\mathrm{op}} & =\int_{\mathsf{x}}\Big\{ r_{0}\left|\boldsymbol{d}^{\mathrm{op}}\right|^{2}+u\left|\boldsymbol{d}^{\mathrm{op}}\right|^{4}+u_{6}\left|\boldsymbol{d}^{\mathrm{op}}\right|^{6}+v_{6}\left|\boldsymbol{d}^{\mathrm{op}}\right|^{6}\cos\left(6\gamma_{\mathrm{op}}\right)\Big\},\label{eq:S_hex_op}
\end{align}
and corresponds to the $Z_{6}$-clock model of Eq.~(\ref{eq:S_axial_clock}).
Thus, we find an important difference between the in-plane $\boldsymbol{d}^{\mathrm{ip}}$
and out-of-plane $\boldsymbol{d}^{\mathrm{op}}$ nematic order parameters
in the hexagonal lattice: whereas the former undergoes a first-order
transition, the latter is expected to undergo a continuous 3D-XY transition,
at which the discrete nature of the nematic angle is irrelevant (in
the renormalization group sense).

The mean-field ground state of $\boldsymbol{d}^{\mathrm{op}}$ is
therefore sixfold degenerate and characterized by the nematic angles
\begin{align}
\gamma_{\mathrm{op}}^{0} & =\frac{\pi}{6}\left(\frac{1+\mathrm{sign}\,v_{6}}{2}\right)+\frac{2\pi}{6}n, & n & \in\{0,1,\dots,5\}.\label{eq:angles_op_Z6}
\end{align}
The distorted Fermi surface associated with this ground state is shown
in Fig.~\ref{fig:FS_hex_trig}(b). In terms of the $(\boldsymbol{n}\boldsymbol{m}\boldsymbol{l})$-representation
derived in Eq.~(\ref{eq:D_ih_op_gs_nml}), we see that, regardless
of the value of the trivial nematic component $d_{1}$, the nematic
axis $\boldsymbol{l}$ is always in-plane and aligned with either
$\mathcal{V}_{1}^{\mathrm{hex}}$ (purple) or $\mathcal{V}_{2}^{\mathrm{hex}}$
(light-blue) defined in Eq.~(\ref{eq:V_hex}), depending on the sign
of the Landau coefficient $v_{6}$. The long ($\boldsymbol{m}$) and
short ($\boldsymbol{n}$) nematic axes are rotated out of the plane
by the tilt angle $\eta_{\mathrm{op}}$ (\ref{eq:eta_op}), which
is exactly $\pm45^{\circ}$ in the limiting case $d_{1}=0$. To visualize
the sixfold degeneracy of this state, it is useful to invoke the symmetry
elements $C_{2z}$ or $IC_{2z}$, one of which is present in every
hexagonal point group. They imply the existence of a non-identical
ground state configuration that emerges from Fig.~\ref{fig:FS_hex_trig}(b)
through a $180^{\circ}$ rotation about the $k_{z}$-axis. The key
point is that, under such a rotation, $\boldsymbol{l}\rightarrow-\boldsymbol{l}$
but $\boldsymbol{m}$, $\boldsymbol{n}$ are not mapped onto $-\boldsymbol{m}$,
$-\boldsymbol{n}$. This implies that, for each of the three choices
for the $\boldsymbol{l}$ axis in $\mathcal{V}_{1}^{\mathrm{hex}}$
or $\mathcal{V}_{2}^{\mathrm{hex}}$, there are two different sets
of possible $\boldsymbol{m}$, $\boldsymbol{n}$ values, resulting
in a sixfold degeneracy. In contrast, such a sixfold degeneracy is
absent for the in-plane state $\boldsymbol{d}^{\mathrm{ip}}$ of Fig.~\ref{fig:FS_hex_trig}(a),
since $\boldsymbol{l}$ remains invariant and $\boldsymbol{m}$, $\boldsymbol{n}$
are mapped onto $-\boldsymbol{m}$, $-\boldsymbol{n}$ under a $C_{2z}$
or $IC_{2z}$ operation.

As we mentioned earlier, the same results derived here for $\mathsf{D_{6h}}$
hold for \{$\mathsf{D_{6}}$, $\mathsf{D_{3h}}$, $\mathsf{C_{6v}}$\}
and \{$\mathsf{C_{6h}}$, $\mathsf{C_{3h}}$, $\mathsf{C_{6}}$\},
as shown in Table~\ref{tab:classification}. The main difference
in the latter set is that an offset angle appears in the anisotropic
terms of Eqs.~(\ref{eq:S_hex_ip}) and (\ref{eq:S_hex_op}), see
Appendix~\ref{sec:hexagonal_Appendix}, which we indicate as $Z_{3}^{*}$-clock
and $Z_{6}^{*}$-clock models in Table~\ref{tab:classification}.

\subsection{Trigonal crystals\label{subsec:Nematicity-trigonal} }

The axial point groups $\mathsf{D_{3}}$, $\mathsf{D_{3d}}$, $\mathsf{C_{3v}}$,
$\mathsf{S_{6}}$ and $\mathsf{C_{3}}$ form the trigonal crystal
system. In what concerns nematicity, its key distinction with respect
to the hexagonal crystal system, Eq.~(\ref{eq:d_hexagonal}), is
that the two nematic doublets, in-plane $\boldsymbol{d}^{\mathrm{ip}}=\left(d_{2},\,d_{5}\right)^{T}=\big(d_{x^{2}-y^{2}},\,d_{2xy}\big)^{T}$
and out-of-plane $\boldsymbol{d}^{\mathrm{op}}=\left(d_{3},\,d_{4}\right)^{T}=\left(d_{2yz},\,d_{2xz}\right)^{T}$,
transform as the same two-dimensional IR. We thus represent the decomposition
of the five-component nematic order parameter $\boldsymbol{d}$ (\ref{eq:d_vector})
as

\begin{align}
\boldsymbol{d} & \quad\overset{\mathrm{trigonal}}{\longrightarrow}\quad\Big(\,\underline{d_{1}},\,\big\{\boldsymbol{d}^{\mathrm{ip}},\,\boldsymbol{d}^{\mathrm{op}}\big\}\,\Big)^{T},\label{eq:d_trigonal}
\end{align}
where the curly brackets are used to indicate degeneracy, i.e. nematic
components that transform as the same IR. Analogously to the hexagonal
case, we must also distinguish the trigonal group sets \{$\mathsf{D_{3}}$,
$\mathsf{D_{3d}}$, $\mathsf{C_{3v}}$\} and \{$\mathsf{S_{6}}$,
$\mathsf{C_{3}}$\}, as the latter lacks the in-plane symmetry directions
encompassed by $\mathcal{V}_{1}^{\mathrm{hex}}$, $\mathcal{V}_{2}^{\mathrm{hex}}$
in Eq.~(\ref{eq:V_hex}). This case, for which $\boldsymbol{d}^{\mathrm{ip}}$
and $\boldsymbol{d}^{\mathrm{op}}$ transform as a complex IR, is
discussed in Appendix~\ref{sec:trigonal_appendix}. Here, as a representative
of the group set \{$\mathsf{D_{3}}$, $\mathsf{D_{3d}}$, $\mathsf{C_{3v}}$\},
we focus our analysis on the point group $\mathsf{C_{3v}}$, for which
$\boldsymbol{d}^{\mathrm{ip}}$ and $\boldsymbol{d}^{\mathrm{op}}$
transform as the IR $E$. Repeating the same steps as in the other
sections, we perform the symmetrized product decomposition:

\begin{align}
\big[\otimes_{j=1}^{2}E\big]_{s} & =A_{1}\oplus E, & \big[\otimes_{j=1}^{3}E\big]_{s} & =A_{1}\oplus A_{2}\oplus E,\label{eq:trig_E3_decomp}\\
\big[\otimes_{j=1}^{4}E\big]_{s} & =A_{1}\oplus2E.\label{eq:trig_E4_decomp}
\end{align}
Clearly, this decomposition is analogous to that of the $E_{2g}$
IR in the hexagonal case, see Eqs.~(\ref{eq:E2g_bil_hex})-(\ref{eq:S_hex_ip}).
This suggests that, individually, each doublet $\boldsymbol{d}^{\mathrm{ip}}$
and $\boldsymbol{d}^{\mathrm{op}}$ would behave as a $Z_{3}$-clock
order parameter. Here, however, we must focus on the combined four-component
vector $\boldsymbol{d}^{E}=\left(\boldsymbol{d}^{\mathrm{ip}},\boldsymbol{d}^{\mathrm{op}}\right)$,
which transforms as $\left(E\oplus E\right)$. Its symmetrized decomposition,
in turn, gives $13$ invariants up to fourth-order in the Landau expansion:

\begin{align}
\big[\otimes_{j=1}^{2}\left(E\oplus E\right)\big]_{s} & =3A_{1}\oplus A_{2}\oplus3E,\label{eq:trig_EE2_decomp}\\
\big[\otimes_{j=1}^{3}\left(E\oplus E\right)\big]_{s} & =4A_{1}\oplus4A_{2}\oplus6E,\label{eq:trig_EE3_decomp}\\
\big[\otimes_{j=1}^{4}\left(E\oplus E\right)\big]_{s} & =6A_{1}\oplus3A_{2}\oplus13E.\label{eq:trig_EE4_decomp}
\end{align}
To systematically construct them, we use the $7$ bilinears that appear
in the decomposition (\ref{eq:trig_EE2_decomp}). These include the
$4$ bilinears that transform as one-dimensional IRs
\begin{align}
D_{\mathrm{ip}}^{A_{1}} & =\left|\boldsymbol{d}^{\mathrm{ip}}\right|^{2}, & D_{\mathrm{io}}^{A_{1}} & =\left|\boldsymbol{d}^{\mathrm{ip}}\right|\left|\boldsymbol{d}^{\mathrm{op}}\right|\cos\left(\gamma_{\mathrm{ip}}-\gamma_{\mathrm{op}}\right),\nonumber \\
D_{\mathrm{op}}^{A_{1}} & =\left|\boldsymbol{d}^{\mathrm{op}}\right|^{2}, & D_{\mathrm{io}}^{A_{2}} & =\left|\boldsymbol{d}^{\mathrm{ip}}\right|\left|\boldsymbol{d}^{\mathrm{op}}\right|\sin\left(\gamma_{\mathrm{ip}}-\gamma_{\mathrm{op}}\right),\label{eq:trig_DA2}
\end{align}
and the $3$ bilinears that transform as two-dimensional IRs
\begin{align}
\boldsymbol{D}_{\mathrm{ip}}^{E} & =\left|\boldsymbol{d}^{\mathrm{ip}}\right|^{2}\left(\begin{smallmatrix}\cos\left(2\gamma_{\mathrm{ip}}\right)\\
-\sin\left(2\gamma_{\mathrm{ip}}\right)
\end{smallmatrix}\right), & \boldsymbol{D}_{\mathrm{io}}^{E} & =\left|\boldsymbol{d}^{\mathrm{ip}}\right|\left|\boldsymbol{d}^{\mathrm{op}}\right|\left(\begin{smallmatrix}\cos\left(\gamma_{\mathrm{ip}}+\gamma_{\mathrm{op}}\right)\\
-\sin\left(\gamma_{\mathrm{ip}}+\gamma_{\mathrm{op}}\right)
\end{smallmatrix}\right),\nonumber \\
\boldsymbol{D}_{\mathrm{op}}^{E} & =\left|\boldsymbol{d}^{\mathrm{op}}\right|^{2}\left(\begin{smallmatrix}\cos\left(2\gamma_{\mathrm{op}}\right)\\
-\sin\left(2\gamma_{\mathrm{op}}\right)
\end{smallmatrix}\right).\label{eq:trig_DE_op}
\end{align}
It is now straightforward to construct the $13$ invariants, each
associated with a Landau coefficient. The three quadratic bilinears
are given by $D_{\mathrm{ip}}^{A_{1}}$, $D_{\mathrm{op}}^{A_{1}}$,
and $D_{\mathrm{io}}^{A_{1}}$, whereas the four cubic invariants
are $\boldsymbol{d}^{\mathrm{ip}}\cdot\boldsymbol{D}_{\mathrm{ip}}^{E}$,
$\boldsymbol{d}^{\mathrm{op}}\cdot\boldsymbol{D}_{\mathrm{op}}^{E}$,
$\boldsymbol{d}^{\mathrm{ip}}\cdot\boldsymbol{D}_{\mathrm{io}}^{E}$,
and $\boldsymbol{d}^{\mathrm{op}}\cdot\boldsymbol{D}_{\mathrm{io}}^{E}$.
As for the quartic invariants, they are given by the squares of the
quadratic invariants $\big(D_{\mathrm{ip}}^{A_{1}}\big)^{2}$ and
$\big(D_{\mathrm{op}}^{A_{1}}\big)^{2}$, the cross terms $D_{\mathrm{ip}}^{A_{1}}D_{\mathrm{op}}^{A_{1}}$,
$D_{\mathrm{ip}}^{A_{1}}D_{\mathrm{io}}^{A_{1}}$ and $D_{\mathrm{op}}^{A_{1}}D_{\mathrm{io}}^{A_{1}}$,
as well as $\big(D_{\mathrm{io}}^{A_{1}}\big)^{2}-\big(D_{\mathrm{io}}^{A_{2}}\big)^{2}$.
Writing the action as $\mathcal{S}=\mathcal{S}_{2}+\mathcal{S}_{3}+\mathcal{S}_{4}$,
we thus have: 
\begin{align}
\mathcal{S}_{2} & =\int_{\mathsf{x}}\Big\{ r_{\mathrm{ip}}\left(\boldsymbol{d}^{\mathrm{ip}}\right)^{2}+r_{\mathrm{op}}\left(\boldsymbol{d}^{\mathrm{op}}\right)^{2}+r_{\mathrm{io}}\left(\boldsymbol{d}^{\mathrm{ip}}\cdot\boldsymbol{d}^{\mathrm{op}}\right)\Big\},\label{eq:trig_S2_real}\\
\mathcal{S}_{3} & =\int_{\mathsf{x}}\Big\{ g_{\mathrm{ip}}\left|\boldsymbol{d}^{\mathrm{ip}}\right|^{3}\mathsf{c}_{3\gamma_{\mathrm{ip}}}+g_{1}\left|\boldsymbol{d}^{\mathrm{ip}}\right|^{2}\left|\boldsymbol{d}^{\mathrm{op}}\right|\mathsf{c}_{2\gamma_{\mathrm{ip}}+\gamma_{\mathrm{op}}}\nonumber \\
 & +g_{\mathrm{op}}\left|\boldsymbol{d}^{\mathrm{op}}\right|^{3}\mathsf{c}_{3\gamma_{\mathrm{op}}}+g_{2}\left|\boldsymbol{d}^{\mathrm{ip}}\right|\left|\boldsymbol{d}^{\mathrm{op}}\right|^{2}\mathsf{c}_{\gamma_{\mathrm{ip}}+2\gamma_{\mathrm{op}}}\Big\},\label{eq:trig_S3_real}\\
\mathcal{S}_{4} & =\int_{\mathsf{x}}\Big\{ u_{\mathrm{ip}}\left|\boldsymbol{d}^{\mathrm{ip}}\right|^{4}+\left|\boldsymbol{d}^{\mathrm{ip}}\right|\left|\boldsymbol{d}^{\mathrm{op}}\right|\left[u_{1}\left|\boldsymbol{d}^{\mathrm{ip}}\right|^{2}+u_{2}\left|\boldsymbol{d}^{\mathrm{op}}\right|^{2}\right]\mathsf{c}_{\gamma_{\mathrm{ip}}-\gamma_{\mathrm{op}}}\nonumber \\
 & +u_{\mathrm{op}}\left|\boldsymbol{d}^{\mathrm{op}}\right|^{4}+\left|\boldsymbol{d}^{\mathrm{ip}}\right|^{2}\left|\boldsymbol{d}^{\mathrm{op}}\right|^{2}\left[u_{\mathrm{io}}^{0}+u_{\mathrm{io}}^{\mathrm{c}}\mathsf{c}_{2\gamma_{\mathrm{ip}}-2\gamma_{\mathrm{op}}}\right]\Big\}.\label{eq:trig_S4_real}
\end{align}
To avoid cumbersome notations, we defined $\mathsf{c}_{\gamma}\equiv\cos\gamma$.
As expected, the degeneracy between $\boldsymbol{d}^{\mathrm{ip}}$
and $\boldsymbol{d}^{\mathrm{op}}$ is manifested in the quadratic
mixing term $\boldsymbol{d}^{\mathrm{ip}}\cdot\boldsymbol{d}^{\mathrm{op}}$. 

To determine the mean-field ground state, we first diagonalize the
quadratic part $\mathcal{S}_{2}$ (\ref{eq:trig_S2_real}) to determine
the combination of $\left(\boldsymbol{d}^{\mathrm{ip}},\boldsymbol{d}^{\mathrm{op}}\right)$
that orders first. We find:
\begin{align}
\mathcal{S}_{2} & =\int_{\mathsf{x}}\Big\{\lambda_{+}\left|\boldsymbol{d}^{+}\right|^{2}+\lambda_{-}\left|\boldsymbol{d}^{-}\right|^{2}\Big\},\label{eq:trig_app_S2_diag-1}
\end{align}
with the eigenvalues 
\begin{align}
\lambda_{\pm} & =\frac{1}{2}\left(r_{\mathrm{ip}}+r_{\mathrm{op}}\pm\sqrt{\left(r_{\mathrm{ip}}-r_{\mathrm{op}}\right)^{2}+r_{\mathrm{io}}^{2}}\,\right),\label{eq:trig_lambdapm_App-2}
\end{align}
and the eigenvectors
\begin{align}
\left(\begin{array}{c}
\boldsymbol{d}^{+}\\
\boldsymbol{d}^{-}
\end{array}\right) & =U^{T}\left(\begin{array}{c}
\boldsymbol{d}^{\mathrm{ip}}\\
\boldsymbol{d}^{\mathrm{op}}
\end{array}\right)=\left(\begin{array}{c}
\mathrm{sign}\left(r_{\mathrm{io}}\right)\beta_{+}\boldsymbol{d}^{\mathrm{ip}}+\beta_{-}\boldsymbol{d}^{\mathrm{op}}\\
-\mathrm{sign}\left(r_{\mathrm{io}}\right)\beta_{-}\boldsymbol{d}^{\mathrm{ip}}+\beta_{+}\boldsymbol{d}^{\mathrm{op}}
\end{array}\right).\label{eq:trig_dpm_App-1}
\end{align}
In these expressions, we defined

\begin{align}
\beta_{\pm} & =\frac{1}{\sqrt{2}}\sqrt{1\pm\frac{r_{\mathrm{ip}}-r_{\mathrm{op}}}{\sqrt{\left(r_{\mathrm{ip}}-r_{\mathrm{op}}\right)^{2}+r_{\mathrm{io}}^{2}}}}\,,\label{eq:beta_pm_tri_app-1}
\end{align}
and the unitary matrix 
\begin{align}
U & =\left(\begin{smallmatrix}\beta_{+}\mathrm{sign}r_{\mathrm{io}} & 0 & -\beta_{-}\mathrm{sign}r_{\mathrm{io}} & 0\\
0 & \beta_{+}\mathrm{sign}r_{\mathrm{io}} & 0 & -\beta_{-}\mathrm{sign}r_{\mathrm{io}}\\
\beta_{-} & 0 & \beta_{+} & 0\\
0 & \beta_{-} & 0 & \beta_{+}
\end{smallmatrix}\right).\label{eq:trig_app_unitary_mat-1}
\end{align}
By construction, $\lambda_{-}<\lambda_{+}$, which implies that the
order parameter combination $\boldsymbol{d}^{-}$ is the one that
condenses. The order parameter combination $\boldsymbol{d}^{+}$,
on the other hand, is a fluctuating field that primarily renormalizes
the action for $\boldsymbol{d}^{-}$. Therefore, to proceed, we consider
only terms that are quadratic or linear in $\boldsymbol{d}^{+}$ (Gaussian
approximation). Introducing the parametrization $\boldsymbol{d}^{\pm}=\left|\boldsymbol{d}^{\pm}\right|\left(\cos\gamma_{\pm},\sin\gamma_{\pm}\right)$,
we can rewrite the action as $\mathcal{S}=\mathcal{S}_{-}+\mathcal{S}_{+-}$
with 
\begin{align}
\mathcal{S}_{-} & =\int_{\mathsf{x}}\Big\{\lambda_{-}\left|\boldsymbol{d}^{-}\right|^{2}+g_{-}\left|\boldsymbol{d}^{-}\right|^{3}\cos\left(3\gamma_{-}\right)+u_{-}\left|\boldsymbol{d}^{-}\right|^{4}\Big\},\label{eq:trig_SM}\\
\mathcal{S}_{+-} & =\int_{\mathsf{x}}\Big\{\lambda_{+}\left|\boldsymbol{d}^{+}\right|^{2}+\tilde{g}_{2}\left|\boldsymbol{d}^{+}\right|\left|\boldsymbol{d}^{-}\right|^{2}\cos\left(\gamma_{+}+2\gamma_{-}\right)\Big\}.\label{eq:trig_SP}
\end{align}
Here, the Landau coefficients $g_{-}$, $\tilde{g}_{2}$, and $u_{-}$
can in principle be expressed in terms of the original Landau coefficients
of Eqs.~(\ref{eq:trig_S2_real})-(\ref{eq:trig_S4_real}) by applying
the unitary transformation $U$ of Eq.~(\ref{eq:trig_app_unitary_mat-1}).
Since the $\boldsymbol{d}^{+}$ action is Gaussian, the minimization
of $\mathcal{S}_{+}$ is straightforward and gives $\boldsymbol{d}^{+}$
in terms of $\boldsymbol{d}^{-}$:
\begin{align}
\left|\boldsymbol{d}^{+}\right| & =\frac{\left|\tilde{g}_{2}\right|}{2\lambda_{+}}\left|\boldsymbol{d}^{-}\right|^{2}, & \gamma_{+} & =-2\gamma_{-}+\left(\frac{1+\mathrm{sign}\,\tilde{g}_{2}}{2}\right)\pi.\label{eq:trig_dP_GS}
\end{align}
Substituting it back in $\mathcal{S}$, we find the Landau expansion
in terms of the ordering field $\boldsymbol{d}^{-}$ alone:

\begin{equation}
\mathcal{S}=\int_{\mathsf{x}}\Big\{\lambda_{-}\left|\boldsymbol{d}^{-}\right|^{2}+g_{-}\left|\boldsymbol{d}^{-}\right|^{3}\cos\left(3\gamma_{-}\right)+\tilde{u}_{-}\left|\boldsymbol{d}^{-}\right|^{4}\Big\},\label{eq:trig_S_final}
\end{equation}
with the reduced quartic coefficient $\tilde{u}_{-}=u_{-}-\frac{\tilde{g}_{2}^{2}}{4\lambda_{+}}$.

The central result is that the nematic action (\ref{eq:trig_S_final})
has the same form as the Landau expansion of the $Z_{3}$-clock model,
see Eq.~(\ref{eq:S_axial_clock}) and Ref.~\citep{Hecker2022}.
Therefore, in all trigonal crystals, the nematic transition belongs
to the 3D $Z_{3}$-Potts/clock ``universality class'', which actually
corresponds to a mean-field first-order transition. Indeed, minimization
of Eq.~(\ref{eq:trig_S_final}) reveals three degenerate ground states:
\begin{align}
\gamma_{-}^{0} & =\frac{\pi}{3}\left(\frac{1+\mathrm{sign}\,g_{-}}{2}\right)+\frac{2\pi}{3}n, & n & \in\{0,1,2\}.\label{eq:trig_gammaM0}
\end{align}
Substitution in Eq.~(\ref{eq:trig_dP_GS}) shows that, in the ground
state, $\boldsymbol{d}^{+}$ and $\boldsymbol{d}^{-}$ are collinear:

\begin{equation}
\left(\begin{array}{c}
\cos\gamma_{+}^{0}\\
\sin\gamma_{+}^{0}
\end{array}\right)=\mathrm{sign}\left(g_{-}\tilde{g}_{2}\right)\left(\begin{array}{c}
\cos\gamma_{-}^{0}\\
\sin\gamma_{-}^{0}
\end{array}\right).
\end{equation}
Therefore, using Eq.~(\ref{eq:trig_dpm_App-1}), we find that the
nematic in-plane and out-of-plane doublets are also collinear, and
thus can be parametrized as:
\begin{align}
\boldsymbol{d}^{\mathrm{ip}} & =\left|\boldsymbol{d}^{E}\right|\cos\delta_{E}\left(\begin{array}{c}
\cos\gamma_{-}^{0}\\
\sin\gamma_{-}^{0}
\end{array}\right), & \boldsymbol{d}^{\mathrm{op}} & =\left|\boldsymbol{d}^{E}\right|\sin\delta_{E}\left(\begin{array}{c}
\cos\gamma_{-}^{0}\\
\sin\gamma_{-}^{0}
\end{array}\right),\label{eq:d_ip_A_fin_tri}
\end{align}
where $\delta_{E}\in[0,\pi]$ and $\left|\boldsymbol{d}^{E}\right|=\sqrt{\left|\boldsymbol{d}^{\mathrm{ip}}\right|^{2}+\left|\boldsymbol{d}^{\mathrm{op}}\right|^{2}}$
are determined by the Landau coefficients. 

To express the trigonal nematic ground state (\ref{eq:d_ip_A_fin_tri})
in the $(\boldsymbol{n}\boldsymbol{m}\boldsymbol{l})$-representation
(\ref{eq:d_nml}), it is convenient to define two new quantities (recall
the $d_{1}$ is the non-zero symmetry-preserving nematic component)
\begin{align}
K_{r} & =-\frac{\sqrt{3}\,\mathrm{sign}(g_{-})}{2}\cos\left(\delta_{E}\right)\frac{\left|\boldsymbol{d}^{E}\right|}{\left|\boldsymbol{d}\right|}-\frac{d_{1}}{2\left|\boldsymbol{d}\right|} & \in & \left[-1,1\right],\label{eq:Kr_tri}\\
K_{r}^{\prime} & =-\frac{\mathrm{sign}(g_{-})}{2}\cos\left(\delta_{E}\right)\frac{\left|\boldsymbol{d}^{E}\right|}{\left|\boldsymbol{d}\right|}+\frac{\sqrt{3}}{2}\frac{d_{1}}{\left|\boldsymbol{d}\right|} & \in & \left[-1,1\right],\label{eq:KrP_tri}
\end{align}
as well as the tilt angle $\eta_{0}\in\left[0,\frac{\pi}{2}\right]$
\begin{align}
\eta_{0} & =\frac{\pi}{2}\frac{1+\mathrm{sign}K_{r}^{\prime}}{2}-\frac{\mathrm{sign}K_{r}^{\prime}}{2}\arccos\Bigg(\frac{\left|K_{r}^{\prime}\right|}{\sqrt{1-\left|K_{r}\right|^{2}}}\Bigg),\label{eq:eta_pm_tri}
\end{align}
and the three orthonormal vectors 
\begin{align}
\boldsymbol{e}_{\parallel} & =\left(\begin{smallmatrix}\cos\gamma_{-}^{0}\\
-\sin\gamma_{-}^{0}\\
0
\end{smallmatrix}\right), & \boldsymbol{e}_{\perp}^{A} & =\left(\begin{smallmatrix}\sin\gamma_{-}^{0}\cos\eta_{0}\\
\cos\gamma_{-}^{0}\cos\eta_{0}\\
\sin\eta_{0}
\end{smallmatrix}\right), & \boldsymbol{e}_{\perp}^{B} & =\left(\begin{smallmatrix}\sin\gamma_{-}^{0}\sin\eta_{0}\\
\cos\gamma_{-}^{0}\sin\eta_{0}\\
-\cos\eta_{0}
\end{smallmatrix}\right).\label{eq:trig_vectors}
\end{align}
Note that $\boldsymbol{e}_{\parallel}$ points along one of the six
high-symmetry directions of the hexagonal point groups, given by $\mathcal{V}_{1}^{\mathrm{hex}}$
and $\mathcal{V}_{2}^{\mathrm{hex}}$ in Eq.~(\ref{eq:V_hex}). Indeed,
the trigonal nematic angle $\gamma_{-}^{0}$ in Eq.~(\ref{eq:trig_gammaM0})
assumes the same values as the in-plane hexagonal nematic angles $\gamma_{\mathrm{ip}}^{0}$
in Eq.~(\ref{eq:angles_ip_Z3}).

In terms of these quantities, the trigonal nematic state (\ref{eq:d_ip_A_fin_tri})
can be conveniently expressed in the $(\boldsymbol{n}\boldsymbol{m}\boldsymbol{l})$-notation
via
\begin{align}
\left|K_{r}\right|\le\frac{1}{2} & \!: & \!\!\alpha & =\frac{\pi}{6}+\arcsin\left(K_{r}\right), & \!\!\boldsymbol{n} & =\boldsymbol{e}_{\perp}^{A},\;\;\boldsymbol{m}=\boldsymbol{e}_{\perp}^{B},\;\;\boldsymbol{l}=\boldsymbol{e}_{\parallel},\nonumber \\
K_{r}<\frac{\text{-}1}{2} & \!: & \!\!\alpha & =\arcsin\left(\left|K_{r}\right|\right)-\frac{\pi}{6}, & \!\!\boldsymbol{n} & =\boldsymbol{e}_{\perp}^{A},\;\;\boldsymbol{m}=\boldsymbol{e}_{\parallel},\;\;\boldsymbol{l}=\boldsymbol{e}_{\perp}^{B},\nonumber \\
K_{r}>\frac{1}{2} & \!: & \!\!\alpha & =\frac{\pi}{2}-\arcsin\left(\left|K_{r}\right|\right), & \!\!\boldsymbol{n} & =\boldsymbol{e}_{\parallel},\;\;\boldsymbol{m}=\boldsymbol{e}_{\perp}^{B},\;\;\boldsymbol{l}=\boldsymbol{e}_{\perp}^{A}.\label{eq:sol_tri}
\end{align}
The similarity to the in-plane nematic state of the 2D isotropic system,
Eq.~(\ref{eq:D_ih_ip_gs_nml}), is apparent \textendash{} indeed,
one can verify that upon setting $\delta_{E}=0$ (pure in-plane nematicity)
one recovers the exact same equations. Conversely, upon setting $\delta_{E}=\pi/2$
(pure out-of-plane nematicity), the out-of-plane isotropic nematic
solution (\ref{eq:D_ih_op_gs_nml}) is recovered. Therefore, the trigonal
nematic state is generally biaxial with $\alpha\neq0,\,\pi/3$, except
for the special point $\left|K_{r}\right|=\frac{1}{2}$, where it
becomes uniaxial. In Figs.~\ref{fig:FS_hex_trig}(c)-(e), we plot
the Fermi surface distortion in the trigonal nematic state (\ref{eq:sol_tri})
for three representative values of $K_{r}$, encompassing each of
the three regimes above. For $\left|K_{r}\right|<\frac{1}{2}$, shown
in Fig.~\ref{fig:FS_hex_trig}(e), the nematic distortion is qualitatively
similar to that caused by an out-of-plane hexagonal order parameter
$\boldsymbol{d}^{\mathrm{op}}$ {[}Fig.~\ref{fig:FS_hex_trig}(b){]},
since the nematic axis $\boldsymbol{l}=\boldsymbol{e}_{\parallel}$
aligns with one of the three in-plane symmetry axes, represented by
the purple lines in the figure {[}corresponding to $\mathcal{V}_{1}^{\mathrm{hex}}$
in Eq.~(\ref{eq:V_hex}){]}. The key difference is that the trigonal
groups do not possess the symmetry elements $C_{2z}$ or $IC_{2z}$,
such that a $180^{\circ}$ degree rotation about the $k_{z}$-axis
does not lead to a degenerate state. As a result, what used to be
a $Z_{6}$-degenerate state in the hexagonal case {[}Fig.~\ref{fig:FS_hex_trig}(b){]}
splits into two sets of $Z_{3}$-degenerate states in the trigonal
case, one for each sign of $g_{-}$ {[}Eq.~(\ref{eq:trig_gammaM0}){]}.
In the regime $\left|K_{r}\right|>\frac{1}{2}$, plotted in Fig.~\ref{fig:FS_hex_trig}(c),
the trigonal nematic state has either the long ($\boldsymbol{m}$)
or the short ($\boldsymbol{n}$) nematic axis aligned with an in-plane
symmetry axis $\boldsymbol{e}_{\parallel}$ , which for $g_{-}<0$
corresponds to $\mathcal{V}_{1}^{\mathrm{hex}}$. This state resembles
the in-plane nematic state of the hexagonal lattice depicted in Fig.~\ref{fig:FS_hex_trig}(a),
albeit slightly rotated out-of-plane. Exactly at $\left|K_{r}\right|=\frac{1}{2}$,
$\alpha=0,\,\pi/3$ and the nematic state is uniaxial, as shown in
Fig.~\ref{fig:FS_hex_trig}(d). The properties of the nematic state
of all trigonal groups are summarized in Table~\ref{tab:classification}. 

\subsection{Tetragonal crystals\label{subsec:Nematicity-tetragonal}}

In any of the seven tetragonal point groups, the five-component nematic
order parameter $\boldsymbol{d}$ (\ref{eq:d_vector}) is decomposed
as: 
\begin{align}
\boldsymbol{d} & \quad\overset{\mathrm{tetragonal}}{\longrightarrow}\quad\Big(\,\underline{d_{1}},\,d_{2},\,d_{5},\,\boldsymbol{d}^{\mathrm{op}}\,\Big)^{T}.\label{eq:d_D4h}
\end{align}
Hence, in contrast to the 2D isotropic system (\ref{eq:d_Dih}), as
well as to the hexagonal and trigonal lattices, in tetragonal crystals
the nematic in-plane doublet $\boldsymbol{d}^{\mathrm{ip}}=\left(d_{2},d_{5}\right)^{T}$
is further decomposed into the two channels $d_{2}=d_{x^{2}-y^{2}}$
and $d_{5}=d_{2xy}$, which transform as non-trivial one-dimensional
IRs. Thus, the tetragonal crystal system is the highest-symmetry crystal
for which a non-trivial nematic component exists that does not transform
as a multi-dimensional IR. 

There are seven different axial point groups in the tetragonal crystal
system: \{$\mathsf{D_{4h}}$, $\mathsf{D_{4}}$, $\mathsf{C_{4v}}$,
$\mathsf{D_{2d}}$\}, which possess in-plane twofold rotational symmetry
axes, and \{$\mathsf{C_{4h}}$, $\mathsf{C_{4}}$, $\mathsf{S_{4}}$\},
which do not possess such axes. As in the previous subsections, the
two sets need to be treated slightly differently, since in the latter
the out-of-plane nematic order parameter $\boldsymbol{d}^{\mathrm{op}}$
transforms as a complex IR and $\left\{ d_{2},d_{5}\right\} $ transform
as the same one-dimensional IR. In this section we will focus on the
first set, illustrating the results for the point group $\mathsf{D_{4h}}$,
while leaving the discussion of the second set of point groups to
Appendix~\ref{sec:tetragonal_Appendix}. 

In $\mathsf{D_{4h}}$, the single-component order parameters $d_{2}=d_{x^{2}-y^{2}}$
and $d_{5}=d_{2xy}$ transform according to the IRs $B_{1g}$ and
$B_{2g}$, respectively, while the out-of-plane doublet $\boldsymbol{d}^{\mathrm{op}}=\left(d_{3},\,d_{4}\right)^{T}=\left(d_{2yz},\,d_{2xz}\right)^{T}$
transforms as the IR $E_{g}$. As single-component real-valued order
parameters, both $d_{2}$ and $d_{5}$ are Ising variables that undergo
a $Z_{2}$-Ising transition described by the action
\begin{align}
\mathcal{S}_{\nu} & =\int_{\mathsf{x}}\Big\{ r_{\nu}\left|d_{\nu}\right|^{2}+u_{\nu}\left|d_{\nu}\right|^{4}\Big\},\label{eq:S_D4h_ip}
\end{align}
with $\nu=\{2,5\}$. We note that this Landau expansion could also
be recast in terms of the in-plane nematic doublet $\boldsymbol{d}^{\mathrm{ip}}=\left(d_{2},\,d_{5}\right)^{T}=\big(d_{x^{2}-y^{2}},\,d_{2xy}\big)^{T}$
as an effective $Z_{2}$-clock model, see Eq.~(\ref{eq:S_axial_clock}):

\begin{equation}
\mathcal{S}_{\mathrm{ip}}=\int_{\mathsf{x}}\Big\{ r_{0}|\boldsymbol{d}^{\mathrm{ip}}|^{2}+h_{2}|\boldsymbol{d}^{\mathrm{ip}}|^{2}\cos\left(2\gamma_{\mathrm{ip}}\right)+\mathcal{O}\left(|\boldsymbol{d}^{\mathrm{ip}}|^{4}\right)\Big\}.\label{eq:S_D4h_ip_aux}
\end{equation}
The second term is nothing but $h_{2}\big(d_{2}^{2}-d_{5}^{2}\big)$,
which implies that the transitions toward $d_{2}$ and $d_{5}$ electronic
nematic order take place at different values of the control parameter,
$r_{0}=-h_{2}$ for $d_{2}$ and $r_{0}=h_{2}$ for $d_{5}$. The
symmetry here also allows for an additional biquadratic term $d_{2}^{2}d_{5}^{2}$.
One could in principle integrate out the fluctuations of the sub-leading
channel, whose effect on the leading channel will only be important
if $h_{2}$ is small. Thus, the Landau coefficients $r_{\nu}$ and
$u_{\nu}$ in Eq.~(\ref{eq:S_D4h_ip}) should be understood as renormalized
Landau coefficients of the ``original'' in\textendash plane nematic
action.

\begin{figure}
\begin{centering}
\includegraphics[width=1\columnwidth]{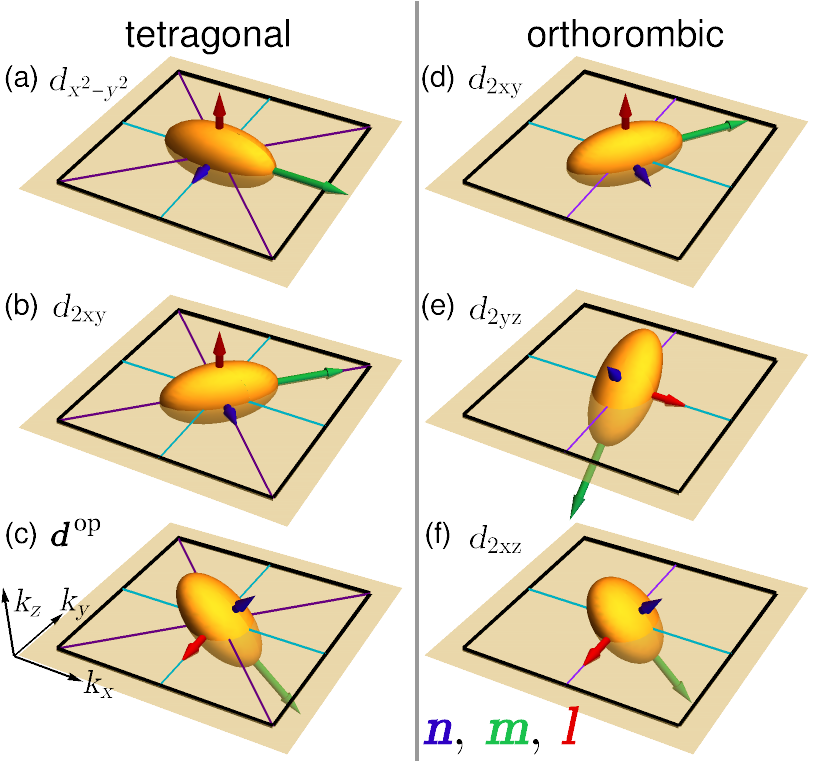}
\par\end{centering}
\caption{Distorted Fermi surface in the nematic phase, as given by Eq.~(\ref{eq:kF_khat}),
plotted together with the nematic axes $\boldsymbol{n}$,$\boldsymbol{m}$,$\boldsymbol{l}$
for the cases of: (a)-(b) tetragonal in-plane $d_{2}=d_{x^{2}-y^{2}}$
and $d_{5}=d_{2xy}$ states (\ref{eq:S_D4h_ip}), (c) tetragonal out-of-plane
doublet state $\boldsymbol{d}^{\mathrm{op}}$ (\ref{eq:angles_op_Z4}),
(d)-(f) orthorhombic nematic states $d_{5}=d_{2xy}$, $d_{3}=d_{2yz}$
and $d_{4}=d_{2xz}$. The unit vectors $\left\{ \boldsymbol{n},\boldsymbol{m},\boldsymbol{l}\right\} $
are rescaled by $2.25\,k_{F}\left(\left\{ \boldsymbol{n},\boldsymbol{m},\boldsymbol{l}\right\} \right)$,
respectively. For the tetragonal and orthorhombic cases we chose the
trivial components as $d_{1}/\left|\boldsymbol{d}\right|=0.1$ and
$\left(\left|\boldsymbol{d}^{e}\right|/\left|\boldsymbol{d}\right|,\gamma_{e}\right)=\left(0.3,1.1\right)$,
respectively. \label{fig:FS_tetra_ortho}}
\end{figure}

The $d_{2}=d_{x^{2}-y^{2}}$ and $d_{5}=d_{2xy}$ Ising-nematic order
parameters are the types of nematic order most widely studied in the
literature \citep{Carlson2006,Xu2008,Fang2008,Raghu2009,Fernandes2010,Metlitski2010,Fischer2011,Fernandes2012,Nie2014,Kontani2019,Wang2019}.
Whereas one often describes them in terms of an effective 2D nematic
order parameter, see Eq.~(\ref{eq:aux_Q_2D}), it is straightforward
to describe them in the $(\boldsymbol{n}\boldsymbol{m}\boldsymbol{l})$-representation
by employing Eq.~(\ref{eq:D_ih_ip_gs_nml}) with the additional constraint
of $\gamma_{\mathrm{ip}}=\{0,\pi\}$ for $d_{2}$ {[}corresponding
to $h_{2}<0$ in Eq.~(\ref{eq:S_D4h_ip_aux}){]} and $\gamma_{\mathrm{ip}}=\{\frac{\pi}{2},\frac{3\pi}{2}\}$
for $d_{5}$ {[}corresponding to $h_{2}>0$ in Eq.~(\ref{eq:S_D4h_ip_aux}){]}.
The corresponding Fermi surface distortions are shown in Figs.~\ref{fig:FS_tetra_ortho}(a),(b).
For the $d_{2}=d_{x^{2}-y^{2}}$ state, the in-plane nematic axes
align with the in-plane principal axes $\mathcal{V}_{1}^{\mathrm{tet}}$
(light-blue) while for $d_{5}=d_{2xy}$ they align with the in-plane
diagonal axes $\mathcal{V}_{2}^{\mathrm{tet}}$ (purple), with
\begin{align}
\mathcal{V}_{1}^{\mathrm{tet}} & =\left\{ \left(\begin{smallmatrix}1\\
0\\
0
\end{smallmatrix}\right),\left(\begin{smallmatrix}0\\
1\\
0
\end{smallmatrix}\right)\right\} , & \mathcal{V}_{2}^{\mathrm{tet}} & =\left\{ \frac{1}{\sqrt{2}}\left(\begin{smallmatrix}1\\
1\\
0
\end{smallmatrix}\right),\frac{1}{\sqrt{2}}\left(\begin{smallmatrix}1\\
-1\\
0
\end{smallmatrix}\right)\right\} .\label{eq:V_tet}
\end{align}

Therefore, the $Z_{2}$ character of the Ising-nematic order parameter
is associated with the $C_{4z}$ operation ($90^{\circ}$ rotation
about the $k_{z}$-axis). Recall that the alignment of the $\boldsymbol{n}$,
$\boldsymbol{m}$, $\boldsymbol{l}$ axes in Eq.~(\ref{eq:D_ih_ip_gs_nml}),
which determines which nematic axes are in-plane, depends on the ratio
$d_{1}/\left|\boldsymbol{d}\right|$ between the trivial (i.e. symmetry-preserving)
nematic component $d_{1}=d_{\frac{1}{\sqrt{3}}(2z^{2}-x^{2}-y^{2})}$
and $\left|\boldsymbol{d}\right|=\sqrt{d_{\nu}^{2}+d_{1}^{2}}$, with
$\nu=\{2,5\}$.

To derive the Landau invariants for the out-of-plane nematic order
parameter $\boldsymbol{d}^{\mathrm{op}}$ we compute, once again,
the decomposition of the symmetrized product for each expansion order,
\begin{align}
\big[\otimes_{j=1}^{2}E_{g}\big]_{s} & =A_{1g}\oplus B_{1g}\oplus B_{2g},\label{eq:decomp_tet_op2}\\
\big[\otimes_{j=1}^{3}E_{g}\big]_{s} & =2E_{g},\label{eq:decomp_tet_op3}\\
\big[\otimes_{j=1}^{4}E_{g}\big]_{s} & =2A_{1g}\!\oplus A_{2g}\oplus B_{1g}\oplus B_{2g}.\label{eq:decomp_tet_op4}
\end{align}
From the bilinears associated with Eq.~(\ref{eq:decomp_tet_op2}),
$D_{\mathrm{op}}^{A_{1g}}=\left|\boldsymbol{d}^{\mathrm{op}}\right|^{2}$,
and 
\begin{align}
D_{\mathrm{op}}^{B_{1g}} & =\left|\boldsymbol{d}^{\mathrm{op}}\right|^{2}\cos\left(2\gamma_{\mathrm{op}}\right), & D_{\mathrm{op}}^{B_{2g}} & =\left|\boldsymbol{d}^{\mathrm{op}}\right|^{2}\sin\left(2\gamma_{\mathrm{op}}\right),\label{eq:tet_bilinears_op}
\end{align}
we can readily identify the three Landau invariants as $\left|\boldsymbol{d}^{\mathrm{op}}\right|^{2}$,
$\left|\boldsymbol{d}^{\mathrm{op}}\right|^{4}$, and $\big(D_{\mathrm{op}}^{B_{1g}}\big)^{2}-\big(D_{\mathrm{op}}^{B_{2g}}\big)^{2}$.
We then obtain the Landau expansion
\begin{align}
\mathcal{S}_{\mathrm{op}} & =\int_{\mathsf{x}}\Big\{ r_{0}\left|\boldsymbol{d}^{\mathrm{op}}\right|^{2}+u\left|\boldsymbol{d}^{\mathrm{op}}\right|^{4}+v_{4}\left|\boldsymbol{d}^{\mathrm{op}}\right|^{4}\cos\left(4\gamma_{\mathrm{op}}\right)\Big\},\label{eq:S_D4h_op}
\end{align}
which corresponds to the $Z_{4}$-clock model of Eq.~(\ref{eq:S_axial_clock}).
The fourfold degenerate ground-state is parametrized by 
\begin{align}
\gamma_{\mathrm{op}}^{0} & =\frac{\pi}{4}\left(\frac{1+\mathrm{sign}v_{4}}{2}\right)+\frac{2\pi}{4}n, & n & \in\{0,1,2,3\}.\label{eq:angles_op_Z4}
\end{align}
Therefore, the out-of-plane nematic transition in a tetragonal crystal
belongs to the 3D-XY universality class; we emphasize that such a
critical behavior may only emerge very close to the transition, depending
on the degree of out-of-plane anisotropy of the system. 

The corresponding representation of Eq.~(\ref{eq:angles_op_Z4})
in the $(\boldsymbol{n}\boldsymbol{m}\boldsymbol{l})$-formalism can
be directly read off from Eq.~(\ref{eq:D_ih_op_gs_nml}). Fig.~\ref{fig:FS_tetra_ortho}(c)
shows the corresponding Fermi surface distortion, with the nematic
axis $\boldsymbol{l}$ aligned with the in-plane high-symmetry axes
of either $\mathcal{V}_{1}^{\mathrm{tet}}$ (light-blue) or $\mathcal{V}_{2}^{\mathrm{tet}}$(purple)
{[}see Eq.~(\ref{eq:V_tet}){]}, depending on the sign of $v_{4}$.
The fourfold degeneracy of the ground state arises from the two axes
in either $\mathcal{V}_{1}^{\mathrm{tet}}$ or $\mathcal{V}_{2}^{\mathrm{tet}}$
with each $\boldsymbol{l}$ can align, combined with the two sets
of directions of $\boldsymbol{n},\,\boldsymbol{m}$ that are related
by a $C_{2z}$ symmetry operation (which is present in all tetragonal
groups). A summary of the results for all tetragonal crystal groups
is presented in Table~\ref{tab:classification}, combining the analysis
of this section with that of Appendix~\ref{sec:tetragonal_Appendix}.

\subsection{Orthorhombic, Monoclinic, and Triclinic crystals\label{subsec:Nematicity-orthormbic}}

In the remaining three crystal systems \textendash{} orthorhombic,
monoclinic, and triclinic \textendash{} all the point groups are Abelian,
and as such do not admit multi-dimensional IRs (real or complex).
As a result, all nematic components transform as one-dimensional IRs,
implying that the non-trivial nematic components behave as $Z_{2}$-Ising
order parameters. Another key difference between these axial groups
and the other ones previously analyzed in this section is that at
least one of the components of $\boldsymbol{d}^{\mathrm{ip}}=\left(d_{2},\,d_{5}\right)^{T}=\big(d_{x^{2}-y^{2}},\,d_{2xy}\big)^{T}$
and $\boldsymbol{d}^{\mathrm{op}}=\left(d_{3},\,d_{4}\right)^{T}=\left(d_{2yz},\,d_{2xz}\right)^{T}$
transform as the trivial IR, like $d_{1}=d_{\frac{1}{\sqrt{3}}(2z^{2}-x^{2}-y^{2})}$.

We start with the orthorhombic crystal system, described by the $3$
point groups \{$\mathsf{D_{2h}}$, $\mathsf{D_{2}}$, $\mathsf{C_{2v}}$\}.
In terms of the IR of these groups, the five nematic components $\boldsymbol{d}$
(\ref{eq:d_vector}) are decomposed according to 
\begin{align}
\boldsymbol{d} & \quad\overset{\mathrm{orthorombic}}{\longrightarrow}\quad\Big(\,\underline{d_{1},\,d_{2}},\,d_{3},\,d_{4},\,d_{5}\,\Big)^{T}.\label{eq:d_D4h-1}
\end{align}
The first two components $\left\{ d_{1},d_{2}\right\} $, highlighted
by the underline, transform trivially under the groups symmetry operations
and thus are generically non-zero. The remaining three components
$d_{3}=d_{2yz}$, $d_{4}=d_{2xz}$ and $d_{5}=d_{2xy}$ correspond
to shear distortions of the Fermi surface, and transform according
to three different one-dimensional IRs, as shown explicitly in Table~\ref{tab:classification}.
As such, they all undergo a $Z_{2}$-Ising transition described by
an action analogous to Eq.~(\ref{eq:S_D4h_ip}). 

These three Ising-nematic states can be expressed in the $(\boldsymbol{n}\boldsymbol{m}\boldsymbol{l})$-representation
provided that one also includes the non-zero trivial components $\boldsymbol{d}^{e}=\left(d_{1},d_{2}\right)=\left|\boldsymbol{d}^{e}\right|\left(\cos\gamma_{e},\sin\gamma_{e}\right)$.
In the case of the nematic order parameter $d_{5}=d_{2xy}$, one can
directly apply the in-plane isotropic parametrization of Eq.~(\ref{eq:D_ih_ip_gs_nml})
with $\left|\boldsymbol{d}\right|=\sqrt{d_{5}^{2}+\left|\boldsymbol{d}^{e}\right|^{2}}$
and $\tan\gamma_{\mathrm{ip}}=d_{5}/d_{2}$. The resulting Fermi surface
distortion, calculated via Eq.~(\ref{eq:kF_khat}), is shown in Fig.~\ref{fig:FS_tetra_ortho}(d).
Since $d_{2}$ is always non-zero, the in-plane nematic axes $\boldsymbol{e}_{\parallel}^{A}$
and $\boldsymbol{e}_{\parallel}^{B}$ defined in Eq.~(\ref{eq:Dih_ip_vectors})
are never aligned with $\mathcal{V}_{2}^{\mathrm{tet}}$ in Eq.~(\ref{eq:V_tet}),
which is consistent with the fact that $\mathcal{V}_{2}^{\mathrm{tet}}$
are not high-symmetry directions in orthorhombic crystals. Note that
one of the nematic axes is aligned with the $k_{z}$-axis. The $Z_{2}$
degeneracy is associated with the $C_{2x}$ ($C_{2y}$) or $IC_{2x}$
($IC_{2y}$) symmetry operations, corresponding to in-plane twofold
rotations or vertical mirror reflections of the nematic axes in Fig.~\ref{fig:FS_tetra_ortho}(d).

As for the nematic order parameters $d_{3}$ and $d_{4}$, one cannot
just apply the out-of-plane parametrization of Eq.~(\ref{eq:D_ih_op_gs_nml}),
since $d_{2}$, which is a component of $\boldsymbol{d}^{\mathrm{ip}}$,
is always non-zero. To express $d_{3}$ and $d_{4}$ in the $(\boldsymbol{n}\boldsymbol{m}\boldsymbol{l})$-representation,
it is convenient to define the auxiliary variables.
\begin{align}
c_{k}^{e} & =\frac{\left|\boldsymbol{d}^{e}\right|}{\left|\boldsymbol{d}\right|}\cos\left(\gamma_{e}-\frac{k\pi}{3}\right), & s_{k}^{e} & =\frac{\left|\boldsymbol{d}^{e}\right|}{\left|\boldsymbol{d}\right|}\sin\left(\gamma_{e}-\frac{k\pi}{3}\right),\label{eq:ortho_c_s}
\end{align}
with $k=\{2,4\}$ and $\left|\boldsymbol{d}\right|=\sqrt{\left|\boldsymbol{d}^{e}\right|^{2}+(d_{3})^{2}+(d_{4})^{2}}$.
We then define the angles

\begin{align}
\eta_{3}^{\pm} & =\frac{1}{2}\frac{d_{3}}{\left|d_{3}\right|}\arccos\Bigg(\frac{\left|s_{2}^{e}\right|}{\sqrt{1-\left(c_{2}^{e}\right)^{2}}}\Bigg)\pm\frac{1+\mathrm{sign}\left(s_{2}^{e}\right)}{2}\frac{\pi}{2},\label{eq:ortho_eta3}\\
\eta_{4}^{\pm} & =\frac{1}{2}\frac{d_{4}}{\left|d_{4}\right|}\arccos\Bigg(\frac{\left|s_{4}^{e}\right|}{\sqrt{1-\left(c_{4}^{e}\right)^{2}}}\Bigg)\pm\frac{1+\mathrm{sign}\left(s_{4}^{e}\right)}{2}\frac{\pi}{2},\label{eq:ortho_eta4}
\end{align}
and, from them, the four unit vectors
\begin{align}
\boldsymbol{e}_{A}^{yz} & =\left(0,\sin\eta_{3}^{+},\cos\eta_{3}^{-}\right)^{T}\!\!, & \boldsymbol{e}_{B}^{yz} & =\left(0,\cos\eta_{3}^{-},\text{-}\sin\eta_{3}^{+}\right)^{T}\!\!,\nonumber \\
\boldsymbol{e}_{A}^{zx} & =\left(\cos\eta_{4}^{-},0,\sin\eta_{4}^{+}\right)^{T}\!\!, & \boldsymbol{e}_{B}^{zx} & =\left(\text{-}\sin\eta_{4}^{+},0,\cos\eta_{4}^{-}\right)^{T}\!\!.\label{eq:ortho_e_perps}
\end{align}
Now, the nematic order parameter $d_{3}=d_{2yz}$ can be conveniently
expressed in the $(\boldsymbol{n}\boldsymbol{m}\boldsymbol{l})$-representation
via:
\begin{align}
\left|c_{2}^{e}\right|\le\frac{1}{2} & : & \!\alpha & =\frac{\pi}{6}+\arcsin\left(c_{2}^{e}\right), & \!\boldsymbol{n} & =\boldsymbol{e}_{A}^{yz},\;\boldsymbol{m}=\boldsymbol{e}_{B}^{yz},\;\boldsymbol{l}=\boldsymbol{e}_{x},\nonumber \\
c_{2}^{e}<\frac{\text{-}1}{2} & : & \!\alpha & =\arcsin\left(\left|c_{2}^{e}\right|\right)-\frac{\pi}{6}, & \!\boldsymbol{n} & =\boldsymbol{e}_{A}^{yz},\;\boldsymbol{m}=\boldsymbol{e}_{x},\;\boldsymbol{l}=\boldsymbol{e}_{B}^{yz},\nonumber \\
c_{2}^{e}>\frac{1}{2} & : & \!\alpha & =\frac{\pi}{2}-\arcsin\left(\left|c_{2}^{e}\right|\right), & \!\boldsymbol{n} & =\boldsymbol{e}_{x},\;\boldsymbol{m}=\boldsymbol{e}_{B}^{yz},\;\boldsymbol{l}=\boldsymbol{e}_{A}^{yz}.\label{eq:ortho_yz_nml}
\end{align}
Analogously, the nematic order parameter $d_{4}=d_{2xz}$ becomes:
\begin{align}
\left|c_{4}^{e}\right|\le\frac{1}{2} & : & \!\alpha & =\frac{\pi}{6}+\arcsin\left(c_{4}^{e}\right), & \!\boldsymbol{n} & =\boldsymbol{e}_{A}^{zx},\;\boldsymbol{m}=\boldsymbol{e}_{B}^{zx},\;\boldsymbol{l}=\boldsymbol{e}_{y},\nonumber \\
c_{4}^{e}<\frac{\text{-}1}{2} & : & \!\alpha & =\arcsin\left(\left|c_{4}^{e}\right|\right)-\frac{\pi}{6}, & \!\boldsymbol{n} & =\boldsymbol{e}_{A}^{zx},\;\boldsymbol{m}=\boldsymbol{e}_{y},\;\boldsymbol{l}=\boldsymbol{e}_{B}^{zx},\nonumber \\
c_{4}^{e}>\frac{1}{2} & : & \!\alpha & =\frac{\pi}{2}-\arcsin\left(\left|c_{4}^{e}\right|\right), & \!\boldsymbol{n} & =\boldsymbol{e}_{y},\;\boldsymbol{m}=\boldsymbol{e}_{B}^{zx},\;\boldsymbol{l}=\boldsymbol{e}_{A}^{zx}.\label{eq:ortho_zx_nml}
\end{align}
Note that the $(\boldsymbol{n}\boldsymbol{m}\boldsymbol{l})$-representations
(\ref{eq:ortho_yz_nml}) and (\ref{eq:ortho_zx_nml}) are very similar
to the in-plane isotropic $(\boldsymbol{n}\boldsymbol{m}\boldsymbol{l})$-representation,
Eq.~(\ref{eq:D_ih_ip_gs_nml}). In fact, the latter can be recovered
from Eq.~(\ref{eq:ortho_c_s}) by setting $k=0$ as $c_{0}^{e}=d_{1}/\left|\boldsymbol{d}\right|$,
and the three cases support a fine-tuned uniaxial nematic order parameter
for $\left|c_{4}^{e}\right|=\left|c_{2}^{e}\right|=\left|c_{0}^{e}\right|=\frac{1}{2}$.
In Figs.~\ref{fig:FS_tetra_ortho}(e)-(f), we show the Fermi surface
distortions corresponding, respectively, to the condensation of the
nematic order parameters $d_{3}=d_{2yz}$ and $d_{4}=d_{2xz}$. In
the former, one of the nematic in-plane axis always aligns with $\boldsymbol{e}_{x}$
while for $d_{4}=d_{2xz}$, it aligns with $\boldsymbol{e}_{y}$,
in accordance with Eqs.~(\ref{eq:ortho_yz_nml}) and (\ref{eq:ortho_zx_nml}).
In both cases, the twofold degeneracy is associated with the symmetry
element $C_{2z}$, i.e. a $180^{\circ}$ rotation about the $k_{z}$-axis.

We now analyze the case of the monoclinic crystal system, encompassed
by the $3$ axial point groups \{$\mathsf{C_{2h}}$, $\mathsf{C_{1h}}=\mathsf{C_{s}}$,
$\mathsf{C_{2}}$\}. In all cases, the decomposition of the five nematic
components $\boldsymbol{d}$ (\ref{eq:d_vector}) is
\begin{align}
\boldsymbol{d} & \quad\overset{\mathrm{monoclinic}}{\longrightarrow}\quad\Big(\,\underline{d_{1},\,d_{2},\,d_{5}},\,\big\{ d_{3},\,d_{4}\big\}\,\Big)^{T}.\label{eq:d_C2h}
\end{align}
As indicated by the underline, the three components $d_{1}=d_{\frac{1}{\sqrt{3}}(2z^{2}-x^{2}-y^{2})}$,
$d_{2}=d_{x^{2}-y^{2}}$, and $d_{5}=d_{2xy}$ transform as the trivial
IR and are thus always non-zero. On the other hand, the nematic order
parameters $d_{3}=d_{2yz}$ and $d_{4}=d_{2xz}$ transform according
to the same one-dimensional IR, as indicated by the curly brackets.
Consequently, the analysis of the Landau expansion is analogous to
that performed in Appendix~\ref{sec:tetragonal_Appendix} for the
tetragonal groups \{$\mathsf{C_{4h}}$, $\mathsf{C_{4}}$, $\mathsf{S_{4}}$\}.
Indeed, we obtain the same form of the nematic action $\mathcal{S}=\mathcal{S}_{2}+\mathcal{S}_{4}$:

\begin{align}
\mathcal{S}_{2} & =\int_{\mathsf{x}}\Big\{ r_{1}\left(d_{3}\right)^{2}+r_{2}\left(d_{4}\right)^{2}+r_{3}d_{3}d_{4}\Big\},\\
\mathcal{S}_{4} & =\int_{\mathsf{x}}\Big\{ u_{1}\left(d_{3}\right)^{4}+u_{2}\left(d_{4}\right)^{4}+u_{3}\left(d_{3}\right)^{2}\left(d_{4}\right)^{2}\nonumber \\
 & +u_{4}d_{3}\left(d_{4}\right)^{3}+u_{5}\left(d_{3}\right)^{3}d_{4}\Big\}.
\end{align}
The outcome, as explained in Appendix~\ref{sec:tetragonal_Appendix},
is that a linear combination of $d_{3}$ and $d_{4}$, which is enforced
by the Landau coefficients, undergoes a $Z_{2}$-Ising transition.
The visualization of this nematic order parameter in the $(\boldsymbol{n}\boldsymbol{m}\boldsymbol{l})$-representation
offers little insight, as the nematic axes $\boldsymbol{n}$, $\boldsymbol{m}$,
$\boldsymbol{l}$ can point anywhere in space. 

The last crystal system is the triclinic one, described by the $2$
axial point groups \{$\mathsf{C_{1}}$, $\mathsf{C_{i}}=\mathsf{S_{2}}$\}.
The symmetry of these crystals is so low that all five nematic components
$\boldsymbol{d}$ transform as the trivial IR, which is a direct consequence
of the absence of rotational symmetry axes:
\begin{align}
\boldsymbol{d} & \quad\overset{\mathrm{triclinic}}{\longrightarrow}\quad\Big(\,\underline{d_{1},\,d_{2},\,d_{3},\,d_{4},\,d_{5}}\Big)^{T}.\label{eq:d_S2}
\end{align}
Therefore, nematic phase transitions cannot occur in triclinic crystals.

\section{Electronic nematicity in quasicrystalline axial point groups \label{sec:Quasicrystal}}

While a large number of known quasicrystals is described by the icosahedral
(i.e., polyhedral) point groups \{$\mathsf{I_{h}}$, $\mathsf{I}$\},
whose electronic nematic properties were analyzed in Sec.~\ref{sec:Nematicity-Icosahedral},
there are also quasicrystalline materials with eightfold, tenfold,
and twelvefold symmetry \citep{Goldman1991}. In contrast to the icosahedral
quasicrystals, the latter are quasiperiodic in two directions and
periodic along an axial direction; consequently, they are described
by non-crystallographic axial point groups \citep{Socolar1989,Rabson1991,Rokhsar1987,Lifshitz1996}.
In this subsection, we investigate the properties of the electronic
nematic order parameter in the class of octagonal, decagonal, and
dodecagonal point groups. Importantly, these point groups describe
not only quasicrystalline materials, but also artificial quasicrystals
obtained from twisting two crystalline 2D materials. The latter will
be discussed in more depth in Subsection~\ref{subsec:twisted-quasicrystals}. 

In all cases studied here, the symmetry-decomposition of the five
nematic components of $\boldsymbol{d}$ (\ref{eq:d_vector}) have
the same form as that for 2D isotropic systems:

\begin{equation}
\boldsymbol{d}\quad\overset{\mathrm{axial\:\:quasicrystals}}{\longrightarrow}\quad\Big(\,\underline{d_{1}},\,\boldsymbol{d}^{\mathrm{ip}},\,\boldsymbol{d}^{\mathrm{op}}\,\Big)^{T}.
\end{equation}
Following the discussion in Sec.~\ref{subsec:Dihedral-group_Dih}
for 2D isotropic systems, our goal is to obtain the first invariant
in the Landau expansion of $\boldsymbol{d}^{\alpha}$ that is not
isotropic, i.e. that is not of the form $\left|\boldsymbol{d}^{\alpha}\right|^{2n}$
with integer $n$ (here, $\alpha$ refers to either $\mathrm{ip}$
or $\mathrm{op}$). To accomplish this in a systematic way, we evaluate
the symmetrized decomposition of the $N^{\mathrm{th}}$-order product
$\big[\otimes_{j=1}^{N}E_{\Gamma}\big]_{s}$, where $E_{\Gamma}$
is the two-dimensional IR according to which $\boldsymbol{d}^{\alpha}$
transforms. The leading-order anisotropic term is obtained from the
product with the smallest $N$ for which the decomposition gives either
two invariants (if $N$ is even) or one invariant (if $N$ is odd).
We note that Ref.~\citep{Liu2024nematic} performed a related analysis
for the $d_{2xy}$ and $d_{x^{2}-y^{2}}$ superconducting order parameters
in $\mathsf{D_{n}}$ point groups.

A summary of all the results presented in this section, as well as
in Sec.~\ref{sec:Nematicity-Icosahedral} for the icosahedral quasicrystal,
is contained in Table~\ref{tab:classification_quasicrystals}.

\subsection{Dodecagonal quasicrystals\label{subsec:Dodecagonal-quasicrystals}}

The non-crystallographic point groups that possess dodecagonal symmetry
are \{$\mathsf{D_{12h}}$, $\mathsf{D_{12}}$, $\mathsf{C_{12v}}$,
$\mathsf{D_{6d}}$, $\mathsf{C_{12h}}$, $\mathsf{C_{12}}$, $\mathsf{S_{12}}$\}.
For concreteness, hereafter we consider $\mathsf{D_{12h}}$, for which
$\boldsymbol{d}^{\mathrm{ip}}$ and $\boldsymbol{d}^{\mathrm{op}}$
transform as the $E_{2g}$ and $E_{1g}$ IR, respectively. We will
discuss later how the results generalize to the other dodecagonal
groups.

To derive the Landau expansion of the doublets, we follow the procedure
outlined in the beginning of this section. Focusing first on the in-plane
doublet $\boldsymbol{d}^{\mathrm{ip}}$, we compute the symmetrized-product
decomposition of the corresponding IR:
\begin{align}
\big[\otimes_{j=1}^{2}E_{2g}\big]_{s} & =A_{1g}\oplus E_{4g},\label{eq:E2g_symmprod_D12h_2}\\
\big[\otimes_{j=1}^{3}E_{2g}\big]_{s} & =B_{1g}\oplus B_{2g}\oplus E_{2g},\label{eq:E2g_symmprod_D12h_3}\\
\big[\otimes_{j=1}^{4}E_{2g}\big]_{s} & =A_{1g}\!\oplus2E_{4g},\label{eq:E2g_symmprod_D12h_4}\\
\big[\otimes_{j=1}^{5}E_{2g}\big]_{s} & =B_{1g}\oplus B_{2g}\oplus2E_{2g},\label{eq:E2g_symmprod_D12h_5}\\
\big[\otimes_{j=1}^{6}E_{2g}\big]_{s} & =2A_{1g}\!\oplus A_{2g}\oplus2E_{4g}.\label{eq:E2g_symmprod_D12h_6}
\end{align}
Therefore, since there are two $A_{1g}$ invariants in (\ref{eq:E2g_symmprod_D12h_6}),
whereas all other even orders have only one invariant, we conclude
that the leading order anisotropic term in the Landau expansion of
$\boldsymbol{d}^{\mathrm{ip}}$ appears at sixth-order. It can be
constructed from the product of the $B_{1g}$ and $B_{2g}$ trilinears,
\begin{align}
B_{1g}: & \left|\boldsymbol{d}^{\mathrm{ip}}\right|^{3}\cos\left(3\gamma_{\mathrm{ip}}\right), & B_{2g}: & \left|\boldsymbol{d}^{\mathrm{ip}}\right|^{3}\sin\left(3\gamma_{\mathrm{ip}}\right),\label{eq:D12h_ip_trilinears}
\end{align}
and thus, it becomes $\left|\boldsymbol{d}^{\mathrm{ip}}\right|^{6}\cos\left(6\gamma_{\mathrm{ip}}\right)$,
resulting in the action:
\begin{align}
\mathcal{S}_{\mathrm{ip}} & =\int_{\mathsf{x}}\Big\{ r_{0}\left|\boldsymbol{d}^{\mathrm{ip}}\right|^{2}+u\left|\boldsymbol{d}^{\mathrm{ip}}\right|^{4}+v_{6}\left|\boldsymbol{d}^{\mathrm{ip}}\right|^{6}\cos\left(6\gamma_{\mathrm{ip}}\right)\Big\}.\label{eq:S_D12h_ip}
\end{align}
The Landau expansion (\ref{eq:S_D12h_ip}) is equivalent to that of
the $Z_{6}$-clock model, displaying a sixfold-degenerate ground state
\begin{align}
\gamma_{\mathrm{ip}}^{0} & =\frac{\pi}{6}\left(\frac{1+\mathrm{sign}v_{6}}{2}\right)+\frac{2\pi}{6}n, & n & \in\{0,1,\dots,5\}.\label{eq:angles_ip_Z6_D12h}
\end{align}

As for the out-of-plane doublet $\boldsymbol{d}^{\mathrm{op}}$, which
transforms as the $E_{1g}$ IR, the symmetrized-product decomposition
up to sixth-order yields only isotropic invariants of the form $\left|\boldsymbol{d}^{\mathrm{op}}\right|^{2n}$:
\begin{align}
\big[\otimes_{j=1}^{2}E_{1g}\big]_{s} & =A_{1g}\oplus E_{2g},\label{eq:E1g_symmprod_D12h_2}\\
\big[\otimes_{j=1}^{3}E_{1g}\big]_{s} & =E_{1g}\oplus E_{3g},\label{eq:E1g_symmprod_D12h_3}\\
\big[\otimes_{j=1}^{4}E_{1g}\big]_{s} & =A_{1g}\oplus E_{2g}\oplus E_{4g},\label{eq:E1g_symmprod_D12h_4}\\
\big[\otimes_{j=1}^{5}E_{1g}\big]_{s} & =E_{1g}\oplus E_{3g}\oplus E_{5g},\label{eq:E1g_symmprod_D12h_5}\\
\big[\otimes_{j=1}^{6}E_{1g}\big]_{s} & =A_{1g}\oplus B_{1g}\oplus B_{2g}\oplus E_{2g}\oplus E_{4g}.\label{eq:E1g_symmprod_D12h_6}
\end{align}
It turns out that an anisotropic invariant can only be constructed
at twelfth-order, from the product of the $B_{1g}$ and $B_{2g}$
IRs that appear in the sixth-order product decomposition:
\begin{align}
B_{1g}: & \left|\boldsymbol{d}^{\mathrm{op}}\right|^{6}\cos\left(6\gamma_{\mathrm{op}}\right), & B_{2g}: & \left|\boldsymbol{d}^{\mathrm{op}}\right|^{6}\sin\left(6\gamma_{\mathrm{op}}\right).\label{eq:D12h_op_sixthorder}
\end{align}
We thus obtain the nematic action: 
\begin{align}
\mathcal{S}_{\mathrm{op}} & =\int_{\mathsf{x}}\Big\{ r_{0}\left|\boldsymbol{d}^{\mathrm{op}}\right|^{2}+u\left|\boldsymbol{d}^{\mathrm{op}}\right|^{4}+v_{12}\left|\boldsymbol{d}^{\mathrm{op}}\right|^{12}\cos\left(12\gamma_{\mathrm{op}}\right)\Big\},\label{eq:S_D12h_op}
\end{align}
which has the same form as the Landau expansion of the $Z_{12}$-clock
model. The ground-state is parametrized by
\begin{align}
\gamma_{\mathrm{op}}^{0} & =\frac{\pi}{12}\left(\frac{1+\mathrm{sign}v_{12}}{2}\right)+\frac{2\pi}{12}n, & n & \in\{0,1,\dots,11\}.\label{eq:angles_op_Z12_D12h}
\end{align}

The in-plane and out-of-plane nematic states given by (\ref{eq:angles_ip_Z6_D12h}),(\ref{eq:angles_op_Z12_D12h})
can be expressed in the $(\boldsymbol{n}\boldsymbol{m}\boldsymbol{l})$-representation
in a straightforward way via the relationships (\ref{eq:D_ih_ip_gs_nml})
and (\ref{eq:D_ih_op_gs_nml}). We show the corresponding Fermi surface
distortions in Figs.~\ref{fig:FS_octa_deca}(a),(b). For $\boldsymbol{d}^{\mathrm{ip}}$,
both in-plane nematic axes point along the high-symmetry directions
associated with twofold in-plane rotations, whereas the third axis
is parallel to the $k_{z}$-axis. In contrast, for $\boldsymbol{d}^{\mathrm{op}}$,
only the nematic axis $\boldsymbol{l}$ aligns with a high-symmetry
direction. 

The same results obtained for $\mathsf{D_{12h}}$ also hold for the
dodecagonal groups \{$\mathsf{D_{12}}$, $\mathsf{C_{12v}}$, $\mathsf{D_{6d}}$\}.
On the other hand, for \{$\mathsf{C_{12h}}$, $\mathsf{C_{12}}$,
$\mathsf{S_{12}}$\}, due to the lack of in-plane symmetry axes, the
nematic doublets transform as complex IRs, resulting in an offset
angle in the clock term of the type (\ref{eq:S_axial_clock_mod}),
similarly to what is shown in Appendices~\ref{sec:hexagonal_Appendix}-\ref{sec:tetragonal_Appendix}
for crystalline axial point groups.

\subsection{Decagonal quasicrystals\label{subsec:Decagonal-quasicrystals}}

The are seven non-crystallographic point groups that possess tenfold
symmetry, \{$\mathsf{D_{10h}}$, $\mathsf{D_{10}}$, $\mathsf{D_{5h}}$,
$\mathsf{C_{10v}}$, $\mathsf{C_{10h}}$, $\mathsf{C_{5h}}$, $\mathsf{C_{10}}$\}.
Similarly to the previous subsection, we need to distinguish the groups
\{$\mathsf{D_{10h}}$, $\mathsf{D_{10}}$, $\mathsf{D_{5h}}$, $\mathsf{C_{10v}}$\}
from the groups \{$\mathsf{C_{10h}}$, $\mathsf{C_{5h}}$, $\mathsf{C_{10}}$\},
as the latter do not have in-plane symmetry axes. The only effect
of this lack of in-plane rotational symmetry is that the clock term
of the nematic action acquires an offset angle, see Eq.~(\ref{eq:S_axial_clock_mod}). 

We consider here the case of $\mathsf{D_{10h}}$. Starting with the
in-plane doublet $\boldsymbol{d}^{\mathrm{ip}}$, which transforms
as the $E_{2g}$ IR, we obtain the symmetrized-product decomposition:

\begin{align}
\big[\otimes_{j=1}^{2}E_{2g}\big]_{s} & =A_{1g}\oplus E_{4g},\label{eq:E2g_symmprod_D10h_2}\\
\big[\otimes_{j=1}^{3}E_{2g}\big]_{s} & =E_{2g}\oplus E_{4g},\label{eq:E2g_symmprod_D10h_3}\\
\big[\otimes_{j=1}^{4}E_{2g}\big]_{s} & =A_{1g}\!\oplus E_{2g}\oplus E_{4g},\label{eq:E2g_symmprod_D10h_4}\\
\big[\otimes_{j=1}^{5}E_{2g}\big]_{s} & =A_{1g}\!\oplus A_{2g}\oplus E_{2g}\oplus E_{4g}.\label{eq:E2g_symmprod_D10h_5}
\end{align}
According to what was explained in the introduction of this section,
the fifth-order invariant must be the leading-order term in the Landau
expansion that is not proportional to the square of the doublet. Directly
employing a fifth-order transformation condition, as an extension
to Eq.~(\ref{eq:SO3_trafo_condition-1}), one finds this invariant
to be $\left|\boldsymbol{d}^{\mathrm{ip}}\right|^{5}\cos\left(5\gamma_{\mathrm{ip}}\right)$,
and thus the nematic action becomes:
\begin{align}
\mathcal{S}_{\mathrm{ip}} & =\int_{\mathsf{x}}\Big\{ r_{0}\left|\boldsymbol{d}^{\mathrm{ip}}\right|^{2}+u\left|\boldsymbol{d}^{\mathrm{ip}}\right|^{4}+v_{5}\left|\boldsymbol{d}^{\mathrm{ip}}\right|^{5}\cos\left(5\gamma_{\mathrm{ip}}\right)\Big\},\label{eq:S_D10h_ip}
\end{align}
which maps onto the $Z_{5}$-clock model. Minimization of the clock
term gives with the fivefold-degenerate ground state 
\begin{align}
\gamma_{\mathrm{ip}}^{0} & =\frac{\pi}{5}\left(\frac{1+\mathrm{sign}v_{5}}{2}\right)+\frac{2\pi}{5}n, & n & \in\{0,1,2,3,4\}.\label{eq:angles_ip_Z5_D10h}
\end{align}

Moving on to the out-of-plane doublet $\boldsymbol{d}^{\mathrm{op}}$,
we note that it transforms as the $E_{1g}$ IR, whose symmetrized
product decomposition is:
\begin{align}
\big[\otimes_{j=1}^{2}E_{1g}\big]_{s} & =A_{1g}\oplus E_{2g},\label{eq:E1g_symmprod_D10h_2}\\
\big[\otimes_{j=1}^{3}E_{1g}\big]_{s} & =E_{1g}\oplus E_{3g},\label{eq:E1g_symmprod_D10h_3}\\
\big[\otimes_{j=1}^{4}E_{1g}\big]_{s} & =A_{1g}\oplus E_{2g}\oplus E_{4g},\label{eq:E1g_symmprod_D10h_4}\\
\big[\otimes_{j=1}^{5}E_{1g}\big]_{s} & =B_{1g}\oplus B_{2g}\oplus E_{1g}\oplus E_{3g},\label{eq:E1g_symmprod_D10h_5}\\
\big[\otimes_{j=1}^{6}E_{1g}\big]_{s} & =A_{1g}\oplus E_{2g}\oplus2E_{4g}.\label{eq:E1g_symmprod_D10h_6}
\end{align}
Similarly to the case of the dodecagonal point group $\mathsf{D_{12h}}$,
only powers of $\left|\boldsymbol{d}^{\mathrm{op}}\right|^{2}$ occur
up to sixth-order. Following the same steps as in that case, we obtain
the clock term from the product of the fifth-order product-decomposition
elements $B_{1g}$ and $B_{2g}$: 
\begin{align*}
B_{1g}: & \left|\boldsymbol{d}^{\mathrm{op}}\right|^{5}\cos\left(5\gamma_{\mathrm{op}}\right), & B_{2g}: & \left|\boldsymbol{d}^{\mathrm{op}}\right|^{5}\sin\left(5\gamma_{\mathrm{op}}\right),
\end{align*}
yielding:
\begin{align}
\mathcal{S}_{\mathrm{op}} & =\int_{\mathsf{x}}\Big\{ r_{0}\left|\boldsymbol{d}^{\mathrm{op}}\right|^{2}+u\left|\boldsymbol{d}^{\mathrm{op}}\right|^{4}+v_{10}\left|\boldsymbol{d}^{\mathrm{op}}\right|^{10}\cos\left(10\gamma_{\mathrm{op}}\right)\Big\}.\label{eq:S_D10h_op}
\end{align}
We therefore find $\boldsymbol{d}^{\mathrm{op}}$ to be a $Z_{10}$-clock
nematic order parameter, whose tenfold degenerate ground states are
given by: 
\begin{align}
\gamma_{\mathrm{op}}^{0} & =\frac{\pi}{10}\left(\frac{1+\mathrm{sign}v_{10}}{2}\right)+\frac{2\pi}{10}n, & n & \in\{0,1,\dots,9\}.\label{eq:angles_op_Z10_D10h}
\end{align}

\begin{figure}
\begin{centering}
\includegraphics[width=1\columnwidth]{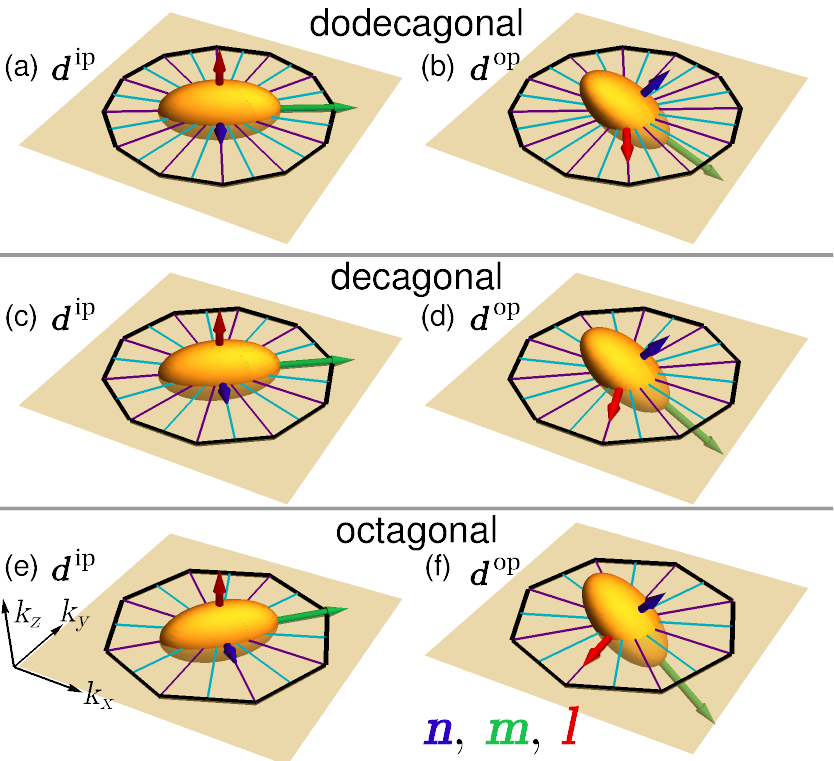}
\par\end{centering}
\caption{Distorted Fermi surface in the nematic phase, as given by Eq.~(\ref{eq:kF_khat}),
plotted together with the nematic axes $\boldsymbol{n}$,$\boldsymbol{m}$,$\boldsymbol{l}$
for the cases of dodecagonal in-plane and out-of-plane nematicity
(\ref{eq:angles_ip_Z6_D12h}),(\ref{eq:angles_op_Z12_D12h}) {[}panels
(a)-(b){]}, decagonal in-plane and out-of-plane nematicity (\ref{eq:angles_ip_Z5_D10h}),(\ref{eq:angles_op_Z10_D10h})
{[}panels (c)-(d){]}, and octagonal in-plane and out-of-plane nematicity
(\ref{eq:angles_ip_Z4_D8h}),(\ref{eq:S_D8h_op}) {[}panels (e)-(f){]}.
The unit vectors $\left\{ \boldsymbol{n},\boldsymbol{m},\boldsymbol{l}\right\} $
are rescaled by $2.25\,k_{F}\left(\left\{ \boldsymbol{n},\boldsymbol{m},\boldsymbol{l}\right\} \right)$,
respectively. We chose the trivial component such that $d_{1}/\left|\boldsymbol{d}\right|=0.1$.
\label{fig:FS_octa_deca}}
\end{figure}

In Figs.~\ref{fig:FS_octa_deca}(c)-(d), we show the distorted Fermi
surfaces obtained from expressing Eqs.~(\ref{eq:angles_ip_Z5_D10h}),(\ref{eq:angles_op_Z10_D10h})
in the $(\boldsymbol{n}\boldsymbol{m}\boldsymbol{l})$-representation.
The Fermi surface properties are analogous to the dodecagonal case,
except that the in-plane high-symmetry directions are those associated
with tenfold rotational symmetry.

\subsection{Octagonal quasicrystals\label{subsec:Octagonal-quasicrystals}}

The analysis of the octagonal point groups \{$\mathsf{D_{8h}}$, $\mathsf{D_{8}}$,
$\mathsf{C_{8v}}$, $\mathsf{D_{4d}}$, $\mathsf{C_{8h}}$, $\mathsf{C_{8}}$,
$\mathsf{S_{8}}$\} mirrors the analyses of the previous two subsections
for the dodecagonal and decagonal quasicrystals. Considering the point
group $\mathsf{D_{8h}}$ for concreteness, we note that $\boldsymbol{d}^{\mathrm{ip}}$
and $\boldsymbol{d}^{\mathrm{op}}$ transform as $E_{2g}$ and $E_{1g}$,
respectively. Performing the symmetrized product decomposition for
$E_{2g}$ gives:

\begin{align}
\big[\otimes_{j=1}^{2}E_{2g}\big]_{s} & =A_{1g}\oplus B_{1g}\oplus B_{2g},\label{eq:E2g_symmprod_octa2}\\
\big[\otimes_{j=1}^{3}E_{2g}\big]_{s} & =2E_{2g},\label{eq:E2g_symmprod_octa3}\\
\big[\otimes_{j=1}^{4}E_{2g}\big]_{s} & =2A_{1g}\!\oplus A_{2g}\oplus B_{1g}\oplus B_{2g}.\label{eq:E2g_symmprod_octa4}
\end{align}
Using the bilinears 
\begin{align}
D^{B_{1g}} & =\left|\boldsymbol{d}^{\mathrm{ip}}\right|^{2}\cos\left(2\gamma_{\mathrm{ip}}\right), & D^{B_{2g}} & =\left|\boldsymbol{d}^{\mathrm{ip}}\right|^{2}\sin\left(2\gamma_{\mathrm{ip}}\right),\label{eq:D8h_ip_bilinears}
\end{align}
it is straightforward to construct the three invariants up to quartic
order, $\left|\boldsymbol{d}^{\mathrm{ip}}\right|^{2}$, $\left|\boldsymbol{d}^{\mathrm{ip}}\right|^{4}$
and $(D^{B_{1g}})^{2}-(D^{B_{2g}})^{2}$. We therefore obtain the
Landau expansion of the $Z_{4}$-clock model:
\begin{align}
\mathcal{S}_{\mathrm{ip}} & =\int_{\mathsf{x}}\Big\{ r_{0}\left|\boldsymbol{d}^{\mathrm{ip}}\right|^{2}+u\left|\boldsymbol{d}^{\mathrm{ip}}\right|^{4}+v_{4}\left|\boldsymbol{d}^{\mathrm{ip}}\right|^{4}\cos\left(4\gamma_{\mathrm{ip}}\right)\Big\},\label{eq:S_D8h_ip}
\end{align}
which has the fourfold-degenerate ground state: 
\begin{align}
\gamma_{\mathrm{ip}}^{0} & =\frac{\pi}{4}\left(\frac{1+\mathrm{sign}v_{4}}{2}\right)+\frac{2\pi}{4}n, & n & \in\{0,1,2,3\},\label{eq:angles_ip_Z4_D8h}
\end{align}
see also Eq.~(\ref{eq:S_D4h_op}). 

Considering now the out-of-plane doublet $\boldsymbol{d}^{\mathrm{op}}$,
the symmetrized product decomposition of $E_{1g}$ to sixth-order
only gives invariants that are powers of $\left|\boldsymbol{d}^{\mathrm{op}}\right|^{2}$:
\begin{align}
\big[\otimes_{j=1}^{2}E_{1g}\big]_{s} & =A_{1g}\oplus E_{2g},\label{eq:E1g_symmprod_octa2}\\
\big[\otimes_{j=1}^{3}E_{1g}\big]_{s} & =E_{1g}\oplus E_{3g},\label{eq:E1g_symmprod_octa3}\\
\big[\otimes_{j=1}^{4}E_{1g}\big]_{s} & =A_{1g}\oplus B_{1g}\oplus B_{2g}\oplus E_{2g},\label{eq:E1g_symmprod_octa4}\\
\big[\otimes_{j=1}^{5}E_{1g}\big]_{s} & =E_{1g}\oplus2E_{3g},\label{eq:E1g_symmprod_octa5}\\
\big[\otimes_{j=1}^{6}E_{1g}\big]_{s} & =A_{1g}\oplus B_{1g}\oplus B_{2g}\oplus2E_{2g}.\label{eq:E1g_symmprod_octa6}
\end{align}
Analogously to the case of the dodecagonal and decagonal quasicrystals,
it is straightforward to find the leading-order anisotropic term,
this time from the product of the quartic combinations: 
\begin{align*}
B_{1g}: & \left|\boldsymbol{d}^{\mathrm{op}}\right|^{4}\cos\left(4\gamma_{\mathrm{op}}\right), & B_{2g}: & \left|\boldsymbol{d}^{\mathrm{op}}\right|^{4}\sin\left(4\gamma_{\mathrm{op}}\right).
\end{align*}
Hence, we find the nematic action 
\begin{align}
\mathcal{S}_{\mathrm{op}} & =\int_{\mathsf{x}}\Big\{ r_{0}\left|\boldsymbol{d}^{\mathrm{op}}\right|^{2}+u\left|\boldsymbol{d}^{\mathrm{op}}\right|^{4}+v_{8}\left|\boldsymbol{d}^{\mathrm{op}}\right|^{8}\cos\left(8\gamma_{\mathrm{op}}\right)\Big\},\label{eq:S_D8h_op}
\end{align}
corresponding to the $Z_{8}$-clock model. The directions of $\boldsymbol{d}^{\mathrm{op}}$
that minimize the action are eightfold degenerate:
\begin{align}
\gamma_{\mathrm{op}}^{0} & =\frac{\pi}{8}\left(\frac{1+\mathrm{sign}v_{8}}{2}\right)+\frac{2\pi}{8}n, & n & \in\{0,1,\dots,7\}.\label{eq:angles_op_Z8_D8h}
\end{align}
The same results hold for \{$\mathsf{D_{8}}$, $\mathsf{C_{8v}}$,
$\mathsf{D_{4d}}$\}, whereas in the groups \{$\mathsf{C_{8h}}$,
$\mathsf{C_{8}}$, $\mathsf{S_{8}}$\}, the clock terms in the nematic
actions of $\boldsymbol{d}^{\mathrm{ip}}$ and $\boldsymbol{d}^{\mathrm{op}}$
acquire an offset angle of the form (\ref{eq:S_axial_clock_mod}). 

The $(\boldsymbol{n}\boldsymbol{m}\boldsymbol{l})$-representation
of the in-plane and out-of-plane nematic doublets (\ref{eq:angles_ip_Z4_D8h})
and (\ref{eq:angles_op_Z8_D8h}) is represented in Figs.~\ref{fig:FS_octa_deca}(e)-(f)
via the corresponding distortions of the Fermi surface. The high-symmetry
in-plane directions associated with the eightfold symmetry of the
quasicrystal are highlighted in the figure.

\subsection{Twisted quasicrystals \label{subsec:twisted-quasicrystals}}

The rapid advances in the field of twistronics opened a new path to
investigate the properties of emergent quasicrystals created by twisting
2D materials. In some cases, the twisted structure is an incommensurate
lattice that, however, retains a crystallographic point group. This
is the case of twisted bilayer graphene with a non-commensurate twist
angle, although the effects of quasi-periodicity are dramatically
enhanced when three graphene layers are twisted by two different angles,
as shown recently in Ref.~\citep{Uri2023}. Interestingly, this work
reported the emergence of robust superconductivity in this quasi-periodic
structure. In what concerns electronic nematicity, as long as the
point group is the same as a crystallographic one, the results derived
in Sec.~\ref{sec:Nematicity-Axial} can be directly applied.

An alternative route to realize twisted quasicrystals, as shown in
Refs.~\citep{Yang2023_a,Yang2023_b}, is to start with two identical
2D materials described by the (planar) crystallographic point group
$\mathsf{D_{n}}$ (with $n=2,3,4,\,6$). Upon twisting the two crystals
by a relative angle of $\pi/n$ with respect to their common $z$-axis,
the resulting structure becomes invariant under an $n$-fold rotation
about $z$ followed by a horizontal mirror reflection (which maps
one layer onto the other). This is nothing but the improper rotation
element $S_{2nz}$. Comparing Eqs.~(\ref{eq:group_Dn}) and (\ref{eq:group_Dnd}),
we conclude that the twisted bilayer is actually a quasi-periodic
``lattice'' with point group $\mathsf{D_{nd}}$, i.e. $\mathsf{D_{n}}\overset{\mathrm{twist}}{\longrightarrow}\mathsf{D_{nd}}$.
Experimentally, both twisted bilayer graphene with a twist angle of
$30^{\circ}$ and twisted bilayer cuprate with a twist angle of $45^{\circ}$
have been realized \citep{Ahn2018,Kim2023}. 

While much of the recent interest in these constructions have focused
on realizing time-reversal symmetry-breaking superconductivity \citep{Haenel2022,Yang2023_a,Yang2023_b},
it has been recently pointed out that electronic nematicity can be
strongly impacted by the enhanced symmetry of the twisted bilayer
\citep{Gali2024}. Here, we further explore this idea by discussing
the properties of the in-plane nematic order parameter $\boldsymbol{d}^{\mathrm{ip}}$
in the four cases of $\pi/n$-twisted bilayers ($n=2,3,4,\,6$). 

The point group in the cases $n=2$ and $n=3$ remain crystallographic,
$\mathsf{D_{2}\overset{\mathrm{twist}}{\longrightarrow}\mathsf{D_{2d}}}$
and $\mathsf{D_{3}\overset{\mathrm{twist}}{\longrightarrow}\mathsf{D_{3d}}}$.
There is, however, one important difference. While $\mathsf{D_{3}}$
and $\mathsf{D_{3d}}$ both belong to the trigonal crystal system,
$\mathsf{D_{2}}$ and $\mathsf{D_{2d}}$ belong to different systems
\textendash{} monoclinic and orthorhombic, respectively. In terms
of the components of $\boldsymbol{d}^{\mathrm{ip}}=\left(d_{2},\,d_{5}\right)^{T}$,
this means that while in $\mathsf{D_{2}}$ only the shear component
$d_{5}$ transforms non-trivially, in $\mathsf{D_{2d}}$ both components
transform non-trivially as two different one-dimensional IRs \textendash{}
see Table~\ref{tab:classification}. Therefore, twisting two monoclinic
layers by $90^{\circ}$ opens up a new nematic instability channel. 

For two tetragonal layers twisted by $45^{\circ}$ ($n=4$), the twisted
bilayer has a non-crystallographic octagonal point group, $\mathsf{D_{4}\overset{\mathrm{twist}}{\longrightarrow}\mathsf{D_{4d}}}$
\citep{Haenel2022}. Using the results above, we see that the two
components of the in-plane nematic doublet $\boldsymbol{d}^{\mathrm{ip}}$
change from transforming as the one-dimensional IRs $B_{1}$ and $B_{2}$
of $\mathsf{D_{4}}$ in each layer to transforming as a single two-dimensional
IR $E_{2}$ of $\mathsf{D_{4d}}$ in the twisted bilayer. Therefore,
the character of the nematic transition changes from $Z_{2}$-Ising
for the individual tetragonal layers to $Z_{4}$-clock for the coupled
$45^{\circ}$-twisted bilayer. As discussed in the end of Sec.~\ref{subsec:Dihedral-group_Dih},
the 2D $4$-state clock model (which has the same properties as the
2D Ashkin-Teller model) undergoes a transition from the disordered
to the ordered phase that can be understood as the merging of the
two BKT transitions of the $q$-state clock model as $q\rightarrow4^{+}$.
Because of its unique character, the exponents of this transition
are non-universal and depend on $g_{\mathrm{ip}}$, except for the
anomalous exponent $\eta=1/4$. Since electronic nematicity is observed
in tetragonal cuprates and iron pnictides/chalcogenides \citep{Fradkin2010,Fernandes2014},
this setting offers an interesting path to realize Ashkin-Teller nematicity.

Finally, the case of two hexagonal layers twisted by $30^{\circ}$
($n=6$) corresponds to $\mathsf{D_{6}\overset{\mathrm{twist}}{\longrightarrow}\mathsf{D_{6d}}}$
and was investigated in Ref.~\citep{Gali2024}. The outcome, which
follows from the nematic properties of dodecagonal quasicrystals,
is that while the uncoupled layers undergo a second-order nematic
transition in the 2D $3$-state clock/Potts universality class (see
Table~\ref{tab:classification}), the coupled $30^{\circ}$-twisted
bilayer undergoes two BKT transitions: the higher one from the disordered
to the critical phase (where there is only quasi-long-range nematic
order) and the lower one from the critical phase to the long-range
ordered nematic phase.

\begin{table*}[t!]  
\setlength{\tabcolsep}{2.1pt}  	 
\setlength{\arrayrulewidth}{.1em} 	 
\renewcommand{\arraystretch}{1.65}  	 
\newcommand{\colA}[1]{\multicolumn{1}{ >{\columncolor{TabGray!100}[\tabcolsep]} c}{#1}}   
\newcommand{\colB}[1]{\multicolumn{1}{ >{\columncolor{TabCyan!100}[\tabcolsep]} c}{#1}}   
\newcommand{\colC}[1]{\multicolumn{1}{ >{\columncolor{TabMagenta!100}[\tabcolsep]} c}{#1}}   
\newcommand{\colD}[1]{\multicolumn{1}{ >{\columncolor{colorD!100}[\tabcolsep]} c}{#1}}
\begin{tabular}{cccccc}  
\hline  
\colA{cubic} & \colA{IR} & \colA{nem.$\!$ basis} & \colA{universality} & \colA{res.$\!$ PG} & \colA{transf.$\!$ matrix} 
\\[0.1em]  \hline    
\colC{$\mathsf{O_{h}}$} & \colC{$E_{g}$} & \colC{$\boldsymbol{d}^{e}$} & \colC{$Z_{3}$-Potts} & \colC{$\mathsf{D_{4h}}$} & \colC{}  
\\[0.1em]  \hline 
\colC{} & \colC{$T_{2g}$} & \colC{$\boldsymbol{d}^{t}$} & \colC{$Z_{4}$-Potts} & \colC{$\mathsf{D_{3d}}$} &  \colC{$\mathcal{R}^{A_{2u}}_g\mathcal{R}^{(3)}_g$} 
\\[0.1em]  \hline 

\colB{$\mathsf{O}$} & \colB{$E$} & \colB{$\boldsymbol{d}^{e}$} & \colB{$Z_{3}$-Potts} & \colB{$\mathsf{D_{4}}$} & \colB{}  
\\[0.1em]  \hline    

\colB{} & \colB{$T_{2}$} & \colB{$\boldsymbol{d}^{t}$} & \colB{$Z_{4}$-Potts} & \colB{$\mathsf{D_{3}}$} & \colB{$\mathcal{R}^{A_{2}}_g \mathcal{R}^{(3)}_g$} 
\\[0.1em]  \hline 

\colC{$\mathsf{T_{d}}$} & \colC{$E$} & \colC{$\boldsymbol{d}^{e}$} & \colC{$Z_{3}$-Potts} & \colC{$\mathsf{D_{2d}}$} & \colC{}  
\\[0.1em]  \hline 

\colC{} & \colC{$T_{2}$} & \colC{$\boldsymbol{d}^{t}$} & \colC{$Z_{4}$-Potts} & \colC{$\mathsf{C_{3v}}$} & \colC{$\mathcal{R}^{(3)}_g$} 
\\[0.1em]  \hline 

\colB{$\mathsf{T_{h}}$} & \colB{$\bar{E}_{g}$} & \colB{$\boldsymbol{d}^{e}$} & \colB{$Z_{3}^{*}$-Potts} & \colB{$\mathsf{D_{2h}}$} & \colB{}  
\\[0.1em]  \hline    

\colB{} & \colB{$T_{g}$} & \colB{$\boldsymbol{d}^{t}$} & \colB{$Z_{4}$-Potts} & \colB{$\mathsf{S_{6}}$} & \colB{$\mathcal{R}^{A_{u}}_g \mathcal{R}^{(3)}_g$} 
\\[0.1em]  \hline 

\colC{$\mathsf{T}$} & \colC{$\bar{E}$} & \colC{$\boldsymbol{d}^{e}$} & \colC{$Z_{3}^{*}$-Potts} & \colC{$\mathsf{D_{2}}$} & \colC{}  
\\[0.1em]  \hline 

\colC{} & \colC{$T$} & \colC{$\boldsymbol{d}^{t}$} & \colC{$Z_{4}$-Potts} & \colC{$\mathsf{C_{3}}$} & \colC{$\mathcal{R}^{(3)}_g$} 

\\[0.1em]  \hline  	   
\colA{hexagonal} & \colA{IR} & \colA{nem.$\!$ basis} & \colA{universality} & \colA{res.$\!$ PG} & \colA{transf.$\!$ matrix} 
\\[0.1em]  \hline 

\colB{$\mathsf{D_{6h}}$ } &  \colB{$E_{2g}$ } & \colB{ $\boldsymbol{d}^{\mathrm{ip}}$ } & \colB{ $Z_{3}$-clock } & \colB{ $\mathsf{D_{2h}}$ } & \colB{ $\mathcal{R}_{g}^{B_{1u}}\sigma^{z}\mathcal{R}_{g}^{(2)}\sigma^{z}$ } 
\\[0.1em]  \hline    \colB{ } & \colB{ $E_{1g}$ } & \colB{ $\boldsymbol{d}^{\mathrm{op}}$ } & \colB{ $Z_{6}$-clock } & \colB{ $\mathsf{C_{2h}}$ } & \colB{$\mathcal{R}_{g}^{A_{1u}}\sigma^{z}\mathcal{R}_{g}^{(2)}\sigma^{z}$} 
\\[0.1em]  \hline    

\colC{$\mathsf{D_{6}}$ } & \colC{$E_{2}$} & \colC{$\boldsymbol{d}^{\mathrm{ip}}$} & \colC{$Z_{3}$-clock}  & \colC{$\mathsf{D_{2}}$} & \colC{$\mathcal{R}_{g}^{B_{1}}\sigma^{z}\mathcal{R}_{g}^{(2)}\sigma^{z}$} 
\\[0.1em]  \hline 

\colC{} & \colC{$E_{1}$} & \colC{$\boldsymbol{d}^{\mathrm{op}}$} & \colC{$Z_{6}$-clock} & \colC{$\mathsf{C_{2}}$} & \colC{$\sigma^{z}\mathcal{R}_{g}^{(2)}\sigma^{z}$} 
\\[0.1em]  \hline 

\colB{$\mathsf{D_{3h}}$ } & \colB{$E^{\prime}$} & \colB{$\boldsymbol{d}^{\mathrm{ip}}$} & \colB{$Z_{3}$-clock} & \colB{$\mathsf{C_{2v}}$} & \colB{$\sigma^{z}\mathcal{R}_{g}^{(2)}\sigma^{z}$} 
\\[0.1em]  \hline    

\colB{} & \colB{$E^{\prime\prime}$} & \colB{$\boldsymbol{d}^{\mathrm{op}}$} & \colB{$Z_{6}$-clock} & \colB{$\mathsf{C_{2}}$} & \colB{$\mathcal{R}_{g}^{A_{1}^{\prime\prime}}\sigma^{z}\mathcal{R}_{g}^{(2)}\sigma^{z}$} 
\\[0.1em]  \hline 

\colC{$\mathsf{C_{6v}}$} & \colC{$E_{2}$} & \colC{$\boldsymbol{d}^{\mathrm{ip}}$} & \colC{$Z_{3}$-clock} & \colC{$\mathsf{C_{2v}}$} & \colC{$\mathcal{R}_{g}^{B_{2}}\sigma^{x}\mathcal{R}_{g}^{(2)}\sigma^{x}$} 
\\[0.1em]  \hline    

\colC{} & \colC{$E_{1}$} & \colC{$\boldsymbol{d}^{\mathrm{op}}$} & \colC{$Z_{6}$-clock} & \colC{$\mathsf{C_{s}}$} & \colC{$\sigma^{x}\mathcal{R}_{g}^{(2)}\sigma^{x}$} 
\\[0.1em]  \hline 

\colB{ $\mathsf{C_{6h}}$} & \colB{$\bar{E}_{2g}$} & \colB{$\boldsymbol{d}^{\mathrm{ip}}$} & \colB{$Z_{3}^{*}$-clock} & \colB{$\mathsf{C_{2h}}$} & \colB{$\mathcal{R}_{g}^{B_{u}}\sigma^{z}\mathcal{R}_{g}^{(2)}\sigma^{z}$} 
\\[0.1em]  \hline 

\colB{} & \colB{$\bar{E}_{1g}$} & \colB{$\boldsymbol{d}^{\mathrm{op}}$} & \colB{$Z_{6}^{*}$-clock} & \colB{$\mathsf{C_{i}}$} & \colB{$\mathcal{R}_{g}^{A_{u}}\sigma^{z}\mathcal{R}_{g}^{(2)}\sigma^{z}$} 
\\[0.1em]  \hline    

\colC{$\mathsf{C_{3h}}$} & \colC{$\bar{E}^{\prime}$} & \colC{$\boldsymbol{d}^{\mathrm{ip}}$} & \colC{$Z_{3}^{*}$-clock} & \colC{$\mathsf{C_{s}}$} & \colC{$\sigma^{z}\mathcal{R}_{g}^{(2)}\sigma^{z}$} 
\\[0.1em]  \hline    

\colC{} & \colC{$\bar{E}^{\prime\prime}$} & \colC{$\boldsymbol{d}^{\mathrm{op}}$} & \colC{$Z_{6}^{*}$-clock} & \colC{$\mathsf{C_{1}}$} & \colC{$\mathcal{R}_{g}^{A^{\prime\prime}}\sigma^{z}\mathcal{R}_{g}^{(2)}\sigma^{z}$} 
\\[0.1em]  \hline    

\colB{$\mathsf{C_{6}}$} & \colB{$\bar{E}_{2}$} & \colB{$\boldsymbol{d}^{\mathrm{ip}}$} & \colB{$Z_{3}^{*}$-clock} & \colB{$\mathsf{C_{2}}$} & \colB{$\mathcal{R}_{g}^{B}\sigma^{z}\mathcal{R}_{g}^{(2)}\sigma^{z}$} 
\\[0.1em]  \hline    

\colB{} & \colB{$\bar{E}_{1}$} & \colB{$\boldsymbol{d}^{\mathrm{op}}$} & \colB{$Z_{6}^{*}$-clock} & \colB{$\mathsf{C_{1}}$} & \colB{$\sigma^{z}\mathcal{R}_{g}^{(2)}\sigma^{z}$} 
\\[0.1em]  \hline

\colA{trigonal} & \colA{IR} & \colA{nem.$\!$ basis} & \colA{universality} & \colA{res.$\!$ PG} & \colA{transf.$\!$ matrix} 
\\[0.1em]  \hline    

\colC{$\mathsf{D_{3d}}$} & \colC{$E_{g}$} & \colC{$\left\{ \boldsymbol{d}^{\mathrm{ip}},\boldsymbol{d}^{\mathrm{op}}\right\} $} & \colC{$Z_{3}$-clock} & \colC{$\mathsf{C_{2h}}$} & \colC{$\mathcal{R}_{g}^{A_{1u}}\sigma^{z}\mathcal{R}_{g}^{(2)}\sigma^{z}$} 
\\[0.1em]  \hline    

\colB{$\mathsf{D_{3}}$} & \colB{$E$} & \colB{$\left\{ \boldsymbol{d}^{\mathrm{ip}},\boldsymbol{d}^{\mathrm{op}}\right\} $} & \colB{$Z_{3}$-clock} & \colB{$\mathsf{C_{2}}$} & \colB{$\sigma^{z}\mathcal{R}_{g}^{(2)}\sigma^{z}$} 
\\[0.1em]  \hline    

\colC{$\mathsf{C_{3v}}$} & \colC{$E$} & \colC{$\left\{ \boldsymbol{d}^{\mathrm{ip}},\boldsymbol{d}^{\mathrm{op}}\right\} $} & \colC{$Z_{3}$-clock} & \colC{$\mathsf{C_{s}}$} & \colC{$\sigma^{x}\mathcal{R}_{g}^{(2)}\sigma^{x}$} 
\\[0.1em]  \hline 

\colB{$\mathsf{S_{6}}$} & \colB{$\bar{E}_{g}$} & \colB{$\left\{ \boldsymbol{d}^{\mathrm{ip}},\boldsymbol{d}^{\mathrm{op}}\right\} $} & \colB{$Z_{3}^{*}$-clock} & \colB{$\mathsf{C_{i}}$} & \colB{$\mathcal{R}_{g}^{A_{u}}\sigma^{z}\mathcal{R}_{g}^{(2)}\sigma^{z}$} 
\\[0.1em]  \hline    

\colC{$\mathsf{C_{3}}$} & \colC{$\bar{E}$} & \colC{$\left\{ \boldsymbol{d}^{\mathrm{ip}},\boldsymbol{d}^{\mathrm{op}}\right\} $} & \colC{$Z_{3}^{*}$-clock} & \colC{$\mathsf{C_{1}}$} & \colC{$\sigma^{z}\mathcal{R}_{g}^{(2)}\sigma^{z}$} 
\\[0.1em]  \hline 
 
\end{tabular} 
$\hfill$ 
\setlength{\tabcolsep}{1.6pt}  	 
\renewcommand{\arraystretch}{1.59}  	 
\begin{tabular}{cccccc} 	
\hline  	
\colA{tetragonal} & \colA{IR} & \colA{nem.$\!$ basis} & \colA{universality} & \colA{res.$\!$ PG} & \colA{transf.$\!$ matrix} 	
\\[0.1em]  \hline 

	\colB{$\mathsf{D_{4h}}$} & \colB{$B_{1g}$} & \colB{$d_{x^{2}\text{-}y^{2}}$} & \colB{$Z_{2}$-Ising} & \colB{$\mathsf{D_{2h}}$} & \colB{}  	
\\[0.1em]  \hline 

	\colB{} & \colB{$B_{2g}$} & \colB{$d_{2xy}$} & \colB{$Z_{2}$-Ising} & \colB{$\mathsf{D_{2h}}$} & \colB{}  	
\\[0.1em]  \hline 

	\colB{} & \colB{$E_{g}$} & \colB{$\boldsymbol{d}^{\mathrm{op}}$} & \colB{$Z_{4}$-clock} & \colB{$\mathsf{C_{2h}}$} & \colB{$\mathcal{R}_{g}^{A_{1u}}\sigma^{z}\mathcal{R}_{g}^{(2)}\sigma^{z}$} 	
\\[0.1em]  \hline 

	\colC{$\mathsf{D_{4}}$} & \colC{$B_{1}$} & \colC{$d_{x^{2}\text{-}y^{2}}$} & \colC{$Z_{2}$-Ising} & \colC{$\mathsf{D_{2}}$} & \colC{}  	
\\[0.1em]  \hline 

	\colC{} & \colC{$B_{2}$} & \colC{$d_{2xy}$} & \colC{$Z_{2}$-Ising} & \colC{$\mathsf{D_{2}}$} & \colC{} 	
\\[0.1em]  \hline 

	\colC{} & \colC{$E$} & \colC{$\boldsymbol{d}^{\mathrm{op}}$} & \colC{$Z_{4}$-clock} & \colC{$\mathsf{C_{2}}$} & \colC{$\sigma^{z}\mathcal{R}_{g}^{(2)}\sigma^{z}$}
\\[0.1em]  \hline 

	\colB{$\mathsf{C_{4v}}$} & \colB{$B_{1}$} & \colB{$d_{x^{2}\text{-}y^{2}}$} & \colB{$Z_{2}$-Ising} & \colB{$\mathsf{C_{2v}}$} & \colB{} 	
\\[0.1em]  \hline 

	\colB{} & \colB{$B_{2}$} & \colB{$d_{2xy}$} & \colB{$Z_{2}$-Ising} & \colB{$\mathsf{C_{2v}}$} & \colB{} 	
\\[0.1em]  \hline 

	\colB{} & \colB{$E$} & \colB{$\boldsymbol{d}^{\mathrm{op}}$} & \colB{$Z_{4}$-clock} & \colB{$\mathsf{C_{s}}$} & \colB{$\sigma^{x}\mathcal{R}_{g}^{(2)}\sigma^{x}$} 	\\[0.1em]  \hline 
	\colC{$\mathsf{D_{2d}}$} & \colC{$B_{1}$} & \colC{$d_{x^{2}\text{-}y^{2}}$} & \colC{$Z_{2}$-Ising} & \colC{$\mathsf{D_{2}}$} & \colC{}  	
\\[0.1em]  \hline 

	\colC{} & \colC{$B_{2}$} & \colC{$d_{2xy}$} & \colC{$Z_{2}$-Ising} & \colC{$\mathsf{C_{2v}}$} & \colC{}  	
\\[0.1em]  \hline 

	\colC{} & \colC{$E$} & \colC{$\boldsymbol{d}^{\mathrm{op}}$} & \colC{$Z_{4}$-clock} & \colC{$C_{2}/\mathsf{C_{s}}$} & \colC{$\mathcal{R}_{g}^{(2)}$} 	
\\[0.1em]  \hline 

	\colB{$\mathsf{C_{4h}}$} & \colB{$B_{g}$} & \colB{$\big\{ d_{x^{2}\text{-}y^{2}},d_{2xy}\big\} $} & \colB{$Z^*_{2}$-Ising} & \colB{$\mathsf{C_{2h}}$} & \colB{} 	
\\[0.1em]  \hline 

	\colB{} & \colB{$\bar{E}_{g}$} & \colB{$\boldsymbol{d}^{\mathrm{op}}$} & \colB{$Z_{4}^{*}$-clock} & \colB{$\mathsf{C_{i}}$} & \colB{$\mathcal{R}_{g}^{A_{u}}\sigma^{x}\mathcal{R}_{g}^{(2)}\sigma^{x}$} 	
\\[0.1em]  \hline 

	\colC{$\mathsf{C_{4}}$} & \colC{$B$} & \colC{$\big\{ d_{x^{2}\text{-}y^{2}},d_{2xy}\big\} $} & \colC{$Z^*_{2}$-Ising} & \colC{$\mathsf{C_{2}}$} & \colC{} 	
\\[0.1em]  \hline 

	\colC{} & \colC{$\bar{E}$} & \colC{$\boldsymbol{d}^{\mathrm{op}}$} & \colC{$Z_{4}^{*}$-clock} & \colC{$\mathsf{C_{1}}$} & \colC{$\sigma^{x}\mathcal{R}_{g}^{(2)}\sigma^{x}$} 	
\\[0.1em]  \hline 

	\colB{$\mathsf{S_{4}}$} & \colB{$B$} & \colB{$\big\{ d_{x^{2}\text{-}y^{2}},d_{2xy}\big\} $} & \colB{$Z^*_{2}$-Ising} & \colB{$\mathsf{C_{2}}$} & \colB{}  	
\\[0.1em]  \hline 

	\colB{} & \colB{$\bar{E}$} & \colB{$\boldsymbol{d}^{\mathrm{op}}$} & \colB{$Z_{4}^{*}$-clock} & \colB{$\mathsf{C_{1}}$} & \colB{$\mathcal{R}_{g}^{(2)}$} 	
\\[0.1em]  \hline 

	\colA{orthorombic} & \colA{IR} & \colA{nem.$\!$ basis} & \colA{universality} & \colA{res.$\!$ PG} & \colA{transf.$\!$ matrix} 	
\\[0.1em]  \hline 
	\colC{$\mathsf{D_{2h}}$} & \colC{$B_{1g}$} & \colC{$d_{2xy}$} & \colC{$Z_{2}$-Ising} & \colC{$\mathsf{C_{2h}}$} & \colC{} 	
\\[0.1em]  \hline 
	\colC{} & \colC{$B_{2g}$} & \colC{$d_{2xz}$} & \colC{$Z_{2}$-Ising} & \colC{$\mathsf{C_{2h}}$} & \colC{} 	
\\[0.1em]  \hline 
	\colC{} & \colC{$B_{3g}$} & \colC{$d_{2yz}$} & \colC{$Z_{2}$-Ising} & \colC{$\mathsf{C_{2h}}$} & \colC{} 	
\\[0.1em]  \hline 
	\colB{$\mathsf{D_{2}}$} & \colB{$B_{1}$} & \colB{$d_{2xy}$} & \colB{$Z_{2}$-Ising} & \colB{$\mathsf{C_{2}}$} & \colB{} 	
\\[0.1em]  \hline 
	\colB{} & \colB{$B_{2}$} & \colB{$d_{2xz}$} & \colB{$Z_{2}$-Ising} & \colB{$\mathsf{C_{2}}$} & \colB{} 	
\\[0.1em]  \hline 
	\colB{} & \colB{$B_{3}$} & \colB{$d_{2yz}$} & \colB{$Z_{2}$-Ising} & \colB{$\mathsf{C_{2}}$} & \colB{} 	
\\[0.1em]  \hline 
	\colC{$\mathsf{C_{2v}}$} & \colC{$A_{2}$} & \colC{$d_{2xy}$} & \colC{$Z_{2}$-Ising} & \colC{$\mathsf{C_{2}}$} & \colC{} 	
\\[0.1em]  \hline 
	\colC{} & \colC{$B_{1}$} & \colC{$d_{2xz}$} & \colC{$Z_{2}$-Ising} & \colC{$\mathsf{C_{s}}$} & \colC{} 	
\\[0.1em]  \hline 
	\colC{} & \colC{$B_{2}$} & \colC{$d_{2yz}$} & \colC{$Z_{2}$-Ising} & \colC{$\mathsf{C_{s}}$} & \colC{} 	
\\[0.1em]  \hline 
	\colA{monoclinic} & \colA{IR} & \colA{nem.$\!$ basis} & \colA{universality} & \colA{res.$\!$ PG} & \colA{transf.$\!$ matrix} 
	\\[0.1em]  \hline 
	\colB{$\mathsf{C_{2h}}$} & \colB{$B_{g}$} & \colB{$\left\{ d_{2xz},d_{2yz}\right\} $} & \colB{$Z_{2}$-Ising} & \colB{$\mathsf{C_{i}}$} & \colB{} 	
\\[0.1em]  \hline 
	\colC{$\mathsf{C_{1h}}\equiv\mathsf{C_{s}}$} & \colC{$A^{\prime\prime}$} & \colC{$\left\{ d_{2xz},d_{2yz}\right\} $} & \colC{$Z_{2}$-Ising} &  \colC{$\mathsf{C_{1}}$} & \colC{} 	
\\[0.1em]  \hline 
	\colB{$\mathsf{C_{2}}$} & \colB{$B$} & \colB{$\left\{ d_{2xz},d_{2yz}\right\} $} & \colB{$Z_{2}$-Ising} & \colB{$\mathsf{C_{1}}$} & \colB{} 	
\\[0.1em]  \hline 

\end{tabular}
\caption{Symmetry properties of the allowed nematic order parameters in all crystallographic point groups (PG).  The columns list the corresponding irreducible representation (IR),  the relevant nematic basis vector  (see main text for the definition),  the universality class of the corresponding Landau expansion,  the residual PG,  and the transformation matrices (see main text).  Note that the $Z_3$-clock and $Z_3$-Potts models have the same properties.  The $Z_{q}^*$-clock model has the same form as the $Z_q$-clock model, except that the clock term has a non-universal offset angle.  The overbar over the IR denotes a complex IR.\label{tab:classification}}  \end{table*}

\begin{table*}[t] 
\setlength{\tabcolsep}{2.1pt}  	  
\setlength{\arrayrulewidth}{.1em} 	  
\renewcommand{\arraystretch}{1.65}     	  
\newcommand{\colA}[1]{\multicolumn{1}{ >{\columncolor{TabGray!100}[\tabcolsep]} c}{#1}}    \newcommand{\colB}[1]{\multicolumn{1}{ >{\columncolor{TabCyan!100}[\tabcolsep]} c}{#1}}    \newcommand{\colC}[1]{\multicolumn{1}{ >{\columncolor{TabMagenta!100}[\tabcolsep]} c}{#1}}    
\newcommand{\colD}[1]{\multicolumn{1}{ >{\columncolor{TabGray!40}[\tabcolsep]} c}{#1}} 

\begin{tabular}{cccccc} 
\hline 
\colA{octagonal} & \colA{IR} & \colA{nem.$\!$ basis} & \colA{universality} & \colA{res.$\!$ PG} & \colA{transf.$\!$ matrix}  
\\[0.1em]  
\hline 
\colC{$\mathsf{D_{8h}}$} & \colC{$E_{2g}$} & \colC{$\boldsymbol{d}^{\mathrm{ip}}$} & \colC{$Z_{4}$-clock} & \colC{$\mathsf{D_{2h}}$} & \colC{} 
\\[0.1em]  \hline     
\colC{} & \colC{$E_{1g}$} & \colC{$\boldsymbol{d}^{\mathrm{op}}$} & \colC{$Z_{8}$-clock} & \colC{$\mathsf{C_{2h}}$} & \colC{$\mathcal{R}_{g}^{A_{1u}}\sigma^{z}\mathcal{R}_{g}^{(2)}\sigma^{z}$}

\\[0.1em]  \hline    
\colB{$\mathsf{D_{8}}$} & \colB{$E_{2}$} & \colB{$\boldsymbol{d}^{\mathrm{ip}}$} & \colB{$Z_{4}$-clock} & \colB{$\mathsf{D_{2}}$} &\colB{}  
\\[0.1em]  \hline    
\colB{} & \colB{$E_{1}$} & \colB{$\boldsymbol{d}^{\mathrm{op}}$} & \colB{$Z_{8}$-clock} & \colB{$\mathsf{C_{2}}$} & \colB{$\sigma^{z}\mathcal{R}_{g}^{(2)}\sigma^{z}$}   
\\[0.1em]  \hline     
\colC{$\mathsf{C_{8v}}$} & \colC{$E_{2}$} & \colC{$\boldsymbol{d}^{\mathrm{ip}}$} & \colC{$Z_{4}$-clock} & \colC{$\mathsf{C_{2v}}$} &\colC{} 

\\[0.1em]  \hline    
\colC{} & \colC{$E_{1}$} & \colC{$\boldsymbol{d}^{\mathrm{op}}$} & \colC{$Z_{8}$-clock} & \colC{$\mathsf{C_{s}}$} & \colC{$\sigma^{x}\mathcal{R}_{g}^{(2)}\sigma^{x}$}   
\\[0.1em]  \hline    
\colB{$\mathsf{D_{4d}}$} & \colB{$E_{2}$} & \colB{$\boldsymbol{d}^{\mathrm{ip}}$} & \colB{$Z_{4}$-clock} & \colB{$\mathsf{D_{2}}/\mathsf{C_{2v}}$} &  \colB{}

\\[0.1em]  \hline 

\colB{} & \colB{$E_{3}$} & \colB{$\boldsymbol{d}^{\mathrm{op}}$} & \colB{$Z_{8}$-clock} & \colB{$\mathsf{C_{2}}/\mathsf{C_{s}}$} & \colB{$\mathcal{R}_{g}^{B_{1}}\sigma^{z}\mathcal{R}_{g}^{(2)}\sigma^{z}$}   
\\[0.1em]  \hline 

\colC{$\mathsf{C_{8h}}$} & \colC{$\bar{E}_{2g}$} & \colC{$\boldsymbol{d}^{\mathrm{ip}}$} & \colC{$Z_{4}^{*}$-clock} & \colC{$\mathsf{C_{2h}}$} & \colC{} 

\\[0.1em]  \hline     
\colC{} & \colC{$\bar{E}_{1g}$} & \colC{$\boldsymbol{d}^{\mathrm{op}}$} & \colC{$Z_{8}^{*}$-clock} & \colC{$\mathsf{C_{i}}$} & \colC{$\mathcal{R}_{g}^{A_{u}}\sigma^{z}\mathcal{R}_{g}^{(2)}\sigma^{z}$} 
\\[0.1em]  \hline    

\colB{$\mathsf{C_{8}}$} & \colB{$\bar{E}_{2}$} & \colB{$\boldsymbol{d}^{\mathrm{ip}}$} & \colB{$Z_{4}^{*}$-clock} & \colB{$\mathsf{C_{2}}$} & \colB{} 
\\[0.1em]  \hline    

\colB{} & \colB{$\bar{E}_{1}$} & \colB{$\boldsymbol{d}^{\mathrm{op}}$} & \colB{$Z_{8}^{*}$-clock} & \colB{$\mathsf{C_{1}}$} & \colB{$\sigma^{z}\mathcal{R}_{g}^{(2)}\sigma^{z}$} 
\\[0.1em]  \hline 

\colC{$\mathsf{S_{8}}$} & \colC{$\bar{E}_{2}$} & \colC{$\boldsymbol{d}^{\mathrm{ip}}$} & \colC{$Z_{4}^{*}$-clock} & \colC{$\mathsf{C_{2}}$} & \colC{} 
\\[0.1em]  \hline 

\colC{} & \colC{$\bar{E}_{3}$} & \colC{$\boldsymbol{d}^{\mathrm{op}}$} & \colC{$Z_{8}^{*}$-clock} & \colC{$\mathsf{C_{1}}$} & \colC{$\mathcal{R}_{g}^{B}\sigma^{z}\mathcal{R}_{g}^{(2)}\sigma^{z}$}
\\[0.1em]  \hline  	    

\colA{decagonal} & \colA{IR} & \colA{nem.$\!$ basis} & \colA{universality} & \colA{res.$\!$ PG} & \colA{transf.$\!$ matrix}  
\\[0.1em]  \hline 

\colB{$\mathsf{D_{10h}}$} & \colB{$E_{2g}$} & \colB{$\boldsymbol{d}^{\mathrm{ip}}$} & \colB{$Z_{5}$-clock} & \colB{$\mathsf{D_{2h}}$} & \colB{} 
\\[0.1em]  \hline 

\colB{} & \colB{$E_{1g}$} & \colB{$\boldsymbol{d}^{\mathrm{op}}$} & \colB{$Z_{10}$-clock} & \colB{$\mathsf{C_{2h}}$} & \colB{$\mathcal{R}_{g}^{A_{1u}}\sigma^{z}\mathcal{R}_{g}^{(2)}\sigma^{z}$} 
\\[0.1em]  \hline 

\colC{$\mathsf{D_{10}}$} & \colC{$E_{2}$} & \colC{$\boldsymbol{d}^{\mathrm{ip}}$} & \colC{$Z_{5}$-clock} & \colC{$\mathsf{D_{2}}$} & \colC{}  
\\[0.1em]  \hline 

\colC{} & \colC{$E_{1}$} & \colC{$\boldsymbol{d}^{\mathrm{op}}$} & \colC{$Z_{10}$-clock} & \colC{$\mathsf{C_{2}}$} & \colC{$\sigma^{z}\mathcal{R}_{g}^{(2)}\sigma^{z}$} 
\\[0.1em]  \hline 

\colB{$\mathsf{D_{5h}}$} & \colB{$E_{2}^{\prime}$} & \colB{$\boldsymbol{d}^{\mathrm{ip}}$} & \colB{$Z_{5}$-clock} & \colB{$\mathsf{C_{2v}}$} & \colB{}  
\\[0.1em]  \hline 

\colB{} & \colB{$E_{1}^{\prime\prime}$} & \colB{$\boldsymbol{d}^{\mathrm{op}}$} & \colB{$Z_{10}$-clock} & \colB{$\mathsf{C_{2}}/\mathsf{C_{s}}$} & \colB{$\mathcal{R}_{g}^{A_{1}^{\prime\prime}}\sigma^{z}\mathcal{R}_{g}^{(2)}\sigma^{z}$} 
\\[0.1em]  \hline 

\colC{$\mathsf{C_{10v}}$} & \colC{$E_{2}$} & \colC{$\boldsymbol{d}^{\mathrm{ip}}$} & \colC{$Z_{5}$-clock} & \colC{$\mathsf{C_{2v}}$} & \colC{}  
\\[0.1em]  \hline 

\colC{} & \colC{$E_{1}$} & \colC{$\boldsymbol{d}^{\mathrm{op}}$} & \colC{$Z_{10}$-clock} & \colC{$\mathsf{C_{s}}$} & \colC{$\sigma^{x}\mathcal{R}_{g}^{(2)}\sigma^{x}$} 
\\[0.1em]  \hline 

\colD{} & \colD{} & \colD{} & \colD{} & \colD{} & \colD{} 

\\[0.1em]  \hline

\end{tabular}  
$\hfill$   	  
\begin{tabular}{cccccc}  
\hline 
	 
\colD{} & \colD{} & \colD{} & \colD{} & \colD{} & \colD{} 

\\[0.1em]  \hline    

\colB{$\mathsf{C_{10h}}$} & \colB{$\bar{E}_{2g}$} & \colB{$\boldsymbol{d}^{\mathrm{ip}}$} & \colB{$Z_{5}^{*}$-clock} & \colB{$\mathsf{C_{2h}}$} & \colB{}  
\\[0.1em]  \hline 

\colB{} & \colB{$\bar{E}_{1g}$} & \colB{$\boldsymbol{d}^{\mathrm{op}}$} & \colB{$Z_{10}^{*}$-clock} & \colB{$\mathsf{C_{i}}$} & \colB{$\mathcal{R}_{g}^{A_{u}}\sigma^{z}\mathcal{R}_{g}^{(2)}\sigma^{z}$} 
\\[0.1em]  \hline 

\colC{$\mathsf{C_{5h}}$} & \colC{$\bar{E}_{2}^{\prime}$} & \colC{$\boldsymbol{d}^{\mathrm{ip}}$} & \colC{$Z_{5}^{*}$-clock} & \colC{$\mathsf{C_{s}}$} & \colC{}  
\\[0.1em]  \hline 

\colC{} & \colC{$\bar{E}_{1}^{\prime\prime}$} & \colC{$\boldsymbol{d}^{\mathrm{op}}$} & \colC{$Z_{10}^{*}$-clock} & \colC{$\mathsf{C_{1}}$} & \colC{$\mathcal{R}_{g}^{A^{\prime\prime}}\sigma^{z}\mathcal{R}_{g}^{(2)}\sigma^{z}$} 
\\[0.1em]  \hline 

\colB{$\mathsf{C_{10}}$} & \colB{$\bar{E}_{2}$} & \colB{$\boldsymbol{d}^{\mathrm{ip}}$} & \colB{$Z_{5}^{*}$-clock} & \colB{$\mathsf{C_{2}}$} & \colB{} 
\\[0.1em]  \hline 

\colB{} & \colB{$\bar{E}_{1}$} & \colB{$\boldsymbol{d}^{\mathrm{op}}$} & \colB{$Z_{10}^{*}$-clock} & \colB{$\mathsf{C_{1}}$} & \colB{$\sigma^{z}\mathcal{R}_{g}^{(2)}\sigma^{z}$}

\\[0.1em]  \hline  	    

\colA{dodecagonal} & \colA{IR} & \colA{nem.$\!$ basis} & \colA{universality} & \colA{res.$\!$ PG} & \colA{transf.$\!$ matrix}  
\\[0.1em]  \hline 

\colC{$\mathsf{D_{12h}}$} & \colC{$E_{2g}$} & \colC{$\boldsymbol{d}^{\mathrm{ip}}$} & \colC{$Z_{6}$-clock} & \colC{$\mathsf{D_{2h}}$} & \colC{} 
\\[0.1em]  \hline 

\colC{} & \colC{$E_{1g}$} & \colC{$\boldsymbol{d}^{\mathrm{op}}$} & \colC{$Z_{12}$-clock} & \colC{$\mathsf{C_{2h}}$} & \colC{$\mathcal{R}_{g}^{A_{1u}}\sigma^{z}\mathcal{R}_{g}^{(2)}\sigma^{z}$} 
\\[0.1em]  \hline 

\colB{$\mathsf{D_{12}}$} & \colB{$E_{2}$} & \colB{$\boldsymbol{d}^{\mathrm{ip}}$} & \colB{$Z_{6}$-clock} & \colB{$\mathsf{D_{2}}$} & \colB{}  
\\[0.1em]  \hline 

\colB{} & \colB{$E_{1}$} & \colB{$\boldsymbol{d}^{\mathrm{op}}$} & \colB{$Z_{12}$-clock} & \colB{$\mathsf{C_{2}}$} & \colB{$\sigma^{z}\mathcal{R}_{g}^{(2)}\sigma^{z}$} 
\\[0.1em]  \hline 

\colC{$\mathsf{C_{12v}}$} & \colC{$E_{2}$} & \colC{$\boldsymbol{d}^{\mathrm{ip}}$} & \colC{$Z_{6}$-clock} & \colC{$\mathsf{C_{2v}}$} & \colC{}  
\\[0.1em]  \hline 

\colC{} & \colC{$E_{1}$} & \colC{$\boldsymbol{d}^{\mathrm{op}}$} & \colC{$Z_{12}$-clock} & \colC{$\mathsf{C_{s}}$} & \colC{$\sigma^{x}\mathcal{R}_{g}^{(2)}\sigma^{x}$} 
\\[0.1em]  \hline 

\colB{$\mathsf{D_{6d}}$} & \colB{$E_{2}$} & \colB{$\boldsymbol{d}^{\mathrm{ip}}$} & \colB{$Z_{6}$-clock} & \colB{$\mathsf{D_{2}}/\mathsf{C_{2v}}$} & \colB{}  
\\[0.1em]  \hline 

\colB{} & \colB{$E_{5}$} & \colB{$\boldsymbol{d}^{\mathrm{op}}$} & \colB{$Z_{12}$-clock} & \colB{$\mathsf{C_{2}}/\mathsf{C_{s}}$} & \colB{$\mathcal{R}_{g}^{B_{1}}\sigma^{z}\mathcal{R}_{g}^{(2)}\sigma^{z}$} 
\\[0.1em]  \hline 

\colC{$\mathsf{C_{12h}}$} & \colC{$\bar{E}_{2g}$} & \colC{$\boldsymbol{d}^{\mathrm{ip}}$} & \colC{$Z_{6}^{*}$-clock} & \colC{$\mathsf{C_{2h}}$} & \colC{}  
\\[0.1em]  \hline 

\colC{} & \colC{$\bar{E}_{1g}$} & \colC{$\boldsymbol{d}^{\mathrm{op}}$} & \colC{$Z_{12}^{*}$-clock} & \colC{$\mathsf{C_{i}}$} & \colC{$\mathcal{R}_{g}^{A_{u}}\sigma^{z}\mathcal{R}_{g}^{(2)}\sigma^{z}$} 
\\[0.1em]  \hline 

\colB{$\mathsf{C_{12}}$} & \colB{$\bar{E}_{2}$} & \colB{$\boldsymbol{d}^{\mathrm{ip}}$} & \colB{$Z_{6}^{*}$-clock} & \colB{$\mathsf{C_{2}}$} & \colB{}  
\\[0.1em]  \hline 

\colB{} & \colB{$\bar{E}_{1}$} & \colB{$\boldsymbol{d}^{\mathrm{op}}$} & \colB{$Z_{12}^{*}$-clock} & \colB{$\mathsf{C_{1}}$} & \colB{$\sigma^{z}\mathcal{R}_{g}^{(2)}\sigma^{z}$}   
\\[0.1em]  \hline 

\colC{$\mathsf{S_{12}}$} & \colC{$\bar{E}_{2}$} & \colC{$\boldsymbol{d}^{\mathrm{ip}}$} & \colC{$Z_{6}^{*}$-clock} & \colC{$\mathsf{C_{2}}$} & \colC{} 
\\[0.1em]  \hline 

\colC{} & \colC{$\bar{E}_{5}$} & \colC{$\boldsymbol{d}^{\mathrm{op}}$} & \colC{$Z_{12}^{*}$-clock} & \colC{$\mathsf{C_{1}}$} & \colC{$\mathcal{R}_{g}^{B}\sigma^{z}\mathcal{R}_{g}^{(2)}\sigma^{z}$} 
\\[0.1em]  \hline 

\colA{icosahedral} & \colA{IR} & \colA{nem.$\!$ basis} & \colA{universality} & \colA{res.$\!$ PG} & \colA{transf.$\!$ matrix}  
\\[0.1em]  \hline 

\colB{$\mathsf{I_{h}}$} & \colB{$H_{g}$} & \colB{$\boldsymbol{d}$} & \colB{}  &  \colB{$\mathsf{D_{3d}}/\mathsf{D_{5d}}$} & \colB{}

\\[0.1em]  \hline 

\colC{$\mathsf{I}$} & \colC{$H$} & \colC{$\boldsymbol{d}$} & \colC{}  & \colC{$\mathsf{D_{3}}/\mathsf{D_{5}}$} & \colC{}  
\\[0.1em]  \hline 

\end{tabular}
\caption{Symmetry properties of the allowed nematic order parameters in non-crystallographic point groups (PG) describing quasicrystals. The columns represent the same quantities as those in Table ~\ref{tab:classification}.  \label{tab:classification_quasicrystals}} 
\end{table*}

\section{Discussion and summary \label{sec:All-crystals}}

In summary, in this paper we derived the properties of electronic
nematic order in all 32 crystallographic point groups and in the non-crystallographic
point groups associated with quasicrystals. We expressed the Fermi
surface distortion patterns caused by nematic order in terms of a
general three-dimensional nematic order parameter expressed in tensorial
form in Eq.~(\ref{eq:Q_alt_parameterization}) and in five-component
vectorial form in Eq.~(\ref{eq:d_nml}). We also established the
critical properties of the allowed nematic transitions in these point
groups. These latter results are summarized in two tables: Table~\ref{tab:classification},
for crystallographic point groups, and Table~\ref{tab:classification_quasicrystals},
for the point groups of quasicrystals.

The crystallographic point groups shown in Table~\ref{tab:classification}
are organized in six blocks corresponding to the cubic, hexagonal,
trigonal, tetragonal, orthorhombic and monoclinic crystal systems;
we do not show the triclinic crystal system since it does not support
any nematic transition . For each point group in the first column,
we list all nematic order parameters that transform as a non-trivial
irreducible representation (IR, second column), together with the
corresponding basis for that order parameter (third column). The possible
basis are: the full five-component nematic order parameter $\boldsymbol{d}=\left(d_{1},\,d_{2},\,d_{3},\,d_{4},\,d_{5}\right)^{T}$,
whose elements correspond, respectively, to the usual charge quadrupolar
order parameters $d_{\frac{1}{\sqrt{3}}(2z^{2}-x^{2}-y^{2})}$, $d_{x^{2}-y^{2}}$,
$d_{2yz}$, $d_{2xz}$, and $d_{2xy}$; the $E_{g}$ and $T_{2g}$
cubic bases $\boldsymbol{d}^{e}=\left(d_{1},\,d_{2}\right)^{T}$ and
$\boldsymbol{d}^{t}=\left(d_{3},\,d_{4},\,d_{5}\right)^{T}$; and
the in-plane and out-of-plane axial bases $\boldsymbol{d}^{\mathrm{ip}}=\left(d_{2},\,d_{5}\right)^{T}$,
and $\boldsymbol{d}^{\mathrm{op}}=\left(d_{3},\,d_{4}\right)^{T}$.
Curly brackets indicate that the enclosed nematic order parameters
transform as the same IR. The fourth column lists the universality
class of the respective nematic Landau expansion: $Z_{2}$-Ising {[}e.g.,
Eq.~(\ref{eq:S_D4h_ip}){]}; $Z_{3}$-clock, which is equivalent
to $Z_{3}$-Potts {[}e.g., Eq.~(\ref{eq:S_Eg}) and (\ref{eq:S_hex_ip}){]};
$Z_{4}$-Potts {[}e.g., Eq.~(\ref{eq:S_T2g}){]}; $Z_{4}$-clock
{[}e.g., Eq.~(\ref{eq:S_D4h_op}){]}; and $Z_{6}$-clock {[}e.g.,
Eq.~(\ref{eq:S_hex_op}){]}. The asterisk notation in $Z_{q}^{*}$-clock
indicates that the clock term has an offset angle {[}e.g., Eq.~(\ref{eq:S_hex_ip_app}){]},
which is always the case when the group lacks twofold in-plane rotation
axes and the nematic order parameter transforms as a complex IR (which
is indicated by an overbar in the second column). The fifth column
shows the residual point group (PG) after the onset of nematic order,
which indicates the set of symmetry elements that remain intact after
nematicity is established. 

Finally, in the last column of Table~\ref{tab:classification}, we
list the transformation matrices of the nematic order parameters expressed
in terms of the transformation matrices of the coordinate vector.
To understand what this entails, consider first the five-component
nematic order parameter $\boldsymbol{d}$. It transforms according
to the IR $\Gamma_{j=2}^{+}$ of the fully-isotropic orthogonal group
$\mathsf{O(3)}$, and therefore transforms with the $5\times5$ matrices
$\mathcal{R}_{+,j=2}(g)=\mathcal{R}_{j=2}(\vartheta,\hat{\boldsymbol{\ell}})=\exp\left(-\mathsf{i}\vartheta\,\boldsymbol{J}^{(2)}\cdot\hat{\boldsymbol{\ell}}\right)$,
where $g=\big(\vartheta,\hat{\boldsymbol{\ell}},\mathcal{I}\big)$
are the symmetry elements of $\mathsf{O(3)}$, see Appendix~\ref{sec:SO3_properties}
for more details. Moving from isotropic space to point groups has
two effects: first, it restricts the symmetry elements $g$ to the
discrete set of operations that characterize the group. Second, it
causes the nematic order parameter to be decomposed into different
irreducible channels. Accordingly, in the point groups, the $5\times5$
matrices $\mathcal{R}_{+,j=2}(g)$ assume a block-diagonal form, where
each block is labeled by the IR characterizing the nematic order parameter.
For one-dimensional IRs, one simply recovers the characters of the
respective IR. For multi-dimensional IRs, the transformation matrices
appear as blocks constructed from $\mathcal{R}_{+,j=2}(g)$. In particular,
for the four multi-component nematic bases $\boldsymbol{d}^{e}$,
$\boldsymbol{d}^{t}$, $\boldsymbol{d}^{\mathrm{ip}}$, and $\boldsymbol{d}^{\mathrm{op}}$
used here, we have:
\begin{align}
\mathcal{R}_{\boldsymbol{d}^{e}}(g) & =\left(\begin{array}{cc}
[\mathcal{R}_{+,j=2}(g)]_{11} & [\mathcal{R}_{+,j=2}(g)]_{12}\\{}
[\mathcal{R}_{+,j=2}(g)]_{21} & [\mathcal{R}_{+,j=2}(g)]_{22}
\end{array}\right),\label{eq:R_de_summary}\\
\mathcal{R}_{\boldsymbol{d}^{t}}(g) & =\left(\begin{smallmatrix}[\mathcal{R}_{+,j=2}(g)]_{33} & [\mathcal{R}_{+,j=2}(g)]_{34} & [\mathcal{R}_{+,j=2}(g)]_{35}\\{}
[\mathcal{R}_{+,j=2}(g)]_{43} & [\mathcal{R}_{+,j=2}(g)]_{44} & [\mathcal{R}_{+,j=2}(g)]_{45}\\{}
[\mathcal{R}_{+,j=2}(g)]_{53} & [\mathcal{R}_{+,j=2}(g)]_{54} & [\mathcal{R}_{+,j=2}(g)]_{55}
\end{smallmatrix}\right),\label{eq:R_dt_summary}\\
\mathcal{R}_{\boldsymbol{d}^{\mathrm{ip}}}(g) & =\left(\begin{array}{cc}
[\mathcal{R}_{+,j=2}(g)]_{22} & [\mathcal{R}_{+,j=2}(g)]_{25}\\{}
[\mathcal{R}_{+,j=2}(g)]_{52} & [\mathcal{R}_{+,j=2}(g)]_{55}
\end{array}\right),\label{eq:R_dip_summary}\\
\mathcal{R}_{\boldsymbol{d}^{\mathrm{op}}}(g) & =\left(\begin{array}{cc}
[\mathcal{R}_{+,j=2}(g)]_{33} & [\mathcal{R}_{+,j=2}(g)]_{34}\\{}
[\mathcal{R}_{+,j=2}(g)]_{43} & [\mathcal{R}_{+,j=2}(g)]_{44}
\end{array}\right).\label{eq:R_dop_summary}
\end{align}

In principle, we could simply use these matrices to define the transformation
properties of the nematic order parameters listed in the table. However,
to construct invariants involving the coupling between the nematic
order parameter and other physical quantities, it is often more convenient
to express these four nematic transformation matrices (\ref{eq:R_de_summary})-(\ref{eq:R_dop_summary})
in terms of transformation matrices of the coordinate vector. The
latter transforms according to the IR $\Gamma_{j=1}^{-}$ of $\mathsf{O(3)}$.
As a result, its transformation matrices for any symmetry element
$g=\big(\vartheta,\hat{\boldsymbol{\ell}},\mathcal{I}\big)$ of $\mathsf{O(3)}$
are given by:
\begin{align}
\mathcal{R}_{g}^{(3)} & =\mathcal{R}_{-,j=1}(g),\label{eq:R3_summary}\\
\mathcal{R}_{g}^{(2)} & =\left(\begin{array}{cc}
[\mathcal{R}_{-,j=1}(g)]_{11} & [\mathcal{R}_{-,j=1}(g)]_{12}\\{}
[\mathcal{R}_{-,j=1}(g)]_{21} & [\mathcal{R}_{-,j=1}(g)]_{22}
\end{array}\right),\label{eq:R2_summary}
\end{align}
where $\mathcal{R}_{-,j=1}(g)=\mathcal{I}\mathcal{R}_{j=1}(\vartheta,\hat{\boldsymbol{\ell}})$
and $\mathcal{R}_{j=1}(\vartheta,\hat{\boldsymbol{\ell}})=\exp\left(-\mathsf{i}\vartheta\,\boldsymbol{J}^{(1)}\cdot\hat{\boldsymbol{\ell}}\right)$,
see Eq.~(\ref{eq:matrices_J1}) in Appendix~\ref{sec:SO3_properties}.
Here, we introduced two different matrices to distinguish between
the cases of a 3D-vector, which transforms with $\mathcal{R}_{g}^{(3)}$
(\ref{eq:R3_summary}) and is relevant for the polyhedral point groups,
and a 2D in-plane vector, which transforms with $\mathcal{R}_{g}^{(2)}$
(\ref{eq:R2_summary}) and is relevant for the axial point groups.
The key point illustrated by the last column of Table~\ref{tab:classification}
is that, for most of the nematic order parameters, the transformation
matrices (\ref{eq:R_de_summary})-(\ref{eq:R_dop_summary}) can be
directly related to either $\mathcal{R}_{g}^{(3)}$ or $\mathcal{R}_{g}^{(2)}$.
For instance, for the nematic order parameter that transforms as the
$T_{2g}$ IR of $\mathsf{O_{h}}$, the last entry on the row means
that $\mathcal{R}_{\boldsymbol{d}^{t}}(g)=\mathcal{R}_{g}^{A_{2u}}\mathcal{R}_{g}^{(3)}$,
where $\mathcal{R}_{g}^{A_{2u}}$ are the characters of the $A_{2u}$
IR. Similarly, for the nematic order parameter transforming as the
$E$ IR of $\mathsf{C_{4v}}$, we have $\mathcal{R}_{\boldsymbol{d}^{\mathrm{op}}}(g)=\sigma^{x}\mathcal{R}_{g}^{(2)}\sigma^{x}$,
where $\sigma^{j}$ are Pauli matrices.

To illustrate the usefulness of this representation of the transformation
matrices, let us construct invariants involving the coupling between
the nematic order parameters and the electric polarization $\boldsymbol{P}$,
which transforms like a vector. To make the example more transparent,
we consider the hexagonal point group $\mathsf{C_{3h}}$ and the nematic
in-plane order parameter $\boldsymbol{d}^{\mathrm{ip}}$. This doublet
transforms according to the IR $E^{\prime}$ and via the transformation
matrix $\mathcal{R}_{\boldsymbol{d}^{\mathrm{ip}}}(g)=\sigma^{z}\mathcal{R}_{g}^{(2)}\sigma^{z}$,
as shown in Table~\ref{tab:classification}. The in-plane polarization
$\boldsymbol{P}_{\mathrm{ip}}=(P_{x},P_{y})$ , on the other hand,
transforms according to the same IR $E^{\prime}$, but via the matrices
$\mathcal{R}^{\boldsymbol{P}_{\mathrm{ip}}}(g)=\mathcal{R}_{g}^{(2)}$
that characterize a vector. Indeed, $\mathcal{R}^{\boldsymbol{P}_{\mathrm{ip}}}(g)$
is related to $\mathcal{R}_{\boldsymbol{d}^{\mathrm{ip}}}(g)$ through
a similarity transformation, the Pauli matrix $\sigma^{z}$. Since
both $\boldsymbol{d}^{\mathrm{ip}}$ and $\boldsymbol{P}_{\mathrm{ip}}$
transform as $E^{\prime}$, there must exist a bilinear invariant.
However, due to the difference in transformation matrices, this invariant
is not simply $(\boldsymbol{d}^{\mathrm{ip}})^{T}\cdot\boldsymbol{P}_{\mathrm{ip}}$,
but instead it is $(\boldsymbol{d}^{\mathrm{ip}})^{T}\sigma^{z}\boldsymbol{P}_{\mathrm{ip}}$,
where the additional Pauli matrix $\sigma^{z}$ is needed to compensate
for the similarity transformation relating $\mathcal{R}^{\boldsymbol{P}_{\mathrm{ip}}}(g)$
and $\mathcal{R}_{\boldsymbol{d}^{\mathrm{ip}}}(g)$. As a second
example, consider the coupling between the nematic out-of-plane doublet
$\boldsymbol{d}^{\mathrm{op}}$ and the out-of-plane polarization
$P_{z}$ within the same point group $\mathsf{C_{3h}}$. The nematic
order parameter transforms according to the IR $E^{\prime\prime}$
and via the transformation matrix $\mathcal{R}_{\boldsymbol{d}^{\mathrm{op}}}(g)=\mathcal{R}_{g}^{A^{\prime\prime}}\big(\sigma^{z}\mathcal{R}_{g}^{(2)}\sigma^{z}\big)$,
see Table~\ref{tab:classification}. The out-of-plane polarization
$P_{z}$, on the other hand, transforms as $A^{\prime\prime}$ and
via the $1\times1$ ``matrix'' $\mathcal{R}^{P_{z}}(g)=\mathcal{R}_{g}^{A^{\prime\prime}}$.
Using the fact that $\boldsymbol{P}_{\mathrm{ip}}$ transforms according
to $\mathcal{R}_{g}^{(2)}$, we can readily construct the invariant
$P_{z}\,\left(\boldsymbol{d}^{\mathrm{op}}\right)^{T}\sigma^{z}\boldsymbol{P}_{\mathrm{ip}}$. 

Table~\ref{tab:classification_quasicrystals} has the same layout
as Table~\ref{tab:classification}, but refers to the quasicrystalline
point groups considered in this work. They correspond to octagonal,
decagonal, dodecagonal, and icosahedral quasicrystals.

The group-theoretical classification presented here can be valuable
in searching for new nematic systems in material databases. More broadly,
our results offer interesting insights into which types of lattice
may realize exotic electronic nematic phenomena. Indeed, for any of
the nematic doublets that behave as a $Z_{q}$-clock order parameter
with $q\geq4$, the nematic transition in 3D belongs to the XY universality
class, which in turn also describes the nematic transition of a 2D
isotropic model. On the other hand, in a 2D system with a $Z_{q}$-clock
nematic order parameter, a critical nematic phase with quasi-long-range
order precedes the onset of long-range order for $q\geq5$, whereas
for $q=4$ one obtains an Ashkin-Teller nematic model, whose critical
properties are described by non-universal critical exponents. From
Tables~\ref{tab:classification} and \ref{tab:classification_quasicrystals},
we see that this condition is satisfied by $\left(d_{2yz},\,d_{2xz}\right)$
nematicity in hexagonal and tetragonal crystals, as well as $\left(d_{2yz},\,d_{2xz}\right)$
and $\big(d_{x^{2}-y^{2}},\,d_{2xy}\big)$ nematicity in octagonal,
decagonal, and dodecagonal quasicrystals. In this regard, it will
be interesting to investigate the properties of the Landau expansion
(\ref{eq:S_Hg}) of the icosahedral nematic order parameter, since
the mean-field ground states are either sixfold or tenfold degenerate. 

Experimentally, these properties should be directly accessible via
probes that measure the nematic susceptibility in different channels,
such as elasto-resistance \citep{Chu12} and Raman spectroscopy \citep{Gallais2016}.
Theoretically, one expects that nematic fluctuations will be enhanced
near the 3D-XY nematic transition, since there will be not only soft
longitudinal (i.e., amplitude) fluctuations but also soft transverse
(i.e., phase) fluctuations associated with the emergent continuous
symmetry. Given that nematic fluctuations have been proposed as potential
drivers of non-Fermi liquid behavior \citep{Metzner2006,Zacharias2009,Maslov2010,Metlitski2010}
and pairing \citep{Lederer2017,Klein2018}, the investigation of emergent
XY-nematicity could provide further insights into these problems as
well. As for the exotic 2D behavior associated with $q\geq4$ clock
in-plane nematicity, the most promising realizations would be in $30^{\circ}$-twisted
hexagonal bilayers ($Z_{6}$-clock) and $45^{\circ}$-twisted tetragonal
bilayers ($Z_{4}$-clock). Both settings have been recently realized
experimentally by twisting graphene \citep{Ahn2018} and monolayers
of cuprate BSCCO \citep{Kim2023}. Interestingly, nematic order has
been observed in underdoped BSCCO \citep{Lawler2010}.

One important ingredient that was not included in this paper and that
deserves further investigation is the role of strain, both extrinsic
and intrinsic. For instance, it was recently shown that the impact
of uniaxial strain on in-plane $Z_{3}$-Potts/clock nematics is fundamentally
different from the well-studied case of Ising-nematics \citep{Chakraborty2023}.
This raises the broader question of how external strain can be used
to probe and modify the behavior of $Z_{q}$-clock nematic order parameters.
Moreover, long-wavelength quantized strain fluctuations, which are
manifested as acoustic phonons, are known to significantly change
the critical properties of $Z_{2}$ and $Z_{3}$ electronic nematic
transitions in tetragonal, hexagonal, and trigonal crystals by promoting
long-range nematic interactions \citep{Xu2009,Karahasanovic2016,Paul2017,Carvalho2019,Fernandes2020,Hecker2022}.
It will be important to perform similar analyses in cubic crystals
as well as quasicrystals, which also host phason modes on top of phonon
modes \citep{Wang1993,Ochoa2019}. Finally, crystalline defects such
as dislocations, vacancies, and interstitials create inhomogeneous
strain distributions, which act as random nematic conjugate fields
\citep{Carlson2006,Dahmen2010,Meese2022}. While most studies have
focused on a random-field Ising-model description of this rich phenomenon,
it will be important to develop models that can capture the long-range
and correlated nature of the various components of the strain fields
generated by these defects.
\begin{acknowledgments}
We thank T. Birol and J. Vi\~nals for fruitful discussions. This
work was primarily supported by the U. S. Department of Energy, Office
of Science, Basic Energy Sciences, Materials Sciences and Engineering
Division, under Award No. DE-SC0020045 (M.H., A.R., and R.M.F.). D.F.A.
was supported by Department of Energy, Office of Basic Energy Science,
Division of Materials Sciences and Engineering under Award No. DE-SC0021971.
\end{acknowledgments}

\appendix

\section{Generators of SO$(3)$ and parametrization of $\boldsymbol{d}$ \label{sec:SO3_properties}}

In this Appendix, we give further details about the generators of
$\mathsf{SO(3)}$. Within the full rotation group $\mathsf{SO(3)}$,
the transformation matrices associated with an irreducible representation
(IR) $\Gamma_{j}$ are given by $\mathcal{R}_{j}(\vartheta,\hat{\boldsymbol{\ell}})=\exp\left(-\mathsf{i}\vartheta\,\boldsymbol{J}^{(j)}\cdot\hat{\boldsymbol{\ell}}\right)$,
parametrized by a rotation angle $\vartheta$ around a unit-length
rotation axis $\hat{\boldsymbol{\ell}}$. The three generators $J_{\mu}^{(j)}$
satisfy the Lie algebra 
\begin{align}
\left[J_{\mu_{1}}^{(j)},J_{\mu_{2}}^{(j)}\right] & =\mathsf{i}\,\epsilon_{\mu_{1}\mu_{2}\mu_{3}}J_{\mu_{3}}^{(j)},\label{eq:Lie_algebra}
\end{align}
with the anti-symmetric Levi-Civita tensor $\epsilon_{\mu_{1}\mu_{2}\mu_{3}}$.
In principle, the matrices $J_{\mu}^{(j)}$ can be defined in the
``Lie basis'' using the standard relationships 
\begin{align}
\hat{J}_{z}^{(j)}\,|j,m\rangle & =m\,|j,m\rangle,\label{eq:J_z}\\
\hat{J}_{\pm}^{(j)}\,|j,m\rangle & =\sqrt{j(j+1)-m(m\pm1)}\,|j,m\pm1\rangle,\label{eq:Jpm}
\end{align}
where $\hat{J}_{\pm}^{(j)}=\hat{J}_{x}^{(j)}\pm\mathsf{i}\hat{J}_{y}^{(j)}$.
However, for the transformation matrices to be orthogonal, i.e. $\mathcal{R}_{j}^{T}(\vartheta,\hat{\boldsymbol{\ell}})=\mathcal{R}_{j}(-\vartheta,\hat{\boldsymbol{\ell}})=\mathcal{R}_{j}^{-1}(\vartheta,\hat{\boldsymbol{\ell}})$,
one needs to impose the requirement $\big(J_{\mu}^{(j)}\big)^{T}=\big(J_{\mu}^{(j)}\big)^{*}=-J_{\mu}^{(j)}$,
i.e. the generators need to be anti-symmetric, and thus, a similarity
transformation has to be applied on Eqs.~(\ref{eq:J_z})-(\ref{eq:Jpm}).
We now list the resulting generators. For $j=0$ the three generators
are $J_{x}^{(0)}=J_{y}^{(0)}=J_{z}^{(0)}=0$, which ensures that a
$\Gamma_{j=0}$ scalar is invariant under all rotations. For $j=1$
the three generators are the well-known anti-symmetric Gell-Mann matrices
\begin{align}
J_{x}^{(1)} & =\mathsf{i}\left(\begin{smallmatrix}0 & 0 & 0\\
0 & 0 & -1\\
0 & 1 & 0
\end{smallmatrix}\right), & J_{y}^{(1)} & =\mathsf{i}\left(\begin{smallmatrix}0 & 0 & 1\\
0 & 0 & 0\\
-1 & 0 & 0
\end{smallmatrix}\right), & J_{z}^{(1)} & =\mathsf{i}\left(\begin{smallmatrix}0 & -1 & 0\\
1 & 0 & 0\\
0 & 0 & 0
\end{smallmatrix}\right),\label{eq:matrices_J1}
\end{align}
which can also be expressed in terms of the anti-symmetric Levi-Cevita
tensor $\big(J_{\mu_{1}}^{(1)}\big)_{\mu_{2}\mu_{3}}=-\mathsf{i}\epsilon_{\mu_{1}\mu_{2}\mu_{3}}$.
The three generators for $j=2$ are given by

\begin{align}
J_{x}^{(2)} & =\mathsf{i}\left(\begin{smallmatrix}0 & 0 & \sqrt{3} & 0 & 0\\
0 & 0 & 1 & 0 & 0\\
-\sqrt{3} & -1 & 0 & 0 & 0\\
0 & 0 & 0 & 0 & 1\\
0 & 0 & 0 & -1 & 0
\end{smallmatrix}\right), & J_{y}^{(2)} & =\mathsf{i}\left(\begin{smallmatrix}0 & 0 & 0 & -\sqrt{3} & 0\\
0 & 0 & 0 & 1 & 0\\
0 & 0 & 0 & 0 & -1\\
\sqrt{3} & -1 & 0 & 0 & 0\\
0 & 0 & 1 & 0 & 0
\end{smallmatrix}\right),\nonumber \\
J_{z}^{(2)} & =\mathsf{i}\left(\begin{smallmatrix}0 & 0 & 0 & 0 & 0\\
0 & 0 & 0 & 0 & -2\\
0 & 0 & 0 & 1 & 0\\
0 & 0 & -1 & 0 & 0\\
0 & 2 & 0 & 0 & 0
\end{smallmatrix}\right),\label{eq:J2_generators}
\end{align}
and the ones for $j=4$ are
\begin{align}
J_{x}^{(4)} & =\frac{\mathsf{i}}{\sqrt{2}}\left(\begin{smallmatrix}0 & 0 & 0 & 0 & 2\sqrt{5} & 0 & 0 & 0 & 0\\
0 & 0 & 2 & 0 & 0 & 0 & 0 & 0 & 0\\
0 & -2 & 0 & -\sqrt{7} & 0 & 0 & 0 & 0 & 0\\
0 & 0 & \sqrt{7} & 0 & 3 & 0 & 0 & 0 & 0\\
-2\sqrt{5} & 0 & 0 & -3 & 0 & 0 & 0 & 0 & 0\\
0 & 0 & 0 & 0 & 0 & 0 & 3 & 0 & 0\\
0 & 0 & 0 & 0 & 0 & -3 & 0 & -\sqrt{7} & 0\\
0 & 0 & 0 & 0 & 0 & 0 & \sqrt{7} & 0 & 2\\
0 & 0 & 0 & 0 & 0 & 0 & 0 & -2 & 0
\end{smallmatrix}\right),\label{eq:J4x_generator}\\
J_{y}^{(4)} & =\frac{\mathsf{i}}{\sqrt{2}}\left(\begin{smallmatrix}0 & 0 & 0 & 0 & 0 & -2\sqrt{5} & 0 & 0 & 0\\
0 & 0 & 0 & 0 & 0 & 0 & 0 & 2 & 0\\
0 & 0 & 0 & 0 & 0 & 0 & \sqrt{7} & 0 & -2\\
0 & 0 & 0 & 0 & 0 & 3 & 0 & -\sqrt{7} & 0\\
0 & 0 & 0 & 0 & 0 & 0 & -3 & 0 & 0\\
2\sqrt{5} & 0 & 0 & -3 & 0 & 0 & 0 & 0 & 0\\
0 & 0 & -\sqrt{7} & 0 & 3 & 0 & 0 & 0 & 0\\
0 & -2 & 0 & \sqrt{7} & 0 & 0 & 0 & 0 & 0\\
0 & 0 & 2 & 0 & 0 & 0 & 0 & 0 & 0
\end{smallmatrix}\right),\label{eq:J4y_generator}\\
J_{z}^{(4)} & =\mathsf{i}\,\left(\begin{smallmatrix}0 & 0 & 0 & 0 & 0 & 0 & 0 & 0 & 0\\
0 & 0 & 0 & 0 & 0 & 0 & 0 & 0 & -4\\
0 & 0 & 0 & 0 & 0 & 0 & 0 & 3 & 0\\
0 & 0 & 0 & 0 & 0 & 0 & -2 & 0 & 0\\
0 & 0 & 0 & 0 & 0 & 1 & 0 & 0 & 0\\
0 & 0 & 0 & 0 & -1 & 0 & 0 & 0 & 0\\
0 & 0 & 0 & 2 & 0 & 0 & 0 & 0 & 0\\
0 & 0 & -3 & 0 & 0 & 0 & 0 & 0 & 0\\
0 & 4 & 0 & 0 & 0 & 0 & 0 & 0 & 0
\end{smallmatrix}\right).\label{eq:J4z_generator}
\end{align}
Clearly, the generators are never unique in the sense that they can
always be rotated via an orthogonal matrix $O^{(j)}$ according to
$[O^{(j)}]^{T}\boldsymbol{J}^{(j)}O^{(j)}$, which corresponds to
a basis rotation. Throughout this work, the generators are fixed according
to Eqs.~(\ref{eq:matrices_J1})-(\ref{eq:J4z_generator}).

Next, we determine the bilinears $\boldsymbol{D}^{j=2}$ and $\boldsymbol{D}^{j=4}$
that occur in the decomposition (\ref{eq:symm_decomp1}) and which
were used in the derivation of the Landau expansion of the 3D isotropic
nematic system. The transformation conditions are structurally similar
to Eq.~(\ref{eq:SO3_trafo_condition}). For example, for the $\Gamma_{j=4}^{+}$
bilinear $D^{j=4,\nu}=d_{i}\Lambda_{i,i^{\prime}}^{j=4,\nu}d_{i^{\prime}}$,
one needs to solve the equation
\begin{align}
\mathcal{R}_{+,j=2}^{T}(g)\Lambda^{j=4,\nu}\mathcal{R}_{+,j=2}(g) & =\mathcal{R}_{+,j=4}(g)_{\nu\nu^{\prime}}\Lambda^{j=4,\nu^{\prime}},\label{eq:SO3_trafo_condition-1}
\end{align}
where $\Lambda^{j=4,\nu}$ are $5\times5$-dimensional matrices with
$i,i^{\prime}=1,\dots,5$ and $\nu,\nu^{\prime}=1,\dots,9$. Recall
that the $\Gamma_{j=2}^{+}$ nematic order parameter $\boldsymbol{d}$
transforms via the matrices $\mathcal{R}_{+,j=2}(g)=\mathcal{R}_{j=2}(\vartheta,\hat{\boldsymbol{\ell}})$
for each element $g=\big(\vartheta,\hat{\boldsymbol{\ell}},\mathcal{I}\big)$. 

A straightforward but tedious calculation then yields $D^{j=0}=\left|\boldsymbol{d}\right|^{2}$,
$\boldsymbol{D}^{j=2}$ given by Eq.~(\ref{eq:bilinears_j02}), and
\begin{align}
\boldsymbol{D}^{j=4} & =\frac{\sqrt{5}}{6}\left(\begin{smallmatrix}\frac{1}{\sqrt{5}}\left[6d_{1}^{2}-4\left(d_{3}^{2}+d_{4}^{2}\right)+\left(d_{2}^{2}+d_{5}^{2}\right)\right]\\
\sqrt{7}\left(d_{2}^{2}-d_{5}^{2}\right)\\
\sqrt{14}\left(d_{2}d_{3}+d_{4}d_{5}\right)\\
2\sqrt{3}d_{1}d_{2}-2\left(d_{3}^{2}-d_{4}^{2}\right)\\
\sqrt{2}\left[\left(2\sqrt{3}d_{1}+d_{2}\right)d_{3}-d_{4}d_{5}\right]\\
\sqrt{2}\left[\left(2\sqrt{3}d_{1}-d_{2}\right)d_{4}-d_{3}d_{5}\right]\\
2\sqrt{3}d_{1}d_{5}+4d_{3}d_{4}\\
\sqrt{14}\left(d_{2}d_{4}-d_{3}d_{5}\right)\\
2\sqrt{7}d_{2}d_{5}
\end{smallmatrix}\right).\label{eq:Dj4}
\end{align}
Importantly, all three bilinears have the same magnitude, $\left(D^{j=0}\right)^{2}=\left(\boldsymbol{D}^{j=2}\right)^{2}=\left(\boldsymbol{D}^{j=4}\right)^{2}=\left|\boldsymbol{d}\right|^{4}$.
From the condition (\ref{eq:SO3_trafo_condition-1}) we see that a
basis rotation $O^{(j=4)}$ on the generators $\boldsymbol{J}^{(j=4)}$
would rotate the basis state $\boldsymbol{D}^{j=4}$ (\ref{eq:Dj4})
into  $O^{(j=4)}\boldsymbol{D}^{j=4}$.

To finish this Appendix, we further discuss the parameterization of
the five-component nematic order parameter within the $(\boldsymbol{n}\boldsymbol{m}\boldsymbol{l})$-representation
(\ref{eq:d_nml}). Recall that:

\begin{align}
\boldsymbol{d} & =\frac{|\boldsymbol{d}|}{\sqrt{3}}\Big\{\cos\left(\alpha\right)\left(n_{\mu}\boldsymbol{\lambda}_{\mu\mu^{\prime}}^{j=2}n_{\mu^{\prime}}\right)+\cos\left(\alpha+\frac{2\pi}{3}\right)\left(m_{\mu}\boldsymbol{\lambda}_{\mu\mu^{\prime}}^{j=2}m_{\mu^{\prime}}\right)\nonumber \\
 & \quad+\cos\left(\alpha+\frac{4\pi}{3}\right)\left(l_{\mu}\boldsymbol{\lambda}_{\mu\mu^{\prime}}^{j=2}l_{\mu^{\prime}}\right)\Big\}.
\end{align}
The three angles describing the three orthonormal eigenvectors $\boldsymbol{n}$,
$\boldsymbol{m}$, $\boldsymbol{l}$ can be chosen in several ways.
The parameterization employed in this work makes use of the spherical
unit vectors 
\begin{align}
\hat{\boldsymbol{e}}_{r} & =\left(\begin{smallmatrix}\cos\varphi\sin\theta\\
\sin\varphi\sin\theta\\
\cos\theta
\end{smallmatrix}\right), & \hat{\boldsymbol{e}}_{\varphi} & =\left(\begin{smallmatrix}\sin\varphi\\
-\cos\varphi\\
0
\end{smallmatrix}\right), & \hat{\boldsymbol{e}}_{\theta} & =\left(\begin{smallmatrix}\cos\varphi\cos\theta\\
\sin\varphi\cos\theta\\
-\sin\theta
\end{smallmatrix}\right).\label{eq:SO3_app_spheric_vectors}
\end{align}
One of the eigenvectors, e.g. $\boldsymbol{l}$, can always be chosen
to be aligned with the radial vector $\hat{\boldsymbol{e}}_{r}$,
i.e. $\boldsymbol{l}=\hat{\boldsymbol{e}}_{r}$. Conversely, the remaining
two vectors can be arbitrarily rotated about this axis by an angle
$\eta$:
\begin{align}
\boldsymbol{l} & =\hat{\boldsymbol{e}}_{r}, & \boldsymbol{n} & =\cos\eta\,\hat{\boldsymbol{e}}_{\varphi}+\sin\eta\,\hat{\boldsymbol{e}}_{\theta}, & \boldsymbol{m} & =-\sin\eta\,\hat{\boldsymbol{e}}_{\varphi}+\cos\eta\,\hat{\boldsymbol{e}}_{\theta}.\label{eq:nml_vectors}
\end{align}
Thus, within the $(\boldsymbol{n}\boldsymbol{m}\boldsymbol{l})$-representation
the five degrees of freedom are encoded in the magnitude $|\boldsymbol{d}|$
and the four angles $\alpha,\varphi,\theta,\eta$. To ensure a one-to-one
mapping between the two representations, (\ref{eq:Q_tensor}) and
(\ref{eq:Q_alt_parameterization}), the parameter ranges have to be
restricted according to \footnote{For the limiting values of $\theta$ the rotation angles are actually
more restricted: For $\theta=0$ we constrain $\varphi\in[0,0]$ and
$\eta\in[0,\pi]$, and for $\theta=\frac{\pi}{2}$ we constrain $\varphi\in[0,2\pi]$
and $\eta\in[0,\frac{\pi}{2}]$.} 
\begin{align}
|\boldsymbol{d}| & \geq0, & \alpha & \in\left[0,\frac{\pi}{3}\right], & \theta & \in\left[0,\frac{\pi}{2}\right], & \varphi & \in\left[0,2\pi\right], & \eta & \in\left[0,\pi\right].\label{eq:parameter_restriction}
\end{align}

\section{Symmetrized products\label{sec:Symmetrized-products}}

In this Appendix, we discuss in more detail the product decomposition
associated with a generic order parameter $\boldsymbol{\eta}=(\eta_{1},\dots,\eta_{\dim n})$
that transforms according to an IR $n$ of some group. Because the
order parameter components commute, i.e. $\eta_{i}\eta_{j}=\eta_{j}\eta_{i}$,
the number of product components is reduced. For example, for a two-component
order parameter $\boldsymbol{\eta}=(\eta_{1},\eta_{2})$ one would
naively expect $2\times2=4$ product components, while in fact, there
are only $3$: $\eta_{1}^{2}$, $\eta_{2}^{2}$ and $\eta_{1}\eta_{2}$,
since the antisymmetric combination $\eta_{1}\eta_{2}-\eta_{2}\eta_{1}$
vanishes. Here, we use group theory to derive expressions for the
product decompositions in the symmetric channel, see e.g. Ref.~\citep{Hamermesh1989}.
Doing so automatically excludes vanishing contributions, such as the
antisymmetric combination $\eta_{1}\eta_{2}-\eta_{2}\eta_{1}$ in
the above example, and removes redundancies associated with double-counting.
To illustrate the last point, consider the symmetrized bilinears in
the above example: $N_{1}=\eta_{1}^{2}+\eta_{2}^{2}$ and $\boldsymbol{N}_{2}=\left(\eta_{1}^{2}-\eta_{2}^{2},2\eta_{1}\eta_{2}\right)$.
Then, a straightforward decomposition would suggest that two invariants
exist in fourth order, $N_{1}^{2}$ and $\boldsymbol{N}_{2}\cdot\boldsymbol{N}_{2}$,
while in fact they are identical since $N_{1}^{2}=\boldsymbol{N}_{2}\cdot\boldsymbol{N}_{2}=\left(\eta_{1}^{2}+\eta_{2}^{2}\right)^{2}$.
Such double-countings are removed by using the symmetrized decomposition.
As we are interested in Landau expansions, we consider products up
to sixth order, but extension to higher orders is straightforward.

Let us begin with the second-order product $P_{2,ij}=\eta_{i}\eta_{j}$.
Using the order parameter transformation $\eta_{i}^{\prime}=\mathcal{R}_{n}(g)_{ii^{\prime}}\eta_{i^{\prime}}$
under a symmetry element $g$, one finds the product to transform
as 
\begin{align}
P_{2,ij}^{\prime} & =\mathcal{R}^{\eta^{2}}(g)_{ij,i^{\prime}j^{\prime}}\,P_{2,i^{\prime}j^{\prime}},\label{eq:Pprime_2}
\end{align}
where $\mathcal{R}^{\eta^{2}}(g)_{ij,i^{\prime}j^{\prime}}=\mathcal{R}_{n}(g)_{ii^{\prime}}\mathcal{R}_{n}(g)_{jj^{\prime}}$.
The transformation matrices $\mathcal{R}^{\eta^{2}}(g)$ themselves
form a representation of the group, i.e. it holds that $\mathcal{R}^{\eta^{2}}(g_{1}g_{2})=\mathcal{R}^{\eta^{2}}(g_{1})\mathcal{R}^{\eta^{2}}(g_{1})$,
see e.g. Ref.~\citep{Hamermesh1989} for details. Correspondingly,
the characters of this representation result from the trace,
\begin{align}
\chi^{\eta^{2}}(g) & =\sum_{i,j}\mathcal{R}^{\eta^{2}}(g)_{ij,ij}=\sum_{i,j}\mathcal{R}_{n}(g)_{ii}\mathcal{R}_{n}(g)_{jj}=\chi_{n}^{2}(g).\label{eq:chi2_full}
\end{align}
The formula (\ref{eq:chi2_full}) contains all symmetry channels,
including the ones that eventually vanish due to the commutativity
of the components of $\boldsymbol{\eta}$. To have these removed,
we symmetrize the transformation matrix $\mathcal{R}^{\eta^{2}}(g)$
before computing its character. Symmetrization involves the addition
of all permutations in Eq.~(\ref{eq:Pprime_2}) with respect to $(i^{\prime},j^{\prime})$,
divided by the number of permutations, i.e. 
\begin{align}
\mathcal{R}_{s}^{\eta^{2}}(g)_{ij,i^{\prime}j^{\prime}} & =\frac{1}{2!}\left[\mathcal{R}^{\eta^{2}}(g)_{ij,i^{\prime}j^{\prime}}+\mathcal{R}^{\eta^{2}}(g)_{ij,j^{\prime}i^{\prime}}\right].\label{eq:R2_s}
\end{align}
Then, the characters for the symmetrized product (\ref{eq:R2_s})
become 
\begin{align}
\chi_{s}^{\eta^{2}}(g) & =\sum_{i,j}\mathcal{R}_{s}^{\eta^{2}}(g)_{ij,ij}=\frac{1}{2}\left(\chi_{n}^{2}(g)+\chi_{n}(g^{2})\right).\label{eq:chi2_s}
\end{align}
Thus, the decomposition of a symmetrized product (\ref{eq:chi2_s})
only contains a subset of the original symmetry channels associated
with the bare product (\ref{eq:chi2_full}). The symmetrized decomposition
is accomplished by using $\mathcal{R}_{s}^{\eta^{2}}(g)$ rather than
$\mathcal{R}^{\eta^{2}}(g)$.

For the higher-order products such as $P_{3,ijk}=\eta_{i}\eta_{j}\eta_{k}$
and $P_{4,ijkl}=\eta_{i}\eta_{j}\eta_{k}\eta_{l}$, one proceeds analogously.
As in the case above, the corresponding transformation matrices 
\begin{align}
\mathcal{R}^{\eta^{3}}(g)_{ijk,i^{\prime}j^{\prime}k^{\prime}} & =\mathcal{R}_{n}(g)_{ii^{\prime}}\mathcal{R}_{n}(g)_{jj^{\prime}}\mathcal{R}_{n}(g)_{kk^{\prime}},\label{eq:Reta3_app}\\
\mathcal{R}^{\eta^{4}}(g)_{ijkl,i^{\prime}j^{\prime}k^{\prime}l^{\prime}} & =\mathcal{R}_{n}(g)_{ii^{\prime}}\mathcal{R}_{n}(g)_{jj^{\prime}}\mathcal{R}_{n}(g)_{kk^{\prime}}\mathcal{R}_{n}(g)_{ll^{\prime}},\label{eq:Reta4_app}
\end{align}
form representations of the group with characters
\begin{align}
\chi^{\eta^{3}}(g) & =\sum_{ijk}\mathcal{R}^{\eta^{3}}(g)_{ijk,ijk}=\chi_{n}^{3}(g),\label{eq:chi_eta3_app}\\
\chi^{\eta^{4}}(g) & =\sum_{ijkl}\mathcal{R}^{\eta^{4}}(g)_{ijkl,ijkl}=\chi_{n}^{4}(g).\label{eq:chi_eta4_app}
\end{align}
The symmetrized transformation matrices are 
\begin{align}
\mathcal{R}_{s}^{\eta^{3}}\!\!(g)_{ijk,i^{\prime}j^{\prime}k^{\prime}} & =\frac{1}{3!}\left[\mathcal{R}^{\eta^{3}}\!\!(g)_{ijk,i^{\prime}j^{\prime}k^{\prime}}+5\left(i^{\prime}\leftrightarrow j^{\prime}\leftrightarrow k^{\prime}\right)\right],\label{eq:Reta3_s_app}\\
\mathcal{R}_{s}^{\eta^{4}}\!\!(g)_{ijkl,i^{\prime}j^{\prime}k^{\prime}l^{\prime}} & =\frac{1}{4!}\Big[\mathcal{R}^{\eta^{4}}\!\!(g)_{ijkl,i^{\prime}j^{\prime}k^{\prime}l^{\prime}}\nonumber \\
 & \qquad+23\left(i^{\prime}\leftrightarrow j^{\prime}\leftrightarrow k^{\prime}\leftrightarrow l^{\prime}\right)\Big],\label{eq:Reta4_s_app}
\end{align}
with the number of added permutation terms indicated explicitly {[}cf.
Eq.~(\ref{eq:R2_s}){]}. One finds the characters associated with
these symmetrized products to be 
\begin{align}
\chi_{s}^{\eta^{3}}\!\!(g) & =\!\sum_{ijk}\!\mathcal{R}_{s}^{\eta^{3}}\!\!(g)_{ijk,ijk}\!\nonumber \\
 & =\big[\chi_{n}^{3}(g)+2\chi_{n}(g^{3})+3\chi_{n}(g)\chi_{n}(g^{2})\big]\big/6,\label{eq:chi_dec_3}\\
\chi_{s}^{\eta^{4}}\!\!(g) & =\!\sum_{ijkl}\mathcal{R}_{s}^{\eta^{4}}\!\!(g)_{ijkl,ijkl}=\frac{1}{24}\Big[\chi_{n}^{4}(g)+6\chi_{n}^{2}(g)\chi_{n}(g^{2})\nonumber \\
 & \quad+8\chi_{n}(g)\chi_{n}(g^{3})+3\chi_{n}^{2}(g^{2})+6\chi_{n}(g^{4})\Big].\label{eq:chis_4}
\end{align}
Similarly, one finds the fifth- and sixth-order expressions
\begin{align}
\chi_{s}^{\eta^{5}}\!\!(g) & =\frac{1}{5!}\Big[\chi_{n}^{5}(g)+10\chi_{n}^{3}(g)\chi_{n}(g^{2})+20\chi_{n}^{2}(g)\chi_{n}(g^{3})\nonumber \\
 & +24\chi_{n}(g^{5})+30\chi_{n}(g)\chi_{n}(g^{4})+20\chi_{n}(g^{2})\chi_{n}(g^{3})\nonumber \\
 & +15\chi_{n}(g)\chi_{n}^{2}(g^{2})\Big],\label{eq:chi_dec_5}\\
\chi_{s}^{\eta^{6}}\!\!(g) & =\frac{1}{6!}\Big[15\chi_{n}^{4}(g)\chi_{n}(g^{2})+120\chi_{n}(g)\chi_{n}(g^{2})\chi_{n}(g^{3})\nonumber \\
 & +\chi_{n}^{6}(g)+15\chi_{n}^{3}(g^{2})+40\chi_{n}^{2}(g^{3})+45\chi_{n}^{2}(g)\chi_{n}^{2}(g^{2})\nonumber \\
 & +144\chi_{n}(g)\chi_{n}(g^{5})+40\chi_{n}^{3}(g)\chi_{n}(g^{3})+120\chi_{n}(g^{6})\nonumber \\
 & +90\chi_{n}(g^{2})\chi_{n}(g^{4})+90\chi_{n}^{2}(g)\chi_{n}(g^{4})\Big].\label{eq:chi_dec_6}
\end{align}

The presented derivation is not restricted to $\eta$ transforming
according to a single IR. In fact, we can equally assume that $\eta$
transforms according to the representation $n_{1}+n_{2}+\ldots+n_{\nu}$.
For example, in the case of a complex IR, we choose $n_{1}=E$ and
$n_{2}=\bar{E}$; in the case of a degeneracy such as in a trigonal
system, we have $n_{1}=E_{g}$ and $n_{2}=E_{g}$, or even combinations
thereof such as in $\mathsf{C_{3}}$ with $n_{1}=n_{3}=E$ and $n_{2}=n_{4}=\bar{E}$.
For clarity, we let $\tilde{\mathcal{R}}_{n}(g)$ be the transformation
matrices associates with the IR $n$, and $\tilde{\chi}_{n}(g)$ the
corresponding character. The nematic order parameter $\boldsymbol{\eta}=(\eta_{1},\eta_{2}\dots,\eta_{\dim n_{1}+\ldots+\dim n_{\nu}})$,
on the other hand, transforms with the matrices $\mathcal{R}_{n}(g)=\tilde{\mathcal{R}}_{n_{1}}(g)\oplus\ldots\oplus\tilde{\mathcal{R}}_{n_{\nu}}(g)$.
Clearly, its characters are given by $\chi_{n}(g)=\tilde{\chi}_{n_{1}}(g)+\ldots+\tilde{\chi}_{n_{\nu}}(g)$.
More importantly, since the transformation matrices $\mathcal{R}_{n}(g)$
are block-diagonal, we find 
\begin{align}
\left[\mathcal{R}_{n}(g)\right]^{\ell} & =\left[\tilde{\mathcal{R}}_{n_{1}}(g)\right]^{\ell}\oplus\ldots\oplus\left[\tilde{\mathcal{R}}_{n_{\nu}}(g)\right]^{\ell},\label{eq:transf_mat_mixed_eta}
\end{align}
with $\ell=\{0,1,2,\dots\}$. As a result, the characters for any
power $\ell$ become 
\begin{align}
\chi_{n}(g^{\ell}) & =\tilde{\chi}_{n_{1}}(g^{\ell})+\ldots+\tilde{\chi}_{n_{\nu}}(g^{\ell}).\label{eq:chi_complex_SM}
\end{align}
We now note that the above formalism equally applies to the ``mixed''
$\boldsymbol{\eta}=(\eta_{1},\eta_{2}\dots,\eta_{\dim n_{1}+\ldots+\dim n_{\nu}})$
with transformation matrices $\mathcal{R}_{n}(g)$. The only difference
is that the characters (\ref{eq:chi_complex_SM}) have to be inserted
into the formulae (\ref{eq:chi2_s})-(\ref{eq:chi_dec_6}).

\section{Minimization of the icosahedral nematic Landau expansion \label{sec:App_Ih}}

In this Appendix, we derive the mean-field phase diagram of the nematic
instability in an icosahedral group, which is presented in Fig.~\ref{fig:phase_diag}.
Starting from the symmetrized product decomposition (\ref{eq:Hg2_decomp})-(\ref{eq:Hg4_decomp})
we first determine the three bilinear combinations $D^{A_{g}}$, $\boldsymbol{D}^{H_{g},1}$
and $\boldsymbol{D}^{H_{g},2}$. The trivial one is given by $D^{A_{g}}=|\boldsymbol{d}|^{2}$.
For the two degenerate $H_{g}$-bilinears we choose a representation
where both bilinears have equal amplitude $\left|\boldsymbol{D}^{H_{g},1}\right|=\left|\boldsymbol{D}^{H_{g},2}\right|=\left|\boldsymbol{d}\right|^{2}$.
This condition indeed fixes the two bilinears and one finds $\boldsymbol{D}^{H_{g},1}=\boldsymbol{D}^{j=2}$
to be identical to the $\mathsf{O(3)}$ bilinear (\ref{eq:bilinears_j02}).
With the amplitudes being equal, it is clear that the obtained $\boldsymbol{D}^{H_{g},2}$
must be related to $\boldsymbol{D}^{H_{g},1}$ through a rotation
of the kind $\boldsymbol{D}^{H_{g},2}=\mathcal{R}^{A}\tilde{\boldsymbol{D}}^{H_{g},1}$
where $\tilde{\boldsymbol{D}}^{H_{g},1}=\boldsymbol{D}^{H_{g},1}\big|_{\boldsymbol{d}\rightarrow\tilde{\boldsymbol{d}}=\mathcal{R}^{B}\boldsymbol{d}}$
with rotation matrices $\mathcal{R}^{A,B}$. Clearly, this choice
preserves the magnitude as $(\boldsymbol{D}^{H_{g},2})^{T}\boldsymbol{D}^{H_{g},2}=(\tilde{\boldsymbol{D}}^{H_{g},1})^{T}\tilde{\boldsymbol{D}}^{H_{g},1}=\left((\tilde{\boldsymbol{d}})^{T}\tilde{\boldsymbol{d}}\right)^{2}=\left((\boldsymbol{d})^{T}\boldsymbol{d}\right)^{2}$.
We find this relation to be satisfied for $\mathcal{R}^{A}=\mathcal{R}_{5}(-\phi_{0})$
and $\mathcal{R}^{B}=\mathcal{R}_{5}(\phi_{0})$, as presented in
the main text in Eqs.~(\ref{eq:D_Hg2_ico})-(\ref{eq:R5-1}). Introducing
the unit vectors $\hat{\boldsymbol{d}}=\boldsymbol{d}/|\boldsymbol{d}|$
and $\hat{\boldsymbol{D}}^{H_{g},1/2}=\boldsymbol{D}^{H_{g},1/2}/|\boldsymbol{d}|^{2}$,
we rewrite the Landau expansion (\ref{eq:S_Hg}) as
\begin{align}
\mathcal{S} & =\int_{\mathsf{x}}|\boldsymbol{d}|^{2}\Big\{ r_{0}+|\boldsymbol{d}|f_{\alpha,\boldsymbol{n},\boldsymbol{m},\boldsymbol{l}}^{(3)}+|\boldsymbol{d}|^{2}f_{\alpha,\boldsymbol{n},\boldsymbol{m},\boldsymbol{l}}^{(4)}\Big\},\label{eq:S_App}
\end{align}
with the direction-dependent functions 
\begin{align}
f_{\alpha,\boldsymbol{n},\boldsymbol{m},\boldsymbol{l}}^{(3)} & =g_{1}\hat{\boldsymbol{d}}\cdot\hat{\boldsymbol{D}}^{H_{g},1}+g_{2}\hat{\boldsymbol{d}}\cdot\hat{\boldsymbol{D}}^{H_{g},2},\label{eq:f3_App}\\
f_{\alpha,\boldsymbol{n},\boldsymbol{m},\boldsymbol{l}}^{(4)} & =u_{1}+u_{2}\,\hat{\boldsymbol{D}}^{H_{g},1}\cdot\hat{\boldsymbol{D}}^{H_{g},2}\,.\label{eq:f4_App}
\end{align}

Within a mean-field analysis, the cubic term in the action (\ref{eq:S_App})
triggers a first-order transition at a reduced temperature $r_{0}>0$.
To derive this solution, one first solves the Landau equations for
$|\boldsymbol{d}|$:
\begin{align}
|\boldsymbol{d}|_{0} & =\frac{3\big|f_{\alpha,\boldsymbol{n},\boldsymbol{m},\boldsymbol{l}}^{(3)}\big|}{8f_{\alpha,\boldsymbol{n},\boldsymbol{m},\boldsymbol{l}}^{(4)}}\left[1+\sqrt{1-\frac{32r_{0}f_{\alpha,\boldsymbol{n},\boldsymbol{m},\boldsymbol{l}}^{(4)}}{9[f_{\alpha,\boldsymbol{n},\boldsymbol{m},\boldsymbol{l}}^{(3)}]^{2}}}\right].\label{eq:d_0}
\end{align}
where we implicitly assumed that $f_{\alpha,\boldsymbol{n},\boldsymbol{m},\boldsymbol{l}}^{(3)}<0$;
we later verified that this is indeed the case for the mean-field
solution. The first-order phase transition occurs at the reduced temperature
$r_{0}^{c}$ where the expansion (\ref{eq:S_App}) evaluated at the
solution (\ref{eq:d_0}) vanishes. We find: 
\begin{align}
r_{0}^{c} & =\max_{\alpha,\boldsymbol{n},\boldsymbol{m},\boldsymbol{l}}\frac{[f_{\alpha,\boldsymbol{n},\boldsymbol{m},\boldsymbol{l}}^{(3)}]^{2}}{4f_{\alpha,\boldsymbol{n},\boldsymbol{m},\boldsymbol{l}}^{(4)}},\label{eq:r0c_App}
\end{align}
where the maximum function determines the optimal direction of the
nematic direction parameters $\alpha,\boldsymbol{n},\boldsymbol{m},\boldsymbol{l}$.
This optimization with respect to the four degrees of freedom has
been conducted numerically to derive the phase diagram in Fig.~\ref{fig:phase_diag}.

An analytical solution is readily available in two regions of the
$(g_{2}/g_{1},\,u_{2}/u_{1})$ parameter-space. First, it is convenient
to parameterize the rotated nematic vector $\tilde{\boldsymbol{d}}=\mathcal{R}_{5}(\phi_{0})\boldsymbol{d}$
as
\begin{align}
\tilde{\boldsymbol{d}} & =\frac{|\boldsymbol{d}|}{\sqrt{3}}\Big\{\cos\left(\tilde{\alpha}\right)\left(\tilde{n}_{\mu}\boldsymbol{\lambda}_{\mu\mu^{\prime}}^{j=2}\tilde{n}_{\mu^{\prime}}\right)+\cos\left(\tilde{\alpha}+\frac{2\pi}{3}\right)\left(\tilde{m}_{\mu}\boldsymbol{\lambda}_{\mu\mu^{\prime}}^{j=2}\tilde{m}_{\mu^{\prime}}\right)\nonumber \\
 & \quad+\cos\left(\tilde{\alpha}+\frac{4\pi}{3}\right)\left(\tilde{l}_{\mu}\boldsymbol{\lambda}_{\mu\mu^{\prime}}^{j=2}\tilde{l}_{\mu^{\prime}}\right)\Big\},\label{eq:d_tilde_nml_app}
\end{align}
similar to the $(\boldsymbol{n}\boldsymbol{m}\boldsymbol{l})$-representation
(\ref{eq:d_nml}). Here, ($\tilde{\alpha}$,$\tilde{\boldsymbol{n}}$,$\tilde{\boldsymbol{m}}$,$\tilde{\boldsymbol{l}}$)
are functions of ($\alpha$,$\boldsymbol{n}$,$\boldsymbol{m}$,$\boldsymbol{l}$)
defined implicitly through $\tilde{\boldsymbol{d}}=\mathcal{R}_{5}(\phi_{0})\boldsymbol{d}$.
This allows us to rewrite the cubic terms in Eq.~(\ref{eq:S_App})
as 
\begin{align}
\boldsymbol{d}\cdot\boldsymbol{D}^{H_{g},1} & =|\boldsymbol{d}|^{3}\cos\left(3\alpha\right),\label{eq:dDHg1}\\
\boldsymbol{d}\cdot\boldsymbol{D}^{H_{g},2} & =\tilde{\boldsymbol{d}}\cdot\tilde{\boldsymbol{D}}^{H_{g},1}=|\boldsymbol{d}|^{3}\cos\left(3\tilde{\alpha}\right).\label{eq:dDHg2}
\end{align}

The two parameter regimes where we can analytically derive the nematic
ground state are spanned by (i) $\mathrm{sign}g_{1}=\mathrm{sign}g_{2}$
and $u_{2}<0$; and by (ii) $\mathrm{sign}g_{1}=-\mathrm{sign}g_{2}$
and $u_{2}>0$. In both regimes, all direction-dependent terms in
the expansion (\ref{eq:S_App}) can be simultaneously minimized.

In the first parameter regime (i), the cubic terms are minimized by
$\hat{\boldsymbol{d}}\cdot\hat{\boldsymbol{D}}^{H_{g},1}=\hat{\boldsymbol{d}}\cdot\hat{\boldsymbol{D}}^{H_{g},2}=-\mathrm{sign}g_{1}$,
which corresponds to $\cos\left(3\alpha\right)=\cos\left(3\tilde{\alpha}\right)=-\mathrm{sign}g_{1}$,
see Eqs.~(\ref{eq:dDHg1})-(\ref{eq:dDHg2}). For concreteness, we
choose $g_{1}<0$, such that the minimization gives $\alpha=\tilde{\alpha}=0$;
a similar procedure can be carried out for $g_{1}>0$. Technically,
one still needs to demonstrate that eigenvectors $\boldsymbol{n}$,$\boldsymbol{m}$,$\boldsymbol{l}$
exist for these values. Setting $\alpha=\tilde{\alpha}=0$ in the
definition $\tilde{\boldsymbol{d}}=\mathcal{R}_{5}(\phi_{0})\boldsymbol{d}$
and in expression (\ref{eq:d_tilde_nml_app}), we find two equations
\begin{align}
\tilde{\boldsymbol{d}} & =|\boldsymbol{d}|\frac{\sqrt{3}}{2}\;n_{\mu}\left(\mathcal{R}_{5}(\phi_{0})\boldsymbol{\lambda}^{j=2}\right)_{\mu\mu^{\prime}}n_{\mu^{\prime}},\label{eq:d_tilde_regA_1}\\
\tilde{\boldsymbol{d}} & =|\boldsymbol{d}|\frac{\sqrt{3}}{2}\;\tilde{n}_{\mu}\boldsymbol{\lambda}_{\mu\mu^{\prime}}^{j=2}\tilde{n}_{\mu^{\prime}},\label{eq:d_tilde_regA_2}
\end{align}
which have to be identical. Equating the two lines (\ref{eq:d_tilde_regA_1})
and (\ref{eq:d_tilde_regA_2}) leads to the five equations 
\begin{align}
2\tilde{n}_{z}^{2}-\tilde{n}_{x}^{2}-\tilde{n}_{y}^{2} & =\frac{1+3\sqrt{5}}{4}n_{x}^{2}+\frac{1-3\sqrt{5}}{4}n_{y}^{2}-\frac{1}{2}n_{z}^{2},\label{eq:regA_eq1}\\
\tilde{n}_{x}^{2}-\tilde{n}_{y}^{2} & =\frac{-1+\sqrt{5}}{4}n_{x}^{2}+\frac{1+\sqrt{5}}{4}n_{y}^{2}-\frac{2\sqrt{5}}{4}n_{z}^{2},\label{eq:regA_eq2}\\
\tilde{n}_{x}\tilde{n}_{z} & =n_{x}n_{z},\label{eq:regA_eq3}\\
\tilde{n}_{y}\tilde{n}_{z} & =n_{y}n_{z},\label{eq:regA_eq4}\\
\tilde{n}_{x}\tilde{n}_{y} & =n_{x}n_{y},\label{eq:regA_eq5}
\end{align}
that need to be simultaneously satisfied by the four degrees of freedom
comprised in $\boldsymbol{n}$ and $\tilde{\boldsymbol{n}}$ (recall
that each unit vector has two independent degrees of freedom). The
set of equations (\ref{eq:regA_eq1})-(\ref{eq:regA_eq5}) has a total
of $10$ solutions. The first four solutions correspond to the equal
amplitude condition $\left|n_{\mu}\right|=\left|\tilde{n}_{\mu}\right|=1/\sqrt{3}$,
which automatically satisfies Eqs.~(\ref{eq:regA_eq1})-(\ref{eq:regA_eq2}).
Eqs.~(\ref{eq:regA_eq3})-(\ref{eq:regA_eq5}) are solved by appropriately
choosing the relative signs between the components of each vector,
yielding $\boldsymbol{n}_{i}=\tilde{\boldsymbol{n}}_{i}=\mathcal{V}_{i}^{111}$
with $\mathcal{V}^{111}$ given by Eq.~(\ref{eq:V111}) and $i=1,2,3,4$.
Recall that changing the sign of $\boldsymbol{n}_{i}$ or $\tilde{\boldsymbol{n}}_{i}$
does not change the corresponding nematic order parameter $\boldsymbol{d}$,
which allows us to discard the solutions $\boldsymbol{n}_{i}=-\tilde{\boldsymbol{n}}_{i}$.

The other six solutions are obtained upon setting one component $n_{\mu}=0$
to zero. This necessarily requires $\tilde{n}_{\mu}=0$ in order to
solve two of the Eqs.~(\ref{eq:regA_eq3})-(\ref{eq:regA_eq5}).
The remaining three equations can then be solved in a straightforward
way. For example, for $n_{z}=0$ we find the two solutions 
\begin{align}
\boldsymbol{n}_{5} & =\left(\alpha_{-}^{(3)},\alpha_{+}^{(3)},0\right), & \tilde{\boldsymbol{n}}_{5} & =\left(\alpha_{+}^{(3)},\alpha_{-}^{(3)},0\right),\label{eq:regA_sol5}\\
\boldsymbol{n}_{6} & =\left(\alpha_{-}^{(3)},-\alpha_{+}^{(3)},0\right), & \tilde{\boldsymbol{n}}_{6} & =\left(\alpha_{+}^{(3)},-\alpha_{-}^{(3)},0\right),\label{eq:regA_sol6}
\end{align}
with $\alpha_{\pm}^{(3)}=\frac{1}{\sqrt{6}}\sqrt{3\pm\sqrt{5}}$.
The other four solutions are found analogously and are presented in
Eqs.~(\ref{eq:d_ico3_2}), (\ref{eq:n_C3s}). Note that the $\tilde{\boldsymbol{n}}$
directions in Eqs.~(\ref{eq:regA_sol5})-(\ref{eq:regA_sol6}) do
not carry any information, they are merely a mathematical construct
used to find the solution.

In the second parameter regime (ii), the cubic terms are minimized
by $\hat{\boldsymbol{d}}\cdot\hat{\boldsymbol{D}}^{H_{g},1}=-\hat{\boldsymbol{d}}\cdot\hat{\boldsymbol{D}}^{H_{g},1}=-\mathrm{sign}g_{1}$
or, equivalently, $\cos\left(3\alpha\right)=-\cos\left(3\tilde{\alpha}\right)=-\mathrm{sign}g_{1}$.
Setting again $g_{1}<0$ for concreteness, we search for solutions
with $\alpha=0$ and $\tilde{\alpha}=\pi/3$. As a result, one obtains
Eq.~(\ref{eq:d_tilde_regA_1}) while the second equation (\ref{eq:d_tilde_regA_2})
is replaced by
\begin{align}
\tilde{\boldsymbol{d}} & =-\frac{\sqrt{3}}{2}\,|\boldsymbol{d}|\;\tilde{m}_{\mu}\boldsymbol{\lambda}_{\mu\mu^{\prime}}^{j=2}\tilde{m}_{\mu^{\prime}}.\label{eq:dtilde_regB_2}
\end{align}
Equating Eqs.~(\ref{eq:dtilde_regB_2}) and (\ref{eq:d_tilde_regA_1})
gives once again five equations 
\begin{align}
\tilde{m}_{x}^{2}+\tilde{m}_{y}^{2}-2\tilde{m}_{z}^{2} & =\frac{1+3\sqrt{5}}{4}n_{x}^{2}+\frac{1-3\sqrt{5}}{4}n_{y}^{2}-\frac{1}{2}n_{z}^{2},\label{eq:regB_eq1}\\
-\tilde{m}_{x}^{2}+\tilde{m}_{y}^{2} & =\frac{-1+\sqrt{5}}{4}n_{x}^{2}+\frac{1+\sqrt{5}}{4}n_{y}^{2}-\frac{2\sqrt{5}}{4}n_{z}^{2},\label{eq:regB_eq2}\\
-\tilde{m}_{x}\tilde{m}_{z} & =n_{x}n_{z},\label{eq:regB_eq3}\\
-\tilde{m}_{y}\tilde{m}_{z} & =n_{y}n_{z},\label{eq:regB_eq4}\\
-\tilde{m}_{x}\tilde{m}_{y} & =n_{x}n_{y}.\label{eq:regB_eq5}
\end{align}
In contrast to the previous case, Eqs.~(\ref{eq:regB_eq1})-(\ref{eq:regB_eq5})
do not allow for solutions with equal amplitudes due to the relative
minus signs in Eqs.~(\ref{eq:regB_eq3})-(\ref{eq:regB_eq5}). The
six existing solutions are obtained by imposing a vanishing component
$n_{\mu}=0$, which directly implies $\tilde{m}_{\mu}=0$, leading
to three remaining equations. Those can be solved in a similar fashion
as the previous case. For instance, for $n_{z}=0$, we find the two
solutions 
\begin{align}
\boldsymbol{n}_{1} & =\left(\alpha_{+}^{(5)},-\alpha_{-}^{(5)},0\right), & \tilde{\boldsymbol{m}}_{1} & =\left(\alpha_{-}^{(5)},\alpha_{+}^{(5)},0\right),\label{eq:regB_sol5}\\
\boldsymbol{n}_{2} & =\left(\alpha_{+}^{(5)},\alpha_{-}^{(5)},0\right), & \tilde{\boldsymbol{m}}_{2} & =\left(\alpha_{-}^{(5)},-\alpha_{+}^{(5)},0\right).\label{eq:regB_sol6}
\end{align}
A similar procedure can be applied for $n_{x}=0$ and $n_{y}=0$,
resulting in the solutions presented in Eqs.~(\ref{eq:d_ico5_2}),(\ref{eq:n_C5s}).

\section{Landau expansion for the tetrahedral groups $\mathsf{T_{h}}$ and
$\mathsf{T}$ \label{sec:cubic_appendix}}

In this Appendix, we derive the Landau expansion of the nematic doublet
order parameter $\boldsymbol{d}^{e}$ in the case of the cubic point
groups \{$\mathsf{T_{h}}$, $\mathsf{T}$\}. The main difference with
respect to the derivation presented in Sec.~\ref{sec:Nematicity-Cubic},
which applied to the cubic point groups \{$\mathsf{O_{h}}$, $\mathsf{O}$,
$\mathsf{T_{d}}$\}, is that here the nematic doublet $\boldsymbol{d}^{e}$
transforms according to a complex IR. We emphasize that, for the \{$\mathsf{T_{h}}$,
$\mathsf{T}$\} point groups, the nematic triplet order parameter
$\boldsymbol{d}^{t}$ still transforms as a real IR, so the results
in Sec.~\ref{sec:Nematicity-Cubic} apply directly to those groups
as well. 

To keep the notation transparent, we focus on the point group $\mathsf{T_{h}}$;
the results apply equally to the group $\mathsf{T}$. Within $\mathsf{T_{h}}$,
the complex combination $\Delta^{e}=\left|\boldsymbol{d}^{e}\right|e^{\mathsf{i}\gamma_{e}}$
transforms as the $E_{g}$ IR while its complex conjugate $\bar{\Delta}^{e}$
transforms as the $\bar{E}_{g}$ IR, such that the two-component doublet
$\boldsymbol{d}^{e}$ transforms according to the representation $\left(E_{g}\oplus\bar{E}_{g}\right)$.
To derive the corresponding Landau expansion, we first compute the
decomposition of the symmetrized product for each expansion order,
following the procedure outlined in Appendix~\ref{sec:Symmetrized-products}:
\begin{align}
\big[\!\otimes_{j=1}^{2}\left(E_{g}\oplus\bar{E}_{g}\right)\big]_{s} & =A_{g}\oplus\left(E_{g}\oplus\bar{E}_{g}\right),\label{eq:cubic_E2_decomp_app}\\
\big[\!\otimes_{j=1}^{3}\left(E_{g}\oplus\bar{E}_{g}\right)\big]_{s} & =2A_{g}\oplus\left(E_{g}\oplus\bar{E}_{g}\right),\label{eq:cubic_E3_decomp_app}\\
\big[\!\otimes_{j=1}^{4}\left(E_{g}\oplus\bar{E}_{g}\right)\big]_{s} & =A_{g}\oplus2\left(E_{g}\oplus\bar{E}_{g}\right).\label{eq:cubic_E4_decomp_app}
\end{align}
We note the existence of one additional cubic invariant in this case,
when compared to the cases in which $\boldsymbol{d}^{e}$ transforms
as a real IR, see Eq.~(\ref{eq:symm_decomp_Eg3-1}). To construct
the four $A_{g}$ invariants in (\ref{eq:cubic_E2_decomp_app})-(\ref{eq:cubic_E4_decomp_app}),
we first determine the bilinears associated with Eq.~(\ref{eq:cubic_E2_decomp_app}):
\begin{align}
D^{A_{g}} & =\left|\boldsymbol{d}^{e}\right|^{2}, & D^{E_{g}} & =\left|\boldsymbol{d}^{e}\right|^{2}e^{-\mathsf{i}2\gamma_{e}}, & D^{\bar{E}_{g}} & =\left|\boldsymbol{d}^{e}\right|^{2}e^{\mathsf{i}2\gamma_{e}}.\label{eq:bils_cubic_E_app}
\end{align}
Then, the four invariants can be written as $\left|\boldsymbol{d}^{e}\right|^{2}$
, $\left|\boldsymbol{d}^{e}\right|^{4}$, $\bar{\Delta}^{e}D^{E_{g}}$
and $\Delta^{e}D^{\bar{E}_{g}}$. From the latter two, we construct
two real combinations $\left(\bar{\Delta}^{e}D^{E_{g}}+\Delta^{e}D^{\bar{E}_{g}}\right)$
and $\mathsf{i}\left(\bar{\Delta}^{e}D^{E_{g}}-\Delta^{e}D^{\bar{E}_{g}}\right)$,
such that the Landau expansion becomes 
\begin{align}
\mathcal{S} & =\int_{\mathsf{x}}\Big\{ r_{0}\left|\boldsymbol{d}^{e}\right|^{2}+\left|\boldsymbol{d}^{e}\right|^{3}\Big[g^{\mathrm{c}}\cos\left(3\gamma_{e}\right)+g^{\mathrm{s}}\sin\left(3\gamma_{e}\right)\!\Big]+u\left|\boldsymbol{d}^{e}\right|^{4}\!\Big\}.\label{eq:S_cubic_E_app-1}
\end{align}
The presence of the second cubic term proportional to $\sin\left(3\gamma_{e}\right)$
can be traced back to the fact that the cubic groups \{$\mathsf{T_{h}}$,
$\mathsf{T}$\} lack an axis of (proper or improper) fourfold rotational
symmetry compared to the cubic groups \{$\mathsf{O_{h}}$, $\mathsf{O}$,
$\mathsf{T_{d}}$\}. To proceed, it is instructive to rewrite the
cubic terms as 
\begin{align}
g^{\mathrm{c}}\cos\left(3\gamma_{e}\right)+g^{\mathrm{s}}\sin\left(3\gamma_{e}\right) & =g\,\cos\left(3\gamma_{e}-\delta_{0}\right),\label{eq:cossin_identity}
\end{align}
with 
\begin{align}
g & =\mathrm{sign}\left(g^{\mathrm{c}}\right)\sqrt{(g^{\mathrm{c}})^{2}+(g^{\mathrm{s}})^{2}}, & \delta_{0} & =\arctan\left(g^{\mathrm{s}}\big/g^{\mathrm{c}}\right).\label{eq:cossin_idenity_aux}
\end{align}
Then, the action (\ref{eq:S_cubic_E_app-1}) becomes
\begin{align}
\mathcal{S} & =\int_{\mathsf{x}}\Big\{ r_{0}\left|\boldsymbol{d}^{e}\right|^{2}+g\left|\boldsymbol{d}^{e}\right|^{3}\,\cos\left(3\gamma_{e}-\delta_{0}\right)+u\left|\boldsymbol{d}^{e}\right|^{4}\Big\},\label{eq:S_cubic_E_app}
\end{align}
illustrating the emergence of an offset angle $\delta_{0}$ for the
nematic director that is non-universal, i.e. it is determined by the
Landau coefficients, which in turn depend on the microscopic model.
As discussed above, this is due to the lack of proper/improper fourfold
rotational symmetry. This Landau expansion thus behaves effectively
as a $Z_{3}$-Potts model whose threefold-degenerate ground-state
directions,

\begin{align}
\gamma_{e}^{0} & =\frac{1}{3}\delta_{0}+\frac{m\pi}{3},\label{eq:angles_tetrahedral_e_Z3_App-1}
\end{align}
are offset by an angle $\delta_{0}/3$. Here, $m=\left\{ 1,3,5\right\} $
for $g>0$ and $m=\left\{ 0,2,4\right\} $ for $g<0$. In Table~\ref{tab:classification},
we signal this modified form of the $Z_{3}$-Potts model via $Z_{3}^{*}$. 

To further visualize the effect of the offset angle, we rewrite the
order parameter $\boldsymbol{d}^{e}=\left|\boldsymbol{d}^{e}\right|\left(\cos\gamma_{e},\sin\gamma_{e}\right)$
in the $(\boldsymbol{n}\boldsymbol{m}\boldsymbol{l})$-representation
of Eq.~(\ref{eq:d_nml}). To accomplish that, we define $n_{e}=\mathrm{div}\left(\gamma_{e},\frac{\pi}{3}\right)$
and $\tilde{\gamma}_{e}=\mathrm{mod}\left(\gamma_{e},\frac{\pi}{3}\right)=\gamma_{e}-n_{e}\frac{\pi}{3}\in[0,\frac{\pi}{3}]$
with $\mathrm{div}$, $\mathrm{mod}$ denoting integer division and
modulo, respectively. Since the definition in Eq.~(\ref{eq:cossin_idenity_aux})
implies $\delta_{0}\in\left[-\pi/2,\,\pi/2\right]$, it follows that
$n_{e}=m-\left(\frac{1-\mathrm{sign}(\delta_{0})}{2}\right)$ and
$\tilde{\gamma}_{e}=\frac{\delta_{0}}{3}+\frac{\pi}{3}\left(\frac{1-\mathrm{sign}(\delta_{0})}{2}\right)\in\left[0,\,\frac{\pi}{3}\right]$.
Then, for even $n_{e}$, the $\boldsymbol{d}^{e}$ order parameter
in the $(\boldsymbol{n}\boldsymbol{m}\boldsymbol{l})$-representation
of Eq.~(\ref{eq:d_nml}) becomes 
\begin{align}
\alpha & =\tilde{\gamma}_{e}, & \boldsymbol{n} & =\boldsymbol{e}_{\tilde{n}_{e}+2}, & \boldsymbol{m} & =\boldsymbol{e}_{\tilde{n}_{e}+1}, & \boldsymbol{l} & =\boldsymbol{e}_{\tilde{n}_{e}},\label{eq:de_state1_app}
\end{align}
where $\tilde{n}_{e}=\frac{1}{2}n_{e}+1\in\{1,2,3\}$ and the summation
in (\ref{eq:de_state1_app}) is understood as modulo $3$. Conversely,
one obtains for odd $n_{e}$,
\begin{align}
\alpha & =\frac{\pi}{3}-\tilde{\gamma}_{e}, & \boldsymbol{n} & =\boldsymbol{e}_{\tilde{n}_{e}}, & \boldsymbol{m} & =\boldsymbol{e}_{\tilde{n}_{e}+1}, & \boldsymbol{l} & =\boldsymbol{e}_{\tilde{n}_{e}+2},\label{eq:de_state2_app}
\end{align}
with $\tilde{n}_{e}=\frac{1}{2}n_{e}+\frac{1}{2}\in\{1,2,3\}$. Interestingly,
in this case of a $Z_{3}^{*}$-Potts transition, the offset angle
only affects $\alpha$ but leaves the nematic axes aligned with the
coordinate axes $\boldsymbol{e}_{1}$, $\boldsymbol{e}_{2}$, and
$\boldsymbol{e}_{3}$. Thus, within the point groups $\mathsf{T_{h}}$
and $\mathsf{T}$, the nematic doublet state is generically biaxial
with $\alpha\neq\{0,\pi/3\}$. Only for $\gamma_{e}=n\frac{\pi}{3}$
{[}$n\in\mathbb{N}${]}, which is a symmetry-enforced condition in
the point groups \{$\mathsf{O_{h}}$, $\mathsf{O}$, $\mathsf{T_{d}}$\},
the state is uniaxial with $\alpha=\{0,\pi/3\}$.

\section{Landau expansion for the hexagonal groups $\mathsf{C_{6h}}$, $\mathsf{C_{3h}}$
and $\mathsf{C_{6}}$ \label{sec:hexagonal_Appendix}}

In this Appendix, we derive the nematic Landau expansion for the hexagonal
point groups \{$\mathsf{C_{6h}}$, $\mathsf{C_{3h}}$, $\mathsf{C_{6}}$\}.
In contrast to the cases presented in Sec.~\ref{subsec:Nematicity_hexagonal},
the nematic in-plane and out-of-plane doublets $\boldsymbol{d}^{\mathrm{ip}}$
and $\boldsymbol{d}^{\mathrm{op}}$ transform according to complex
IRs. For concreteness, we focus on the point group $\mathsf{C_{6h}}$,
for which the complex in-plane and out-of-plane nematic order parameters,
$\Delta^{\mathrm{ip}}=\left|\boldsymbol{d}^{\mathrm{ip}}\right|e^{\mathsf{i}\gamma_{\mathrm{ip}}}$
and $\Delta^{\mathrm{op}}=\left|\boldsymbol{d}^{\mathrm{op}}\right|e^{\mathsf{i}\gamma_{\mathrm{op}}}$,
transform according to the IRs $E_{2g}$ and $E_{1g}$, respectively,
whereas their complex conjugates transform as $\bar{E}_{2g}$ and
$\bar{E}_{1g}$. The same results hold for the other two groups.

To derive the Landau expansion for the in-plane nematic doublet, we
first compute the decomposition of the symmetrized product for each
expansion order, 
\begin{align}
\big[\otimes_{j=1}^{2}\left(E_{2g}\oplus\bar{E}_{2g}\right)\big]_{s} & =A_{g}\oplus\left(E_{2g}\oplus\bar{E}_{2g}\right),\label{eq:hex_app_ip2}\\
\big[\otimes_{j=1}^{3}\left(E_{2g}\oplus\bar{E}_{2g}\right)\big]_{s} & =2A_{g}\oplus\left(E_{2g}\oplus\bar{E}_{2g}\right),\label{eq:hex_app_ip3}\\
\big[\otimes_{j=1}^{4}\left(E_{2g}\oplus\bar{E}_{2g}\right)\big]_{s} & =A_{g}\oplus2\left(E_{2g}\oplus\bar{E}_{2g}\right).\label{eq:hex_app_ip4}
\end{align}
The four invariants are written in terms of the bilinears in Eq.~(\ref{eq:hex_app_ip2}),
\begin{align}
D^{A_{g}} & =\left|\boldsymbol{d}^{\mathrm{ip}}\right|^{2}, & D^{E_{2g}} & =\left|\boldsymbol{d}^{\mathrm{ip}}\right|^{2}e^{-\mathsf{i}2\gamma_{\mathrm{ip}}}, & D^{\bar{E}_{2g}} & =\left|\boldsymbol{d}^{\mathrm{ip}}\right|^{2}e^{\mathsf{i}2\gamma_{\mathrm{ip}}}.\label{eq:bils_hex_ip_app}
\end{align}
resulting in $\left|\boldsymbol{d}^{\mathrm{ip}}\right|^{2}$, $\left|\boldsymbol{d}^{\mathrm{ip}}\right|^{4}$
, $\Delta^{\mathrm{ip}}D^{\bar{E}_{2g}}$ and $\bar{\Delta}^{\mathrm{ip}}D^{E_{2g}}$.
Thus, enforcing the invariants to be real-valued, we obtain the nematic
Landau expansion
\begin{align}
\mathcal{S}_{\mathrm{ip}} & =\int_{\mathsf{x}}\Big\{ r_{0}\left|\boldsymbol{d}^{\mathrm{ip}}\right|^{2}+g_{\mathrm{ip}}\left|\boldsymbol{d}^{\mathrm{ip}}\right|^{3}\cos\left(3\gamma_{\mathrm{ip}}-\delta_{0}\right)+u\left|\boldsymbol{d}^{\mathrm{ip}}\right|^{4}\Big\},\label{eq:S_hex_ip_app}
\end{align}
with an angular-shifted cosine term obtained from the relationship
(\ref{eq:cossin_identity}). The Landau expansion (\ref{eq:S_hex_ip_app})
is that of a $Z_{3}$-clock model with an offset angle $\delta_{0}$,
whose minimization gives
\begin{align}
\gamma_{\mathrm{ip}}^{0} & =\frac{1}{3}\delta_{0}+\frac{\pi}{3}\left(\frac{1+\mathfrak{s}_{g}}{2}\right)+\frac{2\pi}{3}n, & n & =\left\{ 0,1,2\right\} ,\label{eq:angles_ip_Z3-1}
\end{align}
where $\mathfrak{s}_{g}=\mathrm{sign}g_{\mathrm{ip}}$. In the nematic
state, the Fermi surface is similar to that shown in Fig.~\ref{fig:FS_hex_trig}(a),
but arbitrarily rotated about the $k_{z}$-axis according to the offset
angle $\delta_{0}$, which is a Landau coefficient. To see this, we
start from the parametrization of $\boldsymbol{d}^{\mathrm{ip}}$
in Eq.~(\ref{eq:Dih_ip_vectors}) and employ the identities 
\begin{align}
\left(\begin{smallmatrix}\cos\frac{\gamma_{\mathrm{ip}}^{0}}{2}\\
\sin\frac{\gamma_{\mathrm{ip}}^{0}}{2}
\end{smallmatrix}\right) & =\left(\text{-}1\right)^{n}R_{z}\left(\frac{\delta_{0}}{6}\right)\Bigg[\left(\begin{smallmatrix}\cos\tilde{\gamma}_{\mathrm{ip}}^{0}\\
\text{-}\sin\tilde{\gamma}_{\mathrm{ip}}^{0}
\end{smallmatrix}\right)\frac{1-\mathfrak{s}_{g}}{2}+\left(\begin{smallmatrix}\sin\tilde{\gamma}_{\mathrm{ip}}^{0}\\
\cos\tilde{\gamma}_{\mathrm{ip}}^{0}
\end{smallmatrix}\right)\frac{1+\mathfrak{s}_{g}}{2}\Bigg],\label{eq:hex_ip_relation}\\
\left(\begin{smallmatrix}\text{-}\sin\frac{\gamma_{\mathrm{ip}}^{0}}{2}\\
\cos\frac{\gamma_{\mathrm{ip}}^{0}}{2}
\end{smallmatrix}\right) & =\left(\text{-}1\right)^{n}R_{z}\left(\frac{\delta_{0}}{6}\right)\Bigg[\left(\begin{smallmatrix}\sin\tilde{\gamma}_{\mathrm{ip}}^{0}\\
\cos\tilde{\gamma}_{\mathrm{ip}}^{0}
\end{smallmatrix}\right)\frac{1-\mathfrak{s}_{g}}{2}-\left(\begin{smallmatrix}\cos\tilde{\gamma}_{\mathrm{ip}}^{0}\\
\text{-}\sin\tilde{\gamma}_{\mathrm{ip}}^{0}
\end{smallmatrix}\right)\frac{1+\mathfrak{s}_{g}}{2}\Bigg],\label{eq:hex_ip_relation2}
\end{align}
with $\tilde{\gamma}_{\mathrm{ip}}^{0}=\frac{\pi}{3}\frac{1+\mathfrak{s}_{g}}{2}+\frac{2\pi}{3}n$
and the rotation matrix 
\begin{align}
R_{z}\left(\delta\right) & =\left(\begin{smallmatrix}\cos\delta & -\sin\delta\\
\sin\delta & \cos\delta
\end{smallmatrix}\right).\label{eq:Rz_hex_App}
\end{align}
The relationships in Eqs.~(\ref{eq:hex_ip_relation})-(\ref{eq:hex_ip_relation2})
explicitly show that the in-plane nematic axes are rotated by an angle
$\delta_{0}/6$ away from the hexagonal high-symmetry directions listed
in Eq.~(\ref{eq:V_hex}).

Repeating the same steps for the out-of-plane doublet order parameter
$\boldsymbol{d}^{\mathrm{op}}$, we obtain the symmetrized-product
decompositions 
\begin{align}
\big[\otimes_{j=1}^{2}\left(E_{1g}\oplus\bar{E}_{1g}\right)\big]_{s} & =A_{g}\oplus\left(E_{2g}\oplus\bar{E}_{2g}\right),\label{eq:hex_app_op2}\\
\big[\otimes_{j=1}^{3}\left(E_{1g}\oplus\bar{E}_{1g}\right)\big]_{s} & =2B_{g}\oplus\left(E_{1g}\oplus\bar{E}_{1g}\right),\label{eq:hex_app_op3}\\
\big[\otimes_{j=1}^{4}\left(E_{1g}\oplus\bar{E}_{1g}\right)\big]_{s} & =A_{g}\oplus2\left(E_{2g}\oplus\bar{E}_{2g}\right),\label{eq:hex_app_op4}\\
\big[\otimes_{j=1}^{5}\left(E_{1g}\oplus\bar{E}_{1g}\right)\big]_{s} & =2B_{g}\oplus2\left(E_{1g}\oplus\bar{E}_{1g}\right),\label{eq:hex_app_op5}\\
\big[\otimes_{j=1}^{6}\left(E_{1g}\oplus\bar{E}_{1g}\right)\big]_{s} & =3A_{g}\oplus2\left(E_{2g}\oplus\bar{E}_{2g}\right),\label{eq:hex_app_op6}
\end{align}
with the bilinears
\begin{align}
D^{A_{g}} & =\left|\boldsymbol{d}^{\mathrm{op}}\right|^{2}, & D^{E_{2g}} & =\left|\boldsymbol{d}^{\mathrm{op}}\right|^{2}e^{\mathsf{i}2\gamma_{\mathrm{op}}}, & D^{\bar{E}_{2g}} & =\left|\boldsymbol{d}^{\mathrm{op}}\right|^{2}e^{-\mathsf{i}2\gamma_{\mathrm{op}}}.\label{eq:bils_hex_ip_app-1}
\end{align}
Compared to the case of $\boldsymbol{d}^{\mathrm{ip}}$, here the
cubic terms $\Delta^{\mathrm{op}}D^{E_{2g}}$ and $\bar{\Delta}^{\mathrm{op}}D^{\bar{E}_{2g}}$
do not transform as $A_{g}$ but as $B_{g}$. Thus, the five invariants
in Eqs.~(\ref{eq:hex_app_op2})-(\ref{eq:hex_app_op6}) are $\left|\boldsymbol{d}^{\mathrm{op}}\right|^{2}$,
$\left|\boldsymbol{d}^{\mathrm{op}}\right|^{4}$, $\left|\boldsymbol{d}^{\mathrm{op}}\right|^{6}$,
$\left(\Delta^{\mathrm{op}}D^{E_{2g}}\right)^{2}$ and $\big(\bar{\Delta}^{\mathrm{op}}D^{\bar{E}_{2g}}\big)^{2}$,
yielding the Landau expansion 
\begin{align}
\mathcal{S}_{\mathrm{op}} & =\int_{\mathsf{x}}\Big\{ r_{0}\left|\boldsymbol{d}^{\mathrm{op}}\right|^{2}+u\left|\boldsymbol{d}^{\mathrm{op}}\right|^{4}+u_{6}\left|\boldsymbol{d}^{\mathrm{op}}\right|^{6}\nonumber \\
 & +v_{6}\left|\boldsymbol{d}^{\mathrm{op}}\right|^{6}\cos\left(6\gamma_{\mathrm{op}}-\delta_{0}\right)\Big\},\label{eq:S_D6h_E1g-3}
\end{align}
with an offset angle $\delta_{0}$ due to the relation (\ref{eq:cossin_identity}).
Minimization leads to
\begin{align}
\gamma_{\mathrm{op}}^{0} & =\frac{1}{6}\delta_{0}+\frac{\pi}{6}\frac{1+\mathrm{sign}v_{6}}{2}+\frac{2\pi}{6}n, & n & \in\{0,1,\dots,5\},\label{eq:angles_op_Z6-2}
\end{align}
 which, from Eq.~(\ref{eq:D_ih_op_gs_nml}), corresponds to a rotation
of the nematic axes by $\delta_{0}/6$ about the $k_{z}$-axis.

\section{Landau expansion for the trigonal groups $\mathsf{C_{3}}$ and $\mathsf{S_{6}}$
\label{sec:trigonal_appendix}}

This Appendix presents the derivation of the Landau expansion for
the trigonal groups \{$\mathsf{S_{6}}$, $\mathsf{C_{3}}$\}, which
lack the in-plane symmetry directions $\mathcal{V}_{1}^{\mathrm{hex}}$,
$\mathcal{V}_{2}^{\mathrm{hex}}$ in Eq.~(\ref{eq:V_hex}), thus
complementing the analysis shown in Sec.~\ref{subsec:Nematicity-trigonal}
for the trigonal groups \{$\mathsf{D_{3}}$, $\mathsf{D_{3d}}$, $\mathsf{C_{3v}}$\}.
We consider the point group $\mathsf{C_{3}}$ for concreteness; in
this case, the combinations $\Delta^{\mathrm{ip}}=\left|\boldsymbol{d}^{\mathrm{ip}}\right|e^{\mathsf{i}\gamma_{\mathrm{ip}}}$
and $\Delta^{\mathrm{op}}=\left|\boldsymbol{d}^{\mathrm{op}}\right|e^{\mathsf{i}\gamma_{\mathrm{op}}}$
transform according to the complex IR $E$, whereas their complex
conjugates $\bar{\Delta}^{\mathrm{ip}}$, $\bar{\Delta}^{\mathrm{op}}$
transform according to $\bar{E}$. Since we are interested in the
Landau expansion of the total nematic order parameter $\boldsymbol{d}^{E}=\big(\boldsymbol{d}^{\mathrm{ip}},\boldsymbol{d}^{\mathrm{op}}\big)$,
we need the symmetrized product decomposition of $\left(E\oplus\bar{E}\right)\oplus\left(E\oplus\bar{E}\right)$:
\begin{align}
\big[\otimes_{j=1}^{2}\left(\left(E\oplus\bar{E}\right)\oplus\left(E\oplus\bar{E}\right)\right)\big]_{s} & =4A\oplus3\left(E\oplus\bar{E}\right),\label{eq:trig_complex_EEEE2_decomp}\\
\big[\otimes_{j=1}^{3}\left(\left(E\oplus\bar{E}\right)\oplus\left(E\oplus\bar{E}\right)\right)\big]_{s} & =8A\oplus6\left(E\oplus\bar{E}\right),\label{eq:trig_complex_EEEE3_decomp}\\
\big[\otimes_{j=1}^{4}\left(\left(E\oplus\bar{E}\right)\oplus\left(E\oplus\bar{E}\right)\right)\big]_{s} & =9A\oplus13\left(E\oplus\bar{E}\right).\label{eq:trig_complex_EEEE4_decomp}
\end{align}
The bilinear combinations obtained from Eq.~(\ref{eq:trig_complex_EEEE2_decomp})
are 
\begin{align}
D_{\mathrm{ip}}^{A_{1}} & =\left|\boldsymbol{d}^{\mathrm{ip}}\right|^{2}, & D_{\mathrm{io}}^{A_{1}} & =\left|\boldsymbol{d}^{\mathrm{ip}}\right|\left|\boldsymbol{d}^{\mathrm{op}}\right|e^{\mathsf{i}\left(\gamma_{\mathrm{ip}}-\gamma_{\mathrm{op}}\right)},\nonumber \\
D_{\mathrm{op}}^{A_{1}} & =\left|\boldsymbol{d}^{\mathrm{op}}\right|^{2}, & D_{\mathrm{ip}}^{E} & =\left|\boldsymbol{d}^{\mathrm{ip}}\right|^{2}e^{-\mathsf{i}2\gamma_{\mathrm{ip}}},\nonumber \\
D_{\mathrm{op}}^{E} & =\left|\boldsymbol{d}^{\mathrm{op}}\right|^{2}e^{-\mathsf{i}2\gamma_{\mathrm{op}}}, & D_{\mathrm{io}}^{E} & =\left|\boldsymbol{d}^{\mathrm{ip}}\right|\left|\boldsymbol{d}^{\mathrm{op}}\right|e^{-\mathsf{i}\left(\gamma_{\mathrm{ip}}+\gamma_{\mathrm{op}}\right)},\label{eq:trig_complex_bil}
\end{align}
as well as the complex conjugates of the bilinears that are not real-valued.
The $8$ cubic invariants transforming as $A$ are given by 
\begin{align}
\bar{\Delta}^{\mathrm{ip}}D_{\mathrm{ip}}^{E} & =\left|\boldsymbol{d}^{\mathrm{ip}}\right|^{3}e^{-\mathsf{i}3\gamma_{\mathrm{ip}}}, & \bar{\Delta}^{\mathrm{ip}}D_{\mathrm{io}}^{E} & =\left|\boldsymbol{d}^{\mathrm{ip}}\right|^{2}\left|\boldsymbol{d}^{\mathrm{op}}\right|e^{-\mathsf{i}\left(2\gamma_{\mathrm{ip}}+\gamma_{\mathrm{op}}\right)},\nonumber \\
\bar{\Delta}^{\mathrm{op}}D_{\mathrm{op}}^{E} & =\left|\boldsymbol{d}^{\mathrm{op}}\right|^{3}e^{-\mathsf{i}3\gamma_{\mathrm{op}}}, & \bar{\Delta}^{\mathrm{op}}D_{\mathrm{io}}^{E} & =\left|\boldsymbol{d}^{\mathrm{ip}}\right|\left|\boldsymbol{d}^{\mathrm{op}}\right|^{2}e^{-\mathsf{i}\left(\gamma_{\mathrm{ip}}+2\gamma_{\mathrm{op}}\right)},
\end{align}
plus their complex conjugates. Similarly, the $9$ quartic invariants
are
\begin{align}
D_{\mathrm{ip}}^{A}D_{\mathrm{ip}}^{A} & =\left|\boldsymbol{d}^{\mathrm{ip}}\right|^{4}, & D_{\mathrm{ip}}^{A}D_{\mathrm{io}}^{A} & =\left|\boldsymbol{d}^{\mathrm{ip}}\right|^{3}\left|\boldsymbol{d}^{\mathrm{op}}\right|e^{\mathsf{i}\left(\gamma_{\mathrm{ip}}-\gamma_{\mathrm{op}}\right)},\nonumber \\
D_{\mathrm{op}}^{A}D_{\mathrm{op}}^{A} & =\left|\boldsymbol{d}^{\mathrm{op}}\right|^{4}, & D_{\mathrm{op}}^{A}D_{\mathrm{io}}^{A} & =\left|\boldsymbol{d}^{\mathrm{ip}}\right|\left|\boldsymbol{d}^{\mathrm{op}}\right|^{3}e^{\mathsf{i}\left(\gamma_{\mathrm{ip}}-\gamma_{\mathrm{op}}\right)},\nonumber \\
D_{\mathrm{ip}}^{A}D_{\mathrm{op}}^{A} & =\left|\boldsymbol{d}^{\mathrm{ip}}\right|^{2}\left|\boldsymbol{d}^{\mathrm{op}}\right|^{2}, & D_{\mathrm{io}}^{A}D_{\mathrm{io}}^{A} & =\left|\boldsymbol{d}^{\mathrm{ip}}\right|^{2}\left|\boldsymbol{d}^{\mathrm{op}}\right|^{2}e^{\mathsf{i}\left(2\gamma_{\mathrm{ip}}-2\gamma_{\mathrm{op}}\right)},
\end{align}
plus the complex conjugates of the terms on the right column. We can
now write down the Landau expansion in terms of the real-valued combinations
of these invariants. The resulting action becomes $\mathcal{S}=\mathcal{S}_{2}+\mathcal{S}_{3}+\mathcal{S}_{4}$,
with
\begin{align}
\mathcal{S}_{2} & =\int_{\mathsf{x}}\Big\{ r_{\mathrm{ip}}\left(\boldsymbol{d}^{\mathrm{ip}}\right)^{2}+r_{\mathrm{op}}\left(\boldsymbol{d}^{\mathrm{op}}\right)^{2}+r_{\mathrm{io}1}\left(\boldsymbol{d}^{\mathrm{ip}}\cdot\boldsymbol{d}^{\mathrm{op}}\right)\nonumber \\
 & +r_{\mathrm{io}2}(\boldsymbol{d}^{\mathrm{ip}})^{T}\left(-\mathsf{i}\sigma^{y}\right)\boldsymbol{d}^{\mathrm{op}}\Big\},\label{eq:trig_S2_complex}\\
\mathcal{S}_{3} & =\int_{\mathsf{x}}\Big\{\left|\boldsymbol{d}^{\mathrm{ip}}\right|^{3}\left[g_{\mathrm{ip}}^{\mathrm{c}}\mathsf{c}_{3\gamma_{\mathrm{ip}}}+g_{\mathrm{ip}}^{\mathrm{s}}\mathsf{s}_{3\gamma_{\mathrm{ip}}}\right]+\left|\boldsymbol{d}^{\mathrm{op}}\right|^{3}\left[g_{\mathrm{op}}^{\mathrm{c}}\mathsf{c}_{3\gamma_{\mathrm{op}}}+g_{\mathrm{op}}^{\mathrm{s}}\mathsf{s}_{3\gamma_{\mathrm{op}}}\right]\nonumber \\
 & +\left|\boldsymbol{d}^{\mathrm{ip}}\right|\left|\boldsymbol{d}^{\mathrm{op}}\right|^{2}\left[g_{2}^{\mathrm{c}}\mathsf{c}_{\gamma_{\mathrm{ip}}+2\gamma_{\mathrm{op}}}+g_{2}^{\mathrm{s}}\mathsf{s}_{\gamma_{\mathrm{ip}}+2\gamma_{\mathrm{op}}}\right]\nonumber \\
 & +\left|\boldsymbol{d}^{\mathrm{ip}}\right|^{2}\left|\boldsymbol{d}^{\mathrm{op}}\right|\left[g_{1}^{\mathrm{c}}\mathsf{c}_{2\gamma_{\mathrm{ip}}+\gamma_{\mathrm{op}}}+g_{1}^{\mathrm{s}}\mathsf{s}_{2\gamma_{\mathrm{ip}}+\gamma_{\mathrm{op}}}\right]\Big\},\label{eq:trig_S3_complex}\\
\mathcal{S}_{4} & =\int_{\mathsf{x}}\Big\{ u_{\mathrm{ip}}\left|\boldsymbol{d}^{\mathrm{ip}}\right|^{4}+\left|\boldsymbol{d}^{\mathrm{ip}}\right|^{2}\left|\boldsymbol{d}^{\mathrm{op}}\right|^{2}\left[u_{\mathrm{io}}^{\mathrm{c}}\mathsf{c}_{2\gamma_{\mathrm{ip}}-2\gamma_{\mathrm{op}}}+u_{\mathrm{io}}^{\mathrm{s}}\mathsf{s}_{2\gamma_{\mathrm{ip}}-2\gamma_{\mathrm{op}}}\right]\nonumber \\
 & +u_{\mathrm{io}}^{0}\left|\boldsymbol{d}^{\mathrm{ip}}\right|^{2}\left|\boldsymbol{d}^{\mathrm{op}}\right|^{2}+\left[u_{2}^{\mathrm{c}}\mathsf{c}_{\gamma_{\mathrm{ip}}-\gamma_{\mathrm{op}}}+u_{2}^{\mathrm{s}}\mathsf{s}_{\gamma_{\mathrm{ip}}-\gamma_{\mathrm{op}}}\right]\left|\boldsymbol{d}^{\mathrm{ip}}\right|\left|\boldsymbol{d}^{\mathrm{op}}\right|^{3}\nonumber \\
 & +u_{\mathrm{op}}\left|\boldsymbol{d}^{\mathrm{op}}\right|^{4}+\left[u_{1}^{\mathrm{c}}\mathsf{c}_{\gamma_{\mathrm{ip}}-\gamma_{\mathrm{op}}}+u_{1}^{\mathrm{s}}\mathsf{s}_{\gamma_{\mathrm{ip}}-\gamma_{\mathrm{op}}}\right]\left|\boldsymbol{d}^{\mathrm{ip}}\right|^{3}\left|\boldsymbol{d}^{\mathrm{op}}\right|\Big\},\label{eq:trig_S4_complex}
\end{align}
with $23$ invariants. For brevity, we defined $\mathsf{c}_{\gamma}\equiv\cos\gamma$
and $\mathsf{s}_{\gamma}\equiv\sin\gamma$; the Pauli matrix $i\sigma^{y}$
in the quadratic action $\mathcal{S}_{2}$ acts on the two-dimensional
subspace of $\boldsymbol{d}^{\mathrm{ip}}$ and $\boldsymbol{d}^{\mathrm{op}}$.

To minimize $\mathcal{S}$, we proceed in the same way as the analysis
carried out in Sec.~\ref{subsec:Nematicity-trigonal}. Upon diagonalizing
$\mathcal{S}_{2}$, we obtain 
\begin{align}
\mathcal{S}_{2} & =\!\int_{\mathsf{x}}\left(\begin{array}{c}
\boldsymbol{d}^{\mathrm{ip}}\\
\boldsymbol{d}^{\mathrm{op}}
\end{array}\right)^{T}\!\!M\left(\begin{array}{c}
\boldsymbol{d}^{\mathrm{ip}}\\
\boldsymbol{d}^{\mathrm{op}}
\end{array}\right)=\!\int_{x}\Big\{\lambda_{+}\left|\boldsymbol{d}^{+}\right|^{2}+\lambda_{-}\left|\boldsymbol{d}^{-}\right|^{2}\Big\},\label{eq:trig_app_S2_diag}
\end{align}
where 
\begin{align}
\lambda_{\pm} & =\frac{1}{2}\left(r_{\mathrm{ip}}+r_{\mathrm{op}}\pm\sqrt{\left(r_{\mathrm{ip}}-r_{\mathrm{op}}\right)^{2}+r_{\mathrm{io}1}^{2}+r_{\mathrm{io}2}^{2}}\,\right).\label{eq:trig_lambdapm_App}
\end{align}
Here, we used the result $U^{T}MU=\mathrm{diag}\left(\lambda_{+},\lambda_{+},\lambda_{-},\lambda_{-}\right)$
and introduced the diagonal basis 
\begin{align}
\left(\begin{array}{c}
\boldsymbol{d}^{+}\\
\boldsymbol{d}^{-}
\end{array}\right) & =U^{T}\left(\begin{array}{c}
\boldsymbol{d}^{\mathrm{ip}}\\
\boldsymbol{d}^{\mathrm{op}}
\end{array}\right)=\left(\begin{array}{c}
\beta_{+}R_{\delta_{\mathrm{io}}}^{T}\boldsymbol{d}^{\mathrm{ip}}+\beta_{-}\boldsymbol{d}^{\mathrm{op}}\\
-\beta_{-}R_{\delta_{\mathrm{io}}}^{T}\boldsymbol{d}^{\mathrm{ip}}+\beta_{+}\boldsymbol{d}^{\mathrm{op}}
\end{array}\right),\label{eq:trig_dpm_App}
\end{align}
with the rotation matrix 
\begin{align}
R_{\gamma} & =\left(\begin{smallmatrix}\cos\gamma & -\sin\gamma\\
\sin\gamma & \cos\gamma
\end{smallmatrix}\right),\label{eq:R_gamma_mat_summ}
\end{align}
and the unitary matrix 
\begin{align}
U & =\left(\begin{smallmatrix}\beta_{+}\cos\delta_{\mathrm{io}} & -\beta_{+}\sin\delta_{\mathrm{io}} & -\beta_{-}\cos\delta_{\mathrm{io}} & \beta_{-}\sin\delta_{\mathrm{io}}\\
\beta_{+}\sin\delta_{\mathrm{io}} & \beta_{+}\cos\delta_{\mathrm{io}} & -\beta_{-}\sin\delta_{\mathrm{io}} & -\beta_{-}\cos\delta_{\mathrm{io}}\\
\beta_{-} & 0 & \beta_{+} & 0\\
0 & \beta_{-} & 0 & \beta_{+}
\end{smallmatrix}\right),\label{eq:trig_app_unitary_mat}
\end{align}
where 
\begin{align}
\beta_{\pm} & =\frac{1}{\sqrt{2}}\sqrt{1\pm\frac{r_{\mathrm{ip}}-r_{\mathrm{op}}}{\sqrt{\left(r_{\mathrm{ip}}-r_{\mathrm{op}}\right)^{2}+r_{\mathrm{io}1}^{2}+r_{\mathrm{io}2}^{2}}}}\,,\label{eq:beta_pm_tri_app}
\end{align}
and
\begin{align}
\cos\delta_{\mathrm{io}} & =\frac{r_{\mathrm{io}1}}{\sqrt{r_{\mathrm{io}1}^{2}+r_{\mathrm{io}2}^{2}}}, & \sin\delta_{\mathrm{io}} & =\frac{r_{\mathrm{io}2}}{\sqrt{r_{\mathrm{io}1}^{2}+r_{\mathrm{io}2}^{2}}}.\label{eq:deltaio_tri_app}
\end{align}
We emphasize that the corresponding expressions shown in Sec.~\ref{subsec:Nematicity-trigonal}
for the trigonal groups \{$\mathsf{D_{3}}$, $\mathsf{D_{3d}}$, $\mathsf{C_{3v}}$\}
can be recovered from the expressions above upon setting $r_{\mathrm{io}2}=0$,
which gives $\delta_{\mathrm{io}}=0,\:\pi$. The cubic and quartic
actions can also be rewritten in the $\boldsymbol{d}^{\pm}=\left|\boldsymbol{d}^{\pm}\right|\left(\cos\gamma_{\pm},\sin\gamma_{\pm}\right)^{T}$
basis. For instance, $\mathcal{S}_{3}$ becomes
\begin{align}
\mathcal{S}_{3} & =\int_{\mathsf{x}}\Big\{\left|\boldsymbol{d}^{+}\right|^{3}\left[g_{+}^{\mathrm{c}}\mathsf{c}_{3\gamma_{+}}+g_{+}^{\mathrm{s}}\mathsf{s}_{3\gamma_{+}}\right]+\left|\boldsymbol{d}^{-}\right|^{3}\left[g_{-}^{\mathrm{c}}\mathsf{c}_{3\gamma_{-}}+g_{-}^{\mathrm{s}}\mathsf{s}_{3\gamma_{-}}\right]\nonumber \\
 & +\left|\boldsymbol{d}^{+}\right|^{2}\left|\boldsymbol{d}^{-}\right|\left[\tilde{g}_{1}^{\mathrm{c}}\mathsf{c}_{2\gamma_{+}+\gamma_{-}}+\tilde{g}_{1}^{\mathrm{s}}\mathsf{s}_{2\gamma_{+}+\gamma_{-}}\right]\nonumber \\
 & +\left|\boldsymbol{d}^{+}\right|\left|\boldsymbol{d}^{-}\right|^{2}\left[\tilde{g}_{2}^{\mathrm{c}}\mathsf{c}_{\gamma_{+}+2\gamma_{-}}+\tilde{g}_{2}^{\mathrm{s}}\mathsf{s}_{\gamma_{+}+2\gamma_{-}}\right]\Big\}.\label{eq:S3_new_basis}
\end{align}
Here, we defined the new Landau coefficients 
\begin{align}
\boldsymbol{g}_{+} & =\beta_{-}^{3}\boldsymbol{g}_{\mathrm{op}}+\beta_{-}^{2}\beta_{+}R_{\delta_{\mathrm{io}}}^{T}\boldsymbol{g}_{2}+\beta_{-}\beta_{+}^{2}R_{2\delta_{\mathrm{io}}}^{T}\boldsymbol{g}_{1}+\beta_{+}^{3}R_{3\delta_{\mathrm{io}}}^{T}\boldsymbol{g}_{\mathrm{ip}},\nonumber \\
\boldsymbol{g}_{-} & =\beta_{+}^{3}\boldsymbol{g}_{\mathrm{op}}-\beta_{-}\beta_{+}^{2}R_{\delta_{\mathrm{io}}}^{T}\boldsymbol{g}_{2}+\beta_{-}^{2}\beta_{+}R_{2\delta_{\mathrm{io}}}^{T}\boldsymbol{g}_{1}-\beta_{-}^{3}R_{3\delta_{\mathrm{io}}}^{T}\boldsymbol{g}_{\mathrm{ip}},\nonumber \\
\tilde{\boldsymbol{g}}_{1} & =3\beta_{-}^{2}\beta_{+}\boldsymbol{g}_{\mathrm{op}}+\beta_{-}\left(2\beta_{+}^{2}-\beta_{-}^{2}\right)R_{\delta_{\mathrm{io}}}^{T}\boldsymbol{g}_{2}\nonumber \\
 & +\beta_{+}\left(\beta_{+}^{2}-2\beta_{-}^{2}\right)R_{2\delta_{\mathrm{io}}}^{T}\boldsymbol{g}_{1}-3\beta_{-}\beta_{+}^{2}R_{3\delta_{\mathrm{io}}}^{T}\boldsymbol{g}_{\mathrm{ip}},\nonumber \\
\tilde{\boldsymbol{g}}_{2} & =3\beta_{-}\beta_{+}^{2}\boldsymbol{g}_{\mathrm{op}}+\beta_{+}\left(\beta_{+}^{2}-2\beta_{-}^{2}\right)R_{\delta_{\mathrm{io}}}^{T}\boldsymbol{g}_{2}\nonumber \\
 & -\beta_{-}\left(2\beta_{+}^{2}-\beta_{-}^{2}\right)R_{2\delta_{\mathrm{io}}}^{T}\boldsymbol{g}_{1}+3\beta_{-}^{2}\beta_{+}R_{3\delta_{\mathrm{io}}}^{T}\boldsymbol{g}_{\mathrm{ip}},\label{eq:new_cubic_paras_tri_app}
\end{align}
where 
\begin{align}
\boldsymbol{g}_{\mathrm{ip}} & =\left(\begin{array}{c}
g_{\mathrm{ip}}^{\mathrm{c}}\\
g_{\mathrm{ip}}^{\mathrm{s}}
\end{array}\right), & \boldsymbol{g}_{\mathrm{op}} & =\left(\begin{array}{c}
g_{\mathrm{op}}^{\mathrm{c}}\\
g_{\mathrm{op}}^{\mathrm{s}}
\end{array}\right), & \boldsymbol{g}_{1,2} & =\left(\begin{array}{c}
g_{1,2}^{\mathrm{c}}\\
g_{1,2}^{\mathrm{s}}
\end{array}\right),\nonumber \\
\boldsymbol{g}_{\pm} & =\left(\begin{array}{c}
g_{\pm}^{\mathrm{c}}\\
g_{\pm}^{\mathrm{s}}
\end{array}\right), &  &  & \tilde{\boldsymbol{g}}_{1,2} & =\left(\begin{array}{c}
\tilde{g}_{1,2}^{\mathrm{c}}\\
\tilde{g}_{1,2}^{\mathrm{s}}
\end{array}\right).\label{eq:new_cubic_paras2_tri_app}
\end{align}

Similarly to the case of the trigonal groups investigated in Sec.~\ref{subsec:Nematicity-trigonal},
we keep only the terms in the action that are linear and quadratic
in the sub-leading order parameter $\boldsymbol{d}^{+}$. This results
in $\mathcal{S}=\mathcal{S}_{-}\left[\boldsymbol{d}^{-}\right]+\mathcal{S}_{+-}\left[\boldsymbol{d}^{+},\boldsymbol{d}^{-}\right]$
with
\begin{align}
\mathcal{S}_{-} & =\int_{\mathsf{x}}\Big\{\lambda_{-}\left|\boldsymbol{d}^{-}\right|^{2}+g_{-}\left|\boldsymbol{d}^{-}\right|^{3}\cos\left(3\gamma_{-}-\delta_{-}\right)+u_{-}\left|\boldsymbol{d}^{-}\right|^{4}\Big\},\label{eq:trig_SM_App}
\end{align}
where 
\begin{align}
g_{-} & =\mathrm{sign}\left(g_{-}^{\mathrm{c}}\right)\sqrt{\left(g_{-}^{\mathrm{c}}\right)^{2}+\left(g_{-}^{\mathrm{s}}\right)^{2}}, & \delta_{-} & =\arctan\frac{g_{-}^{\mathrm{s}}}{g_{-}^{\mathrm{c}}},\label{eq:gm_deltam_tri_app}
\end{align}
and
\begin{align}
\mathcal{S}_{+-} & =\int_{\mathsf{x}}\Big\{\lambda_{+}\left|\boldsymbol{d}^{+}\right|^{2}+\left|\boldsymbol{d}^{+}\right|^{2}\left|\boldsymbol{d}^{-}\right|\tilde{g}_{1}\cos\left(2\gamma_{+}+\gamma_{-}-\tilde{\delta}_{1}\right)\nonumber \\
 & +\left|\boldsymbol{d}^{+}\right|\left|\boldsymbol{d}^{-}\right|^{2}\tilde{g}_{2}\cos\left(\gamma_{+}+2\gamma_{-}-\tilde{\delta}_{2}\right)\Big\},\label{eq:trig_SP_App_2}
\end{align}
with relationships similar to Eq.~(\ref{eq:gm_deltam_tri_app}) holding
for $\tilde{g}_{1,2}$ and $\tilde{\delta}_{1,2}$ in terms of the
vectors $\tilde{\boldsymbol{g}}_{1,2}$ defined in Eq.~(\ref{eq:new_cubic_paras_tri_app}).
Minimizing with respect to $\boldsymbol{d}^{+}$ gives
\begin{align}
\left|\boldsymbol{d}^{+}\right| & \approx\frac{-\tilde{g}_{2}\cos\left(\gamma_{+}+2\gamma_{-}-\tilde{\delta}_{2}\right)}{2\lambda_{+}}\left|\boldsymbol{d}^{-}\right|^{2}+\mathcal{O}\left(\left|\boldsymbol{d}^{-}\right|^{3}\right),\label{eq:dplus_mf_tri_app}
\end{align}
which, when inserted back in Eq.~(\ref{eq:trig_SP_App_2}), results
in an additional quartic term in $\boldsymbol{d}^{-}$:
\begin{align}
\mathcal{S}_{+-} & =\int_{\mathsf{x}}\Big\{-\frac{\left(\tilde{g}_{2}\right)^{2}\cos^{2}\left(\gamma_{+}+2\gamma_{-}-\tilde{\delta}_{2}\right)}{4\lambda_{+}}\left|\boldsymbol{d}^{-}\right|^{4}\Big\}.\label{eq:Splus_add_tri_app}
\end{align}
Combined with the condition that $\left|\boldsymbol{d}^{+}\right|$
in Eq.~(\ref{eq:Splus_add_tri_app}) must be positive, this additional
quartic term is minimized for the angle
\begin{align}
\gamma_{+} & =-2\gamma_{-}+\tilde{\delta}_{2}+\left(\frac{1+\mathrm{sign}\,\tilde{g}_{2}}{2}\right)\pi.\label{eq:gamma_plus_mf_triapp}
\end{align}
Consequently, the effect of integrating out the fluctuations in the
sub-leading $\boldsymbol{d}^{+}$ channel is to renormalize the quartic
Landau coefficient $u_{-}\rightarrow u_{-}-\left(\tilde{g}_{2}\right)^{2}/(4\lambda_{+})$
of the $\boldsymbol{d}^{-}$ action (\ref{eq:trig_SM_App}), analogously
to what we found in Sec.~\ref{subsec:Nematicity-trigonal}. The main
difference of this action with respect to the action derived in Sec.~\ref{subsec:Nematicity-trigonal}
for the trigonal groups \{$\mathsf{D_{3}}$, $\mathsf{D_{3d}}$, $\mathsf{C_{3v}}$\}
is the offset angle $\delta_{-}$. Thus, the nematic order parameter
$\boldsymbol{d}^{-}$ behaves as a $Z_{3}^{*}$-clock order parameter,
characterized by the threefold-degenerate ground state:
\begin{align}
\gamma_{-}^{0} & =\frac{\delta_{-}}{3}+\frac{\pi}{3}\frac{1+\mathrm{sign}\left(g_{-}\right)}{2}+\frac{2\pi}{3}n, & n & =\{0,1,2\}.\label{eq:trig_gammaM_0_SM}
\end{align}
For these values of $\gamma_{-}^{0}$ , the induced $\boldsymbol{d}^{+}$
nematic order parameter becomes 
\begin{align}
\boldsymbol{d}^{+} & =\mathrm{sign}\left(g_{-}\right)\frac{\left|\tilde{g}_{2}\right|}{2\lambda_{+}}\left|\boldsymbol{d}^{-}\right|\mathcal{R}_{\tilde{\delta}_{2}-\delta_{-}}\boldsymbol{d}^{-}.\label{eq:trig_dP_gs_App}
\end{align}
Therefore, $\boldsymbol{d}^{+}$ is rotated against $\boldsymbol{d}^{-}$
by the offset angle $\tilde{\delta}_{2}-\delta_{-}$. Using Eq.~(\ref{eq:trig_dpm_App}),
it is straightforward to obtain the original nematic order parameter
doublets:
\begin{align}
\boldsymbol{d}^{\mathrm{op}} & =\beta_{-}\boldsymbol{d}^{+}+\beta_{+}\boldsymbol{d}^{-}, & \boldsymbol{d}^{\mathrm{ip}} & =R_{-\delta_{\mathrm{io}}}\left(\beta_{+}\boldsymbol{d}^{+}-\beta_{-}\boldsymbol{d}^{-}\right).\label{eq:trig_dOP_dIP_app}
\end{align}

Because $\boldsymbol{d}^{+}$ is not collinear to $\boldsymbol{d}^{-}$,
$\boldsymbol{d}^{\mathrm{op}}$ and $\boldsymbol{d}^{\mathrm{ip}}$
are generally not going to be collinear either. One consequence of
this property is that the nematic axes $\boldsymbol{n}$,$\boldsymbol{m}$,$\boldsymbol{l}$
will be offset from any high-symmetry axes. This is consistent with
the fact that the groups \{$\mathsf{S_{6}}$, $\mathsf{C_{3}}$\}
have no residual symmetry axes in the nematic phase, see also Table~\ref{tab:classification}.
In contrast, for the trigonal groups \{$\mathsf{D_{3}}$, $\mathsf{D_{3d}}$,
$\mathsf{C_{3v}}$\}, the Landau coefficients satisfy $r_{\mathrm{io}2}=0$,
$g_{\mathrm{ip},\mathrm{op},1,2}^{\mathrm{s}}=0$ and $u_{\mathrm{io},1,2}^{\mathrm{s}}=0$.
This causes the rotation matrix in Eq.~(\ref{eq:trig_dP_gs_App})
to become the identity, $\mathcal{R}_{\tilde{\delta}_{2}-\delta_{-}}=\mathbbm{1}$,
such that $\boldsymbol{d}^{+}$ and $\boldsymbol{d}^{-}$ are collinear. 

\section{Landau expansion for the tetragonal groups $\mathsf{C_{4h}}$, $\mathsf{C_{4}}$
and $\mathsf{S_{4}}$ \label{sec:tetragonal_Appendix}}

In this Appendix, we consider the three tetragonal point groups \{$\mathsf{C_{4h}}$,
$\mathsf{C_{4}}$, $\mathsf{S_{4}}$\} that do not possess in-plane
two-fold rotational symmetry axes. This implies not only that the
two in-plane nematic components $\left\{ d_{2},d_{5}\right\} $ transform
as the same one-dimensional IR, but also that the out-of-plane nematic
doublet $\boldsymbol{d}^{\mathrm{op}}$ transforms as a complex IR.
For concreteness, we focus on the group $\mathsf{C_{4h}}$, for which
$d_{2}$ and $d_{5}$ transform as $B_{g}$ while $\Delta^{\mathrm{op}}=\left|\boldsymbol{d}^{\mathrm{op}}\right|e^{\mathsf{i}\gamma_{\mathrm{op}}}$
and $\bar{\Delta}^{\mathrm{op}}=\left|\boldsymbol{d}^{\mathrm{op}}\right|e^{-\mathsf{i}\gamma_{\mathrm{op}}}$
transform as $E_{g}$ and $\bar{E}_{g}$.

Since the in-plane components are degenerate, we consider the full
in-plane doublet $\boldsymbol{d}^{\mathrm{ip}}=\left(d_{2},d_{5}\right)$,
which transforms according to $(B_{g}\oplus B_{g})$. The decomposition
of the symmetrized product is straightforward:
\begin{align}
\big[\otimes_{j=1}^{2}\left(B_{g}\oplus B_{g}\right)\big]_{s} & =3A_{g},\label{eq:decomp_tet_ip2_app}\\
\big[\otimes_{j=1}^{3}\left(B_{g}\oplus B_{g}\right)\big]_{s} & =4B_{g},\label{eq:decomp_tet_ip3_app}\\
\big[\otimes_{j=1}^{4}\left(B_{g}\oplus B_{g}\right)\big]_{s} & =5A_{g}.\label{eq:decomp_tet_ip4_app}
\end{align}
There are thus $8$ Landau invariants up to fourth-order. The resulting
action $\mathcal{S}=\mathcal{S}_{2}+\mathcal{S}_{4}$ is then:
\begin{align}
\mathcal{S}_{2} & =\int_{\mathsf{x}}\Big\{ r_{1}\left(d_{2}\right)^{2}+r_{2}\left(d_{5}\right)^{2}+r_{3}d_{2}d_{5}\Big\},\label{eq:tet_S2_ip_app}\\
\mathcal{S}_{4} & =\int_{\mathsf{x}}\Big\{ u_{1}\left(d_{2}\right)^{4}+u_{2}\left(d_{5}\right)^{4}+u_{3}\left(d_{2}\right)^{2}\left(d_{5}\right)^{2}\nonumber \\
 & +u_{4}d_{2}\left(d_{5}\right)^{3}+u_{5}\left(d_{2}\right)^{3}d_{5}\Big\}.\label{eq:tet_S4_ip_app}
\end{align}
We follow the same procedure as with the trigonal case in Sec.~\ref{subsec:Nematicity-trigonal}.
The diagonalization of the quadratic action, Eq.~(\ref{eq:tet_S2_ip_app}),
is accomplished via the orthogonal matrix 
\begin{align}
U & =\left(\begin{array}{cc}
\beta_{+} & -\frac{r_{3}}{\left|r_{3}\right|}\beta_{-}\\
\frac{r_{3}}{\left|r_{3}\right|}\beta_{-} & \beta_{+}
\end{array}\right), & \beta_{\pm} & =\frac{1}{\sqrt{2}}\sqrt{1\pm\frac{\Delta r}{\sqrt{\left(\Delta r\right)^{2}+r_{3}^{2}}}}\,,\label{eq:U_ortho_tet_app}
\end{align}
with $\Delta r\equiv r_{1}-r_{2}$. We obtain:
\begin{align}
\mathcal{S}_{2} & =\int_{\mathsf{x}}\Big\{\lambda_{+}\left(d_{+}\right)^{2}+\lambda_{-}\left(d_{-}\right)^{2}\Big\},\label{eq:tet_ip_app_S2_diag}
\end{align}
with eigenvalues 
\begin{align}
\lambda_{\pm} & =\frac{1}{2}\left(r_{1}+r_{2}\pm\sqrt{\left(r_{1}-r_{2}\right)^{2}+r_{3}^{2}}\right),\label{eq:trig_lambdapm_App-1}
\end{align}
and eigenvectors 
\begin{align}
\left(d_{+},d_{-}\right)^{T} & =U^{T}\boldsymbol{d}^{\mathrm{ip}}.\label{eq:tet_app_eigenbasis}
\end{align}
Since $\lambda_{-}<\lambda_{+}$ by construction, the combination
$d_{-}$ orders first. We thus rewrite the action $\mathcal{S}=\mathcal{S}_{-}+\mathcal{S}_{+-}$
as
\begin{align}
\mathcal{S}_{-} & =\int_{\mathsf{x}}\Big\{\lambda_{-}\left(d_{-}\right)^{2}+u_{-}\left(d_{-}\right)^{4}\Big\},\label{eq:tet_app_ip_sM}\\
\mathcal{S}_{+-} & =\int_{\mathsf{x}}\Big\{\lambda_{+}\left(d_{+}\right)^{2}+\tilde{u}_{1}d_{+}\left(d_{-}\right)^{3}+\tilde{u}_{2}\left(d_{+}\right)^{2}\left(d_{-}\right)^{2}\Big\},\label{eq:tet_app_ip_SP}
\end{align}
where we kept only terms that are linear or quadratic in the sub-leading
channel $d_{+}$ and defined
\begin{align}
\left(\begin{smallmatrix}u_{-}\\
\tilde{u}_{1}\\
\tilde{u}_{2}
\end{smallmatrix}\right) & =\beta_{-}^{4}\left(\begin{smallmatrix}u_{1}\\
-u_{5}\\
u_{3}
\end{smallmatrix}\right)+\beta_{+}^{4}\left(\begin{smallmatrix}u_{2}\\
u_{4}\\
u_{3}
\end{smallmatrix}\right)+\beta_{-}^{2}\beta_{+}^{2}\left(\begin{smallmatrix}u_{3}\\
3(u_{5}-u_{4})\\
6u_{1}+6u_{2}-4u_{3}
\end{smallmatrix}\right)\nonumber \\
 & +\frac{r_{3}}{\left|r_{3}\right|}\beta_{-}\beta_{+}^{3}\left(\begin{smallmatrix}-u_{4}\\
4u_{2}-2u_{3}\\
3u_{4}-3u_{5}
\end{smallmatrix}\right)-\frac{r_{3}}{\left|r_{3}\right|}\beta_{-}^{3}\beta_{+}\left(\begin{smallmatrix}u_{5}\\
4u_{1}-2u_{3}\\
3u_{4}-3u_{5}
\end{smallmatrix}\right).\label{eq:tet_app_u_rels}
\end{align}
Minimizing Eq.~(\ref{eq:tet_app_ip_SP}) gives $d_{+}\approx-\frac{\tilde{u}_{1}}{2\lambda_{+}}d_{-}^{3}$
, which upon reinsertion into $\mathcal{S}_{+-}$ leads to a sixth-order
term $d_{-}^{6}$. Hence, fluctuations of the sub-leading channel
only renormalize the sixth-order Landau coefficient of the Ising-nematic
action of the leading channel, Eq.~(\ref{eq:tet_app_ip_sM}). To
express the ground state in the in-plane $(\boldsymbol{n}\boldsymbol{m}\boldsymbol{l})$-representation
of Eq.~(\ref{eq:D_ih_op_gs_nml}), we can substitute $d_{+}\approx-\frac{\tilde{u}_{1}}{2\lambda_{+}}d_{-}^{3}$
in Eq.~(\ref{eq:tet_app_eigenbasis}) and perform the inverse transformation
to obtain both $\left|\boldsymbol{d}^{\mathrm{ip}}\right|$ and the
specific angle $\gamma_{\mathrm{ip}}$, both of which will be determined
by the Landau parameters of the original action (\ref{eq:tet_S2_ip_app})-(\ref{eq:tet_S4_ip_app}).

Proceeding to the out-of-plane doublet $\boldsymbol{d}^{\mathrm{op}}$,
the decompositions of the symmetrized products are:
\begin{align}
\big[\otimes_{j=1}^{2}\left(E_{g}\oplus\bar{E}_{g}\right)\big]_{s} & =A_{g}\oplus2B_{g},\label{eq:decomp_tet_op2_app}\\
\big[\otimes_{j=1}^{3}\left(E_{g}\oplus\bar{E}_{g}\right)\big]_{s} & =2\left(E_{g}\oplus\bar{E}_{g}\right),\label{eq:decomp_tet_op3_app}\\
\big[\otimes_{j=1}^{4}\left(E_{g}\oplus\bar{E}_{g}\right)\big]_{s} & =3A_{g}\oplus2B_{g},\label{eq:decomp_tet_op4_app}
\end{align}
with the bilinears $D_{\mathrm{op}}^{A_{g}}=\left|\boldsymbol{d}^{\mathrm{op}}\right|^{2}$,
\begin{align}
D_{\mathrm{op}}^{B_{g},1} & =\left|\boldsymbol{d}^{\mathrm{op}}\right|^{2}\cos\left(2\gamma_{\mathrm{op}}\right), & D_{\mathrm{op}}^{B_{g},2} & =\left|\boldsymbol{d}^{\mathrm{op}}\right|^{2}\sin\left(2\gamma_{\mathrm{op}}\right).\label{eq:tet_bilinears_app}
\end{align}
The four invariants can be expressed as $\left|\boldsymbol{d}^{\mathrm{op}}\right|^{2}$,
$\left|\boldsymbol{d}^{\mathrm{op}}\right|^{4}$, $D_{\mathrm{op}}^{B_{g},1}D_{\mathrm{op}}^{B_{g},2}$
and $\big(D_{\mathrm{op}}^{B_{g},1}\big)^{2}-\big(D_{\mathrm{op}}^{B_{g},2}\big)^{2}$
which, combined with the relationship (\ref{eq:cossin_identity}),
give the Landau expansion:
\begin{align}
\mathcal{S}_{\mathrm{op}} & =\int_{\mathsf{x}}\Big\{ r_{0}\left|\boldsymbol{d}^{\mathrm{op}}\right|^{2}+u\left|\boldsymbol{d}^{\mathrm{op}}\right|^{4}+v_{4}\left|\boldsymbol{d}^{\mathrm{op}}\right|^{4}\cos\left(4\gamma_{\mathrm{op}}-\delta_{0}\right)\Big\}.\label{eq:S_D6h_E1g-2-1}
\end{align}
As in the cases analyzed in the previous Appendices, the offset angle
$\delta_{0}$ should be understood as a Landau coefficient. The Landau
expansion (\ref{eq:S_D6h_E1g-2-1}) has the shape of a modified $4$-state
clock model, $Z_{4}^{*}$-clock, with the fourfold degenerate angles
offset from the high-symmetry tetragonal directions: 
\begin{align}
\gamma_{\mathrm{op}}^{0} & =\frac{1}{4}\delta_{0}+\frac{2\pi}{4}n+\frac{\pi}{4}\left(\frac{1+\mathrm{sign}v_{4}}{2}\right), & n & \in\{0,1,2,3\}.\label{eq:angles_op_Z6-1-1}
\end{align}
Upon employing Eq.~(\ref{eq:D_ih_op_gs_nml}), we conclude that the
offset angle $\delta_{0}$ leads to a rotation of the nematic axes
about the $k_{z}$-axis by an angle $\delta_{0}/4$.

\bibliography{nematic_classification}

\end{document}